\pdfoutput=1
\documentclass[
	affil-it, 
	auth-lg, 
	british, 
	contents, 
	custom-packages={}, 
	references={bigone}, 
]{lib/preprint}

\tikzset{
    aux/.style={dashed},
    dottedline/.style={dotted},
    gluon/.style={
        decorate, 
        draw=black,
        decoration={
            snake,
            post=lineto,
            post length=0pt,
            segment length=4,
            amplitude=0.9
        }
    }
}
\newcommand{\dbl}[1]{\tilde{#1}}
\newcommand{\dcf}[2]{{}_{#1\!\!}\Upsilon_{\!#2}}

\newcommand{\spacee}[3]{\bigoplus_{\phi\,\in\,(#3)}#1_{#2,\,\phi}}
\newcommand{\mapstomap}[1]{\xmapsto{\,#1\,}}
\newcommand{\forw}{\uparrow}
\newcommand{\backw}{\downarrow}
\newcommand{\trans}{\perp}
\newcommand{\cb}{\tte}
\newcommand{\colfac}{\sfc}
\newcommand{\kinfac}{\sfn}
\makecommand{fr}{\mathfrak}{{Kin,Scal}}
\newcommand{\NLSM}{{\rm NLSM}}
\newcommand{\NLSMst}{{\rm NLSM,\,st}}

\newcommand\numberthis{\addtocounter{equation}{1}\tag{\theequation}}

\begin{document}
    
    \date{28th July 2021}
    
    \email{l.borsten@hw.ac.uk,hk55@hw.ac.uk,branislav.jurco@gmail.com,t.macrelli@ surrey.ac.uk,c.saemann@hw.ac.uk,m.wolf@surrey.ac.uk}
    
    \preprint{EMPG--21--03,DMUS--MP--21/04}
    
    \title{Double Copy from Homotopy Algebras} 
    
    \author[a]{Leron~Borsten}
    \author[a]{Hyungrok~Kim}
    \author[b]{Branislav~Jur{\v c}o}
    \author[c]{Tommaso~Macrelli}
    \author[a]{Christian~Saemann}
    \author[c]{Martin~Wolf}
    
    \affil[a]{Maxwell Institute for Mathematical Sciences,\\ Department of Mathematics, Heriot--Watt University,\\ Edinburgh EH14 4AS, United Kingdom}
    \affil[b]{Charles University Prague,\\ Faculty of Mathematics and Physics, Mathematical Institute,\\ Prague 186 75, Czech Republic}
    \affil[c]{Department of Mathematics, University of Surrey,\\ Guildford GU2 7XH, United Kingdom}
    
    \abstract{
        We show that the BRST--Lagrangian double copy construction of $\caN=0$ supergravity as the `square' of Yang--Mills theory finds a natural interpretation in terms of homotopy algebras. We significantly expand on our previous work arguing the validity of the double copy at the loop level, and we give a detailed derivation of the double-copied Lagrangian and BRST operator. Our constructions are very general and can be applied to a vast set of examples.}
    
    \acknowledgements{We gratefully acknowledge stimulating conversations with Johannes Br\"{o}del, Michael Duff, Jan Gutowski, Henrik Johansson, Silvia Nagy, Jim Stasheff, Richard Szabo, and Alessandro Torrielli. We also thank the organisers of and participants at \emph{QCD Meets Gravity VI} for interesting questions and fruitful discussions about our related work~\cite{Borsten:2020zgj}. L.B., H.K., and C.S. were supported by the Leverhulme Research Project Grant RPG--2018--329 \emph{The Mathematics of M5-Branes}. B.J. was supported by the GA\v{C}R Grant EXPRO 19--28268X. T.M. was partially supported by the EPSRC grant EP/N509772. B.J. would also like to thank the MPI Bonn for hospitality.}
    
    \datamanagement{No additional research data beyond the data presented and cited in this work are needed to validate the research findings in this work.} 
    
    \begin{body}
        
        \section{Introduction and results}   
        
        Recent years have witnessed a resurgence of the idea that gravity can, in some sense, be viewed as the product of two gauge theories,
        \begin{equation}\label{eq:gravitygauge2}
            \text{`gravity}\ =\ \text{gauge}\otimes\text{gauge'}~.
        \end{equation}
        This notion goes back at least to the 1960's~\cite{Papini1965,feynman2018feynman} and was realised concretely at the level of tree-level scattering amplitudes via the Kawai--Lewellen--Tye (KLT) relations of string theory~\cite{Kawai:1985xq}: closed string tree amplitudes can be written as sums over products of open string tree amplitudes. The field theory limit of the KLT relations implies a relationship between tree-level Yang--Mills amplitudes and those of $\caN=0$ supergravity, the universal massless sector of closed string theories consisting of Einstein--Hilbert gravity coupled to a Kalb--Ramond two-form and dilaton.
        
        This paradigm was dramatically advanced with the advent of the Bern--Carrasco--Johansson (BCJ) colour--kinematics duality and the double copy prescription~\cite{Bern:2008qj,Bern:2010ue,Bern:2010yg}. Firstly, it was conjectured~\cite{Bern:2008qj} that gluon amplitudes can be recast so as to manifest a duality between their colour and kinematical data. This was quickly established at the tree level~\cite{BjerrumBohr:2009rd,Stieberger:2009hq}, however it remains conjectural at the loop level. Then, given a gluon amplitude in colour--kinematics-dual form, it can be `double-copied' to yield a bona fide amplitude of $\caN=0$ supergravity. 
        
        \newpage  
        
        In our previous work~\cite{Borsten:2020zgj}, we have shown that the double copy can be realised at the level of `off-mass-shell' perturbative quantum field theories. Specifically,
        \begin{enumerate}[(i)]\itemsep-2pt
            \item the Yang--Mills Becchi--Rouet--Stora--Tyutin (BRST) Lagrangian can be made to manifest tree-level colour--kinematics duality for the full BRST-extended Fock space and 
            \item the Yang--Mills BRST Lagrangian \emph{itself} double-copies to yield the perturbative $\caN=0$ supergravity BRST Lagrangian.
        \end{enumerate}
        An immediate corollary is that the double copy of gluon amplitudes yields the amplitudes of $\caN=0$ supergravity to all orders in perturbation theory, at trees and loops.
        
        \paragraph{Goals of this work.}
        In this work, we shall
        \begin{enumerate}[(i)]\itemsep-2pt
            \item give a detailed exposition of the central ideas contained in~\cite{Borsten:2020zgj}. As a warm-up we apply our methodology to the Lagrangian double copy of the non-linear sigma model, which gives a special galileon theory to all orders in perturbation theory;
            \item show that colour--kinematics duality and the BRST--Lagrangian double copy can be elegantly articulated in terms of homotopy algebras. In particular, the Batalin--Vilkovisky (BV) $L_\infty$-algebra of Yang--Mills theory admits a twisted factorisation, and this makes the double copy construction manifest;
            \item address some of the implications of this perspective for generalisations beyond Yang--Mills theory, colour--kinematics duality, the double-copy, and scattering amplitudes.
        \end{enumerate}
        
        Since one of our aims is to cater to both the scattering amplitudes and the homotopy algebra communities, our discussion will be self-contained to a high level. We also provide separate introductory sections for both the double copy (\cref{sec:background_doublecopy}) and homotopy algebras (\cref{ssec:background_homotopy_algebras}) in the following. The reader familiar with both of these areas may want to skip directly to the results presented in \cref{sec:results}, the outlook given in \cref{sec:outlook}, or the reading guide provided in \cref{ssec:reading_guide}.
        
        \subsection{Double copy}\label{sec:background_doublecopy}
        
        \paragraph{Squaring gauge theory.}
        Heuristically, by~\eqref{eq:gravitygauge2} we mean that one can regard the tensor product of two gauge potentials as the field content of a gravitational theory summarised by
        \begin{equation}\label{eq:ym2On-shell}
            `A_\mu\otimes\bar A_\nu\ =\ g_{\mu\nu}\oplus B_{\mu\nu}\oplus\varphi\text{'}~.
        \end{equation}
        Here, $A_\mu$ and $\barA_\nu$ are the gauge potentials of two distinct Yang--Mills theories with two colour or gauge Lie algebras $\frg$ and $\bar \frg$. After stripping off the colour component, the tensor product of $A_\mu$ and $\bar A_\mu$ yields a metric $g_{\mu\nu}$, the Kalb--Ramond Abelian two-form gauge potential $B_{\mu\nu}$, and a scalar field $\varphi$ called the dilaton. The latter form the field content of $\caN=0$ supergravity, the common Neveu--Schwarz-Neveu--Schwarz sector of the $\alpha'\to0$ limit of closed string theories with classical action
        \begin{equation}\label{eq:N0sugra}
            S^{\caN=0}\ \coloneqq\ \int \rmd^dx\,\sqrt{-g}\,\Big\{\!-\tfrac{1}{\kappa^2}R-\tfrac{1}{d-2}\partial_\mu\varphi\partial^\mu\varphi-\tfrac{1}{12}\rme^{-\frac{4\kappa}{d-2}\varphi}H_{\mu\nu\kappa}H^{\mu\nu\kappa}\Big\}\,,
        \end{equation}
        where $2\kappa^2=16\pi G_{\rm N}^{(d)}$ is Einstein's gravitational constant, $R$ the scalar curvature, and $H_{\mu\nu\kappa}$ the curvature of $B_{\mu\nu}$.        
        
        We can refine our interpretation of equation~\eqref{eq:ym2On-shell} to on-shell states of scattering amplitudes, if we regard it as the tensor product of the corresponding space--time little group representations, 
        \begin{equation}
            \arraycolsep=2.5pt
            \begin{array}{cccccccccccccc}
                \IR^{d-2} &\otimes & \IR^{d-2}\ &\cong\ & \bigodot_0^2\IR^{d-2} & \oplus & \bigwedge^2\IR^{d-2} & \oplus & \IR~,
                \\
                A_i & \otimes & \bar A_j\ &\cong\ & g_{(ij)_0} & \oplus & B_{[ij]} &\oplus & \varphi~,
            \end{array}
        \end{equation}
        where $\IR^{d-2}$ is the vector representation of $\sfSO(d-2)$, $\bigodot_0^2\,\IR^{d-2}$ denotes the trace-free symmetric product, and $i,j=1,\ldots,d-2$. In the context of scattering amplitudes, this amounts to the decomposition of the tensor product of two transverse gluon polarisation tensors into the polarisation tensors of the graviton (transverse-traceless), Kalb--Ramond two-form (transverse) and dilaton,\footnote{This expression is meant to be schematic. In particular, the trace piece corresponding to the dilaton must be supplemented by additional terms to render it left- and right-transverse.}
        \begin{equation}
            \eps_\mu\bar\eps_\nu\ =\ \left(\eps_{(\mu}\bar\eps_{\nu)}-\tfrac1d\eta_{\mu\nu}\eps^\rho\bar\eps_\rho\right)+\eps_{[\mu}\bar\eps_{\nu]}+\tfrac1d\eta_{\mu\nu}\eps^\rho\bar\eps_\rho~.
        \end{equation}
        
        Given this identification of on-shell states, it is natural to wonder about a corresponding identification of scattering amplitudes of $\caN=0$ supergravity. The latter relation, however, has to be subtle. In particular, given that the Weinberg--Witten theorem~\cite{Weinberg:1980kq} forbids composite gravitons under the assumption that there exists a Lorentz covariant conserved energy--momentum tensor,  how should one make sense of such a proposal? Moreover, what happens to the gauge groups and from where would the diffeomorphism invariance of~\eqref{eq:N0sugra} arise? 
        
        The earliest non-trivial concrete realisation of~\eqref{eq:ym2On-shell} came from string theory in the guise of the aforementioned KLT relations~\cite{Kawai:1985xq}: the tree-level scattering amplitudes of closed strings are sums of products of open string amplitudes. The intuition is clear as closed string spectra are given by the tensor product of left and right moving open string spectra. The low energy effective field theory limits of closed (open) strings are given by gravity (Yang--Mills) theories, so the graviton, Kalb--Ramond two-form and dilaton states arise as the tensor product of the gluon states, and we should expect precisely the couplings of~\eqref{eq:N0sugra}. 
    
        The KLT relations were deduced directly from the tree-level string vertex operators, tying both them and their field theory descendants intrinsically to the tree level. Nonetheless, early on, the `closed $=$ open $\otimes$ open' picture was used to construct one-loop graviton amplitudes~\cite{Bern:1993wt}, indicating that such relations may extend beyond the semi-classical regime. The key technical development in this regard was a shift to an `on-shell' perspective that dispensed with the familiar Lagrangian starting point and relied instead on gauge-invariant on-shell amplitude structures, such as recursion relations and unitarity cuts. For example, unitarity methods~\cite{Bern:1994zx,Bern:1994cg,Bern:1995db,  Dixon:1996wi,Bern:1996je,Britto:2010xq,Bern:2011qt} were employed to build $d=4$, $\caN=8$ supergravity two-loop amplitudes from $\caN=4$ supersymmetric Yang--Mills theory, by-passing the usual Lagrangian and Feynman diagram prescription entirely~\cite{Bern:1998ug}. Such results motivated a search for a general all-loop amplitude factorisation.

        \paragraph{Colour--kinematics duality.}
        Importantly, the on-shell methodology uncovered properties of amplitudes not readily visible in the underlying Lagrangians. This includes the colour--kinematics duality of Yang--Mills theory~\cite{Bern:2008qj,Bern:2010ue}. Let us briefly summarise this duality here; we will give the details in \cref{ssec:BCJduality}. For pedagogical introductions and further references see~\cite{Elvang:2015rqa,Elvang:2013cua,Carrasco:2015iwa,Bern:2019prr,Borsten:2020bgv}. 
        
        Firstly, one can write any  gluon amplitude entirely  in terms of trivalent graphs by `blowing' up the four-point contact terms, see \cref{sec:review} for details and notation. The resulting trivalent diagrams are thus \emph{not} the Feynman diagrams of the original theory, but of an equivalent theory. Having done so, the colour--kinematics duality conjecture states that there exists a reparametrisation of the amplitude such that
        \begin{enumerate}[(i)]\itemsep-2pt
            \item for any triple of graphs, $(i,j,k)$ with colour factors $\colfac_i$, $\colfac_j$, and $\colfac_k$, which are built entirely from the  structure constants $f_{ab}{}^c$ of the gauge Lie algebra, obeying a Jacobi identity $\colfac_i+\colfac_j+\colfac_k=0$, the corresponding kinematic factors, $\kinfac_i$, $\kinfac_j$, and $\kinfac_k$, which are built from the momenta and polarisation tensors, also obey the same Jacobi-type identity $\kinfac_i+\kinfac_j+\kinfac_k=0$ and
            \item for any diagram, $i$, such that $\colfac_i\rightarrow-\colfac_i$ under the interchange of two legs then $\kinfac_i\rightarrow-\kinfac_i$.
        \end{enumerate}
        Kinematic factors $\kinfac_i$ satisfying the colour--kinematics duality conditions are referred to as the BCJ numerators.

        Whilst the colour factors satisfy Jacobi identities by definition, it is not at all obvious that the kinematic factors  should obey the same rules; it is certainly not evident form the Yang--Mills Lagrangian. A reorganisation admitting this surprising  relationship between colour and kinematic data  exists for all $n$-point tree-level amplitudes, as has been demonstrated from a number of perspectives~\cite{BjerrumBohr:2010hn,Mafra:2011kj,Broedel:2013tta,Du:2016tbc,Mizera:2019blq}. Although there is as yet no proof that the colour--kinematics duality will continue to hold for general loop amplitudes, there are many highly non-trivial examples providing supportive evidence~\cite{Bern:2009kd,Bern:2014sna,Carrasco:2011mn,Oxburgh:2012zr,Bern:2012uf,Du:2012mt, Yuan:2012rg,Boels:2013bi}. 
        
        The kinematical Jacobi identities have important implications for the structure of the scattering amplitudes themselves, such as the existence of BCJ relations amongst colour-ordered partial amplitudes, reducing the number of independent $n$-point partial amplitudes down from $(n-2)!$ to $(n-3)!$~\cite{Bern:2008qj}. The perhaps most important implication is the double copy of tree-level scattering amplitudes.
        
        \paragraph{Double copy.}
        Consider the BCJ double copy prescription~\cite{Bern:2010ue, Bern:2010yg}. Concretely, take the two $n$-point $L$-loop Yang--Mills amplitudes, both written in trivalent form with respective colour and kinematic factors $(\colfac_i,\kinfac_i)$ and $(\tilde\colfac_i,\tilde\kinfac_i)$, at least one of which has been successfully cast in a colour--kinematics-duality respecting form, say $(\colfac_i,\kinfac_i)$. We can construct a corresponding gravitational theory amplitude by simply replacing each colour factor in $({\colfac}_i,\tilde\kinfac_i)$ with the corresponding kinematic factor of $(\colfac_i,\kinfac_i)$, that is, $(\colfac_i,\tilde\colfac_i)\rightarrow(\kinfac_i,\tilde\kinfac_i)$. We have removed all reference to the gauge group and `doubled' the kinematic terms. This addresses the first of our earlier questions: the gauge Lie algebra is replaced by a `kinematic algebra'. The second question concerning diffeomorphisms is more subtle, but also rests on colour--kinematics duality. For example, assuming colour--kinematics duality, the residual gauge invariance of the Yang--Mills amplitudes implies the invariance of the double copy amplitude under residual diffeomorphisms~\cite{Chiodaroli:2017ngp}. For two Yang--Mills theories, again with possibly unrelated gauge Lie algebras, this double copy procedure generates all possible tree amplitudes of $\caN=0$ supergravity, giving precise meaning to the heuristic equation~\eqref{eq:ym2On-shell}, at least at the semi-classical level. 
        
        This prescription generalises to supersymmetric Yang--Mills theory with both unrelated supersymmetry and gauge algebras. For example, we could take the $(\colfac_i,\kinfac_i)$ from $\caN=4$ supersymmetric Yang--Mills amplitudes and the $(\tilde{\colfac}_i,\tilde{\kinfac}_i)$ from purely bosonic Yang--Mills theory and double-copy them to the amplitudes of $\caN=4$ supergravity~\cite{Bern:2010tq}. Alternatively, if both factors are $\caN=4$ supersymmetric Yang--Mills theories, we produce the amplitudes of $\caN=8$ supergravity~\cite{Bern:2010ue}. This can be thought of as the (low energy limit of the) dimensional reduction on a six-dimensional torus of the `type II $=$ type I $\otimes$ type I' relation of $d=1+9$ superstring theory. By varying the left and right factors over all colour--kinematics duality compatible gauge theories, we generate all double-copy constructible gravitational theories. Whilst concrete constructions are complicated, there is nonetheless a rapidly multiplying zoology of double copy constructible gravity theories~\cite{Bern:2009kd,Bern:2010ue,Bern:2010yg, Chiodaroli:2011pp,Bern:2012gh,Carrasco:2012ca,Damgaard:2012fb,Huang:2012wr,Bargheer:2012gv,Carrasco:2012ca,Carrasco:2013ypa,Chiodaroli:2013upa, Johansson:2014zca,Chiodaroli:2014xia,Chiodaroli:2015rdg,Chiodaroli:2015wal,Chiodaroli:2016jqw,Carrasco:2016ygv,Carrasco:2016ldy,Anastasiou:2016csv,Johansson:2017bfl,Johansson:2017srf,Azevedo:2017lkz,Anastasiou:2017nsz, Chiodaroli:2017ngp,Chiodaroli:2017ehv,Chiodaroli:2018dbu,Azevedo:2018dgo}. 
        
        The double copy is clearly conceptually provocative, suggesting a deep relationship between perturbative quantum Yang--Mills theory and gravity. It is also computationally expedient, bringing seemingly intractable calculations within reach. This has advanced our understanding of perturbative quantum gravity~\cite{Bern:2009kd,Bern:2012cd,Bern:2012gh, Bern:2012uf,Bern:2013qca,Bern:2013uka,Bern:2014lha,Bern:2014sna, Bern:2015xsa,Bern:2018jmv}, revealing a number of unexpected features and calling into question hitherto accepted arguments regarding divergences. 

        For instance, the early expectations~\cite{Deser:2015yfa,Howe:1988qz} regarding the onset of divergences were false in the case of the four-point graviton amplitude of $\caN=8$ supergravity, which was shown to be finite to four loops in~\cite{Bern:2009kd}. This four-loop cancellation can be accounted for by supersymmetry and $\sfE_{7(7)}$ U-duality~\cite{Brodel:2009hu,Green:2010sp,Bossard:2010bd,Beisert:2010jx,Bossard:2011tq, Bossard:2012xs}. However, at seven loops any cancellations could not be `consequences of supersymmetry in any conventional sense'~\cite{Green:2010sp} and would be due to `enhanced cancellations', where the terminology reflects the fact that they cannot be explained by any standard symmetry argument\footnote{See~\cite{Freedman:2018mrv} for possible explanations at three loops that nonetheless fail at four loops.}.
        
        The seven loop case has not yet been verified, but there is evidence for enhanced cancellations from theories with less supersymmetry and, correspondingly, less protection against divergences. For example, the four-point amplitude of $d=4$, $\caN=5$ supergravity has been shown to be finite to four loops, contrary to expectations from standard symmetry arguments~\cite{Bern:2014sna}. This casts serious doubt on the divergence of $\caN=8$ supergravity at seven loops. 
        
        Currently, the cutting edge is the $\caN=8$ four-point five-loop amplitude, which was performed using generalised colour--kinematics duality and the double copy~\cite{Bern:2017ucb,Bern:2018jmv}. It was found to be finite, but the degree of finiteness was in agreement with the standard symmetry arguments, a disappointing outcome for anyone looking for enhanced cancellation that might make for a seven-loop miracle. However, this conclusion was reached in $d=\frac{24}{5}$ where the five-loop amplitude first diverges and there is a $\partial^8R^4$ counter-term.
        
        Altogether, without a complete understanding of the amplitudes, including hidden features such as the double copy construction and enhanced cancellations, such questions about divergences of amplitudes remain open in the absence of explicit calculations.   
        
        \paragraph{Why do colour--kinematics duality and the double copy work?}
        Given these remarkable results, we are compelled to ask why the colour--kinematics duality and the double copy prescription work and whether they remain valid in the full quantum perturbation theory. Although colour--kinematics duality and the double copy were arrived at through an on-shell lens, it may prove instructive to step back to an off-shell, or Lagrangian, point of view. In our previous work~\cite{Borsten:2020zgj},  we took the middle road, incorporating elements of the on-shell picture to facilitate a fully off-shell BRST--Lagrangian double copy.\footnote{See also~\cite{Ferrero:2020vww} for recent work on a plain Lagrangian double copy.} This construction incorporated and generalised three key ideas from the existing literature, which we briefly review here. We shall give a more detailed account of each in the main body of this work.
        \begin{enumerate}[(i)]\itemsep-2pt
            \item The tree-level colour--kinematics duality for physical gluon scattering amplitudes can be made manifest at the level of the Yang--Mills Lagrangian through the introduction of an infinite tower of `identically zero terms'~\cite{Bern:2010yg,Tolotti:2013caa}. As we shall explain, this is not special to the colour--kinematics duality of Yang--Mills theory and can be implemented for any theory admitting a generalised (or even trivial) \emph{internal}--kinematics duality. Here `internal' stands for the (possibly trivial) Lie algebra of any internal symmetry, such as colour or flavour symmetries. For example, we shall discuss the \emph{flavour}--kinematics duality of the non-linear sigma model in \cref{sec:NLSM}. Even Maxwell theory has a trivial $\sfU(1)$--kinematics duality. 
            \item The tree-level colour--kinematics duality manifesting action can itself be double-copied~\cite{Bern:2010yg}. By construction, the tree-level amplitudes of the double copy Lagrangian will match exactly those obtained by the double copy of the tree-level colour--kinematics-dual amplitudes themselves. In the Lagrangian double copy the (polynomials of) colour structure constants are replaced by a second copy of the corresponding differential operators, which can be regarded as `kinematic structure constants'. For self-dual Yang--Mills theory it has been shown that there is a corresponding `kinematic algebra' of area preserving diffeomorphisms~\cite{Monteiro:2011pc,BjerrumBohr:2012mg,Monteiro:2013rya} with further generalisations  given in~\cite{Fu:2016plh,Chen:2019ywi}. Note that complementary to the Lagrangian double copy is the idea that gravitational actions\footnote{specifically those gravity theories that derive from the double copy} can be written in a form that factorises order-by-order~\cite{Bern:1999ji,Bern:2010yg,Hohm:2011dz,Cheung:2016say,Cheung:2016prv, Cheung:2017kzx}. 
            \item There is an off-shell field theory `product' of BRST quantised gauge theories, including the ghost fields, that generates the BRST complex of the double copy theory~\cite{Anastasiou:2014qba,Borsten:2017jpt,Anastasiou:2018rdx}. Applying the Lagrangian double copy and truncating out the dilaton and Kalb--Ramond sector, it has been explicitly shown to give Einstein--Hilbert gravity to cubic order (where colour--kinematics duality is trivially satisfied)~\cite{Borsten:2020xbt}. The necessity of the inclusion of BRST ghosts in the context of `closed $=$ open $\otimes$ open' in string theory was stressed some time ago by Siegel~\cite{Siegel:1988qu,Siegel:1995px}. For our purposes, the key observation is that the linear BRST transformations of the resulting gravity theory follow from those of the gauge theory factors~\cite{Borsten:2017jpt,Anastasiou:2018rdx}. A related perspective on the double copy of symmetries has recently been used to derive all-order diffeomorphisms and extended Bondi--Metzner--Sachs symmetries in the self-dual sector~\cite{Campiglia:2021srh} (see e.g.~\cite{Pohlmeyer:1979ya,Dolan:1983bp,Popov:1995qb,Popov:1996uu,Popov:1998pc,Wolf:2004hp,Wolf:2005sd} for the construction of all hidden symmetries in self-dual Yang--Mills theory and gravity). Combined with sufficient global symmetry, this picture can then be used to identify the gravity theory uniquely~\cite{Borsten:2013bp,Anastasiou:2013hba,Anastasiou:2014qba,Nagy:2014jza,Borsten:2015pla,Anastasiou:2015vba,Anastasiou:2016csv,Anastasiou:2017nsz,Anastasiou:2018rdx}, revealing some interesting properties of double copy theories such as the appearance of the Freudenthal--Rosenfeld--Tits magic square and its generalisation, the magic pyramid~\cite{Borsten:2013bp,Anastasiou:2013hba}. Remarkably, this perspective may be used as a guide to identifying new theories that have no perturbative limit to start with, such as the twin superconformal theories of~\cite{Borsten:2018jjm}.
        \end{enumerate} 
        
        In our previous work~\cite{Borsten:2020zgj}, each of these ideas was generalised or reformulated to show that the BRST--Lagrangian double copy holds to all orders, and here, we shall describe their homotopy algebras underpinnings.

        \paragraph{Other aspects of the double copy.} Let us also mention some of the other generalisations and applications of the double copy. At the classical level, one can apply a double-copy-type construction for classical solutions to generate non-perturbative Kerr--Schild solutions in theories of gravity, such as black holes, or bi-adjoint scalar solutions from gauge theory solutions~\cite{Monteiro:2014cda, Luna:2015paa,Ridgway:2015fdl,Luna:2016due,White:2016jzc,Goldberger:2016iau,Goldberger:2017frp,Luna:2017dtq,Bahjat-Abbas:2017htu, Bahjat-Abbas:2018vgo,Berman:2018hwd,CarrilloGonzalez:2019gof,Bah:2019sda,Alfonsi:2020lub}. This classical double-copy can  be used to relate other features of gauge and gravity theories \cite{Alawadhi:2019urr, Banerjee:2019saj,Guevara:2020xjx} and may also be implemented perturbatively~\cite{Luna:2016hge}. There is an elegant formulation of this idea connecting Yang--Mills field strengths to the Weyl tensor, which has expanded the space of amenable solutions~\cite{Luna:2018dpt,Alawadhi:2020jrv,Godazgar:2020zbv,White:2020sfn, Monteiro:2020plf}. The field theory product \cite{Anastasiou:2014qba} can also be used to elucidate  the classical solution double-copy \cite{Luna:2020adi} and to construct, for example, supersymmetric (single/multi-centre) black hole solutions in $\caN=2$ supergravity~\cite{Cardoso:2016ngt,Cardoso:2016amd}, in the weak-field limit. 

        Alternatively, one can bend amplitudes and the double copy to the problem of classical black hole scattering, strongly motivated by the advent of gravity-wave astronomy~\cite{Shen:2018ebu,Plefka:2018dpa, Cheung:2018wkq, Kosower:2018adc,Plefka:2019hmz,   Maybee:2019jus,Johansson:2019dnu,Bern:2019nnu,Arkani-Hamed:2019ymq,Bern:2019crd, Plefka:2019wyg,Bern:2020buy,Bern:2020uwk,Bern:2021dqo}. Another interesting approach is to seek a geometric and/or world-sheet understanding of these relations through string theory~\cite{BjerrumBohr:2009rd,Stieberger:2009hq,BjerrumBohr:2010hn,Schlotterer:2012ny,Ochirov:2013xba,Bjerrum-Bohr:2014qwa,He:2015wgf} or ambitwistor strings and the scattering equations~\cite{Cachazo:2013iea,Mason:2013sva, Adamo:2013tsa,Dolan:2013isa,Cachazo:2014xea,Geyer:2015bja,Geyer:2016wjx, Bjerrum-Bohr:2016axv,Adamo:2017nia,Geyer:2018xgb,Albonico:2020mge}. 
        
        \subsection{Homotopy algebras and quantum field theory}\label{ssec:background_homotopy_algebras}
        
        In physics, we describe infinitesimal symmetries by Lie algebras and their action on field configurations by Lie algebroids. In the case of gauge symmetry, the latter are more familiar in their dual realisation, known as the Chevalley--Eilenberg picture. The Chevalley--Eilenberg differential encoding the Lie algebra of gauge symmetries as well as its action is called the BRST operator. 
        
        \paragraph{\mathversion{bold}$L_\infty$-algebras and the BV formalism.} The classical observables of a field theory are given by field configurations that satisfy the equations of motion. We can --- and sometimes\footnote{for example, in the case of open BRST complexes such as the ones arising in (unadjusted) higher gauge theories} must --- incorporate the equations of motion into the BRST picture, extending the BRST operator to the Batalin--Vilkovisky (BV) operator~\cite{Batalin:1977pb,Batalin:1981jr,Batalin:1984jr,Batalin:1984ss,Batalin:1985qj,Schwarz:1992nx}. This Chevalley--Eilenberg differential, however, no longer describes a mere Lie algebra, but a homotopy generalisation thereof, known as $L_\infty$-algebra or strong homotopy Lie algebra. These algebras first emerged in string field theory~\cite{Zwiebach:1992ie} taking inspiration from the definition of $A_\infty$-algebras~\cite{Stasheff:1963aa,Stasheff:1963ab}, and were further developed in~\cite{Stasheff:1992bb,Lada:1992wc,Lada:1994mn}. Note that in the same sense as $L_\infty$-algebras generalise Lie algebras, $A_\infty$-algebras generalise associative algebras.

        To be somewhat more explicit, an $L_\infty$-algebra is a generalisation of a differential graded Lie algebra in which the Jacobi identity, as well as its nested forms, is satisfied only up to homotopies. This means that besides the differential $\mu_1$ and the binary products $\mu_2$, there are also products of higher arity $\mu_i$ such that, for example, 
        \begin{equation}\label{eq:sampleJacId}
            \begin{aligned}
                &\mu_2(\ell_1,\mu_2(\ell_2,\ell_3))\pm\mu_2(\ell_3,\mu_2(\ell_1,\ell_2))\pm\mu_2(\ell_2,\mu_2(\ell_3,\ell_1))\ =\
                \\
                &~~~\ =\ \mu_1(\mu_3(\ell_1,\ell_2,\ell_3))\pm \mu_3(\mu_1(\ell_1),\ell_2,\ell_3)\pm \mu_3(\ell_1,\mu_1(\ell_2),\ell_3)\pm \mu_3(\ell_1,\ell_2,\mu_1(\ell_3))~,
            \end{aligned}
        \end{equation}
        where the signs depend on the precise $\IZ$-grading of the arguments. The right-hand side is the Jacobiator, measuring the failure of the Jacobi identity to hold, and importantly, it is given by a homotopy. For $\mu_i=0$ for $i\geq 3$, we recover a differential graded Lie algebra and a (graded) Lie algebra if also $\mu_1=0$. We collected more details on $L_\infty$-algebras in \cref{app:hA_L_infinity}. 
        
        Classically, any BV quantisable field theory is fully described by an $L_\infty$-algebra~\cite{Jurco:2018sby,Jurco:2019bvp}, see~\cite{Fisch:1989rp,Barnich:1997ij,Fulp:2002kk,Berends:1984rq,Movshev:2003ib,Movshev:2004aw,Zeitlin:2007vv,Zeitlin:2007vd,Zeitlin:2007yf,Zeitlin:2007fp,Zeitlin:2008cc,IuliuLazaroiu:2009wz,Costello:2011np,Rocek:2017xsj,Hohm:2017pnh} for earlier and partial accounts. The Maurer--Cartan theory of $L_\infty$-algebras, which in itself is a vast generalisation of Chern--Simons theory for Lie algebras, encompasses the action, the field equations, and all symmetries of general observables, such as gauge and Noether symmetries. Most interestingly, Maurer--Cartan theory also describes the (tree-level) scattering amplitudes of the field theory in question.

        In this sense, the $L_\infty$-framework provides a very natural and unifying description of Lagrangians and scattering amplitudes of a field theory, resolving the question about what should be regarded as fundamental:
        \begin{enumerate}[(i)]\itemsep-2pt
            \item The mathematically appropriate notion of equivalence between $L_\infty$-algebras is given by quasi-isomorphisms, and classically equivalent field theories correspond to $L_\infty$-algebras which are quasi-isomorphic; see \cref{ssec:scatterin_amplitudes,app:hA_L_infinity} for details.
            \item Every $L_\infty$-algebra is quasi-isomorphic to an $L_\infty$-algebra in which the differential $\mu_1$ vanishes identically~\cite{kadeishvili1982algebraic,Kajiura:2003ax}, and such $L_\infty$-algebras are known as minimal models. A minimal model and its higher products describe precisely the tree-level scattering amplitudes of the corresponding field theory~\cite{Kajiura:2001ng,Kajiura:2003ax,Jurco:2018sby,Jurco:2019bvp}, and we explain this in \cref{ssec:scatterin_amplitudes}. Notice that minimal models are related to a Feynman diagram expansion in general, following earlier suggestions~\cite{Kontsevich:2000yf}, and this was used in~\cite{Gwilliam:2012jg} to derive Wick's theorem and Feynman rules for finite-dimensional integrals.
            \item The notion of an $L_\infty$-algebra can be generalised to that of a quantum $L_\infty$-algebra~\cite{Zwiebach:1992ie,Markl:1997bj,Doubek:2017naz}, which corresponds to a solution to the quantum master equation in the BV formalism, see \cref{sec:BVFormalismLinfty}. Ultimately, such an quantum homotopy algebra encapsulates the quantum aspects of the corresponding field theory. In particular, quantum $L_\infty$-algebras also come with a (quantum) minimal model~\cite{Doubek:2017naz}, and their higher products describe precisely the full scattering amplitudes of the corresponding field theory~\cite{Jurco:2019yfd,Saemann:2020oyz}, as we shall review in \cref{ssec:scatterin_amplitudes}. For aspects regarding renormalisation in this context, see e.g.~\cite{Zucchini:2017ilk,Zucchini:2017ydo} and in particular~\cite{Costello:2007ei,Costello:2011aa}.
            \item Since both classical and quantum minimal models can be computed recursively by the homological perturbation lemma~\cite{gugenheim1991perturbation,Crainic:0403266}, see \cref{ssec:scatterin_amplitudes} for details, we obtain  Berends--Giele-type recursion relations for amplitudes in any BV quantisable field theory both at the tree and loop levels~\cite{Macrelli:2019afx,Jurco:2019yfd,Jurco:2020yyu,Saemann:2020oyz}. See also~\cite{Arvanitakis:2019ald} for related discussions of the S-matrix in the $L_\infty$-language,~\cite{Lopez-Arcos:2019hvg,Gomez:2020vat} for the tree-level perturbiner expansion,~\cite{Nutzi:2018vkl} for an $L_\infty$-interpretation of tree-level on-shell recursion relations, and~\cite{Reiterer:2019dys} for the construction of a homotopy BV algebra\footnote{See e.g.~\cite{Loday:2012aa} for the definition of homotopy BV algebras.} description of the BCJ relations and BCJ colour--kinematics duality at the tree level.
        \end{enumerate}

        Because homotopy algebras are the key algebraic structure underlying string field theory, it is perhaps not very surprising that they also play an important role in analysing string and field theories. They are vital in approaches to non-perturbatively completing string theory to M-theory, and we refer the interested reader to the recent review~\cite{Jurco:2019woz} (and references therein) which gives a condensed overview about the various applications of homotopy algebras in physics as well as a basic introduction into higher structures. 

        \paragraph{Homotopy algebras and factorisations.}
        Besides the strong homotopy Lie algebras, or $L_\infty$-algebras, there are other homotopy algebras that are important for our purposes. In particular, we will make use of strong homotopy associative algebras, or $A_\infty$-algebras and strong homotopy commutative algebras, or $C_\infty$-algebras (see e.g.~\cite{Kadeishvili:0811.1655} and references therein). A good example to demonstrate their use is colour-stripping of Yang--Mills theory.        
        
        The $L_\infty$-algebra of Yang--Mills theory $\frL^{\rm YM}$ can be obtained as the anti-symmetrisation of an underlying $A_\infty$-algebra $\frA^{\rm YM}$~\cite{Jurco:2019yfd}. As explained there, this $A_\infty$-algebra allows for an interesting factorisation,
        \begin{equation}
            \frA^{\rm YM}\ =\ \frA^{\rm col}\otimes \frA^{\rm kin}~,
        \end{equation}
        where $\frA^{\rm col}$ is a gauge matrix algebra, regarded as an $A_\infty$-algebra concentrated in degree zero which encodes the colour structure, and $\frA^{\rm kin}$ is concentrated in degrees $0,\ldots,3$ and encodes the colour-stripped interactions. 
        
        As we shall show in \cref{sec:colourDecompositionYM}, this interpretation of colour-stripping\footnote{Note that colour-stripping is not automatically possible, even if all fields take values in the adjoint representation: it requires that the colour coefficients in the interaction terms consist exclusively of (contractions of) the Lie algebra structure constants $f_{ab}{}^c$. For example, the non-Abelian Dirac--Born--Infeld action fails this criterion, even though all fields are adjoint, since its interactions also involve the coefficient $d_{abc}\coloneqq\tr(\{\cb_a,\cb_b\}\cb_c)$.} can be improved by factorising the $L_\infty$-algebra of Yang--Mills theory $\frL^{\rm YM}$ as 
        \begin{equation}
            \frL^{\rm YM}\ =\ \frg \otimes \frC^{\rm YM}~,
        \end{equation}
        where $\frg$ is indeed the colour or gauge Lie algebra and $\frC^{\rm YM}$ is the (unique) $C_\infty$-algebra which fully describes the colour  interactions. A related application of $C_\infty$-algebras was given in~\cite{Zeitlin:2008cc}. 
        
        At first glance, the latter factorisation seems suitable for the description of the double copy. On closer inspection, however, we observe that the factorisation of the Yang--Mills scattering amplitudes is really a factorisation into three parts: the colour part, the form kinematics part and an underlying scalar field theory with cubic interactions, which acts as a `skeleton' for the Feynman diagram expansion. 
        
        Since homotopy algebras underlie the string field theory actions and because the double copy prescription linking gauge theory and gravity amplitudes is motivated by `closed $=$ open $\otimes$ open' string duality, it is not surprising that homotopy algebras provide a good framework for understanding this duality.
    
        \subsection{Results and discussion}\label{sec:results} 
        
        In this paper, we provide an explicit account of the BRST--Lagrangian double copy~\cite{Borsten:2020zgj} and its articulation in terms of homotopy algebras. Here we summarise the key results and features of the BRST--Lagrangian double copy, its implications for scattering amplitudes and BCJ numerators, and the `homotopy double copy'. We also list some collateral results on the relation between homotopy algebras and field theories.
        
        \paragraph{BRST--Lagrangian double copy.}
        Our central result is that the Yang--Mills BRST Lagrangian double-copies to give the perturbative $\caN=0$ supergravity BRST Lagrangian to all orders. The logic of the underlying argument, and the key sub-results  entering into it, are summarised here (cf.~\cref{fig:summary}):
        \begin{enumerate}[(i)]\itemsep-2pt
            \item The tree-level Yang--Mills scattering amplitudes with external states from the extended BRST Hilbert space including the physical transverse gluons, the \emph{unphysical} forward/backward polarised gluons, and (anti)ghost states, can be made to satisfy colour--kinematics duality. See \cref{ssec:ck_duality_YM_BRST_extended_Hilbertspace}.
            \item This extended tree-level BRST colour--kinematics duality can be made manifest in the Yang--Mills BRST Lagrangian. Unlike the colour--kinematics duality for physical gluons, this requires the addition of \emph{non-vanishing} vertices  to the Yang--Mills BRST Lagrangian. However, they may be introduced exclusively through the gauge-fixing fermion and so preserve perturbative quantum equivalence. See \cref{ssec:ck_duality_YM_BRST_extended_Hilbertspace}.
            \item The extended tree-level BRST colour--kinematics duality  manifesting  Yang--Mills BRST Lagrangian can be `strictified'  to possess purely cubic interactions in an extended colour--kinematics duality preserving manner through the introduction of an infinite tower of auxiliary fields. See \cref{ssec:strictified_YM}.
            \item The strict (i.e.~cubic) Yang--Mills Lagrangian which manifests tree-level BRST colour--kinematics duality can be double copied to give a putative perturbative $\caN=0$ supergravity BRST Lagrangian. Similarly, the Yang--Mills BRST operator is double copied to give a putative $\caN=0$ supergravity BRST operator. See  \cref{ssec:construction_double_copied_action}.
            \item By construction, the physical tree-level $\caN=0$ supergravity amplitudes of the double copy Lagrangian match those of $\caN=0$ supergravity. 
            \item The double copy BRST charge is valid on-shell due to tree-level BRST colour--kinematics duality, and the linear double copy BRST charge implies that the double copy amplitudes satisfy the BRST Ward identities. This implies perturbative quantum equivalence to $\caN=0$ supergravity. See \cref{ssec:quantum_equivalence}.
        \end{enumerate}

         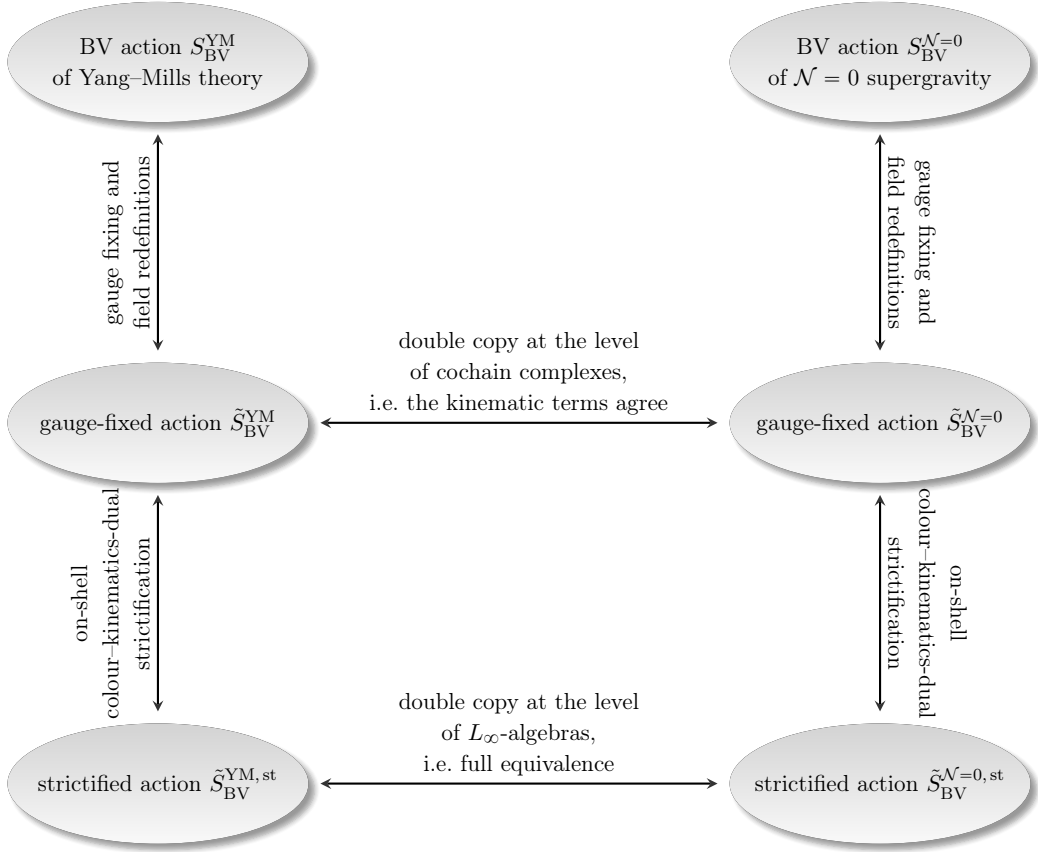
\begin{figure}[ht]
            \begin{center}
                \begin{tikzpicture}[
                    scale=.8,
                    every node/.style={scale=.8}
                    ]
                    \fill [shade,draw=none,top color=gray!35,bottom color=gray!5,blur shadow={shadow blur steps=5}] (-6,-5) ellipse (2.5cm and 1cm);
                    \draw (-6,-5) node [align=center] {BV action $S^{\rm YM}_{\rm BV}$\\[-3pt] of Yang--Mills theory};
                    \draw [thick] (-6,-6.2) edge [out=-90,in=90,<->] node [above,align=center,rotate=90] {gauge fixing and\\[-3pt] field redefinitions} (-6,-9.8);
                    \fill [shade,top color=gray!35,bottom color=gray!5,blur shadow={shadow blur steps=5}] (6,-5) ellipse (2.5cm and 1cm);
                    \draw (6,-5) node [align=center] {BV action $S^{\caN=0}_{\rm BV}$\\[-3pt] of $\caN=0$ supergravity};
                    \draw [thick] (6,-6.2) edge [out=-90,in=90,<->] node [above,align=center,rotate=-90] {gauge fixing and\\[-3pt] field redefinitions} (6,-9.8);
                    \fill [shade,top color=gray!35,bottom color=gray!5,blur shadow={shadow blur steps=5}] (-6,-11) ellipse (2.5cm and 1cm); 
                    \draw (-6,-11) node {gauge-fixed action $\tilde S^{\rm YM}_{\rm BV}$};
                    \draw [thick] (-6,-12.2) edge [out=-90,in=90,<->] node [above,align=center,rotate=90] {on-shell\\[-3pt] colour--kinematics-dual\\[-3pt] strictification} (-6,-15.8);
                    \fill [shade,top color=gray!35,bottom color=gray!5,blur shadow={shadow blur steps=5}] (6,-11) ellipse (2.5cm and 1cm); 
                    \draw (6,-11) node {gauge-fixed action $\tilde S^{\caN=0}_{\rm BV}$};
                    \draw [thick] (6,-12.2) edge [out=-90,in=90,<->] node [above,align=center,rotate=-90] {on-shell\\[-3pt] colour--kinematics-dual\\[-3pt] strictification} (6,-15.8);
                    \draw [thick] (-3.3,-11) edge [out=0,in=180,<->] node [above,align=center] {double copy at the level\\[-3pt] of cochain complexes,\\[-3pt] i.e.~the kinematic terms agree} (3.3,-11);
                    \fill [shade,top color=gray!35,bottom color=gray!5,blur shadow={shadow blur steps=5}] (-6,-17) ellipse (2.5cm and 1cm);
                    \draw (-6,-17) node {strictified action $\tilde S^{\rm YM,\,st}_{\rm BV}$};
                    \fill [shade,top color=gray!35,bottom color=gray!5,blur shadow={shadow blur steps=5}] (6,-17) ellipse (2.5cm and 1cm);
                    \draw (6,-17) node {strictified action $\tilde S^{\caN=0,\,{\rm st}}_{\rm BV}$};
                    \draw [thick] (-3.3,-17) edge [out=0,in=180,<->] node [above,align=center] {double copy at the level\\[-3pt] of $L_\infty$-algebras,\\[-3pt] i.e.~full equivalence} (3.3,-17);
                \end{tikzpicture}
            \end{center}
            \caption{Diagrammatic description of the double copy.}\label{fig:summary}
        \end{figure}
        
        Some comments are in order here. Firstly, we work perturbatively. This implies that, as in~\cite{Bern:2010yg,Tolotti:2013caa}, the BRST colour--kinematics duality manifesting action of~\cite{Borsten:2020zgj} requires an infinite tower of vertices and hence the strictified action contains an infinite tower of auxiliary fields. The intuition is clear: perturbative gravity has all order interactions and these are generated by the double copy of the vertices enforcing BRST colour--kinematics duality. The $n>3$ point interactions of gravity ensure diffeomorphism invariance, and in the BRST framework this  follows from the BRST colour--kinematics duality. Note, however, that perturbatively, i.e.~at  any finite $n$-point, $L$-loop order we only require a finite number of auxiliary fields and terms in the actions.  
        
        \paragraph{Scattering amplitudes and Bern--Carrasco--Johansson numerators.}
        An immediate corollary of this argument is that the Yang--Mills scattering amplitudes double copy into amplitudes of $\caN=0$ supergravity to all orders, tree and loop. The former are computed directly from the tree-level BRST colour--kinematics duality manifesting Lagrangian, which can be used to construct `almost BCJ numerators' that double-copy correctly. Note that to any finite $n$-point, $L$-loop order, deriving this Yang--Mills Lagrangian is a purely algebraic exercise, i.e.~there is no need to solve for \emph{functional} colour--kinematics duality relations.
        
        To be more precise:
        \begin{enumerate}[(i)]\itemsep-2pt
            \item At $n$ points and $L$ loops, one constructs the tree-level BRST colour--kinematics duality manifesting Yang--Mills Lagrangian up to the necessary finite order in auxiliary fields. Being exclusively tree-level, this is a purely algebraic operation. Nonetheless, the number of required auxiliary fields grows quickly, as one needs the largest trivalent tree\footnote{i.e.~a tree with exclusively trivalent vertices} that can be glued into a cubic $n$-point and $L$-loop diagram, i.e.~at one loop and $n$ points one needs $n+2$ point vertices. Already at four points and two loops one needs up to eight points, which requires about $150$ auxiliary fields. We should stress, this is the worst case scenario. It is likely that one can do better by incorporating on-shell methods, in particular generalised unitarity. Also, the process can be automated using computer algebra programmes.
            \item Equipped with such an action, the `almost BCJ numerators' are  given by the  sums the numerators of all Feynman diagrams with the same topology (i.e.~one ignores the distinction amongst the different fields that can sit on the internal lines), which by definition have the same colour numerators and propagators. 
            \item The `almost' qualifier indicates that the numerators so constructed will not necessarily satisfy \emph{perfect} colour--kinematics duality at the loop level. It might be that there are some hidden miracles and they do satisfy perfect colour--kinematics duality, but our arguments do not ensure this, and we have not encountered any reason to think that  this should happen generically. 
            \item Nonetheless, these `almost BCJ numerators' will double copy to yield the corresponding $\caN=0$ supergravity numerators. This gives a bona fide $\caN=0$ supergravity $n$-point and $L$-loop amplitude integrand. 
            \item From a pragmatic point of view, perfect colour--kinematics duality at the loop level is (very probably, i.e.~barring the miracles mentioned above) unnecessarily strong. It would be of practical importance to turn this statement into a precise set of loop integrand `almost colour--kinematics duality' conditions independent of the underlying BRST Lagrangian argument. We intend to address this in future work. The most powerful `almost colour--kinematics duality' conjecture motivated by our construction (i.e.~the conjecture with the weakest condition for loop-level double copy) is that loop integrands with enough internal lines cut to be tree must satisfy perfect colour--kinematics duality.
        \end{enumerate}
        
        \paragraph{Double copy from homotopy algebras.}
        Our central result is that the $L_\infty$-algebra of the strict Yang--Mills Lagrangian which manifests tree-level BRST colour--kinematics duality factorises into a colour  factor, a kinematic vector space, and a scalar theory factor. Schematically,
        \begin{equation}
            \frL^{\rm YM}\ =\ \frg\otimes\frKin\otimes_\tau\frScal~,
        \end{equation}
        where $\frg$ is the colour part, $\frKin$ the kinematic part, $\frScal$ the scalar part. The tensor product between the kinematic and scalar factors is twisted with twist datum $\tau$; see~\cref{ssec:twisted_tp_homotopy_algebras} for details. One can think of this twisting as a form of semi-direct tensor product, and it generates a kinematic algebra acting on the scalar factor. 
        
        Given the factorisation, to double copy is to replace the colour factor with another copy of the twisted kinematic factor. This yields the $L_\infty$-algebra $\frL^{\caN=0}$ of perturbatively BRST quantised $\caN=0$ supergravity, up to a quasi-isomorphism compatible with quantisation. The action, scattering amplitudes and all other features are encoded in the $L_\infty$-algebra.\footnote{Strictly speaking, we also have to provide a path integral measure for the loop amplitudes; we always work with the one arising from canonically quantising all fields.} Perturbatively, it is the complete quantum gravity theory up to the point of renormalisation. Alternatively, one can replace kinematics with colour to give the cubic biadjoint scalar field theory. The scalar factor is common to all three theories. 
        
        Schematically, 
        \begin{equation}
            \begin{array}{cccccccccccc}
                \text{Biadjoint scalar field theory} &\longleftarrow & \text{Yang--Mills theory}&\longrightarrow&\text{$\caN=0$ supergravity}
                \\
                \ \frg\otimes \frg\otimes\frScal&&\ \frg\otimes\frKin\otimes _\tau\frScal&&\ \frKin\otimes_\tau\frKin\otimes_\tau\frScal
            \end{array}
        \end{equation}
        
        Let us expand on the key elements entering this picture. Our starting point is the observation that BV quantised Yang--Mills theory corresponds to an $L_\infty$-algebra, denoted $\frL^{\rm YM}_{\rm BV}$, which upon gauge-fixing yields a BRST $L_\infty$-algebra, denoted $\frL^{\rm YM}_{\rm BRST}$. The BV operator $Q^{\rm YM}_{\rm BV}$ is uniquely determined by the higher products of $\frL^{\rm YM}_{\rm BV}$, i.e.~it is the dual Chevalley--Eilenberg differential, cf.~\cref{sec:BVFormalismLinfty}. The BRST operator $Q^{\rm YM}_{\rm BRST}$ then follows from its gauge-fixing. 
        
        As we saw above, a crucial step in the double copy is the reformulation of scattering amplitudes in terms of cubic interaction vertices. This is particularly natural from the point of view of homotopy algebras, where this is a well-known process known as {\em strictification} or {\em rectification}. The statement of the strictification theorem for homotopy algebras is that any homotopy algebra is quasi-isomorphic to a strict homotopy algebra with higher products that have either one or two inputs and one output, see~\cref{app:structure_theorems}. Field theories with exclusively cubic interaction vertices then simply correspond to strict homotopy algebras. Moreover, quasi-isomorphisms are the proper homotopy algebraic articulation of physical equivalence, since they are isomorphisms on the cohomology and thus preserve the space of physical states while allowing for field redefinitions and for the integrating in and out of auxiliary fields. 
        
        While in general, there are many possible strictifications of a homotopy algebra, the double copy singles out a particular class, namely those corresponding to manifestly BRST colour--kinematics-dual Lagrangians. Each such strict $L_\infty$-algebra $\tilde\frL^{\rm YM,\,st}_{\rm BRST}$ is quasi-isomorphic to $\frL^{\rm YM}_{\rm BRST}$. The non-trivial observation making the double copy manifest is the factorisation of $\tilde\frL^{\rm YM,\,st}_{\rm BRST}$:
        \begin{equation}\label{eq:fact}
            \tilde\frL^{\rm YM,\,st}_{\rm BRST}\ =\ \frg\otimes(\frKin^{\rm st}\otimes_\tau\frScal)~.
        \end{equation}
        Let us provide some further details on the identification~\eqref{eq:fact}:
        \begin{enumerate}[(i)]\itemsep-2pt
            \item $\frg$ is the familiar colour Lie algebra, i.e.~an $L_\infty$-algebra with only $\mu_2$, the Lie bracket, being non-trivial. 
            \item $\frKin^{\rm st}$ is the kinematic algebra. It is a graded vector space of the Poincar\'e representations carried by all the fields of the theory, including the strictification auxiliaries.    Restricting to the familiar BRST fields $A$, $c$, $\bar c$, and $b$,  it is given by
            \begin{equation}\label{eq:exampleKinYM}
                \frKin^{\rm st}\ \coloneqq\ \underbrace{\overset{\ttg}{\IR[1]}}_{\eqqcolon\,\frKin_{-1}}~\oplus~\underbrace{\big(\overset{\ttv^\mu}{\IM^d}\oplus\overset{\ttn}{\IR}\big)}_{\eqqcolon\,\frKin_0}~\oplus~\underbrace{\overset{\tta}{\IR[-1]}}_{\eqqcolon\,\frKin_1}~\oplus~\cdots~,
            \end{equation}
            where we have labelled the basis vectors of the ghost $c$, the gauge potential $A$, the Nakanishi--Lautrup $b$, and anti-ghost $\bar c$ Poincar\'e modules suggestively by $\ttg$, $\ttv^\mu$, $\ttn$, and $\tta$. The ellipses denote the Poincar\'e modules of all the auxiliary fields required for the strictification of the BRST colour--kinematics duality manifesting Lagrangian.  
            \item $\frScal$ is the $L_\infty$-algebra of a cubic scalar field theory. Since it is cubic, this $L_\infty$-algebra has only two higher products, the unary and the binary ones $\mu_1$ and $\mu_2$. Explicitly, $\mu_1$ is simply the wave operator, the unique Lorentz invariant possibility, and $\mu_2$ encodes a cubic scalar interaction, the skeleton of the strictified Yang--Mills interactions.
            \item The map $\tau$ is the twist datum of the tensor product, $\frKin^{\rm st}\otimes_\tau\frScal$. Physically, it encodes the kinematic (differential operator) factors of the Yang--Mills interactions. It is fully determined by the BRST colour--kinematics-dual form of the colour-stripped action of Yang--Mills theory, and it induces the correct tensor product for factorising the $L_\infty$-algebra of $\caN=0$ supergravity.
        \end{enumerate}
        
        For this factorisation of the $L_\infty$-algebra to be sensible, the expected tensor products of homotopy algebras need to exist first. One of our collateral results is that, as mentioned above, colour-stripping in Yang--Mills theory can be regarded as a factorisation of the $L_\infty$-algebra of Yang--Mills theory into a colour or gauge Lie algebra and a kinematical $C_\infty$-algebra. Here, we have 
        \begin{equation}
            \tilde\frL^{\rm YM,\,st}_{\rm BRST}\ =\ \frg\otimes\frC^{\rm YM,\,st}_{\rm BRST}~,
        \end{equation}
        where $\frC^{\rm YM,\,st}_{\rm BRST}$ describes the colour-stripped part of manifestly colour--kinematic-dual, strictified Yang--Mills theory.
        
        The double copy strongly suggests the further factorisation
        \begin{equation}
            \frC^{\rm YM,\,st}_{\rm BRST}\ =\ \frKin^{\rm st}\otimes_\tau\frScal~,
        \end{equation}
        and our notion of twisted tensor product of homotopy algebras is essentially constructed such that this factorisation is possible, see \cref{ssec:tensor_prod_homotopy_algebras}. For example, the action of the differential $\sfm^{\tau_1}_1\,:\,\frKin^{\rm st}\otimes_\tau\frScal\rightarrow\frKin^{\rm st}\otimes_\tau\frScal$ on $\ttv^\mu\otimes \varphi(x)\in \frKin^{\rm st}\otimes_\tau\frScal$ looks schematically like
        \begin{equation}\label{eq:scheme_twisted_differential}
            \sfm^{\tau_1}_1(\ttv^\mu\otimes \varphi(x))\ \sim\ \ttv^\mu\otimes \wave \varphi(x)+\ttn \otimes \partial^\mu \varphi(x)~.
        \end{equation}
    
        Given the full factorisation of the Yang--Mills $L_\infty$-algebra, the double copy prescription becomes manifest. Replace the colour factor $\mathfrak{g}$ with another copy of the kinematics factor $\frKin^{\rm st}$ and twist the tensor product by $\tau$:
        \begin{equation}
            \tilde\frL^{\rm YM,\,st}_{\rm BRST}\ =\ \frg\otimes(\frKin^{\rm st}\otimes_\tau\frScal)\ \xrightarrow{\rm double~copy}\ \frKin^{\rm st}\otimes_{\tau}(\frKin^{\rm st}\otimes_\tau\frScal)\ =\ \tilde\frL^{\caN=0,\,\rm st}_{\rm BRST}~,
        \end{equation}
        where $\tilde\frL^{\caN=0,\,\rm st}_{\rm BRST}$ fully determines the double copy theory. The space of fields in this theory is determined by the tensor product of graded vector spaces, which is an extension of the tensor product~\eqref{eq:ym2On-shell} that includes ghosts, anti-ghosts, Nakanishi--Lautrup fields and the further auxiliary fields arising in the strictification. Since it is constructed from the higher products, the (gauge-fixed) BV differential $Q^{\rm YM}_{\rm BV}$ also factorises and double copies into the (gauge-fixed) BV differential $Q^{\caN=0}_{\rm BV}$. 

        We note that one can also replace kinematics with colour to produce the cubic biadjoint scalar theory, sometimes referred to as the zeroth copy,
        \begin{equation}
            \tilde\frL^{\rm YM,\,st}_{\rm BRST}\ =\ \frg\otimes(\frKin^{\rm st}\otimes_\tau\frScal)\ \xrightarrow{\rm zeroth~copy}\ \frg\otimes\bar\frg\otimes\frScal\ =\ \tilde\frL^{\rm biadj}_{\rm BRST}~.
        \end{equation}
        In this case, the homotopy algebraic discussion becomes straightforward.
        
        To summarise, the homotopy algebraic structure underlying the double copy is the factorisation of the $L_\infty$-algebra $\frL$ of a field theory as
        \begin{equation}\label{eq:summary_generalVectorSpaceFactorisation}
            \frL\ \coloneqq\ \frV\otimes\bar\frV\otimes\frScal~,
        \end{equation}
        where $\frV$ and $\bar\frV$ are two (graded) vector spaces, with the most prominent examples being
        \begin{center}
            \begin{tabular}{|l|c|c|}
                \hline
                & $\frV$ & $\bar\frV$
                \\
                \hline
                Biadjoint scalar field theory & $\frg$ & $\bar\frg$ 
                \\
                Yang--Mills theory & $\frg$ & $\frKin$
                \\
                $\caN=0$ supergravity & $\frKin$ & $\frKin$ 
                \\
                \hline
            \end{tabular}
        \end{center}
        If the factorisation is suitable, which means that it is compatible with colour--kinematics duality, then the double copy is a mapping\footnote{This is not a morphism of $L_\infty$-algebras, which would imply a map between elements of the $L_\infty$-algebras. A simple analogy is the mapping of vector spaces from $V$ to $V\otimes W$ for some fixed vector space $W$.} between $L_\infty$-algebras of classical field theories obtained from substitutions of the factors $\frV$ and $\bar \frV$. The advantage of this {\em homotopy double copy} is that it is fully off-shell and goes beyond on-shell amplitudes. Furthermore, it suggests a lift to homotopy algebraic structures in string field theory.
        
        Let us also list a few secondary results which we obtained collecting the necessary tools for our homotopy algebraic discussion of the double copy.
        \begin{enumerate}[(i)]\itemsep-2pt
            \item We demonstrate in \cref{sec:colourDecompositionYM} that the familiar colour-stripping of Yang--Mills scattering amplitudes corresponds to a factorisation of the $L_\infty$-algebra of Yang--Mills theory into a colour or gauge Lie algebra factor and a kinematical $C_\infty$-algebra. This factorisation extends to the level of actions.
            \item A mathematical argument that we were not able to find in the literature is that the rather evident tensor product between certain strict homotopy algebras guarantees the existence of a tensor product between the corresponding general homotopy algebras by homotopy transfer via the homological perturbation lemma. The full argument is given in \cref{ssec:tensor_prod_homotopy_algebras}.
            \item Finally, we show in \cref{app:hA_L_infinity} that in the homotopy algebraic picture, finite gauge transformations can be regarded as curved morphisms of $L_\infty$-algebra.
        \end{enumerate}

        \subsection{Outlook}\label{sec:outlook} 

        In this work, we focus on the case of Yang--Mills theory and $\caN=0$ supergravity. However, when making the replacement
        \begin{equation}
            \frg\otimes(\frKin\otimes_\tau\frScal)\ \longrightarrow\ \bar\frKin\otimes_{\bar\tau}(\frKin\otimes_\tau\frScal)
        \end{equation}
        there is no reason to restrict to $(\bar\frKin,\bar\tau)\cong(\frKin, \tau)$. We could have taken $(\bar\frKin,\bar\tau)$ from any BV Lagrangian field theory admitting a factorisation $\bar\frKin\otimes_{\bar\tau}\frScal$ and enjoying a generalised notion of tree-level colour--kinematics duality. For example, taking the $(\bar\frKin,\bar\tau)$ of $\caN=4$ super Yang--Mills theory, $\bar\frKin\otimes_{\bar\tau}(\frKin\otimes_\tau \frScal)$ would be the $L_\infty$-algebra of pure $\caN=4$ supergravity. In this example, both theories have vanilla colour--kinematics duality, but this also need not be the case. For instance, the flavour--kinematics duality of the non-linear sigma model is an example of (rather trivially) generalised colour--kinematics duality. One can even consider theories that have \emph{no} `colour' factor at all, such as Maxwell theory. This is not a vacuous statement. For example, the trivial factorisation of Maxwell, $\frL^{\rm Max}=\frKin \otimes_\tau\frScal$, can be  enhanced by introducing graded antisymmetric higher products in $\frScal$ that satisfy kinematic Jacobi identities. Since $\frKin\otimes_\tau\frScal$ is a $C_\infty$-algebra, these higher products do not contribute to the Maxwell action, but if we then tensor with a colour Lie algebra $\frg$ we recover BRST colour--kinematics-dual Yang--Mills theory.  
        
        With these ingredients, there are numerous immediate generalisations. We only require
        \begin{enumerate}[(i)]\itemsep-2pt
            \item tree-level colour--kinematics duality --- there is growing zoology of such theories;
            \item an underlying $L_\infty$-algebra --- this includes all BV quantisable Lagrangian field theories and so is very general;
            \item that the $L_\infty$-algebra factorises in an appropriate manner --- this is essentially the requirement that the gauge and space--time symmetries do not mix.
        \end{enumerate}
        The last condition restricts the apparent vast generality a bit. There are examples, such as the non-Abelian Dirac--Born--Infeld theory, where colour structure constants arise whose compatibility with a factorisation is not apparent. In such cases further work is required, before proceeding directly on the homotopy double copy. Up to this issue, our machinery is powerful enough to derive all order validity of the double copy from the validity at the tree level. Let us summarise  here  some of the possibilities, indicating the outstanding questions that must be addressed to realise this claim.
        
        \paragraph{Supergravity.}
        The first obvious generalisation is the inclusion of supersymmetry. For irreducible super Yang--Mills multiplets, colour--kinematics duality for gluons ensures colour--kinematics duality for the entire multiplet. This can be shown using a supersymmetric Ward identity argument, entirely analogous to the BRST Ward identity argument for BRST colour--kinematics duality given in \cref{sec:BRSTExtension}. The factorisation requirement is obviously satisfied, so in principle, there is no obstruction. For gauge theory factors of $\caN=1$ supersymmetric Yang--Mills theory and $\caN=0$ Yang--Mills theory, which yields $\caN=1$ supergravity minimally coupled to a single chiral multiplet, it is particularly straightforward, since there is a convenient superfield formalism as described in~\cite{Anastasiou:2014qba}. 
        
        It is more subtle and interesting when both factors are supersymmetric and there is a Ramond--Ramond sector. Since the gauginos have no linear gauge (BRST) transformation, their product must be identified with field strengths that are to be regarded as the fundamental fields. The intuition from string theory is clear --- the Ramond--Ramond sector couples to the string world-sheet only through the field strengths and never the bare potentials. Indeed, the type II supergravity Lagrangians can both be written without any bare Ramond--Ramond potentials and the Lagrangian double copy in the Ramond--Ramond sector does indeed generate a Lagrangian that is formulated purely in terms of fundamental field strengths. Of course,  it is perturbatively equivalent to the familiar formulation in term of field strengths of potentials. Interestingly, this is automatically achieved via Sen's mechanism for writing Lagrangians for self-dual field strengths~\cite{Sen:2015nph}\footnote{See also~\cite{Rist:2020uaa} where it is shown that this mechanism arises very directly from a, again, homotopy algebraic perspective.}, but without necessarily imposing self-duality. We shall spell out the details in forthcoming work. 
        
        With these basic ingredients accounted for, the door is then open to the plethora of double-copy constructible theories, such as (almost, cf.~\cite{Anastasiou:2017nsz}) all $\caN \geq 2$ ungauged supergravity theories, (super) Einstein--Yang--Mills--scalar theories~\cite{Chiodaroli:2014xia} and gauged supergravity (with Poincar\'e background)~\cite{Chiodaroli:2017ehv}. Each comes with interesting features that must still be addressed in the BRST--Lagrangian double copy formalism, e.g.~spontaneous symmetry breaking, but none that present an obvious obstruction. 
        
        \paragraph{Abelian Dirac--Born--Infeld theory.} 
        The Abelian Dirac--Born--Infeld (DBI) scattering amplitudes have been double copy constructed~\cite{Cachazo:2014xea,Chen:2013fya,Albonico:2020mge}. They follow from the product of the non-linear sigma model and (super) Yang--Mills amplitudes. Given that we already have both the BRST manifestly colour--kinematics-dual formulations of the non-linear sigma model, Yang--Mills theory and their $L_\infty$-algebras, we can immediately apply the homotopy double copy to obtain the perturbative DBI BRST Lagrangian by replacing the colour factor of Yang--Mills with the kinematics of the non-linear sigma model. 
        
        \paragraph{Conformal gravity.}
        More ambitiously, one can approach conformal gravity by including higher-dimensional operators in the Yang--Mills theories~\cite{Johansson:2017srf,Johansson:2018ues}. Again, if colour--kinematics duality holds to all points, then it may be possible to turn the homotopy double copy handle, although the higher dimensional operators will have to be treated carefully. 
        
        \paragraph{Closed string theory.} 
        Given the KLT origins of the double copy, it is natural to seek an `open $\otimes$ open $=$ closed' stringy extension. Moreover, both open and closed string field theory are built on homotopy algebras~\cite{Zwiebach:1992ie, Kajiura:2003ax}, and so the homotopy double copy is a natural framework. Bosonic open string field theory comes strictified out of the box~\cite{Witten:1986qs,Kajiura:2003ax}, so in this sense it may seem to be ready to be double copied. However, one must still pick a partition of the moduli space consistent with colour--kinematics duality so the double copy is not \emph{a priori} automatic. All choices will be quasi-isomorphic, but there seems to be no reason to suppose that every partition of the moduli space gives rise to a strictification compatible with colour--kinematics duality. In this case, one would have a situation similar to Yang--Mills theory, where only certain strictifications are compatible with colour--kinematics duality. Alternatively, it could be that all quasi-isomorphic choices can be double copied and only on taking the field theory limit is this structure broken. Ambitiously, one could consider the more general (or other) formulations of the open/closed string duality.

        \subsection{Reading guide}\label{ssec:reading_guide}
        
        As stated before, it is our intention to be highly self-contained in our presentation, in order to make the homotopy algebraic perspective on the double copy accessible for readers unfamiliar with either homotopy algebras or the double copy (or both!). This reading guide may provide some further help.
        
        For readers unfamiliar with the double copy, good stating points are \cref{sec:review}, with a concise review of the basics, and \cref{sec:NLSM}, which spells out the details in the case of the related but much simpler double copy of the amplitudes of the non-linear sigma model to those for the special galileon. The explicit details for the factorisation of homotopy algebras involved in the gauge--gravity double copy are then presented in \cref{sec:factorisation_free_field_theories} at the level of free field theories and \cref{sec:full_double_copy} at the level of the full actions. The latter section also contains the proof that translates the double copy of amplitudes from tree to full quantum (i.e.~loop) level.
        
        Readers unfamiliar with the BV formalism and the BV formulation of standard field theories will benefit from the general discussion in \cref{sec:BV_section} as well as the concrete examples presented in \cref{sec:Examples}. Some quantum field theoretic preliminaries that are crucial to extending the double copy of amplitudes to loop level are reviewed or developed in \cref{sec:QFT_preliminaries}.
        
        For readers unfamiliar with homotopy algebras, we have collected the basic definitions and results in \cref{app:homotopyAlgebras}. The link to field theories is gently introduced in \cref{ssec:NLSM_homotopy_picture}, using the example of the non-linear sigma model. The general picture and the link to the BV formalism are then developed in \cref{sec:BV_section}; the correspondence between actions and $L_\infty$-algebras is explained in \cref{sec:BVFormalismLinfty}, while the link between scattering amplitudes and $L_\infty$-algebras is presented in \cref{ssec:scatterin_amplitudes}. Concrete examples of $L_\infty$-algebras for a number of field theories relevant in the gauge--gravity double copy are then given in \cref{sec:Examples}.
        
        \section{Double copy basics}\label{sec:review}
        
        \subsection{Scattering amplitude generalities}\label{ssec:scattering_amplitudes_generalities}
        
        We start with a brief review of gluon scattering amplitudes to set some notation and to be self-contained.
        
        \paragraph{Gluon scattering amplitudes.}
        Consider Yang--Mills theory with a semi-simple compact matrix Lie algebra $\frg$ as gauge algebra. Because the Lie bracket in $\frg$ describes naturally a cubic interaction vertex $[-,-]:\frg\times\frg\rightarrow\frg$, the possibility of relating colour to kinematics relies on writing the amplitude in terms of trivalent diagrams only,
        \begin{equation}\label{eq:GeneralGluonAmplitude}
            \scA_{n,L}\ =\ (-\rmi)^{n-3+3L}g^{n-2+2L}\sum_i\int\prod^L_{l=1}\frac{\rmd^dp_l}{(2\pi)^dS_i}\frac{\colfac_i\kinfac_i}{d_i}~.
        \end{equation}
        Here, $\scA_{n,L}$ is the $n$-point $L$-loop gluon scattering amplitude, and $g$ the Yang--Mills coupling constant. The sum is over all $n$-point $L$-loop diagrams, labelled $i$, with only trivalent vertices (\emph{not} the Feynman diagrams of the original theory). The \uline{colour numerator} or \uline{colour factor} $\colfac_i$ associated to a diagram $i$ is composed of gauge algebra structure constants and can be read off directly from the trivalent diagram. The \uline{kinematic numerator} or \uline{kinematic factor} $\kinfac_i$ associated to diagram $i$ is a polynomial of Lorentz-invariant contractions of polarisation vectors and momenta. The \uline{denominator} $d_i$ associated to a diagram $i$ is the product of the Feynman--'t Hooft propagators, i.e.~the product of the squared momenta of all internal lines of the diagram $i$. Finally,  $S_i\in\mathbb{N}$ is the \uline{symmetry factor} associated to a diagram $i$, defined in the same way as for Feynman diagrams, accounting for any over-counting due to the diagram symmetries. At the tree level, i.e.~for $L=0$,~\eqref{eq:GeneralGluonAmplitude} simplifies to 
        \begin{equation}
            \scA_{n,0}\ =\ (-\rmi)^{n-3}g^{n-2}\sum_{i=1}^{(2n-5)!!}\frac{\colfac_i\kinfac_i}{d_i}~,
        \end{equation}
        since there are $(2n-5)!!$ trivalent tree diagrams at $n$ points.
        
        This trivalent form exists because the four-point contact terms can always be `blown-up' and absorbed into corresponding three-point diagrams:
        \begin{equation}
            \begin{tikzpicture}[
                scale=1,
                every node/.style={scale=1},
                baseline={([yshift=-.5ex]current bounding box.center)}
                ]
                \matrix (m) [
                matrix of nodes,
                ampersand replacement=\&, 
                column sep=0.2cm, 
                row sep=0.2cm
                ]{
                    {} \& {}\& {} \& {} \& {}
                    \\
                    {} \& {} \& {}\& {} \& {}
                    \\
                    {} \& {} \& {}\& {} \& {}
                    \\
                    {} \& {} \& {}\& {} \& {}
                    \\
                    {} \& {} \& {} \& {} \& {}
                    \\
                };
                \draw [gluon] (m-1-1) -- (m-3-3.center);
                \draw [gluon] (m-5-1) -- (m-3-3.center);
                \draw [gluon] (m-1-5) -- (m-3-3.center);
                \draw [gluon] (m-5-5) -- (m-3-3.center);
                \foreach \x in {(m-3-3)}{
                    \fill \x circle[radius=2pt];
                }
            \end{tikzpicture}
            \qquad\longrightarrow\qquad 
            \begin{tikzpicture}[
                scale=1,
                every node/.style={scale=1},
                baseline={([yshift=-.5ex]current bounding box.center)}
                ]
                \matrix (m) [
                matrix of nodes,
                ampersand replacement=\&,
                column sep=0.2cm,
                row sep=0.2cm
                ]{
                    {} \& {}\& {} \& {} \& {}
                    \\
                    {} \& {} \& {}\& {} \& {}
                    \\
                    {} \& {} \& {} \& {} \& {}
                    \\
                    {} \& {} \& {}\& {} \& {}
                    \\
                    {} \& {} \& {} \& {} \& {}
                    \\
                };
                \draw [gluon] (m-1-1) -- (m-3-2.center);
                \draw [gluon] (m-5-1) -- (m-3-2.center);
                \draw [aux] (m-3-2.center) -| node[near start,below] {$s$} (m-3-4.center);
                \draw [gluon] (m-1-5) -- (m-3-4.center);
                \draw [gluon] (m-5-5) -- (m-3-4.center);
                \foreach \x in {(m-3-2), (m-3-4)}{
                    \fill \x circle[radius=2pt];
                }
            \end{tikzpicture}
            +
            \begin{tikzpicture}[
                scale=1,
                every node/.style={scale=1},
                baseline={([yshift=-.5ex]current bounding box.center)}
                ]
                \matrix (m) [
                matrix of nodes,
                ampersand replacement=\&,
                column sep=0.2cm,
                row sep=0.2cm
                ]{
                    {} \& {} \& {} \& {} \& {}
                    \\
                    {} \& {} \& {} \& {} \& {}
                    \\
                    {} \& {} \& {} \& {} \& {}
                    \\
                    {} \& {} \& {} \& {} \& {}
                    \\
                    {} \& {} \& {} \& {} \& {}
                    \\
                };
                \draw [gluon] (m-1-1) -- (m-2-3.center);
                \draw [gluon] (m-5-1) -- (m-4-3.center);
                \draw [aux] (m-2-3.center) -| node[near end,left] {$t$} (m-4-3.center);
                \draw [gluon] (m-1-5) -- (m-2-3.center);
                \draw [gluon] (m-5-5) -- (m-4-3.center);
                \foreach \x in {(m-2-3), (m-4-3)}{
                    \fill \x circle[radius=2pt];
                }
            \end{tikzpicture}
            + 
            \begin{tikzpicture}[
                scale=1,
                every node/.style={scale=1},
                baseline={([yshift=-.5ex]current bounding box.center)}
                ]
                \matrix (m) [
                matrix of nodes,
                ampersand replacement=\&,
                column sep=0.2cm,
                row sep=0.2cm
                ]{
                    {} \& {}\& {} \& {} \& {}
                    \\
                    {} \& {} \& {}\& {} \& {}
                    \\
                    {} \& {} \& {}\& {} \& {}
                    \\
                    {} \& {} \& {}\& {} \& {}
                    \\
                    {} \& {}\& {} \& {} \& {}
                    \\
                };
                \draw [gluon] (m-1-1) -- (m-3-4.center);
                \draw [gluon] (m-5-1) -- (m-3-2.center);
                \draw [aux] (m-3-2.center) -| node[near start,below] {$u$} (m-3-4.center);
                \draw [gluon] (m-1-5) -- (m-3-2.center);
                \draw [gluon] (m-5-5) -- (m-3-4.center);
                \foreach \x in {(m-3-2), (m-3-4)}{
                    \fill \x circle[radius=2pt];
                }
            \end{tikzpicture}
        \end{equation}
        
        \paragraph{Four-point tree-level gluon scattering amplitude from trivalent diagrams.}
        Consider the simplest example of the four-point tree-level scattering amplitude,
        \begin{equation}
            \scA_{4,0}\ =\
            \begin{tikzpicture}[
                scale=1,
                every node/.style={scale=1},
                baseline={([yshift=-.5ex]current bounding box.center)}
                ]
                \matrix (m) [
                matrix of nodes,
                ampersand replacement=\&,
                column sep=0.2cm,
                row sep=0.2cm
                ]{
                    $1$ \& {} \& {} \& {} \& $4$
                    \\
                    {} \& {} \& {} \& {} \& {}
                    \\
                    {} \& {} \& {} \& {} \& {}
                    \\
                    {} \& {} \& {} \& {} \& {}
                    \\
                    $2$ \& {}\& {} \& {} \& $3$
                    \\
                };
                \draw [gluon] (m-1-1) -- (m-3-2.center);
                \draw [gluon] (m-5-1) -- (m-3-2.center);
                \draw [gluon] (m-3-2.center) -| node[near start,below] {$s$} (m-3-4.center);
                \draw [gluon] (m-1-5) -- (m-3-4.center);
                \draw [gluon] (m-5-5) -- (m-3-4.center);
                \foreach \x in {(m-3-2), (m-3-4)}{
                    \fill \x circle[radius=2pt];
                }
            \end{tikzpicture}
            +
            \begin{tikzpicture}[
                scale=1,
                every node/.style={scale=1},
                baseline={([yshift=-.5ex]current bounding box.center)}
                ]
                \matrix (m) [
                matrix of nodes,
                ampersand replacement=\&,
                column sep=0.2cm,
                row sep=0.2cm
                ]{
                    $1$ \& {} \& {} \& {} \& $4$
                    \\
                    {} \& {} \& {} \& {} \& {}
                    \\
                    {} \& {} \& {} \& {} \& {}
                    \\
                    {} \& {} \& {} \& {} \& {}
                    \\
                    $2$ \& {}\& {} \& {} \& $3$
                    \\
                };
                \draw [gluon] (m-1-1) -- (m-2-3.center);
                \draw [gluon] (m-5-1) -- (m-4-3.center);
                \draw [gluon] (m-2-3.center) -| node[near end,left] {$t$} (m-4-3.center);
                \draw [gluon] (m-1-5) -- (m-2-3.center);
                \draw [gluon] (m-5-5) -- (m-4-3.center);
                \foreach \x in {(m-2-3), (m-4-3)}{
                    \fill \x circle[radius=2pt];
                }
            \end{tikzpicture}
            +
            \begin{tikzpicture}[
                scale=1,
                every node/.style={scale=1},
                baseline={([yshift=-.5ex]current bounding box.center)}
                ]
                \matrix (m) [
                matrix of nodes,
                ampersand replacement=\&,
                column sep=0.2cm,
                row sep=0.2cm
                ]{
                    $1$ \& {} \& {} \& {} \& $4$
                    \\
                    {} \& {} \& {} \& {} \& {}
                    \\
                    {} \& {} \& {} \& {} \& {}
                    \\
                    {} \& {} \& {} \& {} \& {}
                    \\
                    $2$ \& {}\& {} \& {} \& $3$
                    \\
                };
                \draw [gluon] (m-1-1) -- (m-3-4.center);
                \draw [gluon] (m-5-1) -- (m-3-2.center);
                \draw [gluon] (m-3-2.center) -| node[near start,below] {$u$} (m-3-4.center);
                \draw [gluon] (m-1-5) -- (m-3-2.center);
                \draw [gluon] (m-5-5) -- (m-3-4.center);
                \foreach \x in {(m-3-2), (m-3-4)}{
                    \fill \x circle[radius=2pt];
                }
            \end{tikzpicture}
            +
            \begin{tikzpicture}[
                scale=1,
                every node/.style={scale=1},
                baseline={([yshift=-.5ex]current bounding box.center)}
                ]
                \matrix (m) [
                matrix of nodes,
                ampersand replacement=\&,
                column sep=0.2cm,
                row sep=0.2cm
                ]{
                    $1$ \& {} \& {} \& {} \& $4$
                    \\
                    {} \& {} \& {} \& {} \& {}
                    \\
                    {} \& {} \& {} \& {} \& {}
                    \\
                    {} \& {} \& {} \& {} \& {}
                    \\
                    $2$ \& {}\& {} \& {} \& $3$
                    \\
                };
                \draw [gluon] (m-1-1) -- (m-3-3.center);
                \draw [gluon] (m-5-1) -- (m-3-3.center);
                \draw [gluon] (m-1-5) -- (m-3-3.center);
                \draw [gluon] (m-5-5) -- (m-3-3.center);
                \foreach \x in {(m-3-3)}{
                    \fill \x circle[radius=2pt];
                }
            \end{tikzpicture}.
        \end{equation}
        Explicitly, with all momenta incoming,
        \begin{subequations}
            \begin{equation}\label{eq:FourPointTreeGluonAmplitude}
                \begin{aligned}
                    \begin{tikzpicture}[
                        scale=1,
                        every node/.style={scale=1},
                        baseline={([yshift=-.5ex]current bounding box.center)}
                        ]
                        \matrix (m) [
                        matrix of nodes,
                        ampersand replacement=\&,
                        column sep=0.2cm,
                        row sep=0.2cm
                        ]{
                            $1$ \& {} \& {} \& {} \& $4$
                            \\
                            {} \& {} \& {} \& {} \& {}
                            \\
                            {} \& {} \& {} \& {} \& {}
                            \\
                            {} \& {} \& {} \& {} \& {}
                            \\
                            $2$ \& {}\& {} \& {} \& $3$
                            \\
                        };
                        \draw [gluon] (m-1-1) -- (m-3-2.center);
                        \draw [gluon] (m-5-1) -- (m-3-2.center);
                        \draw [gluon] (m-3-2.center) -| node[near start,below] {$s$} (m-3-4.center);
                        \draw [gluon] (m-1-5) -- (m-3-4.center);
                        \draw [gluon] (m-5-5) -- (m-3-4.center);
                        \foreach \x in {(m-3-2), (m-3-4)}{
                            \fill \x circle[radius=2pt];
                        }
                    \end{tikzpicture}
                    \ &=\ -\rmi g^2\frac{f^{abe}{f_e}^{cd}\kinfac_s}{s}\ \eqqcolon\ -\rmi g^2\frac{\colfac_s\kinfac_s}{s}~,
                    \\
                    \begin{tikzpicture}[
                        scale=1,
                        every node/.style={scale=1},
                        baseline={([yshift=-.5ex]current bounding box.center)}
                        ]
                        \matrix (m) [
                        matrix of nodes,
                        ampersand replacement=\&,
                        column sep=0.2cm,
                        row sep=0.2cm
                        ]{
                            $1$ \& {} \& {} \& {} \& $4$
                            \\
                            {} \& {} \& {} \& {} \& {}
                            \\
                            {} \& {} \& {} \& {} \& {}
                            \\
                            {} \& {} \& {} \& {} \& {}
                            \\
                            $2$ \& {}\& {} \& {} \& $3$
                            \\
                        };
                        \draw [gluon] (m-1-1) -- (m-2-3.center);
                        \draw [gluon] (m-5-1) -- (m-4-3.center);
                        \draw [gluon] (m-2-3.center) -| node[near end,left] {$t$} (m-4-3.center);
                        \draw [gluon] (m-1-5) -- (m-2-3.center);
                        \draw [gluon] (m-5-5) -- (m-4-3.center);
                        \foreach \x in {(m-2-3), (m-4-3)}{
                            \fill \x circle[radius=2pt];
                        }
                    \end{tikzpicture}
                    \ &=\ -\rmi g^2\frac{f^{aed}{f_e}^{bc}\kinfac_t}{t}\ \eqqcolon\ -\rmi g^2\frac{\colfac_t \kinfac_t}{t}~,
                    \\
                    \begin{tikzpicture}[
                        scale=1,
                        every node/.style={scale=1},
                        baseline={([yshift=-.5ex]current bounding box.center)}
                        ]
                        \matrix (m) [
                        matrix of nodes,
                        ampersand replacement=\&,
                        column sep=0.2cm,
                        row sep=0.2cm
                        ]{
                            $1$ \& {} \& {} \& {} \& $4$
                            \\
                            {} \& {} \& {} \& {} \& {}
                            \\
                            {} \& {} \& {} \& {} \& {}
                            \\
                            {} \& {} \& {} \& {} \& {}
                            \\
                            $2$ \& {}\& {} \& {} \& $3$
                            \\
                        };
                        \draw [gluon] (m-1-1) -- (m-3-4.center);
                        \draw [gluon] (m-5-1) -- (m-3-2.center);
                        \draw [gluon] (m-3-2.center) -| node[near start,below] {$u$} (m-3-4.center);
                        \draw [gluon] (m-1-5) -- (m-3-2.center);
                        \draw [gluon] (m-5-5) -- (m-3-4.center);
                        \foreach \x in {(m-3-2), (m-3-4)}{
                            \fill \x circle[radius=2pt];
                        }
                    \end{tikzpicture}
                    \ &=\ -\rmi g^2\frac{f^{aec}{f_e}^{db}\kinfac_u}{u}\ \eqqcolon\ -\rmi g^2\frac{\colfac_u\kinfac_u}{u}~, 
                    \\
                    \begin{tikzpicture}[
                        scale=1,
                        every node/.style={scale=1},
                        baseline={([yshift=-.5ex]current bounding box.center)}
                        ]
                        \matrix (m) [
                        matrix of nodes,
                        ampersand replacement=\&,
                        column sep=0.2cm,
                        row sep=0.2cm
                        ]{
                            $1$ \& {} \& {} \& {} \& $4$
                            \\
                            {} \& {} \& {} \& {} \& {}
                            \\
                            {} \& {} \& {} \& {} \& {}
                            \\
                            {} \& {} \& {} \& {} \& {}
                            \\
                            $2$ \& {}\& {} \& {} \& $3$
                            \\
                        };
                        \draw [gluon] (m-3-3.center) -- (m-1-1);
                        \draw [gluon] (m-5-1) -- (m-3-3.center);
                        \draw [gluon] (m-1-5) -- (m-3-3.center);
                        \draw [gluon] (m-5-5) -- (m-3-3.center);
                        \foreach \x in {(m-3-3)}{
                            \fill \x circle[radius=2pt];
                        }
                    \end{tikzpicture}
                    \ &=\ -\rmi g^2\left(\colfac_s\kinfac^{(4)}_s-\colfac_t\kinfac^{(4)}_t-\colfac_u\kinfac^{(4)}_u\right).
                \end{aligned}
            \end{equation}
            Here, we have made use of the standard Mandelstam variables $s\coloneqq(p_1+p_2)^2$, $t\coloneqq(p_1+p_4)^2$, and $u\coloneqq(p_1+p_3)^2$ in the trivalent $s$-, $t$-, and $u$-channel diagrams, respectively. Furthermore, upon setting $p_{ij}\coloneqq p_i-p_j$ where $i,j,\ldots=1,\ldots,4$ label the different gluons, the kinematic numerators are given by
            \begin{equation}\label{eq:FourPointKinematicFactors}
                \begin{aligned}
                    \kinfac_s\ &\coloneqq\ 4\big[(\eps_1\cdot p_2)\eps_2-(\eps_2\cdot p_1)\eps_1+\tfrac12(\eps_1\cdot\eps_2)p_{12}\big]\,\cdot
                    \\
                    &\kern2cm\cdot\big[(\eps_3\cdot p_4)\eps_4-(\eps_4\cdot p_3)\eps_3+\tfrac12(\eps_3\cdot\eps_4)p_{34}\big]~,
                    \\
                    \kinfac_t\ &\coloneqq\ -4\big[(\eps_1\cdot p_4)\eps_4-(\eps_4\cdot p_1)\eps_1+\tfrac12(\eps_1\cdot\eps_4)p_{14}\big]\,\cdot
                    \\
                    &\kern2cm\cdot\big[(\eps_2\cdot p_3)\eps_3-(\eps_3\cdot p_3)\eps_3+\tfrac12(\eps_2\cdot\eps_3)p_{23}\big]~,
                    \\
                    \kinfac_u\ &\coloneqq\ 4\big[(\eps_2\cdot p_4)\eps_4-(\eps_4\cdot p_4)\eps_4+\tfrac12(\eps_2\cdot\eps_4)p_{24}\big]\,\cdot
                    \\
                    &\kern2cm\cdot\big[(\eps_1\cdot p_3)\eps_3-(\eps_3\cdot p_1)\eps_1+\tfrac12(\eps_1\cdot\eps_3)p_{13}\big]~,
                \end{aligned}
            \end{equation}
            where $\eps_i$ is the polarisation vector of the $i$-th gluon and $x\cdot y\coloneqq\eta_{\mu\nu}x^\mu x^\nu$. We also have suggestively labelled the kinematic numerators appearing in the four-point contact term in~\eqref{eq:FourPointTreeGluonAmplitude} by $\kinfac^{(4)}_s$, $\kinfac^{(4)}_t$, and $\kinfac^{(4)}_u$. They are given by
            \begin{equation}\label{eq:FourContact}
                \begin{aligned}
                    \kinfac^{(4)}_s\ &\coloneqq\ (\eps_1\cdot\eps_3)(\eps_2\cdot\eps_4)-(\eps_1\cdot\eps_4)(\eps_2\cdot\eps_3)~,
                    \\
                    \kinfac^{(4)}_t\ &\coloneqq\ -(\eps_1\cdot\eps_2)(\eps_3\cdot\eps_4)+(\eps_1\cdot\eps_3)(\eps_2\cdot\eps_4)~,
                    \\
                    \kinfac^{(4)}_u\ &\coloneqq\ -(\eps_1\cdot\eps_4)(\eps_2\cdot\eps_3)+(\eps_1\cdot\eps_2)(\eps_4\cdot\eps_3)~.
                \end{aligned}
            \end{equation}
        \end{subequations}
        
        Upon summing up~\eqref{eq:FourPointTreeGluonAmplitude}, the four-point tree-level scattering amplitude is a sum over the three trivalent diagrams,
        \begin{subequations}
            \begin{equation}\label{eq:FourPointTrivalent}
                \scA_{4,0}\ =\ -\rmi g^2\left(\frac{\colfac_s\kinfac'_s}{s}+\frac{\colfac_t \kinfac'_t}{t}+\frac{\colfac_u\kinfac'_u}{u}\right)
            \end{equation}
            with
            \begin{equation}\label{eq:ns}
                \kinfac'_s\ \coloneqq\ \kinfac_s+s\kinfac_s^{(4)}~,~~~
                \kinfac'_t\ \coloneqq\ \kinfac_t-t\kinfac_t^{(4)}~,~~~
                \kinfac'_u\ \coloneqq\ \kinfac_u-u\kinfac_u^{(4)}~.
            \end{equation}
        \end{subequations}
        Note that any $n$-point $L$-loop diagram with a four-point contact term is accompanied by three diagrams that are identical except that the four-point contact term is replaced by the $s$-, $t$-, and $u$-channel trivalent diagrams, and the above argument can be applied. Of course, this can be realised at the Lagrangian level by introducing an auxiliary field strictifying the action to be cubic~\cite{Bern:2010yg}.
        
        \begin{remark}
            We may introduce the colour-stripped vertex $\tilde F_{\mu\nu\rho}$ in momentum space, intentionally written as `structure constants' analogous to $f_{abc}$,
            \begin{equation}
                \tilde F^{\mu_1\mu_2\mu_3}(p_1,p_2, p_3)\ \coloneqq\!\!
                \begin{tikzpicture}[
                    scale=1,
                    every node/.style={scale=1},
                    baseline={([yshift=-.5ex]current bounding box.center)}
                    ]
                    \matrix (m) [
                    matrix of nodes,
                    ampersand replacement=\&,
                    column sep=0.2cm,
                    row sep=0.2cm
                    ]{
                        $p_1$ \& {} \& {} \& {} \& {}
                        \\
                        {} \& {} \& {} \& {} \& {}
                        \\
                        {} \& {} \& {} \& {} \& {}
                        \\
                        {} \& {} \& {} \& {} \& {}
                        \\
                        $p_3$ \& {}\& {} \& {} \& {}
                        \\
                    };
                    \draw (m-1-1) -- (m-3-3.center);
                    \draw (m-3-5) -- (m-3-3.center);
                    \draw (m-5-1) -- (m-3-3.center);
                    \foreach \x in {(m-3-3)}{
                        \fill \x circle[radius=2pt];
                    }
                \end{tikzpicture}
                \!\!p_2\ \coloneqq\ p_{12}^{\mu_3}\eta^{\mu_1\mu_2}+p_{23}^{\mu_1}\eta^{\mu_2\mu_3}+p_{31}^{\mu_2}\eta^{\mu_3\mu_1}~,
            \end{equation}
            so that
            \begin{equation}
                \begin{gathered}
                    \kinfac_s\ =\ \eps^{\mu_1}_1\eps^{\mu_2}_2\tilde F_{\mu_1\mu_2\rho}\tilde F^\rho{}_{\mu_3\mu_4}\eps^{\mu_3}_3\eps^{\mu_4}_4~,~~~
                    \kinfac_t\ =\ \eps^{\mu_1}_1\eps^{\mu_4}_4\tilde F_{\mu_1\rho\mu_4}\tilde F^\rho{}_{\mu_2\mu_3}\eps^{\mu_2}_2\eps^{\mu_3}_3~,
                    \\
                    \kinfac_u\ =\ \eps^{\mu_1}_1\eps^{\mu_3}_3\tilde F_{\mu_1\rho\mu_3}\tilde F^\rho{}_{\mu_4\mu_2}\eps^{\mu_2}_2\eps^{\mu_4}_4~.
                \end{gathered}
            \end{equation}
            This observation will become important in \cref{ssec:double_copy}.
        \end{remark}
        
        \paragraph{Generalised gauge transformations.}
        We note that the three colour numerators satisfy the Jacobi identity,\footnote{Here, square brackets indicate anti-symmetrisation of the enclosed indices.}
        \begin{equation}
            \colfac_s-\colfac_t-\colfac_u\ =\ 3 f^{ea[b}f_{e}{}^{cd]}\ =\ 0~,
        \end{equation}
        so that a shift of the kinematic numerators by an arbitrary function $\alpha$,
        \begin{equation}
            \kinfac'_s\ \mapsto\ \kinfac'_s-s\alpha~,~~~
            \kinfac'_t\ \mapsto\ \kinfac'_t+t\alpha~,~~~
            \kinfac'_u\ \mapsto\ \kinfac'_u+u\alpha~,
        \end{equation}
        leaves the amplitude~\eqref{eq:FourPointTrivalent} invariant. These shifts, corresponding to an additional freedom in the choice of the kinematic numerators, were referred to as \uline{generalised gauge transformations} in~\cite{Bern:2008qj}. Of course, this applies to any triple of trivalent diagrams $(i,j,k)$ that only differ in a common four-point subdiagram with colour numerators satisfying a Jacobi identity of the form $\colfac_i+\colfac_j+\colfac_k=0$, where the generalised gauge transformation acting on the corresponding kinematic numerators is given by  
        \begin{equation}
            \kinfac_i\ \mapsto\ \kinfac_i+s_i\alpha~,~~~
            \kinfac_j\ \mapsto\ \kinfac_j+s_j\alpha~,~~~
            \kinfac_k\ \mapsto\ \kinfac_k+s_k\alpha~,
        \end{equation}
        and $s_i$, $s_j$, and $s_k$ are the Mandelstam variables of the common four-point subdiagram in which the three diagrams differ.
        
        \subsection{Colour--kinematics duality}\label{ssec:BCJduality}
        
        \paragraph{Bern--Carrasco--Johansson duality.}
        The Bern--Carrasco--Johansson (BCJ) colour--kinematics duality conjecture now states the following.
        
        \begin{conjecture} (Bern--Carrasco--Johansson,~\cite{Bern:2008qj,Bern:2010ue})
            \begin{subequations}
                There exists a choice of kinematic numerators  of the trivalent diagrams entering the scattering amplitude $\scA_{n,L}$ such that
                \begin{enumerate}[(i)]\itemsep-2pt
                    \item whenever a triple of trivalent diagrams $(i,j,k)$ has colour numerators obeying
                    \begin{equation}
                        \colfac_i+\colfac_j+\colfac_k\ =\ 0
                    \end{equation}
                    due to the Jacobi identity, then the corresponding kinematic numerators obey precisely the same identity
                    \begin{equation}
                        \kinfac_i+\kinfac_j+\kinfac_k\ =\ 0~;
                    \end{equation}
                    \item if in any individual diagram, $\colfac_i\mapsto-\colfac_i$ under the interchange of two legs, then $\kinfac_i\mapsto-\kinfac_i$ at the same time. 
                \end{enumerate}
            \end{subequations}
        \end{conjecture}

        \paragraph{Three- and four-point tree-level gluon scattering amplitudes.}
        Evidently, the tree-level three-point scattering amplitude (allowing complex momenta), which consists of a single diagram, trivially satisfies the duality since under interchange of any two edges $\colfac=f^{abc} \mapsto -\colfac$, since $f^{abc}$ is totally anti-symmetric, and the same is true for $\kinfac=\eps_{1}^{\mu_1}\eps_{2}^{\mu_2}\eps_{3}^{\mu_3}F_{\mu_1\mu_2\mu_3}$, since $F_{\mu_1\mu_2\mu_3}$ is totally anti-symmetric.
        
        The tree-level four-point scattering amplitude was known to satisfy colour--kinematics-duality before the notion of this duality had been articulated~\cite{Zhu:1980sz,Goebel:1980es}. Indeed, using momentum conservation $\sum_ip_i=0$ and transversality $\eps_i\cdot p_i=0$,~\eqref{eq:FourPointKinematicFactors},~\eqref{eq:FourContact}, and~\eqref{eq:ns} immediately imply that the kinematic numerators satisfy
        \begin{equation}
            \kinfac'_s-\kinfac'_t-\kinfac'_u\ =\ 0~.
        \end{equation} 
        This agrees with the colour Jacobi identity $\colfac_s-\colfac_t-\colfac_u=0$. Note that this would have failed without the additional contributions from the four-point contact term~\eqref{eq:FourContact}. At higher points, one would also need the on-shell conditions $p_i^2=0$.
        
        The fact that the kinematic identity holds without any intervention besides blowing up the four-point contact term is due to the special kinematics of the four-point amplitude.    At higher points, not all possible choices of ${\sf n}_i$ will satisfy the required kinematic  identities. Already at five points it is non-trivial~\cite{Bern:2008qj}, although there is a particularly nice representation of the colour--kinematics-dual amplitude in this case~\cite{Broedel:2011pd}.
        
        \paragraph{General tree-level gluon scattering amplitudes.}
        Thinking of the $(2n-5)!!$ colour $\colfac_i$ and kinematic numerators $\kinfac_i$ as column vectors, denoted by $\bmc$ and $\bmn$, we can trivially rewrite the $n$-point tree amplitude as 
        \begin{equation}
            \scA_{n,0}\ =\ \bmc^\sfT\bmD\bmn
            \ewith
            \bmD_{ij}\ \coloneqq\ \frac{\delta_{ij}}{d_j}~.
        \end{equation}
        The number of linearly independent (under the Jacobi identities) colour numerators $\colfac_i$ is $(n-2)!$, which, using the multi-peripheral colour decomposition of~\cite{DelDuca:1999rs}, is seen to be the same as the  number of linearly independent partial colour-stripped scattering amplitudes $A_n^{i'}\coloneqq A_{n,0}(12\sigma_{i'}(34\cdots n))$ under the Kleiss--Kuijf relations~\cite{Kleiss:1988ne}, where $\{\sigma_{i'}\}_{i'=1}^{n-2}=S_{n-2}$ and $A_{n,0}(1\cdots n)$ is the colour-ordered $n$-gluon tree amplitude. 
        
        Thus, we can choose a subset consisting of $(n-2)!$ linearly independent colour numerators, also called \uline{primaries}, and put them into a $(n-2)!$-component column vector $\bmc_{\rm m}$. The rest are generated by the $\sum_{k=1}^{\lfloor\frac12(n-2)\rfloor}\frac{1}{2^{2k}}C^{n-2}_{2k}C^{2k}_k(n-2)!$ linearly independent Jacobi identities, 
        \begin{equation}
            \bmc\ =\ \bmJ\bmc_{\rm m}~,
        \end{equation}
        where $\bmJ$ is a $(2n-5)!!\times (n-2)!$ matrix encoding these relations. For example, at four points, in the conventions of~\eqref{eq:FourPointTreeGluonAmplitude},  we can choose $\colfac_t, \colfac_u$ as our primary colour numerators, and then
        \begin{equation}\label{eq:nJn}
            \bmJ\ =\ 
            \begin{pmatrix}
                1 & 1
                \\
                1 & 0
                \\ 
                0 & 1
            \end{pmatrix}.
        \end{equation}
        In this form, colour--kinematics duality requires the existence of kinematic numerators satisfying 
        \begin{equation}
            \bmn\ =\ \bmJ\bmn_{\rm m}~.
        \end{equation}
        We also have the $(n-2)!$ ostensibly linearly independent (prior to applying the BCJ relations) $n$-point partial amplitudes $\bmA_{i'}\coloneqq A_n^{i'}$ which may be written as
        \begin{equation}\label{eq:Aton}
            \bmA\ =\ \bmP\bmn~,
        \end{equation}
        where $\bmP$ is an $(n-2)!\times (2n-5)!!$ matrix of propagators with signs determined by the permutations defining the components of $\bmA$ relative to the colour order of the corresponding graphs. 
        
        If~\eqref{eq:nJn} can be realised then 
        \begin{equation}
            \bmA\ =\ \bmP\bmJ\bmn_{\rm m}~.
        \end{equation}
        Note that, while this relation looks as if it immediately identifies the colour--kinematics duality respecting $\bmn$ in terms of the partial scattering amplitudes, $\bmP\bmJ$ is necessarily singular. 
        
        However, the required relations are purely algebraic, and we can solve for  $(n-3)!$ elements of $\bmn_{\rm m}$ in terms of $(n-3)!$ partial amplitudes and the remaining $(n-2)!-(n-3)!=(n-3)!(n-3)$ elements of $\bmn_{\rm m}$. On substituting this solution back into~\eqref{eq:Aton}, we encounter a surprise: the dependence of the remaining $(n-3)!(n-3)$ partial amplitudes on $\bmn_{\rm m}$  drops out entirely, and we are left with a new set of relations amongst the $(n-2)!$ partial scattering amplitudes. These are known as the \uline{BCJ relations} and were introduced in~\cite{Bern:2008qj}, where they were shown to hold explicitly up to eight points. Assuming the colour--kinematics-duality, the $(n-2)!$ partial amplitudes are in fact an overcomplete basis, which is reduced to $(n-3)!$ linearly independent partial amplitudes by the implied BCJ relations. 
        
        Conversely, given the BCJ relations it is possible to explicitly construct a representation of the total amplitude such that colour--kinematics duality holds~\cite{BjerrumBohr:2010hn,Kiermaier:2010}. The $n$-point BCJ relations were shown to hold in~\cite{BjerrumBohr:2009rd,Stieberger:2009hq} by considering the $\alpha'\rightarrow0$ limit of string theory monodromy relations, confirming the colour--kinematics  duality conjecture at the tree level. The BCJ relations  may also be deduced from pure spinor cohomology~\cite{Mafra:2015vca}. There are a number of powerful stringy perspectives on the BCJ relations, see for example~\cite{Tye:2010dd,Mafra:2012kh,Schlotterer:2012ny,Broedel:2013tta}, including $\alpha'$ deformations respecting colour--kinematics duality~\cite{Broedel:2012rc}.  A purely field theoretic derivation was given in~\cite{Feng:2010my} using only Britto--Cachazo--Feng--Witten recursion~\cite{Britto:2005fq}. They were also established in $\caN=4$ super Yang--Mills theory~\cite{Cachazo:2012uq}, which contains the Yang--Mills case, using the Roiban--Spradlin--Volovich--Witten connected formalism. Recently, it has been shown~\cite{Mizera:2019blq}, via a residue theorem, that the tree-amplitudes written in terms of intersection numbers~\cite{Mizera:2017rqa,Mizera:2019gea} automatically satisfy the colour--kinematics duality. 
        
        \paragraph{Colour--kinematics duality for loops.}
        Our discussion so far has been restricted to the tree level. The statement of the duality for loops is the same as for trees, up to some minor subtleties. In particular, the kinematic numerators are functions of the loop momenta. Moreover, the kinematic Jacobi-type identities are functional identities at the loop level. Hence, one cannot straightforwardly  solve for the kinematic numerators via a pseudo-inverse as in the tree-level case.
        
        The four-point one-loop example in $\caN=4$ supersymmetric Yang--Mills theory is particularly simple, due to the simple structure of one-loop amplitudes~\cite{Brink:1976bc,Ferber:1977qx}. See for example~\cite{Elvang:2015rqa,Elvang:2013cua,Bern:2019prr}. For Yang--Mills theory at one and two loops, see~\cite{Bern:2013yya}. For detailed examples at three loops, see for example~\cite{Bern:2010ue,Carrasco:2011hw}. These simple cases make it clear that colour--kinematics duality can work at the loop level. However, the proof of colour--kinematics duality at the tree level given in~\cite{Kiermaier:2010,BjerrumBohr:2010hn} relied on the Kawai--Lewellen--Tye relations and therefore does not extend to loop level. At the time of writing there is no proof that colour--kinematics duality will hold to all loops, despite an impressive number of highly non-trivial concrete examples~\cite{Bern:2010ue,Bern:2010tq,Carrasco:2011mn,Bern:2011rj,BoucherVeronneau:2011qv,Bern:2012cd,Bern:2012gh,Oxburgh:2012zr,Bern:2012uf,Du:2012mt,Yuan:2012rg,Bern:2013uka,Boels:2013bi,Bern:2013yya,Bern:2013qca,Bern:2014lha,Bern:2014sna,Mafra:2015mja,Johansson:2017bfl}. 
        
        \subsection{Double copy}\label{ssec:double_copy}
        
        \paragraph{Graviton scattering amplitudes.}
        Expanding the Einstein--Hilbert action perturbatively around the Minkowski background $g_{\mu\nu}=\eta_{\mu\nu}+\kappa h_{\mu\nu}$, we can construct graviton scattering amplitudes as pioneered by DeWitt~\cite{DeWitt:1967yk,DeWitt:1967ub,DeWitt:1967uc}. The Feynman diagrams for gravitons include $n$-point vertices for all $n$, and schematically we have
        \begin{equation}
            S^{\rm EH}\ =\ -\tfrac{1}{\kappa^2}\int\rmd^dx\,\sqrt{-g}\,R\ \sim\ \int\rmd^dx\,\sum_{n=0}^\infty\kappa^n\partial\partial h^{n+2}~,
        \end{equation}
        where $2\kappa^2=16\pi G_{\rm N}^{(d)}$ is Einstein's gravitational constant.\footnote{See \cref{ssec:ex:EH} for more details on the perturbative analysis of the Einstein--Hilbert action.} However, just as for the four-point vertex in Yang--Mills theory, these can all be absorbed into the kinematic numerators of the purely trivalent diagrams. For example, consider a purely trivalent diagram $i$, contributing $\frac{N_i}{d_i}$, where $N_i$ is the kinematic numerator, to the amplitude integrand, and another diagram $i_{(4)}$, contributing $\frac{N_{i_{(4)}}}{d_{i_{(4)}}}$, which is identical except that one trivalent four-point sub-diagram with propagator $s$ has been contracted to a four-point vertex. Then, $sd_{i_{(4)}}=d_{i}$ and so,
        \begin{equation}
            \frac{N_i}{d_i}+\frac{N_{i_{(4)}}}{d_{i_{(4)}}}\ =\ \frac{N_i+s N_{i_{(4)}}}{d_i}\ =\ \frac{N'_i}{d_i}~. 
        \end{equation}
        This argument is not affected by the inclusion of the Kalb--Ramond and dilaton sectors.  Consequently, the $\caN=0$ supergravity scattering amplitudes structurally resemble closely the gluon scattering amplitudes~\eqref{eq:GeneralGluonAmplitude},
        \begin{equation}
            \scH_{n,L}\ =\ (-\rmi)^{n-3+3L}\left(\frac{\kappa}{2}\right)^{n-2+2L}\sum_i\int\prod^{L}_{l=1}\frac{\rmd^dp_l}{(2\pi)^dS_i}\frac{N_i}{d_i}~.
        \end{equation}
        
        \paragraph{Double copy.}
        What is less immediately apparent is that the BCJ double copy prescription implies that, given colour--kinematics duality, the $\caN=0$ kinematic numerators, for factorisable  external states, can always be written as a product: $N_i=n_i\tilde{n}_i$. More precisely, let us state the BCJ double copy prescription for Yang--Mills theory~\cite{Bern:2010ue,Bern:2010yg}: given any two $n$-point $L$-loop gluon scattering amplitudes, 
        \begin{equation}
            \begin{aligned}
                \scA_{n,L}\ &=\ (-\rmi)^{n-3+3L}g^{n-2+2L}\sum_{i}\int\prod^{L}_{l=1}\frac{\rmd^dp_l}{(2\pi)^dS_i}\frac{\colfac_i\kinfac_i}{d_i}~,
                \\
                \tilde{\scA}_{n,L}\ &=\ (-\rmi)^{n-3+3L}g^{n-2+2L}\sum_i\int\prod^{L}_{l=1}\frac{\rmd^dp_l}{(2\pi)^dS_i}\frac{\colfac_i\tilde{\kinfac}_i}{d_i}~,
            \end{aligned}
        \end{equation}
        at least one of which respects colour--kinematics duality\footnote{The $\colfac_i$ should not   be explicitly evaluated under the integral (i.e.~internal indices should not be summed when the corresponding momentum is undetermined) in case they accidentally vanish before being replaced by the loop-momenta dependent kinematic numerators.}, let us assume it is $\scA_{n,L}$, we may `double copy' by replacing the colour numerators by kinematic numerators which respect colour--kinematics duality, while sending $g\mapsto\frac{\kappa}{2}$, to generate a new scattering amplitude, 
        \begin{equation}
            \scH_{n, L}\ =\ (-\rmi)^{n-3+3L}\left(\frac{\kappa}{2}\right)^{n-2+2L}\sum_i\int\prod^{L}_{l=1}\frac{\rmd^dp_l}{(2\pi)^dS_i}\frac{\kinfac_i\tilde{\kinfac}_i}{d_i}~,
        \end{equation}
        which is guaranteed to be a bona-fide scattering amplitude of $\caN=0$ supergravity. This remarkable fact was conjectured in~\cite{Bern:2008qj,Bern:2010ue} and shown to be true in~\cite{Bern:2010yg}. Note that this is an all-loop-order statement: if colour--kinematics duality holds to all loop orders, then $\caN=0$ supergravity is the double copy of Yang--Mills theory to all orders. The open question, in this context, is whether or not colour--kinematics duality holds to arbitrary loop order.
        
        We have discussed only pure Yang--Mills theory and $\caN=0$ supergravity. However, as mentioned in \cref{sec:outlook}, there is an ever-growing zoo of colour--kinematics duality respecting and double copy constructible theories. There are also various counterexamples, the most obvious being Yang--Mills theory coupled to adjoint fermions in a non-supersymmetric way~\cite{Chiodaroli:2013upa}. Such a coupling is incompatible with the assumption of colour--kinematics duality, which implies that the fermion couplings obey a Fierz identity that ensures supersymmetry. The latter is also evident from the double copy perspective: the product of gluon and fermion states yields gravitino states, which must couple supersymmetrically.    
        
        \subsection{Manifestly colour--kinematics-dual action}\label{ssec:manifest_bcj_action}
        
        The decomposition of scattering amplitudes into diagrams with trivalent vertices raises the question if there is an action principle for which these diagrams are the genuine Feynman diagrams. The homotopy algebraic perspective which we adopt in the rest of the paper yields the general statement that for any BV quantisable field theory there is a physically equivalent action with only trivalent vertices, cf.~\cref{ssec:strictified_YM}. We call the latter theory a \uline{strictification} of the former, because the homotopy algebras underlying a field theory with trivalent vertices is called \uline{strict}\footnote{Mathematicians would also use the terms {\em rectification} and {\em rectified}.}. An example of a strictification of four-dimensional Yang--Mills theory is the first-order formulation with an additional self-dual two-form~\cite{Okubo:1979gt}. In the case of the double copy, however, manifest colour--kinematics duality requires a different strictification, which we review below.
        
        \paragraph{Reorganisation of general tree-level scattering amplitudes.}
        For each tree $\Gamma$ containing a higher-than-trivalent vertex with $n$ external vertices, consider its contribution $\scA_\Gamma$ to a total scattering amplitude $\caA$. We split $\scA_\Gamma$ into a sum over trivalent trees with the same number of external vertices,
        \begin{equation}
            \scA_\Gamma\ =\ \sum_{\Gamma'\in{\rm Tree}_{3,n}}\scA(\Gamma)_{\Gamma'}~,
        \end{equation}
        where ${\rm Tree}_{3,n}$ is the set of trivalent trees with $n$ external vertices. The value of each summand is to be determined later by the colour--kinematics condition. Doing this for every tree involving a vertex which is higher than trivalent (the case where $\Gamma$ is already trivalent is trivial), we have reorganised the tree-level amplitudes of our theory into a sum over trivalent diagrams:
        \begin{equation}
            \sum_{\Gamma\in{\rm Tree}_n}\scA_\Gamma\ =\ \sum_{\Gamma'\in{\rm Tree}_{3,n}}\scA'_{\Gamma'}~,
        \end{equation}
        where
        \begin{equation}
            \scA'_{\Gamma'}\ =\ \sum_{\Gamma\in{\rm Tree}_n}\scA(\Gamma)_{\Gamma'}~.
        \end{equation}
        
        We require that our reorganisation be \emph{local}: the way each $n$-ary vertex is resolved into a trivalent tree subdiagram is independent of the rest of the diagram. It then suffices to specify, for each $n$-ary vertex with $n\geq4$, how to split it into a sum over trivalent trees. That is, for the tree diagram $T_n$ with $n$ external legs and one internal $n$-ary vertex, i.e.~the $n$-point contact contribution with a single $n$-ary vertex, we specify its decomposition
        \begin{equation}
            \scA_{T_n}\ =\ \sum_{\Gamma'\in{\rm Tree}_{3,n}}\scA'(T_n)_{\Gamma'}\
            =\ \sum_{\Gamma'\in{\rm Tree}_{3,n}}\frac{N(\Gamma')}{D(\Gamma')}~,
        \end{equation}
        where we have chosen a particular ansatz in order to satisfy the colour--kinematics identities:
        \begin{enumerate}[(i)]\itemsep-2pt
            \item $D(\Gamma')$ is a differential\footnote{in fact, pseudo-differential for $n=2$} operator of degree $2(n-3)$ which is the product of the inverse propagators $\wave$ corresponding to the $n-3$ internal edges in $\Gamma'$, i.e.~the kinematic denominator. In particular, $D=\wave^{-1}$ for $n=2$ and $D=1$ for $n=3$.
            \item $N(\Gamma')$ is a differential operator, corresponding in momentum space to a polynomial of the external momenta $p_1,\ldots,p_n$ and the polarisation vectors $\eps_1,\ldots,\eps_n$.
        \end{enumerate}
        Because only vertices of degree at most $n$ contribute to the $n$-point amplitude, we can solve for these decompositions recursively. Concretely, supposing that one knows the decomposition for all vertices of degrees at most $n$, then one simply writes an equation in as many unknowns as there are for all possible ways to decompose the $(n+1)$-ary vertex into trees, and solves for the colour--kinematics duality equations (a system of linear equations) at $n+1$ points. The initial case of the iteration is $n=2$, where the decomposition of trivalent vertices is trivial, and one must verify that the colour--kinematics identities hold for tree-level three-point functions.
        
        A~priori, at each level of the iteration, i.e.~for any given $n$, there may be infinitely many solutions or no solutions; tree-level colour--kinematics duality of the amplitudes of a field theory then amounts to the assertion that the latter is never the case.
        
        \paragraph{Colour--kinematics-dual Yang--Mills action.}
        We now specialise to the case of Yang--Mills theory with the physical field being the gauge field. For simplicity, we shall work in Feynman gauge.\footnote{For a general linear Lorentz-covariant gauge, the kinematic numerator $\sfn$ will instead be a rational function containing terms such as $\frac{p^\mu p^\nu}{p^2}$.} The Yang--Mills action only contains terms up to quartic order in the field, but this does not mean that we cannot split the vanishing quintic and higher-order terms into sums that are attributed to different trivalent trees; indeed, such a procedure is necessary to fulfil the colour--kinematics identities. For Yang--Mills theory, we can be more precise about the ansatz:
        \begin{equation}
            \scA_{T_n}\ =\ \sum_{\Gamma'\in{\rm Tree}_{3,n}}\scA'(T_n)_{\Gamma'}\
            =\ \sum_{\Gamma'\in{\rm Tree}_{3,n}} \frac{\colfac(\Gamma')\kinfac(\Gamma')}{D(\Gamma')}~,
        \end{equation}
        where the denominator $D(\Gamma')$ is as before, but the numerator has been split into the colour numerator $\colfac(\Gamma')$ and the kinematic numerator $\kinfac(\Gamma')$. More explicitly,
        \begin{enumerate}[(i)]\itemsep-2pt
            \item $\colfac(\Gamma)$ is a group theory factor, corresponding to contractions of $n-2$ copies of the colour group structure constants corresponding to the vertices of the trivalent diagram $\Gamma'$.
            \item $\kinfac(\Gamma')$ is a differential operator of degree $n-2$, corresponding to the kinematic numerator; equivalently, in momentum space, a polynomial expression of the external momenta and polarisation vectors, whose degree is homogeneously $n-2$ in terms of the momenta and $k$ in terms of the polarisation vectors.
        \end{enumerate}
        
        The exact form of this splitting was computed in~\cite{Tolotti:2013caa}, with the first few terms contained in~\cite{Bern:2010yg}. In the former paper, the authors present their result in terms of a (non-local) action with the Lagrangian
        \begin{equation}
            \scL^{\rm YM}\ =\ \scL^{\rm YM}_2+\scL^{\rm YM}_3+\cdots~,
        \end{equation}
        in which the $n$-th order term is 
        \begin{equation}\label{eq:higher_terms_L_BCJ}
            \scL^{\rm YM}_n\ =\ \sum_{\Gamma\in{\rm Tree}_{3,n}}O^{\mu_1\cdots\mu_n}_{n,\Gamma}\frac{\tr\Big\{[A_{\mu_{\sigma(1)}},A_{\mu_{\sigma(2)}}]\,[\ldots[A_{\mu_{\sigma(3)}},A_{\mu_{\sigma(4)}}]\ldots,A_{\mu_{\sigma(n)}}]\Big\}}{\wave_{j_{n,\Gamma,1}}\cdots\wave_{j_{n,\Gamma,n-3}}}~,
        \end{equation}
        where the permutation $\sigma$ is determined by the tree-level diagram $\Gamma$ and where $O^{\mu_1\cdots \mu_n}_{n,\Gamma}$ is a sum of polynomials in the inverse Minkowski metric $\eta^{\mu\nu}$ and $n-2$ partial differential operators $\partial_\mu$ acting on one of the $n$ occurrences of the field $A$ in the numerator.
        
        Note that the expressions for the splitting of the higher-arity vertices are simply substituted into the ordinary Yang--Mills action, even though the resulting expression is simply algebraically equal to the original Yang--Mills action (in particular, the higher-order vertices vanish due to colour Jacobi identities), giving an impression reminiscent of the `ghosts of departed quantities' of Newtonian calculus~\cite{Berkely:1734aa}. However, there is nothing mysterious about this action; it simply expresses how higher-order vertices, most of which are zero, are split apart and distributed into trivalent trees. 
        
        The order-by-order calculation of the splitting of the higher-degree Feynman vertices is, in principle, a straightforward exercise, and there is nothing specific to Yang--Mills theory (apart from perhaps the ansatz of the numerator), provided that the tree-level colour--kinematics identities in fact hold. In particular, one can readily compute such a splitting for gravity, except that there the ansatz is of a different form,
        \begin{equation}
            \scH_{T_n}\ =\ \sum_{\Gamma'\in{\rm Tree}_{3,n}}\scH'(T_n)_{\Gamma'}\ =\ \sum_{\Gamma'\in{\rm Tree}_{3,n}}\frac{\kinfac(\Gamma')\kinfac(\Gamma')}{D(\Gamma')}~,
        \end{equation}
        where the colour numerator is replaced by another copy of the kinematic numerator.
        
        In dealing with scattering amplitudes, we can freely use the on-shell condition $p_i^2=0$ and the transversality condition $\eps_i\cdot p_i=0$ for the external momenta $p_i$ and polarisation vectors $\eps_i$ when performing the above manipulations. Thus, the action is colour--kinematics-dual only on shell.

        \section{Non-linear sigma model and special galileons}\label{sec:NLSM}
        
        Before delving into the details of the double copy of Yang--Mills theory to $\caN=0$ supergravity, we first consider the simpler example of the double copy of the non-linear sigma model on a Lie group to the special galileon~\cite{Chen:2013fya,Chen:2014dfa,Du:2016tbc,Carrasco:2016ldy,Cheung:2016prv}. This example is considerably simpler because we can ignore the technicalities due to gauge symmetry and Becchi--Rouet--Stora--Tyutin (BRST) quantisation. 
        
        In this simpler example, the non-linear sigma model enjoys a \uline{flavour--kinematics duality}, the analogue of colour--kinematics duality in Yang--Mills theory. The roles of these two dualities differ slightly: whereas in Yang--Mills theory the colour--kinematics duality ensures the existence of a BRST operator in the double copy, the flavour--kinematics duality in the non-linear sigma model ensures (amongst other things) avoidance of the Ostrogradsky instability. This instability generically arises for Lagrangians involving derivatives of higher order than two, in which the Hamiltonian is unbounded from below, cf.~e.g.~\cite{Motohashi:2014opa}.
        
        \subsection{Review of the essentials}
        
        Let us first recall some of the background material.
        
        \paragraph{Non-linear sigma model.}
        Consider $d$-dimensional Minkowski space $\IM^d\coloneqq\IR^{1,d-1}$ with metric $(\eta_{\mu\nu})=\diag(-1,1,\ldots,1)$ with $\mu,\nu,\ldots=0,1,\ldots,d-1$ and local coordinates $x^\mu$ together with a semi-simple compact matrix Lie group $\sfG$.  
        To define the non-linear sigma model action, we are interested in maps $g:\IM^d\rightarrow\sfG$, or rather their flat current 
        \begin{equation}
            j_\mu\ \coloneqq\ g^{-1}\partial_\mu g~,
        \end{equation}
        which takes values in the Lie algebra $\frg$ of $\sfG$. We take $\cb_a$ as a basis of $\frg$ with $a,b,\ldots=1,2,\ldots,\dim(\frg)$, $[\cb_a,\cb_b]={f_{ab}}^c\cb_c$ with $[-,-]$ the Lie bracket on $\frg$, and $\inner{\cb_a}{\cb_b}\coloneqq-\tr(\cb_a\cb_b)=\delta_{ab}$ with `$\tr$' the matrix trace. The action without a potential term is then given by
        \begin{equation}\label{eq:PCMAction}
            S^\NLSM\ \coloneqq\ \tfrac12\int\rmd^dx\,\tr\{j_\mu j^\mu\}\ =\ -\tfrac12\int\rmd^dx\,(g^{-1}\partial_\mu g)_a(g^{-1}\partial^\mu g)^a~,
        \end{equation}
        and this special case is also called the \uline{principal chiral model}.
        
        Upon setting\footnote{We use here the exponential parametrisation $g=\rme^\phi$. There are other possible parametrisations used in this context in the literature, such as the Cayley parametrisation (see e.g.~\cite{Kostant:0109066}). Our treatment below, however, does not depend on the choice of parametrisation as long as this parametrisation is defined by an equivariant map $\frg\to\sfG$, where both $\frg$ and $\sfG$ are equipped with adjoint actions.} $g\coloneqq\rme^\phi$ and $\ad_\phi(-)\coloneqq[\phi,-]$ for $\phi:\IM^d\rightarrow\frg$, and using the formula
        \begin{equation}
            j_\mu\ =\ \rme^{-\phi}\partial_\mu\rme^\phi\ =\ \frac{1-\rme^{-\ad_\phi}}{\ad_\phi}(\partial_\mu\phi)\ =\ \sum_{n=0}^\infty\frac{(-1)^n}{(n+1)!}\,\ad_\phi^n(\partial_\mu\phi)~,
        \end{equation}
        which follows from the Baker--Campbell--Hausdorff formula, we may rewrite~\eqref{eq:PCMAction} as
        \begin{equation}\label{eq:PCMActionLieAlgebra}
            S^\NLSM\ =\ \tfrac12\int\rmd^dx\,\sum_{n=0}^\infty\frac{(-1)^n}{((n+1)!)^2}\,\tr\big\{\partial_\mu\phi\,\ad_\phi^{2n}(\partial^\mu\phi)\big\}~.
        \end{equation}
        Because of the symmetry $\phi\mapsto-\phi$, corresponding to the symmetry $g\mapsto g^{-1}$ of~\eqref{eq:PCMAction}, there are only Feynman vertices of even degree, each of which contains exactly two derivatives.
        
        \paragraph{Galileons.} As mentioned above, a generic Lagrangian involving derivatives of higher order than two runs into the Ostrogradsky instability. We can avoid this if we carefully select an ansatz such that even though the action contains higher-derivative terms, the corresponding equations of motion are at most of second order in the derivatives. For a scalar field theory, the most general such ansatz is that of the \uline{galileon}~\cite{Fairlie:1992nb,Fairlie:1992zn,Fairlie:1991qe,Dvali:2000hr,Nicolis:2008in}, see also~\cite{Curtright:2012gx} for a review.
        
        The galileon theory is a theory of a scalar field $\phi$ which is invariant under the Galilean-type symmetry
        \begin{equation}
            \phi(x)\ \mapsto\ \phi(x) + c + b_\mu x^\mu~,
        \end{equation}
        where $c$ is a constant and $b_\mu$ is a constant vector on $\IM^d$. In $d$ space--time dimensions, there are $d+1$ possible terms that satisfy the Galilean-type symmetry. Specifically, the general action is of the form~\cite{Nicolis:2008in}
        \begin{equation}
            S^{\rm Gal}\ \coloneqq\ \int\rmd^dx\,\sum_{n=1}^{d+1}\alpha_n\scL^{\rm Gal}_n
            \ewith
            \scL^{\rm Gal}_n\ \coloneqq\ \phi\,\eps^{\mu_1\cdots\mu_d}\eps^{\nu_1\cdots\nu_d}\left(\prod_{i=1}^{n-1}\partial_{\mu_i}\partial_{\nu_i}\phi\right)\left(\prod_{i=n}^d\eta_{\mu_i\nu_i}\right),
        \end{equation}
        where $\eps^{\mu_1\cdots\mu_d}$ is the usual Levi--Civita symbol. This action is parametrised by the $d+1$ coefficients $\alpha_i$, among which $\alpha_1$, corresponding to the tadpole $\scL^{\rm Gal}_1\,\propto\,\phi$, should be set to zero, and $\alpha_2$, corresponding to the kinetic term $\scL^{\rm Gal}_2\,\propto\,\phi\wave\phi$ where $\wave\coloneqq\partial_\mu\partial^\mu$, should be canonically normalised. Thus, one obtains a $(d-1)$-dimensional moduli space of possible galileon theories in $d$ space--time dimensions.
        
        There is a special point in this moduli space called the \uline{special galileon}~\cite{Hinterbichler:2015pqa,Cheung:2014dqa,Cachazo:2014xea},
        \begin{equation}
            \alpha_{2n}\ \coloneqq\ \frac1{2n}\binom d{2n-1}\frac1{M^{2n-2}}
            \eand
            \alpha_{2n+1}\ \coloneqq\ 0~,
        \end{equation}
        where $M$ is a mass scale. At this point, the scattering amplitudes become extremal in a specific sense. In particular, all amplitudes with an odd number of external particles vanish, which is not the case for the generic galileon theory~\cite{Cachazo:2014xea}. Operationally, it may be defined as the galileon theory obtained by double-copying the scattering amplitudes of the non-linear sigma model.
        
        \subsection{Flavour--kinematics duality and double copy}
        
        We will transform the non-linear sigma model action~\eqref{eq:PCMActionLieAlgebra} into that of the special galileon via the following steps:
        \begin{enumerate}[(i)]\itemsep-2pt
            \item Put the non-linear sigma model action into a manifestly flavour--kinematics-dual form.
            \item Introduce infinitely many auxiliary fields to render the action cubic (i.e.~strictify) in a manner compatible with flavour--kinematics duality.
            \item We square the coefficients of the cubic action to obtain a raw double-copied action.
            \item Upon integrating out the infinite tower of auxiliary fields of the double copy and a suitable field redefinition, we are guaranteed to recover the special galileon action.
        \end{enumerate}
        Unlike the case of Yang--Mills theory, we will not need a detailed argument to ensure the existence of a BRST operator.
        
        \paragraph{Manifestly flavour--kinematics-dual action.}
        As explained in \cref{ssec:manifest_bcj_action}, once it is known that flavour--kinematics duality (and hence double copy) holds at the tree level, it is automatic that one can write down a manifestly flavour--kinematics-dual form of the action.
        
        In particular, we may organise the infinitely many terms or the Lagrangian $\scL^\NLSM\coloneqq\sum_{n=0}^\infty\scL_{2n}^\NLSM$ of the non-linear sigma model Lagrangian~\eqref{eq:PCMActionLieAlgebra} into a manifestly flavour--kinematics-dual form as
        \begin{equation}\label{eq:PCMBCJDualLagrangian}
            \scL^\NLSM_{2n}\ =\ \sum_{\Gamma\in{\rm Tree}_{3,2n}}O_{2n,\Gamma}\frac{\tr\Big\{[\phi,\phi]\,[\ldots[\phi,\phi]\ldots,\phi]\Big\}}{\wave_{j_{2n,\Gamma,1}}\cdots\wave_{j_{2n,\Gamma,2n-3}}}~,
        \end{equation}
        and this expression needs to be read as follows. Firstly, $O_{2n,\Gamma}$ is a sum of polynomials in the inverse Minkowski metric $\eta^{\mu\nu}$ and $2(2n-2)$ partial differential operators $\partial_\mu$ acting on one of the $2n$ occurrences of the field $\phi$ in the numerator (and rendering the commutators non-trivial). Secondly, the subscripts on the inverse wave operators similarly indicate which product of fields\footnote{corresponding to an internal propagator in the Feynman diagrams} they act on. Finally, the sum ranges over trivalent trees $\Gamma$ with $2n$ external legs, in which the flavour contraction is determined by the topology of the tree $\Gamma$. In particular, for $n=1$,
        \begin{equation}
            \scL^\NLSM_2\ =\ \tfrac12\tr\{\phi\wave\phi\}
        \end{equation}
        is the canonical kinetic term.
        
        The Lagrangian~\eqref{eq:PCMBCJDualLagrangian} is the analogue of the manifestly colour--kinematics-dual Lagrangian~\eqref{eq:higher_terms_L_BCJ} for Yang--Mills theory. The former perhaps seems less strange than the latter, because unlike Yang--Mills theory, the non-linear sigma model already has an infinite number of terms to begin with. The method of construction of the manifestly dual action, however, is exactly the same in both cases.
        
        \paragraph{Strictification.}
        The flavour--kinematics-dual action~\eqref{eq:PCMBCJDualLagrangian} has two defects: it is neither local nor cubic\footnote{or strict, in homotopy algebraic language; cf.~\cref{app:structure_theorems}}, and thus does not produce exclusively cubic Feynman vertices. We can remedy both defects by introducing an infinite tower of auxiliary fields. We note that to a fixed order in perturbation theory (with a bounded number of external legs and loops), the number of auxiliary fields that enter is always finite, just as in ordinary Yang--Mills theory, cf.~\cref{ob:usual_double_copy} below.
        
        In the case of the non-linear sigma model, we require two scalar auxiliary fields $C^1$ and $C^2$, both in the adjoint representation, at the quartic order; quintic and higher orders will each require multiple auxiliary fields, all of which are Lorentz tensors, but with varying numbers of Lorentz indices, and take values in the adjoint representation. The strictified action is of the form
        \begin{equation}\label{eq:PCMStrictifiedAction}
            S^\NLSMst\ =\ \int\rmd^dx\,\tr\left\{\tfrac12\phi\wave\phi+
            C^1_\mu \wave C^{2,\mu} + \alpha C^1_\mu[\phi,\partial^\mu \phi]+\beta(\wave C^{2,\mu})[\phi, \partial_\mu \phi]+\dotsb\right\}~,
        \end{equation}
        where the dimensionful coefficients $\alpha$ and $\beta$ are tuned so as to give the correct four-point amplitude and to manifest flavour--kinematics duality. Notice that the strictification is not arbitrary, but mostly determined by the form of the manifestly flavour--kinematics-dual form of the action~\eqref{eq:PCMBCJDualLagrangian}. 
        
        \paragraph{Double copy.}
        The next step is to engineer an action which reproduces the double copy of the flavour--kinematics-dual amplitudes on the nose, and it is not hard to see that this consists essentially of the following:
        \begin{enumerate}[(i)]\itemsep-2pt
            \item We take the tensor square of the field content $(\phi,C^1,C^2,\ldots)$, so as to obtain an infinite quadrant of fields,
            \begin{equation}
                \begin{tikzcd}[row sep=5pt,column sep=5pt]
                    \tilde\phi\ \coloneqq\ \phi\otimes\phi & \tilde C^1_{{\rm L},\mu}\ \coloneqq\ C^1_\mu\otimes\phi & \tilde C^{2}_{\rm L,\mu}\ \coloneqq\ C^2_\mu\otimes\phi & \cdots 
                    \\
                    \tilde C^1_{{\rm R},\mu}\ \coloneqq\ \phi\otimes C^1_\mu & \tilde C^{11}_{\rm LR,\mu\nu}\ \coloneqq\ C^1_\mu\otimes C^1_\nu & \tilde C^{12}_{\rm LR,\mu\nu}\ \coloneqq\ C^1_\mu\otimes C_\nu^2 & \cdots
                    \\
                    \tilde C^{2}_{{\rm R,\mu}}\ \coloneqq\ \phi\otimes C^2_\mu & \tilde C^{21}_{\rm LR,\mu\nu}\ \coloneqq\ C_\mu^2\otimes C^1_\nu & \tilde C^{22}_{\rm LR,\mu\nu}\ \coloneqq\ C_\mu^2\otimes C_\nu^2 &\cdots
                    \\
                    \vdots & \vdots & \vdots & \ddots
                \end{tikzcd}
            \end{equation}
            all of which except for $\phi\otimes \phi$ are regarded as auxiliary.
            \item The kinematical terms are given by (off-diagonal) wave operators.
            \item The interaction vertices are simply the products of two of the interaction vertices of the non-linear sigma model.
        \end{enumerate}
        The double-copied action is then
        \begin{equation}\label{eq:PCMDCAction}
            \begin{aligned}
                \dbl{S}^{\rm DC}\ &\coloneqq\ \int\rmd^dx\,\Big\{\big(\tfrac12\tilde\phi\wave\tilde\phi+\tfrac12\tilde C^1_{\rm L\mu}\wave\tilde C^{2,\mu}_{\rm L}+\tfrac12\tilde C^1_{\rm R,\mu}\wave\tilde C^{2,\mu}_{\rm R}+\tfrac12\tilde C^{11}_{\rm LR\mu\nu}\wave\tilde C^{22,\mu\nu}_{\rm LR}+\cdots\big)\ +\,
                \\
                &\kern2.3cm+\Lambda^{\frac{d}{2}-3}\big(\alpha^2\tilde C_{\rm LR,\mu\nu}^{11}(\partial^\mu\partial^\nu\tilde\phi)\tilde\phi+\alpha^2\tilde C_{\rm LR}^{11,\mu\nu}(\partial_\mu\tilde \phi)\partial_\nu\tilde\phi\ +\\
                &\kern3.9cm+\alpha^2(\partial_\nu\tilde C_{\rm L}^{1,\mu})(\partial_\mu\tilde C_{\rm R}^{1,\nu})\tilde\phi+\alpha^2(\partial_\nu\tilde C_{\rm L}^{1,\mu})(\partial_\mu\tilde\phi)\tilde C_{\rm R}^{1,\nu}\ +\\
                &\kern3.9cm+\alpha^2\tilde C_{\rm L}^{1,\mu}(\partial_\mu\partial_\nu\tilde\phi)\tilde C_{\rm R}^{1,\nu}+\alpha^2\tilde C_{\rm L}^{1,\mu}(\partial_\mu\tilde C_{\rm R}^{1,\nu})\partial_\nu\tilde\phi+\dotsb\big)\Big\}~,
            \end{aligned}
        \end{equation}
        where in the interests of brevity we have only shown the terms corresponding to the double copy of \(\phi\) and \(C^1\).
        Note that we have introduced a new mass scale $\Lambda$ in the double copy, in the same way that the Planck scale enters during the double copy of Yang--Mills theory to Einstein gravity.
        
        The six interaction terms shown above, corresponding to the strictification of a quartic galileon interaction, illustrate the six possible ways of double copying the flavour-stripped interaction vertex $C^1_\mu\partial^\mu\phi\phi$ namely, schematically,
        \begin{equation}
            \begin{gathered}
                \left(\begin{smallmatrix}
                    C^1 & \partial\phi & \phi
                    \\
                    &\otimes
                    \\
                    C^1 & \partial\phi & \phi
                \end{smallmatrix}\right)
                ~,~~~
                \left(\begin{smallmatrix}
                    C^1 & \partial\phi & \phi
                    \\
                    &\otimes
                    \\
                    C^1 & \phi & \partial\phi
                \end{smallmatrix}\right)
                ~,~~~
                \left(\begin{smallmatrix}
                    C^1 & \partial\phi & \phi
                    \\
                    &\otimes
                    \\
                    \partial\phi & C^1 & \phi
                \end{smallmatrix}\right)
                ~,\\
                \left(\begin{smallmatrix}
                    C^1 & \partial\phi & \phi
                    \\
                    &\otimes
                    \\
                    \partial\phi & \phi & C^1
                \end{smallmatrix}\right)
                ~,~~~
                \left(\begin{smallmatrix}
                    C^1 & \partial\phi & \phi
                    \\
                    &\otimes
                    \\
                    \phi & \partial\phi & C^1
                \end{smallmatrix}\right)
                ~,~~~
                \left(\begin{smallmatrix}
                    C^1 & \partial\phi & \phi
                    \\
                    &\otimes
                    \\
                    \phi & C^1 & \partial\phi
                \end{smallmatrix}\right)
                ~.
            \end{gathered}
        \end{equation}
        
        By construction, the action~\eqref{eq:PCMDCAction} reproduces the double copy of the amplitudes of the non-linear sigma model and its tree-level amplitudes coincide with those of the special galileon.
        
        \paragraph{Physical equivalence.} To compare against the usual action of the special galileon $S^{\rm sGal}$, we can either integrate out the auxiliaries of $\tilde S^{\rm DC}$ or rewrite $S^{\rm sGal}$ in manifestly flavour--kinematics-dual cubic terms, introducing a tower of auxiliary fields. We will give arguments in both cases; the former involves a small suspension of disbelief, the latter does not have this gap, but is more indirect.
        
        In the first version of the argument, we straightforwardly integrate out all auxiliary fields, i.e.\ all fields except for \(\tilde\phi\). One may worry that the result is non-local. However, since the tree amplitudes are all local, the resulting terms will have to conspire to hide this non-locality. Let us assume, and this is the slight gap in the argument, that this does not happen. That is, we obtain a local action $\dbl{S}^{\rm DC}_2$ of a single scalar field $\tilde\phi$ whose tree amplitudes agree with those of the special galileon. 
        
        The two actions $\dbl{S}^{\rm DC}_2$ and $S^{\rm sGal}$ certainly agree at the quadratic and the cubic levels (for which there are no vertices). They can differ at the quartic level, but only up to terms that are not visible in the four-point tree-level amplitude, because the tree-level amplitudes agree. That is, the difference has to be of the form
        \begin{equation}
            \int\rmd^dx\,\tr\Big\{Z\wave\tilde\phi + \caO(\tilde\phi^5)\Big\},
        \end{equation}
        where $Z$ is some local cubic Lie algebra-valued functional of $\tilde\phi$. In this case, we can perform a field redefinition of either side of the form
        \begin{equation}\label{eq:sGal_field_redef}
            \tilde\phi\ \mapsto\ \tilde\phi+\theta Z
        \end{equation}
        with the coefficient $\theta$ tuned such that $\dbl{S}^{\rm DC}_2[\tilde\phi+\theta Z]$ agrees up to the quartic level with the action of the special galileon.
        
        We can then iterate this argument at the quintic, sextic, etc.~orders. The sceptical reader may still worry about whether this sequence of field redefinitions converges, but this is irrelevant from the perspective of perturbative quantum field theory, since to any desired order in perturbation theory, only finitely many interaction vertices contribute.
        
        Note that for the computation of $n$-loop amplitudes at certain loop orders, the degree of the appearing vertices is bounded from above, and we can conclude that from the perspective of perturbative quantum field theory, the two actions agree up to a local field redefinition.
        
        Notice that we started with an action with infinitely many terms which, after a suitable field redefinition, reduces to only finitely many terms. This is reminiscent of how the infinitely many terms of perturbative gravity reduce to a single term of the Einstein--Hilbert action. Furthermore, a Galilean-type symmetry has appeared that avoids the Ostrogradsky instability. Presumably, if one had started from a generic theory of adjoint scalars without flavour--kinematics duality, or if one started with a strictification of the non-linear sigma model that did not manifest flavour--kinematics duality, then our construction would have yielded a galileon-like theory that, nevertheless, would not avoid the Ostrogradsky instability. The miracle of the Galilean-type symmetry is opaque in our formalism (unlike the BRST symmetry in the gauged case), but it nevertheless occurs.
        
        Upon path-integral quantisation, such field redefinitions produce a Jacobian in the form of local operators, which can be cancelled by local counterterms. Hence, the double copied action and the action of the special Galilean define equivalent perturbative quantum theories up to local counterterms.
        
        We now give the alternative, gap-free but less direct argument. To ensure locality throughout, instead of integrating out auxiliary fields from $\tilde S^{\rm DC}$, we will introduce auxiliaries into $S^{\rm sGal}$. In the strictified action~\eqref{eq:PCMStrictifiedAction}, the auxiliary fields can be formally put on external legs, with well defined, local tree-level scattering amplitudes. The latter are particular collinear limits of $\tilde \phi$-only amplitudes, i.e.~tree-level scattering amplitudes whose external legs are exclusively copies of the field $\tilde\phi$. Similarly, the action of the special galileon can be put into a manifestly flavour--kinematics-dual form, by which we mean the kinematic numerators factorise into kinematic Jacobi identity respecting pieces. This form of the action can then be strictified introducing auxiliary fields. In doing so, we must take care to introduce the special galileon auxiliary fields in a manner compatible with that of~\eqref{eq:PCMStrictifiedAction}: namely, the special galileon auxiliaries' equations of motion are the double copies of the non-linear sigma model auxiliaries. Then tree-level flavour--kinematics duality implies that the double copy relation holds not only between \(\phi\)-only amplitudes of \(S^\NLSM\) and \(\tilde\phi\)-only amplitudes of \(S^{\rm sGal}\), but also between \(S^\NLSM\) amplitudes with external auxiliaries and $S^{\rm sGal}$ amplitudes with external auxiliaries.
        
        Thus, by construction, all tree-level amplitudes agree between $\dbl{S}^{\rm DC}$ and $S^{\rm sGal}$, including those with auxiliaries on external legs; both are local, cubic actions. Now, the field-redefinition argument above applies to this pair of actions as follows. If  $\dbl{S}^{\rm DC}$ and $S^{\rm sGal}$ differ, then the difference must be of the form
        \begin{equation}
            \dbl S^{\rm DC}[\tilde\varphi]-S^{\rm sGal}[\tilde\varphi]\ =\ \sum_iZ^i[\tilde\varphi]\wave\tilde\varphi_i
        \end{equation}
        for some local functionals \(Z^i[\tilde\varphi]\), where $\tilde \varphi_i$ stands for an arbitrary field, physical or auxiliary. Upon a field redefinition in $\dbl S^{\rm DC}$ of the form
        \begin{equation}
            \tilde\varphi_i\ \mapsto\ \tilde\varphi_i +\theta^i Z^i[\tilde\varphi]~,
        \end{equation}
        (no summation implied)
        we can tune the coefficients \(\theta^i\) so that $\dbl S^{\rm DC}$ and $S^{\rm sGal}$ coincide at the cubic level. Of course, this comes at the cost of potentially introducing quartic terms into $\dbl{S}^{\rm DC}$. But the argument continues to work nevertheless: since the quartic terms possibly present in $\dbl{S}^{\rm DC}$ are not visible at four-point tree amplitudes, they must be proportional to \(\wave\tilde\varphi_i\), which in turn can be absorbed by a field redefinition, possibly producing quintic terms in $\dbl{S}^{\rm DC}$. Another field redefinition pushes the quintic terms to sextic ones, the sextic ones to septic ones, and so on ad~infinitum.
        
        Again, the convergence of the field redefinitions is not of interest to us: to any desired order in perturbation theory, we only need finitely many interaction vertices. Thus, the loop amplitudes of $\dbl S^{\rm DC}$ and $S^{\rm sGal}$ agree to any desired order in perturbation theory, up to local counterterms; they define equivalent perturbative quantum theories up to local counterterms.
         
        \paragraph{Relation to prior work.} Let us briefly point out how our approach differs from the related prior work of Cheung--Shen~\cite{Cheung:2016prv}. Inspired by a particular dimensional reduction of Yang--Mills theory in $2d+1$ space--time dimensions~\cite{Cheung:2017yef}, the non-linear sigma model is here effectively embedded as a subsector into a theory of two vector-like fields and a scalar. The action of this larger theory is already in flavour--kinematics duality manifesting form. The special galileon then appears as a particular subsector of the square of this theory.
        
        The way that Cheung--Shen construct the double copy is similar to ours, except that their action is already strictified with a finite number of fields. We have seen that if we drop this restriction and allow ourselves an infinite tower of auxiliary fields, then it is not necessary to add any new degrees of freedom (except auxiliary fields), and the double copy becomes exactly the special galileon upon integrating out the auxiliary fields.
        
        \subsection{Formulation in terms of homotopy algebras}\label{ssec:NLSM_homotopy_picture}
        
        The previous construction can be elegantly formulated using the language of homotopy algebras\footnote{Relevant definitions and results on homotopy algebras are collected in \cref{sec:BVFormalismLinfty,app:homotopyAlgebras}, but we shall not need them yet in this section, which is a gentle motivation of some of these definitions.}, and this reformulation serves again as a simpler example of our perspective on the gauge-gravity double copy. 
        
        \paragraph{\mathversion{bold}$L_\infty$-algebra of the strictified non-linear sigma model.} Feynman diagrams are constructed using $n$-point vertices that, with some bias, can be regarded as taking $n-1$ input fields and combining them into a new field. They also involve propagators, which are the inverses of differential operators mapping a single input field into a new field. Both of these structures can be regarded as `higher products'
        \begin{equation}
            \mu_1\ :\ \frL_1\ \rightarrow\ \frL_1\eand \mu_2\ :\ \frL_1\times \frL_1\ \rightarrow \frL_1
        \end{equation}
        of an $L_\infty$-algebra $\frL$ with underlying graded vector space $\frL_i=\bigoplus_{i\in\IZ} \frL_i$. Here, $\frL_1$ is the vector space of all fields and $\frL_2$ is a second copy of $\frL_1$ shifted by one in degree and identified with the space of BV anti-fields. This is the space in which the `right-hand side' of the equations of motion takes values. Usually one has further nontrivial vector subspaces $\frL_i$ with $i<1$ describing gauge symmetries and $\frL_i$ with $i>2$ describing Noether symmetries; we will come to this later when discussing Yang--Mills theory.
        
        In the case of the strictified non-linear sigma model, we have no gauge symmetry, merely chiral fields and exclusively trivalent vertices. Therefore
        \begin{equation}
            \frL^\NLSMst\ \coloneqq\ \frL^\NLSMst_1\oplus\frL^\NLSMst_2~,
        \end{equation}
        where $\frL^\NLSMst_1$ is the space of all possible field configurations (including the auxiliary fields), $\frL^\NLSMst_2\cong\frL^\NLSMst_1$, and we have only two non-vanishing maps
        \begin{equation}
            \begin{gathered}
                \mu^\NLSMst_1\,:\,\frL^\NLSMst_1\ \rightarrow\ \frL^\NLSMst_2~,
                \\
                \mu^\NLSMst_2\,:\,\frL^\NLSMst_1\times\frL^\NLSMst_1\ \rightarrow\ \frL^\NLSMst_2~.
            \end{gathered}
        \end{equation}
        For example,
        \begin{equation}
            \mu^\NLSMst_1(\phi)\ =\ \wave \phi\in \frL^\NLSMst_{2,\phi^+}~,~~~\mu^\NLSMst_2(\phi,\phi)\ =\ \underbrace{\wave(\phi\partial^\mu\phi)}_{\in \frL^\NLSMst_{2,C^{1+}}}+\ldots~,
        \end{equation}
        where $\frL^\NLSMst_{2,\phi^+}$ and $\frL^\NLSMst_{2,C^{1+}}$ are the subspaces of $\frL^\NLSMst_2$ in which the anti-fields $\phi^+$ and $C^{1+}$ take their values. We note that $\mu^\NLSMst_2$ is graded anti-symmetric (i.e.~symmetric when forgetting about the degree of the fields), and we have the usual polarisation identity
        \begin{equation}
            \begin{aligned}
                \mu^\NLSMst_2(\phi_1,\phi_2)\ =\ \tfrac12\Big(&\mu^\NLSMst_2(\phi_1+\phi_2,\phi_1+\phi_2)-\\
                &\hspace{1cm}-\mu^\NLSMst_2(\phi_1,\phi_1)-\mu^\NLSMst_2(\phi_2,\phi_2)\Big)~.
            \end{aligned}
        \end{equation}
        We also have an inner product $\langle-,-\rangle$ on $\frL^\NLSMst$ of degree $-3$, which pairs elements in $\frL^\NLSMst_1$ with elements in $\frL^\NLSMst_2$. The action of the non-linear sigma model is then given by the homotopy Maurer--Cartan action for $\frL^\NLSMst$,
        \begin{equation}
            S^\NLSMst\ =\ \tfrac12\langle a,\mu^\NLSMst_1(a)\rangle+\tfrac1{3!}\langle a,\mu^\NLSMst_2(a,a)\rangle~,
        \end{equation}
        where $a$ is a generic element in $\frL^\NLSMst_1$.
        
        \paragraph{Factorisation.}
        Since every field (including the auxiliary fields) carries an adjoint representation of the flavour symmetry, we can flavour-strip the fields in the theory in a similar way that one can colour-strip the fields in the case of Yang--Mills theory. In homotopy algebraic language, this corresponds to a factorisation
        \begin{equation}
            \frL^\NLSMst\ =\ \frg\otimes\frC^\NLSMst\ =\ \underbrace{\frg\otimes\frC^\NLSMst_1}_{=\,\frL^\NLSMst_1}~~\oplus~~\underbrace{\frg\otimes\frC^\NLSMst_2}_{=\,\frL^\NLSMst_2}~,
        \end{equation}
        where $\frC^\NLSMst_1\cong\frC^\NLSMst_2$ can be interpreted as the field space of a theory of an uncharged scalar field, together with a tower of auxiliary fields.
        Importantly, also the two maps $\mu^\NLSMst_1$ and $\mu^\NLSMst_2$ factorise,
        \begin{equation}
            \mu_1^\NLSMst\ =\ \sfid\otimes\sfm^\NLSMst_1
            \eand
            \mu_2^\NLSMst\ =\ [-,-]\otimes\sfm_2^\NLSMst~.
        \end{equation}
        Because the map $\sfm_2^\NLSMst$ is now graded symmetric (i.e.~anti-symmetric after forgetting the fields' degrees), 
        \begin{equation}
            \Big(~\frC^\NLSMst,\sfm_1^\NLSMst,\sfm_2^\NLSMst~\Big)
        \end{equation}
        is not an $L_\infty$-algebra but a $C_\infty$-algebra.\footnote{Recall that the tensor product of a Lie algebra and a commutative algebra is a Lie algebra. We have just encountered an example of the corresponding homotopy algebraic generalisation.}
        
        In the double copy of Yang--Mills theory, there is a kinematical factor that is treated on equal footing and interchanged with the colour factor. We therefore should strip off this factor as well. In the non-linear sigma model, this corresponds to factorising the fields into a single scalar field times the vector space of the Lorentz tensors on $\IM^d$ that the auxiliary fields form. 
        
        The graded vector space underlying $\frC^\NLSMst$ factorises as
        \begin{equation}
            \frC^\NLSMst\ =\  \frKin\otimes\frScal\ =\ \underbrace{\frKin\otimes\frScal_1}_{=\,\frL^{\rm scal}_1}~~\oplus~~\underbrace{\frKin\otimes\frScal_2}_{=\,\frL^{\rm scal}_2}~,
        \end{equation}
        where $\frScal_1\cong \frScal_2$ is simply the space of a single scalar (anti)field, while
        \begin{equation}
            \frKin\ \coloneqq\ \underbrace{\IR}_{\phi}\oplus\underbrace{\IM^d}_{C^1}\oplus\underbrace{\IM^d}_{C^2}\oplus\cdots
        \end{equation}
        The higher product $\sfm^\NLSMst_1$ factorises trivially as
        \begin{equation}
            \sfm_1^\NLSMst\ =\ \sfid\otimes\sfm_1^\NLSMst~,
        \end{equation}
        but the factorisation of $\sfm^\NLSMst_2$ is harder. Notice that this higher product does two things:
        \begin{enumerate}[(i)]\itemsep-2pt
            \item It implements the differential operator that appears as the kinematic numerator in the flavour--kinematics-dual form of the amplitude.
            \item It multiplies the two fields pointwise, implementing locality.
        \end{enumerate}
        We can disentangle the two functions by means of the \uline{twisted tensor product}, whose precise definition will be given later in \cref{ssec:twisted_tp_homotopy_algebras}; it  suffices to say that it is custom-made to implement precisely the above separation. In terms of this twisted tensor product, we can factorise
        \begin{equation}
            \sfm_2^\NLSMst\ =\ \sfm_2^\frKin\otimes_\tau\mu_2^\frScal~,
        \end{equation}
        where
        \begin{equation}
            \mu_2^\frScal\,:\,(\phi_1,\phi_2)\ \mapsto\ \phi_1\phi_2
        \end{equation}
        is simply the pointwise product of two scalar fields, and the \uline{twist datum} $\tau$ encodes the differential operators that appear as numerators in flavour--kinematics duality. We further note that $\mu_2^\frScal$ is symmetric in the fields (i.e.~graded-anti-symmetric in the elements of $\frScal$) such that $(\frScal,\mu_1^\frScal,\mu_2^\frScal)$ forms an $L_\infty$-algebra.
        
        \paragraph{Double copy.} Altogether, we have factorised the $L_\infty$-algebra of (the manifestly flavour--kinematics-dual formulation of) the non-linear sigma model as
        \begin{equation}
            \frL^\NLSMst\ =\ \frg\otimes\frL^{\rm scal}\ =\ \frg\otimes(\frKin\otimes_\tau\frScal)~.
        \end{equation}
        In terms of this factorisation, the double copy prescription is straightforward to phrase. The field space and the action of the double-copied form of the special galileon is given by the $L_\infty$-algebra
        \begin{equation}\label{eq:sGal_fac}
            \frL_1^{\rm sGal}\ =\ \frKin\otimes_\tau(\frKin\otimes_\tau\frScal)~,
        \end{equation}
        where the double appearance of the twist datum $\tau$ reflects the fact that the differential operators appearing in the numerator of the flavour--kinematics-dual representation have been squared. 
        
        The validity of the factorisation~\eqref{eq:sGal_fac} is not hard to see, in particular once one understands the link between higher products of $L_\infty$-algebras and tree-level scattering amplitudes of the encoded field theories that we shall explain later in detail.

        \section{Field theories, Batalin--Vilkovisky complexes, and homotopy algebras}\label{sec:BV_section}
        
        In the following, we summarise how perturbative quantum field theory is naturally formulated in the language of homotopy algebras. The bridge between field theories and homotopy algebras is provided by the Batalin--Vilkovisky (BV) formalism~\cite{Batalin:1981jr,Schwarz:1992nx}. Our discussion follows the treatment in~\cite{Jurco:2018sby,Jurco:2019bvp}; see also~\cite{Jurco:2020yyu} for a pedagogical summary and~\cite{Saemann:2020oyz} for a detailed discussion of Feynman diagrams. Basic definitions and results on homotopy algebras and homotopy Maurer--Cartan theory are collected in \cref{app:homotopyAlgebras} for convenience. We start with the Becchi--Rouet--Stora--Tyutin (BRST) formalism for the archetypal example of Yang--Mills theory. This will also prepare our discussion in \cref{sec:Examples}.
        
        \subsection{Motivation}\label{sec:BVMotivation}
        
        \paragraph{Yang--Mills action.}
        As before, we consider $d$-dimensional Minkowski space $\IM^d\coloneqq\IR^{1,d-1}$ with metric $(\eta_{\mu\nu})=\diag(-1,1,\ldots,1)$ with $\mu,\nu,\ldots=0,1,\ldots,d-1$ and local coordinates $x^\mu$. Let $\frg$ be a semi-simple compact matrix Lie algebra with basis $\cb_a$ with $a,b,\ldots=1,2,\ldots,\dim(\frg)$, $[\cb_a,\cb_b]={f_{ab}}^c\cb_c$ with $[-,-]$ the Lie bracket on $\frg$, and $\inner{\cb_a}{\cb_b}\coloneqq-\tr(\cb_a\cb_b)=\delta_{ab}$ with `$\tr$' the matrix trace.
        
        The action for Yang--Mills theory in $R_\xi$-gauge for some real constant $\xi$ in the BRST formalism reads as
        \begin{subequations}
            \begin{equation}\label{eq:YMActionComponents}
                S^{\rm YM}_{\rm BRST}\ \coloneqq\ \int\rmd^dx\,\Big\{\!-\tfrac14 F_{a\mu\nu}F^{a\mu\nu}-\bar c_a\partial^\mu(\nabla_\mu c)^a+\tfrac\xi2b_ab^a+b_a\partial^\mu A^a_\mu\,\Big\}
            \end{equation}
            with 
            \begin{equation}
                F^a_{\mu\nu}\ \coloneqq\ \partial_\mu A_\nu^a-\partial_\nu A_\mu^a+{f_{bc}}^agA^b_\mu A^c_\nu
                \eand
                (\nabla_\mu c)^a\ \coloneqq\ \partial_\mu c^a+g{f_{bc}}^aA^b_\mu c^c~,
            \end{equation}
        \end{subequations}
        where $g$ is the Yang--Mills coupling constant, $A^a_\mu$ are the components of the $\frg$-valued one-form gauge potential on $\IM^d$, and $c^a$, $b^a$, and $\bar c^a$ are the components of $\frg$-valued functions corresponding to the ghost, the Nakanishi--Lautrup field, and the anti-ghost field, respectively.
        
        \paragraph{\mathversion{bold}$\IZ$-graded vector spaces.}
        We note that the fields in the action~\eqref{eq:YMActionComponents} are graded by their \uline{ghost number} as detailed in \cref{tab:ghostNumberYM}.
        Therefore, we should view them as coordinate functions on a \uline{$\IZ$-graded vector space} $\frV=\bigoplus_{k\in\IZ}\frV_k$. Elements of $\frV_k$ are said to be \uline{homogeneous} of degree~$k$, and we shall use the notation $|\ell|_\frV$ to denote the degree of a homogeneous element $\ell\in\frV$.
        \begin{table}[ht]
            \begin{center}
                \begin{tabular}{|l|c|c|c|c|}
                    \hline
                    field $\Phi^I$ & $c^a$ & $A^a_\mu$ & $b^a$ & $\bar c^a$
                    \\
                    \hline
                    ghost number $|\Phi^I|_{\rm gh}$ & $1$ & $0$ & $0$ & $-1$
                    \\
                    \hline
                \end{tabular}
                \vspace{-15pt}
            \end{center}
            \caption{Ghost numbers of the fields in Yang--Mills theory.\label{tab:ghostNumberYM}}
        \end{table}
        
        The \uline{tensor product} of two $\IZ$-graded vector spaces $\frV$ and $\frW$ is defined as
        \begin{equation}\label{eq:tensor_product_graded_vector_spaces}
            \frV\otimes\frW\ =\ \bigoplus_{k\in\IZ}(\frV\otimes\frW)_k
            \ewith
            (\frV\otimes\frW)_k\ \coloneqq\ \bigoplus_{i+j=k}\frV_i\otimes\frW_j~,
        \end{equation}
        and the degree in $\frV\otimes\frW$ is thus the sum of the degrees in $\frV$ and $\frW$. 
        
        We shall denote the \uline{dual} of a $\IZ$-graded vector space $\frV$ by $\frV^*$,\footnote{We will not discuss the analytical subtleties of this construction in the infinite-dimensional case, except to note that the dual spaces will be degree-wise topological duals.} and we have
        \begin{equation}
            \frV^*\ =\ \bigoplus_{k\in\IZ}(\frV^*)_k
            \ewith
            (\frV^*)_k\ \coloneqq\ (\frV_{-k})^*~.
        \end{equation}
        In particular, elements in $\frV_k$ have the opposite degree of elements in $(\frV_k)^*$.
        
        Given a $\IZ$-graded vector space $\frV$, we can introduce the \uline{degree-shifted $\IZ$-graded vector space} $\frV[l]$ for $l\in\IZ$ by
        \begin{equation}\label{eq:shift_notation}
            \frV[l]\ =\ \bigoplus_{k\in\IZ}(\frV[l])_k
            \ewith
            (\frV[l])_k\ \coloneqq\ \frV_{k+l}~.
        \end{equation}
        For an ordinary vector space $\frV\equiv\frV_0$, for instance, $\frV[1]$ consists of elements of degree~$-1$ since only $(\frV[1])_{-1}=\frV_0$ is non-trivial. Note that $(\frV\otimes\frW)[l]=\frV[l]\otimes\frW=\frV\otimes\frW[l]$ and $(\frV[l])^*=\frV^*[-l]$ for all $l\in\IZ$. For convenience, we introduce the notion of a \uline{shift isomorphism}
        \begin{equation}\label{eq:shiftIso}
            \sigma\,:\,\frV\ \rightarrow\ \frV[1]
        \end{equation}
        which lowers the degree of every element of $\frV$, that is, $\sigma:\frV_k\rightarrow(\frV[1])_{k-1}$. 
        
        We note that the action~\eqref{eq:YMActionComponents} is built of polynomial functions and their derivatives. By the algebra of \uline{polynomial functions} on a $\IZ$-graded vector space $\frV$, we mean the $\IZ$-graded symmetric tensor algebra $\scC^\infty(\frV)\coloneqq\bigodot^\bullet\frV^*$. Basis elements of $\frV^*$ can be regarded as the coordinate functions on $\frV$. Explicitly, such a function looks like 
        \begin{equation}\label{eq:functionExpansion}
            \sff\ =\ f+\xi^\alpha f_\alpha+\tfrac12\xi^\alpha\xi^\beta f_{\alpha\beta}+\cdots\ \in\ \scC^\infty(\frV)~, 
        \end{equation}
        where $\xi^\alpha$ are basis elements of $\frV^*$ and $f,f_\alpha,f_{\alpha\beta},\ldots$ are constants. We have $\xi^\alpha\xi^\beta=(-1)^{|\xi^\alpha|_{\frV^*}|\xi^\beta|_{\frV^*}}\xi^\beta\xi^\alpha$. Note that if $\frV$ is a vector space of some suitably smooth functions or, more generally, sections of some vector bundle, then the dual $\frV^*$, being the space of distributions, contains not only the ordinary dual coordinate functions but also all of their derivatives.
        
        \paragraph{BRST operator in Yang--Mills theory.} 
        The reason for introducing ghosts in the first place is the gauge symmetry of Yang--Mills theory, which in the BRST and BV formalisms is captured in a dual formulation as a differential on a differential graded commutative algebra that is called the \uline{Chevalley--Eilenberg algebra}. More specifically, this is the algebra of polynomial functions, and the differential is a nilquadratic vector field $Q:\scC^\infty(\frV)\rightarrow\scC^\infty(\frV)$ of degree one, $Q^2=0$, known as the \uline{homological vector field}. A $\IZ$-graded vector space with such a homological vector field is called a \uline{$Q$-vector space}. 
        
        The prime example of a $Q$-vector space is that of an ordinary vector space $\frg$ with basis $\cb_a$ for $a,b,\ldots=1,\ldots,\dim(\frg)$, regarded as the $\IZ$-graded vector space $\frg[1]$. On $\frg[1]$, we have coordinates $\xi^a$ only in degree one and thus, the most general vector field $Q:\scC^\infty(\frg[1])\rightarrow\scC^\infty(\frg[1])$ of degree one is of the form
        \begin{equation}
            Q\ \coloneqq\ \tfrac12\xi^b\xi^c{f_{cb}}^a\parder{\xi^a}
            \quad\Rightarrow\quad
            Q\xi^a\ =\ \tfrac12\xi^b\xi^c{f_{cb}}^a
        \end{equation}
        for some constants ${f_{ab}}^c=-{f_{ba}}^c$. The condition $Q^2=0$ is equivalent to the Jacobi identity for the ${f_{ab}}^c$ so that $Q$ induces a Lie bracket $[\cb_a,\cb_b]={f_{ab}}^c\cb_c$ on $\frg$. The differential graded algebra $(\scC^\infty(\frg[1]),Q)$ is the Chevalley--Eilenberg algebra of the Lie algebra $(\frg,[-,-])$ to which we alluded above. In order to translate between $Q$ and $[-,-]$, it is useful to define the \uline{contracted coordinate functions}\footnote{These are often used in the string field theory literature, albeit shifted such that $\sfa$ is of degree zero.} 
        \begin{equation}\label{eq:LieAlgContractedCoordinates}
            \sfa\ \coloneqq\ \xi^a\otimes\cb_a\ \in\ (\frg[1])^*\otimes\frg
        \end{equation}
        of degree one in $(\frg[1])^*\otimes\frg$. Consequently,
        \begin{equation}\label{eq:LieAlgQ}
            \begin{aligned}
                Q\sfa\ &\coloneqq\ (Q\xi^a)\otimes\cb_a
                \\
                &\,=\ \tfrac12\xi^b\xi^c{f_{cb}}^a\otimes\cb_a
                \\
                &\,=\ -\tfrac12 \xi^b\xi^c\otimes{f_{bc}}^a\cb_a
                \\
                &\,=\ -\tfrac12 \xi^b\xi^c\otimes[\cb_b,\cb_c]
                \\
                &\,\eqqcolon\ -\tfrac12[\xi^b\otimes\cb_b,\xi^c\otimes\cb_c]
                \\
                &\,=\ -\tfrac12[\sfa,\sfa]~.
            \end{aligned}
        \end{equation}
        
        More general vector fields arise in the Chevalley--Eilenberg algebras of $L_\infty$-algebras and $L_\infty$-algebroids, cf.~e.g.~\cite{Jurco:2018sby} for further details. In the case of Yang--Mills theory, the homological vector field $Q^{\rm YM}_{\rm BRST}$ describing the gauge symmetry acts according to 
        \begin{equation}\label{eq:YMBRST_1}
            \begin{aligned}
                c^a\ &\mapstomap{Q^{\rm YM}_{\rm BRST}}\ -\tfrac12g{f_{bc}}^ac^bc^c~,&~~~
                \bar c^a\ &\mapstomap{Q^{\rm YM}_{\rm BRST}}\ b^a~,
                \\
                A^a_\mu\ &\mapstomap{Q^{\rm YM}_{\rm BRST}}\ (\nabla_\mu c)^a~,~~~&
                b^a\ &\mapstomap{Q^{\rm YM}_{\rm BRST}}\ 0~.
            \end{aligned}
        \end{equation}
        These transformations are known as the \uline{BRST transformations} and $Q^{\rm YM}_{\rm BRST}$ as the \uline{BRST operator}. One readily verifies that $(Q^{\rm YM}_{\rm BRST})^2=0$, that is, $Q_{\rm BRST}$ is a differential. In addition, the action~\eqref{eq:YMActionComponents} is $Q^{\rm YM}_{\rm BRST}$-closed, that is, $Q^{\rm YM}_{\rm BRST}S^{\rm YM}_{\rm BRST}=0$, which ensures gauge choice independence and unitarity of the physical S-matrix.
        
        We shall denote the minimal field space\footnote{This graded vector space is, in fact, the space of sections of a graded vector bundle, and fields and their derivatives are sections of the corresponding jet bundle; but these details would not enlighten our discussion any further so we suppress them.} for gauge-fixed Yang--Mills theory by $\frL^{\rm YM}_{\rm BRST}$, but the ghost number is the degree of coordinate functions on $\frL^{\rm YM}_{\rm BRST}[1]$. Explicitly,
        \begin{equation}
            \begin{gathered}
                \frL^{\rm YM}_{\rm BRST}\ =\ \frL^{\rm YM}_{\rm BRST,\,0}\oplus\frL^{\rm YM}_{\rm BRST,\,1}\oplus\frL^{\rm YM}_{\rm BRST,\,2}~,
                \\
                \frL^{\rm YM}_{\rm BRST,\,0}\ \coloneqq\ \scC^\infty(\IM^d)\otimes\frg
                ~,~~~
                \frL^{\rm YM}_{\rm BRST,\,1}\ \coloneqq\ (\Omega^1(\IM^d)\oplus\scC^\infty(\IM^d))\otimes\frg~,
                \\
                \frL^{\rm YM}_{\rm BRST,\,2}\ \coloneqq\ \scC^\infty(\IM^d)\otimes\frg
            \end{gathered}
        \end{equation}
        and $c$, $A$, $b$, and $\bar c$ are coordinate functions on $\frL^{\rm YM}_{\rm BRST,\,0}[1]=(\frL^{\rm YM}_{\rm BRST}[1])_{-1}$, $\frL^{\rm YM}_{\rm BRST,\,1}[1]=(\frL^{\rm YM}_{\rm BRST}[1])_0$, $\frL^{\rm YM}_{\rm BRST,\,1}[1]=(\frL^{\rm YM}_{\rm BRST}[1])_0$, and $\frL^{\rm YM}_{\rm BRST,\,2}[1]=(\frL^{\rm YM}_{\rm BRST}[1])_1$ and thus of degrees $1$, $0$, $0$, and $-1$, respectively. Moreover, the action~\eqref{eq:YMActionComponents} is a polynomial function $S^{\rm YM}_{\rm BRST}\in\scC^\infty(\frL^{\rm YM}_{\rm BRST}[1])$ on $\frL^{\rm YM}_{\rm BRST}[1]$  of total ghost number zero, $|S^{\rm YM}_{\rm BRST}|_{\scC^\infty(\frL^{\rm YM}_{\rm BRST}[1])}=0$. In the following, we shall write $|-|_{\rm gh}$ as a shorthand for both $|-|_{(\frL^{\rm YM}_{\rm BRST}[1])^*}$ and $|-|_{\scC^\infty(\frL^{\rm YM}_{\rm BRST}[1])}$.
        
        The $Q$-vector space $(\frL^{\rm YM}_{\rm BRST}[1],Q^{\rm YM}_{\rm BRST})$ describes the Lie algebra of gauge transformations as well as its action on the various fields, which together form an action Lie algebroid. This becomes clear when comparing~\eqref{eq:YMBRST_1} to~\eqref{eq:LieAlgQ}; the latter is the evident generalisation, e.g.~to the corresponding formulas for a differential graded Lie algebra.        
        
        We note that gauge-invariant objects are $Q^{\rm YM}_{\rm BRST}$-closed and that gauge-trivial objects are $Q^{\rm YM}_{\rm BRST}$-exact. Therefore, physical observables are in the cohomology of $Q_{\rm BRST}$. The pair of fields $(b,\bar c)$ is known as a \uline{trivial pair}, as $Q^{\rm YM}_{\rm BRST}$ links the two fields by an identity map. They vanish in the $Q^{\rm YM}_{\rm BRST}$-cohomology and thus are not observable.
        
        As in~\eqref{eq:LieAlgContractedCoordinates}, it will turn out useful to define the contracted coordinates
        \begin{subequations}\label{eq:YMsuperfield}
            \begin{equation}
                \begin{aligned}
                    \sfa\ &\coloneqq\ \int\rmd^dx\,\Big\{c^a(x)\otimes(\cb_a\otimes\tts_x)+A^a_\mu(x)\otimes(\cb_a\otimes\ttv^\mu\otimes\tts_x)\,+
                    \\
                    &\kern2.5cm+\,b^a(x)\otimes(\cb_a\otimes\tts_x)+\bar c^a(x)\otimes(\cb_a\otimes\tts_x)\Big\}~,
                \end{aligned}
            \end{equation}
            where $\cb_a$, $\ttv^\mu$, and $\tts_x$ are basis vectors on $\frg$, $T^*_x\IM^d$, and $\scC^\infty(\IM^d)$, respectively (and thus, we have an identification $\ttv^\mu\mathrel{\hat=}\rmd x^\mu$). It should be noted that $\sfa$ is an element of $(\frL^{\rm YM}_{\rm BRST}[1])^*\otimes\frL^{\rm YM}_{\rm BRST}$ of degree one, and it can be regarded as a superfield which contains all the fields of different ghost numbers. The component fields can be recovered by projecting onto the respective ghost numbers. In the following, we will write symbolically
            \begin{equation}\label{eq:YMsuperfieldAbstract}
                \sfa\ =\ \Phi^I\otimes\cb_I
            \end{equation}
            for DeWitt indices $I,J,\ldots$, which contain Lorentz and gauge indices as well as space--time position. A contraction of DeWitt indices involves sums over all discrete indices and evident integrals over the continuous ones.
        \end{subequations}
        
        \subsection{Batalin--Vilkovisky formalism and \texorpdfstring{$L_\infty$}{L-infty}-algebras}\label{sec:BVFormalismLinfty}
        
        The above example of Yang--Mills theory has demonstrated how $\IZ$-graded vector spaces and homological vector fields enter into the description of a gauge field theory in the BRST formalism. In particular, gauge-invariant observables were contained in the cohomology of $Q_{\rm BRST}$. To fully characterise classical observables, however, we also need to impose the equations of motion. This is the purpose of the more general Batalin--Vilkovisky (BV) formalism. As a byproduct, the BV formalism can cater for \uline{open} gauge symmetries which are gauge symmetries for which $Q_{\rm BRST}$ is a differential only on-shell. The BV operator $Q_{\rm BV}$, which generalises the BRST operator $Q_{\rm BRST}$, encodes the Chevalley--Eilenberg description of a cyclic $L_\infty$-algebra (i.e.~an $L_\infty$-algebra with a notion of inner product). The gauge-fixed form of this cyclic $L_\infty$-algebra will be crucial for our formulation of the double copy of amplitudes.
        
        \paragraph{BV operator.} 
        Let $\frL_{\rm BRST}[1]$ be a $\IZ$-graded vector space of fields of a general field theory. Then we have also a correspondence between the fields and the coordinate functions on this space. In order to encode the field equations for all the fields in the action of an operator $Q_{\rm BV}$, we `double' this vector space such that we have for each field $\Phi^I$ of ghost number $|\Phi^I|_{\rm gh}$ an \uline{anti-field} $\Phi^+_I$ of ghost number $|\Phi^+_I|_{\rm gh}\coloneqq-1-|\Phi^I|_{\rm gh}$ so that
        \begin{equation}\label{eq:QBVeom}
            Q_{\rm BV} \Phi^+_I\ \coloneqq\ (-1)^{|\Phi^I|_{\rm gh}}\delder[S_{\rm BRST}]{\Phi^I}+\cdots~.
        \end{equation}
        Here, the ellipsis denotes terms at least linear in the anti-fields. Formally, this doubling amounts to considering the cotangent space
        \begin{equation}
            \frL_{\rm BV}[1]\ \coloneqq\ T^*[-1](\frL_{\rm BRST}[1])~~~\Leftrightarrow~~~\frL_{\rm BV}\ \coloneqq\ T^*[-3]\frL_{\rm BRST}~,
        \end{equation}
        which yields a canonical symplectic form
        \begin{equation}\label{eq:BVOmega}
            \omega\ \coloneqq\ \delta\Phi^+_I\wedge\delta\Phi^I
        \end{equation}
        of ghost number $-1$. This symplectic form $\omega$, in turn, induces a Poisson bracket, also known as the \uline{anti-bracket}. It reads explicitly as\footnote{The signs arise as follows. Hamiltonian vector fields $V_F$ are given by $V_F\intprod\omega=\delta F$ for some function $F$. The Poisson bracket is then given by $\{F,G\}\coloneqq V_F\intprod V_G\intprod\omega=V_F(G)$ from which the signs follow using the explicit form~\eqref{eq:BVOmega} of $\omega$. The signs are often absorbed using left- and right-derivatives; however, for clarity we shall keep them explicitly.}
        \begin{equation}\label{eq:BVBracket}
            \{F,G\}\ =\ (-1)^{|\Phi^I|_{\rm gh}(|F|_{\rm gh}+1)}\delder[F]{\Phi^I}\delder[G]{\Phi^+_I}-(-1)^{(|\Phi^I|_{\rm gh}+1)(|F|_{\rm gh}+1)}\delder[F]{\Phi^+_I}\delder[G]{\Phi^I}~,
        \end{equation}
        and it is of ghost number one so that $\{F,G\}=-(-1)^{(|F|_{\rm gh}+1)(|G|_{\rm gh}+1)}\{G,F\}$. 
        
        The \uline{classical Batalin--Vilkovisky action} is now a function $S_{\rm BV}\in\scC^\infty(\frL_{\rm BV}[1])$ of ghost number zero, which obeys the \uline{classical master equation}
        \begin{subequations}
            \begin{equation}\label{eq:CME}
                \{S_{\rm BV},S_{\rm BV}\}\ =\ 0~,
            \end{equation}
            which extends the original action $S_0$ of the field theory (without ghosts or trivial pairs)\footnote{Here, $|_{\Phi^+_I=0}$ is the restriction to the subspace of BV field space where all anti-fields are zero.}
            \begin{equation}
                S_{\rm BV}|_{\Phi^+_I=0}\ =\ S_0~,
            \end{equation}
            and whose Hamiltonian vector field extends the BRST differential,
            \begin{equation}\label{eq:BRST_differential_definition}
                (Q_{\rm BV} \Phi^I)|_{\Phi^+_J=0}\ =\ Q_{\rm BRST} \Phi^I
            \end{equation}
        \end{subequations}
        with
        \begin{equation}\label{eq:CQ}
            Q_{\rm BV}\ \coloneqq\ \{S_{\rm BV},-\}~.
        \end{equation}
        We note that $Q_{\rm BV}^2=0$ and~\eqref{eq:CME} are equivalent.
        
        The last two conditions fix the terms of $S_{\rm BV}$ which are constant and linear in the anti-fields to read as
        \begin{equation}
            S_{\rm BV}\ =\ S_0+(-1)^{|\Phi^I|_{\rm gh}}\Phi^+_I Q_{\rm BRST}\Phi^I+\cdots~.
        \end{equation}
        General theorems now state that for each action and compatible BRST operator, there is a corresponding BV action and a BV operator, cf.~\cite{Gomis:1994he}.
        
        In a general theory, we will usually have a physical field $a$ of ghost number zero as well as ghosts $c_0$ together with higher ghosts $c_{-k}$ of each ghost number $-k+1$ as coordinate functions on $\frL_{\rm BV}[1]$. Higher ghosts are non-trivial only in theories with higher gauge invariance. All fields come with the corresponding anti-fields $a^+$, $c_0^+$, and $c_{-k}^+$. To accommodate gauge fixing, we will have to expand the field space further by trivial pairs and corresponding anti-fields, as already encountered in the previous section.
        
        The equations of motion generate an ideal $\scI$ in $\scC^\infty(\frL_{\rm BRST}[1])$, and the functions on the solutions space are the quotient $\scC^\infty(\frL_{\rm BRST}[1])/\scI$. Because of~\eqref{eq:CQ},
        \begin{equation}
            Q_{\rm BV}\Phi^+_I\ =\ (-1)^{|\Phi^I|_{\rm gh}}\delder[S_{\rm BV}]{\Phi^I}~,
        \end{equation}
        and the gauge-invariant functions on the solutions space are described by the $Q_{\rm BV}$-cohomology.
        
        \paragraph{\mathversion{bold}$L_\infty$-algebras.} 
        Following~\eqref{eq:YMsuperfield}, we define again a superfield
        \begin{equation}\label{eq:superfield}
            \sfa\ \coloneqq\ \sfa^\bmI\otimes\cb_\bmI\ =\ \Phi^I\otimes\cb_I+\Phi^+_I\otimes\cb^I
        \end{equation}
        of degree one in $(\frL_{\rm BV}[1])^*\otimes\frL_{\rm BV}$, where $\bmI$ runs over all fields, ghosts, ghosts for ghosts and the corresponding anti-fields, as well as space--time and Lie algebra indices. As in~\eqref{eq:LieAlgQ}, we may extend the action of $Q_{\rm BV}$ to elements in $(\frL_{\rm BV}[1])^*\otimes\frL_{\rm BV}$ by left action and write
        \begin{subequations}
            \begin{equation}\label{eq:Qsupermu}
                Q_{\rm BV}\sfa\ =\ \{S_{\rm BV},\sfa\}\ =\ -\sff(\sfa)\ewith\sff(\sfa)\ \eqqcolon\ \sum_{i\geq1}\frac{1}{i!}\mu'_i(\sfa,\ldots,\sfa)~.
            \end{equation}
            The $\mu'_i$ now encode $i$-ary graded anti-symmetric linear maps $\mu_i:\frL_{\rm BV}\times\cdots\times\frL_{\rm BV}\rightarrow\frL_{\rm BV}$, which can be extracted by the formulas
            \begin{equation}\label{eq:ghostProd}
                \begin{aligned}
                    \mu'_1(\sfa)\ &\coloneqq\ (-1)^{|\sfa^\bmI|_{\rm gh}}\sfa^\bmI\otimes\mu_1(\cb_\bmI)~,
                    \\
                    \mu'_i(\sfa,\ldots,\sfa)\ &\coloneqq\ (-1)^{i\sum_{j=1}^i|\sfa^{\bmI_j}|_{\rm gh}+\sum_{j=2}^{i}|\sfa^{\bmI_j}|_{\rm gh}\sum_{k=1}^{j-1}|\cb_{\bmI_k}|_{\frL_{\rm BV}}}\sfa^{\bmI_1}\cdots\sfa^{\bmI_i}\otimes\mu_i(\cb_{\bmI_1},\ldots,\cb_{\bmI_i})~,
                \end{aligned}
            \end{equation}
        \end{subequations}
        see~\cite{Jurco:2018sby} for a much more detailed exposition.\footnote{Note that the $\mu_i'$ define, in fact, an $L_\infty$-structure on $\scC^\infty(\frL_{\rm BV}[1])\otimes\frL_{\rm BV}$.} The condition $Q_{\rm BV}^2=0$ then amounts to the \uline{homotopy Jacobi identities}
        \begin{subequations}\label{eq:homotopyJacobi}
            \begin{equation}
                \sum_{i_1+i_2=i}\sum_{\sigma\in{\rm \overline{Sh}}(i_1;i)}(-1)^{i_2}\chi(\sigma;\ell_1,\ldots,\ell_i)\mu_{i_2+1}(\mu_{i_1}(\ell_{\sigma(1)},\ldots,\ell_{\sigma(i_1)}),\ell_{\sigma(i_1+1)},\ldots,\ell_{\sigma(i)})\ =\ 0
            \end{equation}
            for all $\ell_1,\ldots,\ell_i\in\frL_{\rm BV}$. The sum is over all $(i_1;i)$ \uline{unshuffles} $\sigma$ which consist of permutations $\sigma$ of $\{1,\ldots,i\}$ so that the first $i_1$ and the last $i-i_1$ images of $\sigma$ are ordered. Moreover, $\chi(\sigma;\ell_1,\ldots,\ell_i)$ is the \uline{Koszul sign} given by
            \begin{equation}\label{eq:KoszulSign}
                \ell_1\wedge\ldots\wedge\ell_i\ =\ \chi(\sigma;\ell_1,\ldots,\ell_i)\,\ell_{\sigma(1)}\wedge\ldots\wedge\ell_{\sigma(i)}~.
            \end{equation}
        \end{subequations}
        The pair $(\frL_{\rm BV},\mu_i)$ with products $\mu_i$ subject to~\eqref{eq:homotopyJacobi} is called an \uline{$L_\infty$-algebra}, cf.~\cref{app:hA_L_infinity}. In our present setting, $\frL_{\rm BV}$ is, in fact, a \uline{cyclic $L_\infty$-algebra} because of the presence of the symplectic form $\omega$. Specifically, if we consider the shift isomorphism~\eqref{eq:shiftIso}, then $\omega$ induces the (indefinite) inner product\footnote{We will, in the bulk of our paper, deviate from this sign convention in order to simplify the signs arising in our double copy formalism.}
        \begin{subequations}\label{eq:BVInner}
            \begin{equation}
                \inner{\ell_1}{\ell_2}\ \coloneqq\ (-1)^{|\ell_1|_{\frL_{\rm BV}}}\omega(\sigma(\ell_1),\sigma(\ell_2))
            \end{equation}
            of degree $-3$ in $\frL_{\rm BV}$ and of ghost number zero. It is cyclic in the sense that
            \begin{equation}
                \inner{\ell_1}{\mu_i(\ell_2,\ldots,\ell_{i+1})}\ =\ (-1)^{i+i(|\ell_1|_{\frL_{\rm BV}}+|\ell_{i+1}|_{\frL_{\rm BV}})+|\ell_{i+1}|_{\frL_{\rm BV}}\sum_{j=1}^i|\ell_j|_{\frL_{\rm BV}}}\inner{\ell_{i+1}}{\mu_i(\ell_1,\ldots,\ell_i)}~,
            \end{equation}
        \end{subequations}
        which is a consequence of the vanishing of the Lie derivative of $\omega$ along $Q_{\rm BV}$. This is equivalent to saying that the higher products $\mu_i$, with the first $i-1$ arguments fixed, act as graded derivations on $\inner{-}{-}$.
        
        \paragraph{\mathversion{bold}Correspondence between actions and $L_\infty$-algebras.} 
        Every cyclic $L_\infty$-algebra $(\frL_{\rm BV},\mu_i)$ comes with a \uline{homotopy Maurer--Cartan action}, cf.~\cref{app:homotopyAlgebras}. In particular, the functional
        \begin{equation}
            S^{\rm hMC}\ \coloneqq\ \sum_{i\geq1}\frac{1}{(i+1)!}\inner{a}{\mu_i(a,\ldots,a)}
        \end{equation}
        for $a\in\frL_{\rm BV,1}$ reproduces the action for the physical fields. Using the superfield $\sfa$ defined in~\eqref{eq:superfield}, we can write down a more general homotopy Maurer--Cartan action
        \begin{subequations}
            \begin{equation}\label{eq:CBVAction}
                S^{\rm shMC}\ \coloneqq\ \sum_{i\geq1}\frac{1}{(i+1)!}\inner{\sfa}{\mu'_i(\sfa,\ldots,\sfa)}'~,
            \end{equation}
            where we define 
            \begin{equation}\label{eq:superfield_hMC_action}
                \inner{f^\bmI_1\otimes\cb_\bmI}{f^\bmJ_2\otimes\cb_\bmJ}'\ \coloneqq\ (-1)^{|f_1^\bmI|_{\rm gh}+|f_2^\bmJ|_{\rm gh}+|\cb_\bmI|_{\frL_{\rm BV}}|f_2^\bmJ|_{\rm gh}}~f_1^\bmI f_2^\bmJ~\inner{\cb_\bmI}{\cb_\bmJ}
            \end{equation}
        \end{subequations}
        for $f^\bmI_{1,2}\in\scC^\infty(\frL_{\rm BV}[1])$. This superfield version of the homotopy Maurer--Cartan action is, in fact, the full BV action $S_{\rm BV}$. Put differently,~\eqref{eq:CBVAction} satisfies the quantum master equation~\eqref{eq:QME} if and only if the $\mu_i$ in $\mu'_i$ via~\eqref{eq:ghostProd} satisfy the homotopy Jacobi identities~\eqref{eq:homotopyJacobi}. We shall refer to the action~\eqref{eq:CBVAction} as the \uline{superfield homotopy Maurer--Cartan action} of the $L_\infty$-algebra $(\frL_{\rm BV},\mu_i)$.
        
        In summary, the BV formalism provides an equivalence between classical field theories and cyclic $L_\infty$-algebras, where the BV operator plays the role of the Chevalley--Eilenberg differential of the $L_\infty$-algebra. Clearly, the BV action corresponding to an $L_\infty$-algebra $\frL_{\rm BV}$ is physically only interesting if its degree-one part is non-trivial. To read off the $L_\infty$-algebra from a particular action functional, we note that using~\eqref{eq:superfield_hMC_action} we have
        \begin{subequations}\label{eq:translation_BV_L_infty}
            \begin{equation}
                \begin{aligned}
                    \inner{\sfa}{\mu'_i(\sfa,\ldots,\sfa)}'\ &=\ \inner{\sfa^{\bmI_{i+1}}\otimes\cb_{\bmI_{i+1}}}{\mu'_i(\sfa^{\bmI_1}\otimes\cb_{\bmI_1},\ldots,\sfa^{\bmI_i}\otimes\cb_{\bmI_i})}'
                    \\
                    &=\ \zeta(\bmI_1,\ldots,\bmI_i)\,\sfa^{\bmI_{i+1}}\sfa^{\bmI_1}\cdots\sfa^{\bmI_i}\inner{\cb_{\bmI_{i+1}}}{\mu_i(\cb_{\bmI_1},\ldots,\cb_{\bmI_i})}
                \end{aligned}
            \end{equation}
            with the sign $\zeta(\bmI_1,\ldots,\bmI_i)$ given by
            \begin{equation}\label{eq:action-translation-sign}
                \zeta(\bmI_1,\ldots,\bmI_i)\ \coloneqq\ (-1)^{\sum_{k=1}^i|\sfa^{\bmI_k}|_{\rm gh}(i+k+\sum_{j=k}^i|\sfa^{\bmI_j}|_{\rm gh})}~.
            \end{equation}
        \end{subequations}
        More explicitly,
        \begin{equation}\label{eq:shMC_translation_formulas}
            \begin{aligned}
                \langle \sfa,\mu'_1(\sfa)\rangle'\ &=\ (-1)^{|\sfa^{\bmI_1}|_{\rm gh}}\,\sfa^{\bmI_{2}}\sfa^{\bmI_1}\inner{\cb_{\bmI_{2}}}{\mu_1(\cb_{\bmI_1})}~,
                \\
                \langle \sfa,\mu'_2(\sfa,\sfa)\rangle'\ &=\ (-1)^{(|\sfa^{\bmI_1}|_{\rm gh}+1)|\sfa^{\bmI_2}|_{\rm gh}}\,\sfa^{\bmI_{3}}\sfa^{\bmI_1}\sfa^{\bmI_2}\inner{\cb_{\bmI_3}}{\mu_2(\cb_{\bmI_1},\cb_{\bmI_2})}~,
            \end{aligned}
        \end{equation}
        and we shall make use of these formulas later.
        
        \begin{remark}\label{rem:fields_vs_coordinate_functions}
            The exchange of the coordinate functions on field space with the actual fields can easily lead to confusion. Let us therefore summarise the situation once more. Actual fields (usually sections of a bundle or connections and their generalisations) are elements of a graded vector space $\frL_{\rm BV}$. The $L_\infty$-algebra structure is defined on the vector space $\frL_{\rm BV}$. The symbols appearing in an action $S$ are, technically speaking, not fields but coordinate functions on the grade-shifted field space $\frL_{\rm BV}[1]$, the same way that in differential geometry one writes the metric in terms of the symbols $x^\mu$, which are not points in space--time but rather real-valued coordinate functions defined on space--time. Once we evaluate the action for particular fields, the coordinate functions are replaced by their values. Similarly, the BV operator, the anti-bracket etc.~all act on or take as arguments polynomial functions on $\frL_{\rm BV}[1]$, which are given by polynomial expressions in the coordinate functions as well as their derivatives, which are also contained in $(\frL_{\rm BV}[1])^*$. To simplify notation, the coordinate function for a field (e.g.~in an action) will be denoted by the same symbol as the field (element of the $L_\infty$-algebra), as commonly done in quantum field theory.
        \end{remark}
        
        \begin{remark}\label{rem:Schwartz_issues}
            The integral defining the action $S$ of a classical field theory is mathematically usually not well defined. At a classical level, this does not matter because we are never interested in the value of $S$ itself, and we can treat all integrals as formal expressions. For definiteness, mathematicians often drop the action and work with the Lagrangian instead. This can easily be done in the $L_\infty$-algebra picture, working with graded modules over the ring of functions instead of graded vector spaces.
            
            At quantum level, however, the value of $S$ for particular field configurations does play a role, and one needs to carefully restrict the field space such that all integrals are indeed well-defined, cf.~\cite{Macrelli:2019afx}. One suitable restriction offers itself for the perturbative treatment. We split the field space into interacting fields, $\frF_{\rm int}$, which can simply be identified with Schwartz functions on Minkowski space $\scS(\IM^d)$, and free fields $\frF_{\rm free}$, which can be identified with solutions to the free equations of motion (i.e.~fields in the kernel of $\mu_1$), which are Schwartz type for any fixed time-slice of Minkowski space,
            \begin{equation}\label{eq:properFunctionSpace}
                \frF\ \coloneqq\ \frF_{\rm int}\oplus\frF_{\rm free}
                \ewith
                \frF_{\rm int}\ \coloneqq\ \scS(\IM^d)
                \eand
                \frF_{\rm free}\ \coloneqq\ \ker_\scS(\mu_1)~.
            \end{equation}
            The elements of $\ker_\scS(\mu_1)$ are, of course, the states that label the asymptotic on-shell states in perturbation theory. On the other hand, the fields in $\scS(\IM^d)$ are the propagating degrees of freedom found on internal lines in Feynman diagrams. The decomposition~\eqref{eq:properFunctionSpace} is very much in the spirit of the homological perturbation lemma, which can be used to construct the scattering amplitudes, as we shall discuss below.
            
            We note that the wave operator is invertible on $\scS(\IM^d)$ and the inverse is indeed the propagator $\sfh$, as we shall discuss in more detail below. This allows us to define the operators $\sqrt{\wave}$ and $\frac{1}{\sqrt{\wave}}$ on $\scS(\IM^d)$, which we continue to all of $\frF$ by mapping elements of $\ker_\scS(\mu_1)$ to zero. This fact will play an important role later.
        \end{remark}
        
        \paragraph{Gauge fixing.}
        The next step in the BV formalism is the implementation of gauge fixing. This is achieved by a canonical transformation
        \begin{equation}\label{eq:gaugeFixingCanonicalTrafo}
            S^{\rm gf}_{\rm BV}\big[\Phi^I,\tilde\Phi^+_I\big]\ \coloneqq\ S_{\rm BV}\left[\Phi^I,\Phi^+_I+\delder[\Psi]{\Phi^I}\right]
        \end{equation}
        which is mediated by a choice of \uline{gauge-fixing fermion}, the generating functional for the canonical transformation, which is a function $\Psi\in\scC^\infty(\frL_{\rm BV}[1])$ of ghost number $-1$. The action~\eqref{eq:gaugeFixingCanonicalTrafo} is then gauge-fixed if its Hessian is invertible. This requires a careful choice of~$\Psi$: the trivial choice $\Psi=0$ leads back to the original action. When the classical BV action is only linear in the anti-fields, as is e.g.~the case for Yang--Mills theory and all the field theories we are dealing with, we may set the anti-fields in $S^{\rm gf}_{\rm BV}$ to zero after gauge-fixing, without loss of generality since the BV operator reduces to a BRST operator.
        
        Note that to construct the gauge-fixing fermion $\Psi$ of ghost number $-1$, we will have to introduce additional fields of negative ghost number together with their anti-fields, such as e.g.~the anti-ghost $\bar c$ and the Nakanishi--Lautrup field $b$ in the case of Yang--Mills theory. If we do not change the $Q_{\rm BV}$-cohomology, these new fields do not affect the observables. This can trivially be achieved if $Q_{\rm BV}$ maps one field to another,
        \begin{equation}
            \bar c\ \mapstomap{Q_{\rm BV}}\ b~,~~~
            b\ \mapstomap{Q_{\rm BV}}\ 0~,~~~
            \bar c^+\ \mapstomap{Q_{\rm BV}}\ 0~,~~~
            b^+\ \mapstomap{Q_{\rm BV}}\ -\bar c^+~,
        \end{equation}
        cf.~\eqref{eq:YMBRST_1}. We shall encounter a number of more involved examples in \cref{sec:Examples}.
        
        \paragraph{\mathversion{bold}Quantum master equation and quantum $L_\infty$-algebras.}
        Besides the canonical symplectic form~\eqref{eq:BVOmega}, we also have a canonical second-order differential operator on $\scC^\infty(\frL_{\rm BV}[1])$, called the \uline{Batalin--Vilkovisky Laplacian}, and defined as 
        \begin{equation}\label{eq:BVLaplacian}
            \lap F\ \coloneqq\ (-1)^{|\Phi^I|_{\rm gh}+|F|_{\rm gh}}\frac{\delta^2 F}{\delta\Phi^+_I\delta\Phi^I}
        \end{equation}
        for $F\in\scC^\infty(\frL_{\rm BV}[1])$. 
        
        The BV Laplacian plays a key role in the path integral quantisation of a theory. In particular, the gauge fixing~\eqref{eq:gaugeFixingCanonicalTrafo} is implemented at the path-integral level as
        \begin{equation}\label{eq:gaugeFixingPathIntegral}
            Z_\Psi\ \coloneqq\ \int_{\frL_{\rm BV}}\mu(\Phi^I,\Phi^+_I)\,\delta\left(\Phi^+_I-\delder[\Psi]{\Phi^I}\right)\rme^{\frac{\rmi}{\hbar}S^\hbar_{\rm qBV}[\Phi^I,\Phi^+_I]}~,
        \end{equation}
        where $\mu$ is a measure that is compatible with the symplectic form $\omega$, $\delta$ is a functional delta distribution, $\hbar$ is a formal parameter, and $S^\hbar_{\rm qBV}\in\scC^\infty(\frL_{\rm BV}[1])$ is a functional of ghost number zero with
        \begin{equation}
            S^\hbar_{\rm qBV}|_{\hbar=0}\ =\ S_{\rm BV}~.
        \end{equation} 
        For $Z_\Psi$ to be independent of the choice of gauge-fixing fermion $\Psi$, $S^\hbar_{\rm qBV}$ must satisfy the \uline{quantum master equation}~\cite{Batalin:1981jr}\footnote{Specifically, one requires $Z_{\Psi+\delta\Psi}=Z_\Psi$ for an infinitesimal deformation $\delta\Psi$ of $\Psi$; the space of gauge-fixing fermions $\Psi$ (whose Hessians may not be invertible) is contractible, so $Z_\Psi$ is globally independent of $\Psi$.}
        \begin{equation}\label{eq:QME}
            \lap\rme^{\frac{\rmi}{\hbar}S^\hbar_{\rm qBV}}\ =\ 0
            \qquad\Longleftrightarrow\qquad
            \{S^\hbar_{\rm qBV},S^\hbar_{\rm qBV}\}-2\rmi\hbar\lap S^\hbar_{\rm qBV}\ =\ 0~.
        \end{equation}
        Consequently, we obtain as generalisation of~\eqref{eq:CQ} the quantum BRST--BV operator
        \begin{equation}\label{eq:QQ}
            Q_{\rm qBV}\ \coloneqq\ \{S^\hbar_{\rm qBV},-\}-2\rmi\hbar\lap~,
        \end{equation}
        and the quantum master equation~\eqref{eq:QME} is equivalent to $Q_{\rm qBV}^2=0$. Note that contrary to the classical version, the quantum version~\eqref{eq:QQ} is no longer a derivation. Solutions $S^\hbar_{\rm qBV}$ to~\eqref{eq:QME} are called \uline{quantum Batalin--Vilkovisky actions}. We may now solve~\eqref{eq:QME} order by order in $\hbar$ generalising the products $\mu'_i$ in~\eqref{eq:CBVAction} to products $\mu'_{i,L}$ for $L=0,1,2,\ldots$ to reflect the $\hbar$-dependence with $\mu'_{i,L=0}=\mu'_i$ and $\mu'_{i,L=-1}\coloneqq0$. Consequently, we consider the ansatz
        \begin{equation}\label{eq:QBVAction}
            S^{\rm qshMC}\ \coloneqq\ \sum_{\substack{i\geq1\\ L\geq0}}\frac{\hbar^L}{(i+1)!}\inner{\sfa}{\mu'_{i,L}(\sfa,\ldots,\sfa)}'
        \end{equation}
        for the superfield~\eqref{eq:superfield}. The action~\eqref{eq:QBVAction} satisfies the quantum master equation~\eqref{eq:QME} if and only if the $\mu_{i,L}$ satisfy the \uline{quantum homotopy Jacobi identities}~\cite{Zwiebach:1992ie,Markl:1997bj,Doubek:2017naz}
        \begin{equation}\label{eq:QJacobi}
            \begin{aligned}
                &\sum_{\substack{i_1+i_2=i\\L_1+L_2=L}}\sum_{\sigma\in{\rm \overline{Sh}}(i_1;i)}(-1)^{i_2}\chi(\sigma;\ell_1,\ldots,\ell_i)\mu_{i_2+1,L_2}(\mu_{i_1,L_1}(\ell_{\sigma(1)},\ldots,\ell_{\sigma(i_1)}),\ell_{\sigma(i_1+1)},\ldots,\ell_{\sigma(i)})\,-
                \\[-10pt]
                &\kern3cm-\rmi\mu_{i+2,L-1}(\cb^\bmI,\cb_\bmI,\ell_1,\ldots,\ell_i)\ =\ 0
            \end{aligned}
        \end{equation}
        for $\ell_1,\ldots,\ell_i\in\frL_{\rm BV}$, where the $\mu_{i,L}$ are as in~\eqref{eq:ghostProd} via the $\mu'_{i,L}$. Here $\cb^\bmI\coloneqq\cb_\bmJ\omega^{\bmJ\bmI}$, where $\omega^{\bmI\bmJ}$ is the inverse of the symplectic form~\eqref{eq:BVOmega} when written as $\omega=\frac12\delta\sfa^\bmI\wedge\omega_{\bmI\bmJ}\delta\sfa^\bmJ$. Furthermore,~\eqref{eq:Qsupermu} generalises to
        \begin{equation}\label{eq:QsupermuQuantum}
            Q_{\rm qBV}\sfa\ =\ -\sum_{\substack{i\geq1\\ L\geq0}}\frac{\hbar^L}{i!}\mu'_{i,L}(\sfa,\ldots,\sfa)~.
        \end{equation}
        The tuple $(\frL_{\rm BV},\mu_{i,L},\omega)$ with the products $\mu_{i,L}$ subject to~\eqref{eq:QJacobi} is called a \uline{quantum} or \uline{loop $L_\infty$-algebra}. In the classical limit $\hbar\to0$, the higher products $\mu_{i,L}$ for $L>0$ become trivial, and we recover a cyclic $L_\infty$-algebra. Note that for scalar field theory, Yang--Mills theory, and also Chern--Simons theory, the classical BV action also satisfies the quantum master equation and hence, in those cases, we may set $S^\hbar_{\rm qBV}=S_{\rm BV}$, in which case $\mu_{i,L}=0$ for $L>0$.
        
        \subsection{Scattering amplitudes and \texorpdfstring{$L_\infty$}{L-infty}-algebras}\label{ssec:scatterin_amplitudes}
        
        Above, we saw that actions of field theories are encoded in cyclic $L_\infty$-algebras. The same holds for tree-level scattering amplitudes and loop-level scattering amplitudes are encoded in quantum $L_\infty$-algebras, as we shall explain in this section.
        
        \paragraph{Equivalence of field theories.} Classically, two physical theories are equivalent, if they have an isomorphic space of observables.\footnote{This is weaker than the statement that tree-level scattering amplitudes coincide. To define asymptotic in- and out-states in the same Hilbert space, one needs the additional data of the symplectic form $\omega$. Two classical theories have the same tree-level scattering amplitudes if they are related by a quasi-isomorphisms compatible with the cyclic structures. Again, see~\cite{Jurco:2018sby} for some more details.} Translated to the BV formalism, this implies that classically equivalent physical theories have isomorphic $Q_{\rm BV}$-cohomology. Dually, this implies that two physical theories are classically equivalent, if they have \uline{quasi-isomorphic} $L_\infty$-algebras, which is also mathematically the natural notion of equivalence for $L_\infty$-algebras; see \cref{app:homotopyAlgebras} for more details. 
        
        Given two $L_\infty$-algebras $(\frL_{\rm BV},\mu_i)$ and $(\tilde\frL_{\rm BV},\tilde\mu_i)$ constructed from a BV action, a \uline{morphism} $\phi:\frL_{\rm BV}\rightarrow\tilde \frL_{\rm BV}$ of $L_\infty$-algebras is a collection of $i$-linear totally graded anti-symmetric maps $\phi_i:\frL_{\rm BV}\times\cdots\times\frL_{\rm BV}\rightarrow\tilde\frL_{\rm BV}$ of degree~$1-i$ subject to the conditions~\eqref{eq:morphism_L}. We note that the homotopy Jacobi identities~\eqref{eq:homotopyJacobi} imply that $\mu_1$ and $\tilde\mu_1$ are differentials. Therefore, we may consider their cohomologies $H^\bullet_{\mu_1}(\frL_{\rm BV})\coloneqq\bigoplus_{k\in\IZ}H^k_{\mu_1}(\frL_{\rm BV})$ and $H^\bullet_{\tilde\mu_1}(\tilde\frL_{\rm BV})\coloneqq\bigoplus_{k\in\IZ}H^k_{\tilde\mu_1}(\tilde\frL_{\rm BV})$. We also note that the identity~\eqref{eq:morphism_L} implies that $\phi_1$ is a cochain map, that is, $\tilde\mu_1\circ \phi_1=\phi_1\circ\mu_1$ and thus descends to a map $H^\bullet_{\mu_1}(\frL_{\rm BV})\rightarrow H^\bullet_{\tilde\mu_1}(\tilde\frL_{\rm BV})$ on the cohomologies. Quasi-isomorphisms are those morphisms for which $\phi_1$ induces an isomorphism on cohomology. 
        
        Under quasi-isomorphisms, the physical theory remains unchanged as is explained in \cref{app:homotopyAlgebras}, see also~\cite{Kajiura:2001ng,Kajiura:2003ax,Jurco:2018sby,Jurco:2019bvp,Jurco:2020yyu}. In particular, the BV actions $S_{\rm BV}$ and $\tilde S_{\rm BV}$ for $\frL_{\rm BV}$ and $\tilde\frL_{\rm BV}$ are related as $\tilde S_{\rm BV}=\phi^*S_{\rm BV}$, where we used the pullback $\phi^*:\scC^\infty(\tilde\frL_{\rm BV}[1])\rightarrow\scC^\infty(\frL_{\rm BV}[1])$ dual to the morphism $\phi$. Consequently, quasi-isomorphisms constitute the correct notion of equivalence\footnote{Here, we are a bit cavalier about the inclusion of the cyclic structure; again, see~\cite{Jurco:2018sby} for some more details.}.
        
        Because the $Q_{\rm BV}$-cohomologies in ghost numbers different from zero (i.e.~dual to $L_\infty$-degree one) are not measurable, one may wonder if the notion of a full quasi-isomorphism is not too restrictive. For perturbation theory, agreement in $H^1_{\mu_1}(\frL_{\rm BV})$ is certainly sufficient, and this can often be extended to an agreement in further cohomologies, cf.~e.g.~\cite[Appendix C]{Saemann:2019dsl}. Moreover, some fields in $L_\infty$-degree zero, such as e.g.~anti-fields of anti-ghosts and Nakanishi--Lautrup fields, are often unphysical, and arise only as internal fields in loop diagrams. Therefore their contributions to $H^1_{\mu_1}(\frL_{\rm BV})$ can also be disregarded when identifying physical observables. At a technical level, one can restrict these fields such that the kernel of the differential operator describing their linearised equations of motion vanishes, cf.~\cref{rem:Schwartz_issues}.
        
        \paragraph{Tree-level scattering amplitudes.} 
        There is an $L_\infty$-structure $\mu_i^\circ$ with vanishing differential $\mu^\circ_1$ on the cohomology $\frL_{\rm BV}^\circ\coloneqq H^\bullet_{\mu_1}(\frL_{\rm BV})$ of an $L_\infty$-algebra $(\frL_{\rm BV},\mu_i)$ such that $\frL_{\rm BV}^\circ$ and $\frL_{\rm BV}$ are quasi-isomorphic. This $L_\infty$-algebra $\frL_{\rm BV}^\circ$ is called the \uline{minimal model} of $\frL_{\rm BV}$, cf.~\cref{app:homotopyAlgebras}. The minimal model corresponds to a field theory equivalent to the original field theory, but without any propagating degrees of freedom. Its higher products therefore have to be the tree-level scattering amplitudes~\cite{Zwiebach:1992ie,Jurco:2018sby,Nutzi:2018vkl,Macrelli:2019afx}.
        
        The relation between $\frL_{\rm BV}$ and $\frL_{\rm BV}^\circ$ can be understood as follows. We start from the underlying cochain complexes and the following diagram:
        \begin{subequations}
            \begin{equation}\label{eq:diag_contracting_homotopy_L}
                \begin{tikzcd}
                    \ar[loop,out=160,in= 200,distance=20,"\sfh" left] (\frL_{\rm BV},\mu_1)\arrow[r,twoheadrightarrow,shift left]{}{\sfp} & (\frL_{\rm BV}^\circ,0) \arrow[l,hookrightarrow,shift left]{}{\sfe}~.
                \end{tikzcd}
            \end{equation}
            Here, $\sfp$ is the obvious projection onto the cohomology, and $\sfe$ is a choice of embedding (involving choices, e.g.~a choice of gauge for gauge theories). The quasi-isomorphism also gives rise to a contracting homotopy $\sfh$, which is a linear map of degree~$-1$. The maps $\sfe$ and $\sfh$ can be chosen such that 
            \begin{equation}
                \begin{gathered}
                    \sfid \ =\ \mu_1\circ\sfh+\sfh\circ\mu_1+\sfe\circ\sfp~,
                    \\
                    \sfp\circ\sfe\ =\ \sfid ~,
                    \\
                    \sfp\circ\sfh\ =\ \sfh\circ\sfe\ =\ \sfh\circ\sfh\ =\ 0~,
                    \\
                    \sfp\circ \mu_1\ =\ \mu_1\circ\sfe\ =\ 0~.
                \end{gathered}
            \end{equation}
        \end{subequations}
        Mathematically, this is an abstract \uline{Hodge--Kodaira decomposition}. The map $\sfh$ in $L_\infty$-degree two turns out to be the (Feynman--'t Hooft) propagator of the physical theory in question~\cite{Mnev:2006ch,Mnev:2008sa,Cattaneo:0811.2045}, see also~\cite{Cattaneo:2015vsa} and references therein.
        
        We directly extend the diagram~\eqref{eq:diag_contracting_homotopy_L} to the Chevalley--Eilenberg picture, where we have 
        \begin{subequations}
            \begin{equation}\label{eq:contractingHomotopyZero}
                \begin{gathered}
                    \begin{tikzcd}
                        \ar[loop,out=194,in= 166,distance=20,"\sfH_0"]\kern20pt(\scC^\infty(\frL_{\rm BV}[1]),Q_{\rm BV,0})\arrow[r,twoheadrightarrow,shift left]{}{\sfE_0} & (\scC^\infty(\frL^\circ_{\rm BV}[1]),0) \arrow[l,shift left,hookrightarrow]{}{\sfP_0} 
                    \end{tikzcd}
                    \\
                    \sfid \ =\ \sfP_0\circ\sfE_0+Q_{\rm BV,0}\circ\sfH_0+\sfH_0\circ Q_{\rm BV,0}~,
                    \\
                    \sfE_0\circ\sfP_0\ =\ \sfid~,
                    \\
                    \sfE_0\circ\sfH_0\ =\ \sfH_0\circ\sfP_0\ =\ \sfH_0\circ\sfH_0\ =\ 0~,
                    \\
                    \sfE_0\circ Q_{\rm BV,0}\ =\ Q_{\rm BV,0}\circ\sfP_0\ =\ 0~,
                \end{gathered}
            \end{equation}
            where $Q_{\rm BV,0}$ is the linear part of $Q_{\rm BV}$, which encodes the differential $\mu_1$. The maps $\sfE_0$, $\sfP_0$, and $\sfH_0$ are defined by the `tensor trick'~\cite{gugenheim1991perturbation} as 
            \begin{equation}
                \sfF_0\ =\ \sum_{i\geq1}\frac{1}{i!}(\sfF_0)^i\efor\sfF_0\ \in\ \{\sfE_0,\sfP_0,\sfH_0\}
            \end{equation}
            with
            \begin{equation}\label{eq:E0P0H0}
                (\sfE_0)^i\ \coloneqq\ (\sfe^*)^{\odot i}~,~~
                (\sfP_0)^i\ \coloneqq\ (\sfp^*)^{\odot i}~,~~
                (\sfH_0)^i\ \coloneqq\kern-8pt \sum_{k+l=i-1}1^{\odot k}\odot\sfh^*\odot(\sfp^*\circ\sfe^*)^{\odot l}~.
            \end{equation}
        \end{subequations}
        We can now regard the non-linear part 
        \begin{equation}
            \delta\ \coloneqq\ Q_{\rm BV}-Q_{\rm BV,0}
        \end{equation}
        of $Q_{\rm BV}$ as a perturbation and use the homological perturbation lemma~\cite{gugenheim1991perturbation,Crainic:0403266}, which asserts that there is a contracting homotopy
        \begin{subequations}
            \begin{equation}
                \begin{gathered}
                    \begin{tikzcd}
                        \ar[loop,out=194,in= 166,distance=20,"\sfH"]\kern15pt(\scC^\infty(\frL_{\rm BV}[1]),Q_{\rm BV})\arrow[r,twoheadrightarrow,shift left]{}{\sfE} & (\scC^\infty(\frL^\circ_{\rm BV}[1]),Q^\circ_{\rm BV}) \arrow[l,shift left,hookrightarrow]{}{\sfP} 
                    \end{tikzcd}\\
                    \sfid\ =\ \sfP\circ\sfE+Q_{\rm BV}\circ\sfH+\sfH\circ Q_{\rm BV}~,
                    \\
                    \sfE\circ\sfP\ =\ \sfid~,
                    \\
                    \sfE\circ\sfH\ =\ \sfH\circ\sfP\ =\ \sfH\circ\sfH\ =\ 0,
                    \\
                    \sfE\circ Q_{\rm BV}\ =\ Q^\circ_{\rm BV}\circ\sfE~,~~~Q_{\rm BV}\circ\sfP\ =\ \sfP\circ Q^\circ_{\rm BV}
                \end{gathered}
            \end{equation}
            in the deformed setting. In particular,
            \begin{equation}
                \begin{gathered}
                    \sfE\ =\ \sfE_0\circ(\sfid+\delta\circ\sfH_0)^{-1}~,~~~
                    \sfH\ =\ \sfH_0\circ(\sfid+\delta\circ\sfH_0)^{-1}~,
                    \\
                    \sfP\ =\ \sfP_0-\sfH\circ\delta\circ\sfP_0~,~~~
                    Q^\circ_{\rm BV}\ =\ \sfE\circ\delta\circ\sfP_0~,
                \end{gathered}
            \end{equation}
        \end{subequations}        
        and $Q_{\rm BV}^\circ$ is the Chevalley--Eilenberg differential encoding the higher products of the minimal model and thus its tree-level scattering amplitudes. Note that here, the inverse operators are to be seen as geometric series.\footnote{Because we are interested in perturbation theory, we do not have to concern ourselves with convergence issues.} The formula for $Q^\circ_{\rm BV}$ is then recursive, which has interesting consequences~\cite{Macrelli:2019afx,Jurco:2019yfd}.
        
        Translated to the dual picture, the homological perturbation lemma yields the following formulas for the quasi-isomorphism $\phi:(\frL_{\rm BV},\mu_i)\rightarrow(\frL_{\rm BV}^\circ,\mu_i^\circ)$~\cite{Kajiura:2003ax}:
        \begin{subequations}
            \begin{equation}\label{eq:minimalQuasiIso}
                \begin{aligned}
                    \phi_1(\ell^\circ_1)\ &\coloneqq\ \sfe(\ell^\circ_1)~,
                    \\
                    \phi_2(\ell^\circ_1,\ell^\circ_2)\ &\coloneqq\ - (\sfh\circ\mu_2)(\phi_1(\ell^\circ_1),\phi_1(\ell^\circ_2))~,
                    \\
                    &~~\vdots
                    \\
                    \phi_i(\ell^\circ_1,\ldots,\ell^\circ_i)\ &\coloneqq\ -\sum_{j=2}^i\frac{1}{j!} \sum_{k_1+\cdots+k_j=i}\sum_{\sigma\in{\rm \overline{Sh}}(k_1,\ldots,k_{j-1};i)}\chi(\sigma;\ell^\circ_1,\ldots,\ell^\circ_i)\zeta(\sigma;\ell^\circ_1,\ldots,\ell^\circ_i)\,\times
                    \\
                    &\kern1cm\times(\sfh\circ\mu_j)\big(\phi_{k_1}\big(\ell^\circ_{\sigma(1)},\ldots,\ell^\circ_{\sigma(k_1)}\big),\ldots,\phi_{k_j}\big(\ell^\circ_{\sigma(k_1+\cdots+k_{j-1}+1)},\ldots,\ell^\circ_{\sigma(i)}\big)\big)\,,
                \end{aligned}
            \end{equation}
            and the products $\mu^\circ_i:\frL_{\rm BV}^\circ\times\cdots\times\frL_{\rm BV}^\circ\to\frL_{\rm BV}^\circ$ are constructed recursively as
            \begin{equation}\label{eq:minimalProducts}
                \begin{aligned}
                    \mu^\circ_1(\ell^\circ_1)\ &\coloneqq\ 0~,
                    \\
                    \mu^\circ_2(\ell^\circ_1,\ell^\circ_2)\ &\coloneqq\ (\sfp\circ\mu_2)(\phi_1(\ell^\circ_1),\phi_1(\ell^\circ_2))~,
                    \\
                    &~~\vdots\\
                    \mu^\circ_i(\ell^\circ_1,\ldots,\ell^\circ_i)\ &\coloneqq\ \sum_{j=2}^i\frac{1}{j!} \sum_{k_1+\cdots+k_j=i}\sum_{\sigma\in{\rm \overline{Sh}}(k_1,\ldots,k_{j-1};i)}\chi(\sigma;\ell^\circ_1,\ldots,\ell^\circ_i)\zeta(\sigma;\ell^\circ_1,\ldots,\ell^\circ_i)\,\times
                    \\
                    &\kern1cm\times(\sfp\circ\mu_j)\big(\phi_{k_1}\big(\ell^\circ_{\sigma(1)},\ldots,\ell^\circ_{\sigma(k_1)}\big),\ldots,\phi_{k_j}\big(\ell^\circ_{\sigma(k_1+\cdots+k_{j-1}+1)},\ldots,\ell^\circ_{\sigma(i)}\big)\big)\,,
                \end{aligned}
            \end{equation}
        \end{subequations}
        where $\ell^\circ_1,\ldots,\ell^\circ_i\in\frL_{\rm BV}^\circ$. Here, $\chi$ and $\zeta$ are again the Koszul sign~\eqref{eq:KoszulSign} and the sign factor~\eqref{eq:zeta-sign}, respectively.
        
        Using the higher products of the minimal model, $n$-point tree-level scattering amplitudes of the free fields $a^\circ_1,\ldots,a^\circ_n\in H^1_{\mu_1}(\frL_{\rm BV})$ are then computed using formula~\cite{Macrelli:2019afx} (see also~\cite{Kajiura:2001ng,Kajiura:2003ax} for the case of string field theory) 
        \begin{equation}\label{eq:treeLevelAmplitude}
            \scA_{n,0}(a^\circ_1,\ldots,a^\circ_n)\ =\ \rmi\inner{a^\circ_1}{\mu^\circ_{n-1}(a^\circ_2,\ldots,a^\circ_n)}~.
        \end{equation}
        Furthermore, in~\cite{Macrelli:2019afx} it was shown that the recursion relations~\eqref{eq:minimalQuasiIso} encode the famous Berends--Giele recursion relations~\cite{Berends:1987me} for gluon scattering in Yang--Mills theory. For a related discussion of the S-matrix in the language of $L_\infty$-algebras, see also~\cite{Arvanitakis:2019ald} as well as~\cite{Nutzi:2018vkl,Reiterer:2019dys} for an interpretation of tree-level on-shell recursion relations in terms of $L_\infty$-algebras.
        
        \paragraph{Loop-level scattering amplitudes.}
        In order to extend the above discussion to recursion relations for loop-level amplitudes, we follow~\cite{Doubek:2017naz,Jurco:2019yfd,Jurco:2020yyu}. Recall that in the transition from the classical to the quantum master equation, the classical BV operator is deformed in powers of $\hbar$ according to
        \begin{equation}
            Q_{\rm BV}\ \coloneqq\ \{S_{\rm BV},-\}~~~\rightarrow~~~Q_{\rm qBV}\ \coloneqq\ \{S^\hbar_{\rm qBV},-\}-2\rmi\hbar\lap
            \ewith 
            S^\hbar_{\rm qBV}\ =\ S_{\rm BV}+\caO(\hbar)~.
        \end{equation}
        Consequently, the perturbation 
        \begin{equation}
            \delta\ \coloneqq\ Q_{\rm qBV}-Q_{\rm qBV,0}\ =\ Q_{\rm qBV}-Q_{\rm BV,0}
        \end{equation}
        between the full and linearised part of $Q_{\rm qBV}$ is now also deformed in powers of $\hbar$. Starting again from the diagram~\eqref{eq:contractingHomotopyZero}, we use the homological perturbation lemma to obtain a contracting homotopy 
        \begin{subequations}\label{eq:deformedCH}
            \begin{equation}
                \begin{gathered}
                    \begin{tikzcd}
                        \ar[loop,out=194,in= 166,distance=20,"\sfH"]\kern15pt(\scC^\infty(\frL_{\rm BV}[1]),Q_{\rm qBV})\arrow[r,twoheadrightarrow,shift left]{}{\sfE} & (\scC^\infty(\frL^\circ_{\rm BV}[1]),Q^\circ_{\rm qBV}) \arrow[l,shift left,hookrightarrow]{}{\sfP} 
                    \end{tikzcd}\\
                    \sfid\ =\ \sfP\circ\sfE+Q_{\rm qBV}\circ\sfH+\sfH\circ Q_{\rm qBV}~,
                    \\
                    \sfE\circ\sfP\ =\ \sfid~,
                    \\
                    \sfE\circ\sfH\ =\ \sfH\circ\sfP\ =\ \sfH\circ\sfH\ =\ 0~,
                    \\
                    \sfE\circ Q_{\rm qBV}\ =\ Q^\circ_{\rm qBV}\circ\sfE~,~~~
                    Q_{\rm qBV}\circ\sfP\ =\ \sfP\circ Q^\circ_{\rm qBV}~,
                \end{gathered}
            \end{equation}
            where 
            \begin{equation}
                \begin{gathered}
                    \sfE\ =\ \sfE_0\circ(\sfid+\delta\circ\sfH_0)^{-1}~,~~~
                    \sfH\ =\ \sfH_0\circ(\sfid+\delta\circ\sfH_0)^{-1}~,
                    \\
                    \sfP\ =\ \sfP_0-\sfH\circ\delta\circ\sfP_0~,~~~
                    Q^\circ_{\rm qBV}\ =\ \sfE\circ\delta\circ\sfP_0~.
                \end{gathered}
            \end{equation}
        \end{subequations}
        Note that because $\delta$ contains the second order differential operator $\Delta$, none of the maps will be algebra morphisms in general; this is just a consequence of the fact that $Q^\circ_{\rm qBV}$ defines a loop homotopy algebra.
        
        Importantly, the differential $Q^\circ$ can be written as~\cite{Pulmann:2016aa,Doubek:2017naz}
        \begin{equation}\label{eq:QQmin}
            Q^\circ_{\rm qBV}\ =\ \{W^\hbar_{\rm qBV},-\}^\circ-2\rmi\hbar\lap^\circ~,
        \end{equation}
        where $\{-,-\}^\circ$ and $\lap^\circ$ are the anti-bracket and the BV Laplacian on $\scC^\infty(\frL^\circ_{\rm BV}[1])$, respectively, and $W^\hbar_{\rm qBV}$ is of the form~\eqref{eq:QBVAction} but with $\mu_{1,L=0}^\circ=0$. Altogether, we obtain $(\frL^\circ_{\rm BV}[1],Q^\circ_{\rm qBV})$ which corresponds to a quantum $L_\infty$-structure on $H^\bullet_{\mu_{1,L=0}}(\frL_{\rm BV})$ with a differential that vanishes to zeroth order in $\hbar$.
        
        The quantum BV action $W^\hbar_{\rm qBV}$ is the action that encodes all scattering amplitudes to arbitrary loop order in perturbation theory.\footnote{One should not confuse the quantum BV action with the one-particle-irreducible effective action or the Wilsonian effective action, even though it has the form of $\hbar$-corrections to the classical action.} In particular, for theories for which the classical BV action also satisfies the quantum master equation, which includes scalar field theory, Yang--Mills theory, and Chern--Simons theory, $L$ coincides with the loop expansion order and hence, the products $\mu_{n-1,L}^\circ$ are the $L$-loop integrands for the $n$-point scattering amplitude. Consequently,~\eqref{eq:treeLevelAmplitude} generalises to
        \begin{equation}
            \scA_{n,L}(a^\circ_1,\ldots,a^\circ_n)\ =\ \rmi\,\inner{a^\circ_1}{\mu^\circ_{n-1,L}(a^\circ_2,\ldots,a^\circ_n)}~.
        \end{equation}
        To construct the $\mu_{i,L}$, we note that~\eqref{eq:deformedCH} immediately implies
        \begin{equation}\label{eq:ERecursion}
            \sfE\ =\ \sfE_0-\sfE\circ\delta\circ\sfH_0
        \end{equation}
        which is a recursion relation for $\sfE$. Hence, we can iterate this equation to obtain $\sfE$ recursively and substitute the result into $Q^\circ_{\rm qBV}=\sfE\circ\delta\circ\sfP_0$ from~\eqref{eq:deformedCH} with $\sfP_0$ given in~\eqref{eq:E0P0H0}. We conclude, in analogy with~\eqref{eq:QsupermuQuantum}, that
        \begin{equation}\label{eq:QRecursion}
            Q^\circ_{\rm qBV}\sfa^\circ\ =\ -\sum_{\substack{i\geq1\\ L\geq0}}\frac{\hbar^L}{i!}\mu'^\circ_{i,L}(\sfa^\circ,\ldots,\sfa^\circ)~,
        \end{equation}
        from which the $\mu'^\circ_{i,L}$ and thus the $\mu^\circ_{i,L}$ can be read off. We refer to~\cite{Jurco:2019yfd,Jurco:2020yyu} for full details. It is not difficult to see that for $\hbar\to0$, the recursion relation~\eqref{eq:ERecursion} coincides with the recursion relation~\eqref{eq:minimalQuasiIso} and~\eqref{eq:QRecursion} with that for the products~\eqref{eq:minimalProducts} for the minimal model at the tree level.
        
        \section{Examples of homotopy algebras of field theories}\label{sec:Examples}
        
        In the following, we review the actions, the BV complexes and the dual $L_\infty$-algebra structures of the field theories relevant to our homotopy algebraic treatment of the double copy. We note that none of the theories we discuss in this section requires the BV formalism for quantisation. As explained before, however, it does make the link to homotopy algebras evident and clarifies the freedom we have in choosing gauges, an important aspect in our later discussion.
        
        \subsection{Biadjoint scalar field theory}\label{sec:biadjoint}
        
        We start with the simplest relevant field theory, namely biadjoint scalar field theory with cubic interaction. This theory appeared in the scattering amplitudes and double copy literature in various incarnations~\cite{Hodges:2011wm,Vaman:2010ez,Cachazo:2013iea,Monteiro:2013rya,Cachazo:2014xea,Monteiro:2014cda,Chiodaroli:2014xia,Naculich:2014naa,Luna:2015paa,Naculich:2015zha,Chiodaroli:2015rdg,Luna:2016due,White:2016jzc,Cheung:2016prv,Chiodaroli:2017ngp,Brown:2018wss}. 
        
        In particular, let $\frg$ and $\bar\frg$ be two semi-simple compact matrix Lie algebras. For $(\frg\otimes\bar\frg)$-valued functions on Minkowski space $\IM^d$, we define a symmetric bracket and an inner product by linearly extending
        \begin{equation}
            \begin{aligned}
                [\cb_1\otimes\bar\cb_1,\cb_2\otimes\bar\cb_2]_{\frg\otimes\bar\frg}\ &\coloneqq\ [\cb_1,\cb_2]_\frg\otimes[\bar\cb_1,\bar\cb_2]_{\bar\frg}~,
                \\
                \inner{\cb_1\otimes\bar\cb_1}{\cb_2\otimes\bar\cb_2}_{\frg\otimes\bar\frg}\ &\coloneqq\ \tr_\frg(\cb_1\cb_2)\,\tr_{\bar\frg}(\bar\cb_1\bar\cb_2)
            \end{aligned}
        \end{equation}
        for all $\cb_{1,2}\in\frg$ and $\bar\cb_{1,2}\in\bar\frg$. 
        
        \paragraph{BV action and BV operator.}
        The BV action for biadjoint scalar field theory then reads as 
        \begin{equation}\label{eq:biadjoint:action}
            S^{\rm biadj}\ \coloneqq\ \int\rmd^d x\,\Big\{\tfrac{1}{2}\inner{\varphi}{\wave\varphi}_{\frg\otimes\bar\frg}-\tfrac{\lambda}{3!}\inner{\varphi}{[\varphi,\varphi]_{\frg\otimes\bar\frg}}_{\frg\otimes\bar\frg}\Big\}~,
        \end{equation}
        where $\lambda$ is a coupling constant, $\wave\coloneqq\eta^{\mu\nu}\partial_\mu\partial_\nu$, and $\varphi$ is a scalar field taking values in $\frg\otimes\bar\frg$. We write $\varphi\in(\frg\otimes\bar\frg)\otimes\frF$ where $\frF$ is a suitable function space discussed shortly. Introducing basis vectors $\cb_a$ and $\bar\cb_{\bar a}$ on $\frg$ and $\bar\frg$, respectively, we can rewrite this action in component form
        \begin{subequations}
            \begin{equation}
                S^{\rm biadj}\ =\ \int\rmd^dx\,\Big\{\tfrac12\varphi_{a\bar a}\wave\varphi^{a\bar a}-\tfrac{\lambda}{3!}f_{abc}f_{\bar a\bar b\bar c}\varphi^{a\bar a}\varphi^{b\bar b}\varphi^{c\bar c}\Big\}~,
            \end{equation}
            where 
            \begin{equation}
                \begin{gathered}
                    \tr_\frg(\cb_a\cb_b)\ =\ -\delta_{ab}~,~~~
                    \tr_{\bar\frg}(\bar\cb_{\bar a}\bar\cb_{\bar b})\ =\ -\delta_{\bar a\bar b}~,
                    \\
                    f_{abc}\ \coloneqq\ -\tr_\frg(\cb_a[\cb_b,\cb_c]_\frg)~,~~~
                    f_{\bar a\bar b\bar c}\ \coloneqq\ -\tr_{\bar\frg}(\bar\cb_{\bar a}[\bar\cb_{\bar b},\bar \cb_{\bar c}]_{\bar\frg})~.
                \end{gathered}
            \end{equation}
        \end{subequations}
        Besides the field $\varphi$, we also have the anti-field $\varphi^+$ and the BV operator~\eqref{eq:CQ} acts according to
        \begin{equation}\label{eq:BVOperatorBiadj}
            \varphi^{a\bar a}\ \mapstomap{Q_{\rm BV}}\ 0
            \eand
            \varphi^{+a\bar a}\ \mapstomap{Q_{\rm BV}}\ \wave\varphi^{a\bar a}-\tfrac\lambda2{f_{bc}}^a{f_{\bar b\bar c}}^{\bar a}\varphi^{b\bar b}\varphi^{c\bar c}~.
        \end{equation}
        
        \paragraph{\mathversion{bold}$L_\infty$-algebra.}
        The BV operator~\eqref{eq:BVOperatorBiadj} is the Chevalley--Eilenberg differential of an $L_\infty$-algebra $\frL^{\rm biadj}_{\rm BV}$ which has the underlying cochain complex\footnote{Here, `$*$' denotes the trivial vector space.}
        \begin{equation}\label{eq:biadjointComplex}
            \begin{tikzcd}
                *\arrow[r] & \underbrace{\stackrel{\varphi^{a\bar a}}{(\frg\otimes\bar\frg)\otimes\frF}}_{\frL^{\rm biadj}_{\rm BV,\,1}} \arrow[r,"\wave"] & \underbrace{\stackrel{\varphi^{+a\bar a}}{(\frg\otimes\bar\frg)\otimes\frF}}_{\frL^{\rm biadj}_{\rm BV,\,2}} \arrow[r] & *
            \end{tikzcd}
        \end{equation}
        with cyclic inner product
        \begin{equation}
            \inner{\varphi}{\varphi^+}\ \coloneqq\ \int\rmd^dx\,\varphi^{a\bar a}\varphi^+_{a\bar a}~,
        \end{equation}
        and the only non-trivial higher product is 
        \begin{equation}
            (\varphi^{a\bar{a}},\varphi^{b\bar{b}})\ \mapstomap{\mu_2}\ -\lambda{f_{bc}}^a{f_{\bar b\bar c}}^{\bar a}\varphi^{b\bar b}\varphi^{c\bar c}~.
        \end{equation}
        At this point it is important to recall \cref{rem:fields_vs_coordinate_functions} and that we always use the same symbol for a coordinate function on field space and the corresponding elements of field space.
        
        The field space $\frF$ can roughly be thought of as the smooth functions of Minkowski space $\scC^\infty(\IM^d)$. More precisely, however, the field space is the direct sum of interacting fields and solutions to the (colour-stripped) equations of motion, cf.~\cref{rem:Schwartz_issues}.
        
        \subsection{Yang--Mills theory}\label{ssec:ex:YM}
        
        A key player in the double copy is Yang--Mills theory on $d$-dimensional Minkowski space $\IM^d$ with a semi-simple compact matrix Lie algebra $\frg$ as gauge algebra. The gauge potential $A_\mu^a$ is a one-form on $\IM^d$ taking values in $\frg$. Let $\nabla$ be the connection with respect to $A$. Infinitesimal gauge transformations act according to
        \begin{equation}
            A_\mu^a\ \mapsto\ \tilde A_\mu^a\ \coloneqq\ A_\mu^a+(\nabla_\mu c)^a
            \eforall
            c\ \in\ \scC^\infty(\IM^d)\otimes\frg~.
        \end{equation}
        
        \paragraph{BV action and BV operator.}
        The list of all the fields required in the BV formulation of Yang--Mills theory together with their properties is found in \cref{tab:fields:YM}, and the BV action is~\cite{Batalin:1981jr}
        \begin{equation}\label{eq:BVActionYM}
            S_{\rm BV}^{\rm YM}\ \coloneqq\ \int \rmd^dx\,\Big\{\!-\tfrac14F_{a\mu\nu}F^{a\mu\nu}+A^+_{a\mu}(\nabla^\mu c)^a+\tfrac g2{f_{bc}}^ac^+_ac^bc^c-b^a\bar c^+_a\Big\}~.
        \end{equation}
        As in \cref{sec:BVMotivation}, all the fields are rescaled such that the Yang--Mills coupling constant $g$ appears in all interaction vertices. Consequently, the BV operator~\eqref{eq:CQ} acts as
        \begin{equation}\label{eq:BVOperatorYM}
            \begin{gathered}
                c^a\ \mapstomap{Q_{\rm BV}}\ -\tfrac g2{f_{bc}}^ac^bc^c~,~~~
                c^{+a}\ \mapstomap{Q_{\rm BV}}\ -(\nabla^\mu A^+_\mu)^a-g{f_{bc}}^ac^bc^{+c}~,
                \\
                A_\mu^a\ \mapstomap{Q_{\rm BV}}\ (\nabla_\mu c)^a~,~~~
                A^{+a}_\mu\ \mapstomap{Q_{\rm BV}}\ (\nabla^\nu F_{\nu\mu})^a-g{f_{bc}}^aA^{+b}_\mu c^c~,
                \\
                b^a\ \mapstomap{Q_{\rm BV}}\ 0~,~~~
                b^{+a}\ \mapstomap{Q_{\rm BV}}\ -\bar c^{+a}~,
                \\
                \bar c^a\ \mapstomap{Q_{\rm BV}}\ b^a~,~~~
                \bar c^{+a}\ \mapstomap{Q_{\rm BV}}\ 0~.
            \end{gathered}
        \end{equation}
        
        \begin{table}[ht]
            \begin{center}
                \resizebox{\textwidth}{!}{
                    \begin{tabular}{|c|l|c|c|c|c|c|c|c|}
                        \hline
                        \multicolumn{5}{|c|}{fields} & \multicolumn{4}{c|}{anti-fields}
                        \\
                        \hline
                        & role & $|-|_{\rm gh}$ & $|-|_\frL$ & dim & & $|-|_{\rm gh}$ & $|-|_\frL$ & dim
                        \\
                        \hline
                        $c^a$ & ghost field & 1 & 0 & $\tfrac{d}{2}-2$ & $c^{+a}$ & $-2$ & 3 & $\tfrac{d}{2}+2$ 
                        \\
                        $A_\mu^a$ & physical field & 0 & 1 & $\tfrac{d}{2}-1$ & $A^{+a}_\mu$ & $-1$ & 2 & $\tfrac{d}{2}+1$
                        \\
                        $b^a$ & Nakanishi--Lautrup field & 0 & 1 & $\tfrac{d}{2}$ & $b^{+a}$ & $-1$ & $2$ & $\tfrac{d}{2}$ 
                        \\
                        $\bar c^a$ & anti-ghost field & $-1$ & $2$ & $\tfrac{d}{2}$ & $\bar c^{+a}$ & 0 & 1 & $\tfrac{d}{2}$
                        \\
                        \hline
                    \end{tabular}
                }
            \end{center}
            \caption{The full set of BV fields for Yang--Mills theory on $\IM^d$ with gauge Lie algebra $\frg$, including their ghost numbers, their $L_\infty$-degrees, and their mass dimensions. The mass dimension of the coupling constant $g$ is $2-\frac{d}{2}$.}\label{tab:fields:YM}
        \end{table}
        
        \paragraph{\mathversion{bold}$L_\infty$-algebra.}
        The BV operator~\eqref{eq:BVOperatorYM} is the Chevalley--Eilenberg differential of an $L_\infty$-algebra which we shall denote by $\frL^{\rm YM}_{\rm BV}$. This $L_\infty$-algebra has the underlying complex\footnote{This complex has been rediscovered several times in the literature. For early references, see~\cite{Zeitlin:2007vv,Zeitlin:2007yf}; more detailed historical references are found in~\cite{Jurco:2018sby}.}
        \begin{subequations}\label{eq:YMBVLAlgebra}
            \begin{equation}\label{eq:BVComplexYM}
                \begin{tikzcd}[column sep=50pt]
                    & \stackrel{A_\mu^a}{\Omega^1(\IM^d)\otimes\frg}\arrow[r,"-(\partial_\nu\partial^\mu-\delta^\mu_\nu\wave)"] & \stackrel{A^{+a}_\mu}{\Omega^1(\IM^d)\otimes\frg}\arrow[rdd,"-\partial^\mu"] & 
                    \\
                    &\stackrel{b^a}{\scC^\infty(\IM^d)\otimes\frg}\arrow[start anchor=south east, end anchor= north west,dr,"~\sfid",pos=0.05] & \stackrel{b^{+a}}{\scC^\infty(\IM^d)\otimes\frg} 
                    \\
                    \underbrace{\spacer{2ex}\stackrel{c^a}{\scC^\infty(\IM^d)\otimes\frg}}_{\eqqcolon\,\frL^{\rm YM}_{\rm BV,\,0}}
                    \arrow[ruu,"-\partial_\mu"] & \underbrace{\spacer{2ex}\stackrel{\bar c^{+a}}{\scC^\infty(\IM^d)\otimes\frg}}_{\eqqcolon\,\frL^{\rm YM}_{\rm BV,\,1}}\arrow[start anchor=north east, end anchor= south west,crossing over,ur,"-\sfid",pos=0.05,swap] & \underbrace{\spacer{2ex}\stackrel{\bar c^a}{\scC^\infty(\IM^d)\otimes\frg}}_{\eqqcolon\,\frL^{\rm YM}_{\rm BV,\,2}} & \underbrace{\spacer{2ex}\stackrel{c^{+a}}{\scC^\infty(\IM^d)\otimes\frg}}_{\eqqcolon\,\frL^{\rm YM}_{\rm BV,\,3}}
                \end{tikzcd}
            \end{equation}
            We shall label the subspaces $\frL^{\rm YM}_{{\rm BV},\,i}$ to which the various fields belong by the corresponding subscripts, that is,
            \begin{equation}
                \begin{gathered}
                    \frL^{\rm YM}_{\rm BV,\,0}\ =\ \frL^{\rm YM}_{{\rm BV},\,0,\,c}~,~~~
                    \frL^{\rm YM}_{\rm BV,\,1}\ =\ \spacee{\frL^{\rm YM}}{\rm BV,\,1}{A,\,b,\,\bar c^+}~,
                    \\
                    \frL^{\rm YM}_{\rm BV,\,2}\ =\ \spacee{\frL^{\rm YM}}{\rm BV,\,2}{A^+,\,b^+,\,\bar c}~,~~~
                    \frL^{\rm YM}_{\rm BV,\,3}\ =\ \frL^{\rm YM}_{3,\,c^+}~,
                \end{gathered}
            \end{equation}
            and the non-trivial actions of the differential $\mu_1$ in $\frL^{\rm YM}_{{\rm BV},\,i}$ are
            \begin{equation}
                \begin{aligned}
                    c^a\ &\mapstomap{\mu_1}\ -\partial_\mu c^a\ \in\ \frL^{\rm YM}_{{\rm BV},\,1,\,A}~,
                    \\
                    \colvec{
                        A_\mu^a,
                        b^a,
                        \bar c^{+a}
                    }
                    \ &\mapstomap{\mu_1}\ 
                        \colvec{
                            -(\partial_\mu\partial^\nu-\delta^\nu_\mu\wave)A^a_\nu,
                            -\bar c^{+a},
                            b^a
                        }
                    \ \in\ \spacee{\frL^{\rm YM}}{\rm BV,\,2}{A^+,\,b^+,\,\bar c}~,
                    \\
                    A^{+a}_{\mu}\ &\mapstomap{\mu_1}\ -\partial^\mu A^{+a}_\mu\ \in\ \frL^{\rm YM}_{{\rm BV},\,3,\,c^+}~.
                \end{aligned}
            \end{equation}
            The non-vanishing higher products are
            \begin{equation}
                \begin{aligned}
                    (c^a,c^b)\ &\mapstomap{\mu_2} \ g{f_{bc}}^ac^bc^c\ \in\ \frL^{\rm YM}_{{\rm BV},\,0,\,c}~,
                    \\
                    (A_\mu^a,c^b)\ &\mapstomap{\mu_2}\ -g{f_{bc}}^aA_\mu^bc^c\ \in\ \frL^{\rm YM}_{{\rm BV},\,1,\,A}~,
                    \\
                    (A^{+a}_\mu,c^b)\ &\mapstomap{\mu_2}\ -g{f_{bc}}^aA_\mu^{+b}c^c\ \in\ \frL^{\rm YM}_{{\rm BV},\,2,\,A^+}~,
                    \\
                    (A_\mu^a,A_\nu^b)\ &\mapstomap{\mu_2}\ 2g{f_{bc}}^a\Big(\partial^\nu(A_\nu^bA_\mu^c)+2A^{b\nu}\partial_{[\nu}A^c_{\mu]}\Big)\ \in\ \frL^{\rm YM}_{{\rm BV},\,2,\,A^+}~,
                    \\
                    (c^a,c^{+b})\ &\mapstomap{\mu_2}\ g{f_{bc}}^ac^bc^{+c}\ \in\ \frL^{\rm YM}_{{\rm BV},\,3,\,c^+}~,
                    \\
                    (A_\mu^a,A^{+b}_\nu)\ &\mapstomap{\mu_2}\ -g{f_{bc}}^aA_\mu^bA^{+c\mu}\ \in\ \frL^{\rm YM}_{{\rm BV},\,3,\,c^+}~,
                    \\
                    (A_\mu^a,A_\nu^b,A_\kappa^c)\ &\mapstomap{\mu_3}\ 3!g^2A^{\nu c}A_\nu^d A_\mu^e {f_{ed}}^b{f_{bc}}^a\ \in\ \frL^{\rm YM}_{{\rm BV},\,2,\,A^+}~,
                \end{aligned}
            \end{equation}
        \end{subequations}
        and the general expressions follow from polarisation. One can check that $(\frL^{\rm YM}_{\rm BV},\mu_i)$ forms an $L_\infty$-algebra, and with the inner products
        \begin{equation}\label{eq:innerProductYM}
            \begin{aligned}
                \inner{A}{A^+}\ &\coloneqq\ \int\rmd^dx\,A^a_\mu A^{+\mu}_a~,~~
                &
                \inner{b}{b^+}\ &\coloneqq\ \int\rmd^dx\,b^ab^+_a~,
                \\
                \inner{c}{c^+}\ &\coloneqq\ \int\rmd^dx\,c^ac^+_a~,~~
                &
                \inner{\bar c}{\bar c^+}\ &\coloneqq\ -\int\rmd^dx\,\bar c^a\bar c^+_a~,
            \end{aligned}
        \end{equation}
        it becomes a cyclic $L_\infty$-algebra. Note that the superfield homotopy Maurer--Cartan action~\eqref{eq:CBVAction} reduces to the BV action~\eqref{eq:BVActionYM} when using these higher products and inner products together with~\eqref{eq:translation_BV_L_infty}.
        
        \paragraph{Gauge fixing.}
        We have discussed the general gauge-fixing procedure in the BV formalism in \cref{sec:BVMotivation}. Here, to implement $R_\xi$-gauge for some real parameter $\xi$, we choose the gauge-fixing fermion
        \begin{equation}\label{eq:gaugeFixingFermionYM}
            \Psi\ \coloneqq\ -\int\rmd^dx\,\bar c_a\big(\partial^\mu A^a_\mu+\tfrac\xi2 b^a\big)\,.
        \end{equation}
        Following~\eqref{eq:gaugeFixingCanonicalTrafo} and~\eqref{eq:gaugeFixingPathIntegral}, the Lagrangian of the resulting gauge-fixed BV action is
        \begin{equation}\label{eq:YMBVActionComponents}
            \begin{aligned}
                S_{\rm BV}^{\rm YM,\,gf}\ &= \ \int \rmd^dx\,\Big\{\!-\tfrac14F_{a\mu\nu}F^{a\mu\nu}-\bar c_a\partial^\mu(\nabla_\mu c)^a+\tfrac\xi2b_ab^a+b_a\partial^\mu A^a_\mu\,+
                \\
                &\kern3cm+A^+_{a\mu}(\nabla_\mu c)^a+\tfrac g2{f_{bc}}^ac^+_ac^bc^c-b^a\bar c^+_a\,\Big\}~,
            \end{aligned}
        \end{equation}
        and after putting to zero the anti-fields, we obtain
        \begin{equation}\label{eq:CGFYMA}
            S^{\rm YM}_{\rm BRST}\ =\ \int\rmd^dx\,\Big\{\!-\tfrac14F_{a\mu\nu}F^{a\mu\nu}-\bar c_a\partial^\mu(\nabla_\mu c)^a+\tfrac\xi2b_ab^a+b_a\partial^\mu A^a_\mu\Big\}\,.
        \end{equation}
        This is precisely the action appearing in~\eqref{eq:YMActionComponents}.
        
        \subsection{Free Kalb--Ramond two-form}\label{ssec:ex:KR}

        The next theory which we would like to discuss is that of a free two-form gauge potential $B\in\Omega^2(\IM^d)$. It has a three-form curvature given by
        \begin{equation}
            H_{\mu\nu\kappa}\ \coloneqq\ \partial_\mu B_{\nu\kappa}+\partial_\nu B_{\kappa\mu}+\partial_\kappa B_{\mu\nu}~~\in\Omega^3(\IM^d)
        \end{equation}
        and transforms under the infinitesimal gauge transformations as 
        \begin{equation}
            B_{\mu\nu}\ \mapsto\ \tilde B_{\mu\nu}\ \coloneqq\ B_{\mu\nu}+\partial_\mu\Lambda_\nu-\partial_\nu\Lambda_\mu~,
        \end{equation}
        where $\Lambda\in\Omega^1(\IM^d)$ is the one-form gauge parameter. Note that the gauge parameters themselves transform under a \uline{higher gauge symmetry},
        \begin{equation}
            \Lambda_\mu\ \mapsto\ \tilde\Lambda_\mu\ \coloneqq\ \Lambda_\mu+\partial_\mu\lambda~,
        \end{equation}
        where $\lambda\in\scC^\infty(\IM^d)$ is the (scalar) higher gauge parameter. 
        
        \begin{table}[ht]
            \begin{center}
                \resizebox{\textwidth}{!}{
                    \begin{tabular}{|c|l|c|c|c|c|c|c|c|}
                        \hline
                        \multicolumn{5}{|c|}{fields} & \multicolumn{4}{c|}{anti-fields}
                        \\
                        \hline
                        & role & $|-|_{\rm gh}$ & $|-|_\frL$ & dim & & $|-|_{\rm gh}$ & $|-|_\frL$ & dim 
                        \\
                        \hline
                        $\lambda$ & ghost--for--ghost field & 2 & $-1$ & $\tfrac{d}{2}-3$ & $\lambda^+$ & $-3$ & 4 & $\tfrac{d}{2}+3$
                        \\
                        $\Lambda_\mu$ & ghost field & 1 & 0 & $\tfrac{d}{2}-2$ & $\Lambda^+_\mu$ & $-2$ & 3 & $\tfrac{d}{2}+2$
                        \\
                        $\gamma$ & trivial pair partner of $\eps$ & 1 & 0 & $\tfrac{d}{2}-1$ & $\gamma^+$ & $-2$ & $3$ & $\tfrac{d}{2}+1$
                        \\
                        $B_{\mu\nu}$ & physical field & 0 & 1 & $\tfrac{d}{2}-1$ & $B^+_{\mu\nu}$ & $-1$ & 2 & $\tfrac{d}{2}+1$
                        \\
                        $\alpha_\mu$ & Nakanishi--Lautrup field & 0 & 1 & $\tfrac{d}{2}$ & $\alpha^+_\mu$ & $-1$ & $2$ & $\tfrac{d}{2}$ 
                        \\
                        $\eps$ & trivial pair partner of $\gamma$ & $0$ & $1$ & $\tfrac{d}{2}-1$ & $\eps^+$ & $-1$ & 2 & $\tfrac{d}{2}+1$
                        \\
                        $\bar \Lambda_\mu$ & anti-ghost field & $-1$ & $2$ & $\tfrac{d}{2}$ & $\bar \Lambda^+_\mu$ & 0 & 1 & $\tfrac{d}{2}$
                        \\
                        $\bar \gamma$ & trivial pair partner of $\bar\lambda$ & $-1$ & 2 & $\tfrac{d}{2}+1$ & $\bar \gamma^+$ & $0$ & $1$ & $\tfrac{d}{2}-1$
                        \\
                        $\bar \lambda$ & trivial pair partner of $\bar\gamma$ & $-2$ & $3$ & $\tfrac{d}{2}+1$  & $\bar \lambda^+$ & 1 & 0 & $\tfrac{d}{2}-1$
                        \\
                        \hline
                    \end{tabular}
                }
            \end{center}
            \caption{The full set of BV fields for the free Kalb--Ramond field, including their ghost numbers, their $L_\infty$-degrees, and their mass dimension. Besides the physical field, the ghost field, and ghost--for--ghost field, we also introduced trivial pairs $(\alpha,\bar\Lambda)$, $(\gamma,\eps)$, and $(\bar\gamma,\bar\lambda)$ together with their anti-fields.}\label{tab:fields:KR}
        \end{table}
        
        \paragraph{BV action and BV operator.}
        The full set of fields required for gauge fixing in the BV formalism is given by what is known as the \uline{Batalin--Vilkovisky triangle}~\cite{Batalin:1984jr}, see also ~\cite{Jurco:2018sby} for a recent review in the notation used here. The complete list of BV fields is given in \cref{tab:fields:KR}. Following the discussion of~\cite{Batalin:1984jr}, the BV action reads as
        \begin{equation}\label{eq:BVActionKR}
            S_{\rm BV}^{\rm KR}\ \coloneqq\ \int\rmd^dx\,\Big\{\!-\tfrac{1}{12}H_{\mu\nu\kappa}H^{\mu\nu\kappa}+2B^+_{\mu\nu}\partial^\mu\Lambda^\nu-\Lambda^+_\mu\partial^\mu\lambda-\bar\Lambda^+_\mu\alpha^\mu+\bar\lambda^+\bar\gamma+\eps^+\gamma\Big\}\,,
        \end{equation}
        where the factor of two has been introduced for later convenience. Consequently, the BV operator acts~\eqref{eq:CQ} as
        \begin{equation}\label{eq:BVOperatorKR}
            \begin{aligned}
                \lambda\ &\mapstomap{Q_{\rm BV}}\ 0~,~~~&
                \lambda^+\ &\mapstomap{Q_{\rm BV}}\ \partial^\mu\Lambda^+_\mu~,
                \\
                \Lambda_\mu\ &\mapstomap{Q_{\rm BV}}\ \partial_\mu\lambda~,~~~&
                \Lambda^+_\mu\ &\mapstomap{Q_{\rm BV}}\ -2\partial^\nu B_{\nu\mu}^+~,
                \\
                \gamma\ &\mapstomap{Q_{\rm BV}}\ 0~,~~~&
                \gamma^+\ &\mapstomap{Q_{\rm BV}}\ \eps^+~,
                \\
                B_{\mu\nu}\ &\mapstomap{Q_{\rm BV}}\ \partial_\mu\Lambda_\nu-\partial_\nu\Lambda_\mu~,~~~&
                B^+_{\mu\nu}\ &\mapstomap{Q_{\rm BV}}\ \tfrac12\partial^\kappa H_{\kappa\mu\nu}~,
                \\
                \alpha_\mu\ &\mapstomap{Q_{\rm BV}}\ 0~,~~~&
                \alpha^+_\mu\ &\mapstomap{Q_{\rm BV}}\ \bar\Lambda^+_\mu~,
                \\
                \eps\ &\mapstomap{Q_{\rm BV}}\ \gamma~,~~~&
                \eps^+\ &\mapstomap{Q_{\rm BV}}\ 0~,
                \\
                \bar\Lambda_\mu\ &\mapstomap{Q_{\rm BV}}\ \alpha_\mu~,~~~&
                \bar\Lambda^+_\mu\ &\mapstomap{Q_{\rm BV}}\ 0~,
                \\
                \bar\gamma\ &\mapstomap{Q_{\rm BV}}\ 0~,~~~&
                \bar\gamma^+\ &\mapstomap{Q_{\rm BV}}\ \bar\lambda^+~,
                \\
                \bar\lambda\ &\mapstomap{Q_{\rm BV}}\ \bar\gamma~,~~~&
                \bar\lambda^+\ &\mapstomap{Q_{\rm BV}}\ 0~.
            \end{aligned}
        \end{equation}

        \paragraph{\mathversion{bold}$L_\infty$-algebra.}
        The BV operator~\eqref{eq:BVOperatorKR} is the Chevalley--Eilenberg differential of an $L_\infty$-algebra $\frL^{\rm KR}_{\rm BV}$, which has the underlying complex
        \begin{subequations}
            \begin{equation}
                \begin{tikzcd}[
                    every label/.append style={scale=.95},
                    cells={nodes={scale=.95}},row sep=0.5cm
                    ]
                    \stackrel{\lambda}{\scC^\infty(\IM^d)}\rar["-\partial_\mu"] &\stackrel{\Lambda_\mu}{\Omega^1(\IM^d)}\rar["2\partial_{[\nu}"]&\stackrel{B_{\mu\nu}}{\Omega^2(\IM^d)}\arrow{r}{\mu_1}&\stackrel{B^+_{\mu\nu}}{\Omega^2(\IM^d)} \rar["2\partial^\nu"]&\stackrel{\Lambda_\mu^+}{\Omega^1(\IM^d)}\rar["-\partial^{\mu}"] &\stackrel{\lambda^+}{\scC^\infty(\IM^d)}
                    \\
                    &&\stackrel{\bar\Lambda_\mu^+}{\Omega^1(\IM^d)}\arrow[start anchor=south east, end anchor= north west,dr,"~\sfid",pos=0.05]&\stackrel{\bar\Lambda_\mu}{\Omega^1(\IM^d)} 
                    \\
                    & & \stackrel{\alpha^{\phantom{+}}_\mu}{\Omega^1(\IM^d)}\arrow[start anchor=north east, end anchor= south west,crossing over,ur,"-\sfid",pos=0.05,swap]& \stackrel{\alpha^+_\mu}{\Omega^1(\IM^d)} & &
                    \\
                    & \stackrel{\gamma}{\scC^\infty(\IM^d)}\rar["\sfid"] & \stackrel{\eps}{\scC^\infty(\IM^d)}& \stackrel{\eps^+}{\scC^\infty(\IM^d)}\rar["-\sfid"]& \stackrel{\gamma^+}{\scC^\infty(\IM^d)}
                    \\
                    \underbrace{\spacer{2ex}~~~~~~~~~~}_{\eqqcolon\,\frL^{\rm KR}_{-1}}& \underbrace{\spacer{2ex}\stackrel{\bar\lambda^+}{\scC^\infty(\IM^d)}}_{\eqqcolon\,\frL^{\rm KR}_{{\rm BV},\,0}}\rar["\sfid"] & 
                    \underbrace{\spacer{2ex}\stackrel{\bar\gamma^+}{\scC^\infty(\IM^d)}}_{\eqqcolon\,\frL^{\rm KR}_{{\rm BV},\,1}}
                    & \underbrace{\spacer{2ex}\stackrel{\bar\gamma}{\scC^\infty(\IM^d)}}_{\eqqcolon\,\frL^{\rm KR}_{{\rm BV},\,2}}\rar["-\sfid"]
                    & \underbrace{\spacer{2ex}\stackrel{\bar\lambda}{\scC^\infty(\IM^d)}}_{\eqqcolon\,\frL^{\rm KR}_{{\rm BV},\,3}}
                    & \underbrace{\spacer{2ex}~~~~~~~~~~}_{\eqqcolon\,\frL^{\rm KR}_{{\rm BV},\,4}}
                \end{tikzcd}
            \end{equation}
            with  
            \begin{equation}
                \begin{gathered}
                    \frL^{\rm KR}_{{\rm BV},\,-1}\ =\ \frL^{\rm KR}_{{\rm BV},\,-1,\,\lambda}~,~~~
                    \frL^{\rm KR}_{{\rm BV},\,0}\ =\ \spacee{\frL^{\rm KR}}{{\rm BV},\,0}{\Lambda\,,\gamma\,,\bar\lambda^+}~,
                    \\
                    \frL^{\rm KR}_{{\rm BV},\,1}\ =\ \spacee{\frL^{\rm KR}}{{\rm BV},\,1}{B,\,\bar\Lambda^+,\,\alpha,\,\eps,\,\bar\gamma^+}~,~~~
                    \frL^{\rm KR}_{{\rm BV},\,2}\ =\ \spacee{\frL^{\rm KR}}{{\rm BV},\,2}{B^+,\,\bar\Lambda,\,\alpha^+,\,\eps^+,\,\bar\gamma}~,
                    \\
                    \frL^{\rm KR}_{{\rm BV},\,3}\ =\ \spacee{\frL^{\rm KR}}{{\rm BV},\,3}{\Lambda^+,\,\gamma^+,\,\bar\lambda}~,~~~
                    \frL^{\rm KR}_{{\rm BV},\,4}\ =\ \frL^{\rm KR}_{{\rm BV},\,4,\,\lambda^+}~,
                \end{gathered}
            \end{equation}
            and the non-vanishing action of the differential $\mu_1$ given by
            \begin{equation}
                \begin{aligned}
                    \lambda\ &\mapstomap{\mu_1}\ -\partial_\mu\lambda\ \in\ \frL^{\rm KR}_{{\rm BV},\,0,\,\Lambda}~,
                    \\
                    \colvec{
                        \Lambda_\mu,
                        \gamma,
                        \bar\lambda^+
                    }
                    \ &\mapstomap{\mu_1}\ 
                    \colvec{
                        -2\partial_{[\mu}\Lambda_{\nu]},
                        \gamma,
                        \bar\lambda^+
                    }
                    \ \in\ \spacee{\frL^{\rm KR}}{{\rm BV},\,1}{B,\,\eps,\,\bar\gamma^+}~,
                    \\
                    \colvec{
                        B_{\mu\nu},
                        \bar\Lambda_\mu^+,
                        \alpha_\mu
                    }
                    \ &\mapstomap{\mu_1}\ 
                    \colvec{
                        \tfrac12\partial^\kappa H_{\kappa\mu\nu},
                        \alpha_\mu,
                        -\bar\Lambda^+_\mu
                    }
                    \ \in\ \spacee{\frL^{\rm KR}}{{\rm BV},\,2}{B^+,\,\bar\Lambda,\,\alpha^+}~,
                    \\
                    \colvec{
                        B_{\mu\nu}^+,
                        \eps^+,
                        \bar\gamma
                    }
                    \ &\mapstomap{\mu_1}\ 
                    \colvec{
                        2\partial^\nu B_{\mu\nu}^+,
                        -\eps^+,
                        -\bar\gamma
                    }
                    \ \in\ \spacee{\frL^{\rm KR}}{{\rm BV},\,3}{\Lambda^+,\,\gamma^+,\,\bar\lambda}~,
                    \\
                    \Lambda_{\mu}^+\ &\mapstomap{\mu_1}\ -\partial^\mu\Lambda^+_\mu\ \in\ \frL^{\rm KR}_{{\rm BV},\,4,\,\lambda^+}~,
                \end{aligned}
            \end{equation}
        \end{subequations}
        There are no higher products because the theory is free. The $L_\infty$-algebra $\frL^{\rm KR}_{\rm BV}$ becomes cyclic upon introducing
        \begin{equation}\label{eq:innerProductKR}
            \begin{aligned}
                \inner{\lambda}{\lambda^+}\ &\coloneqq\ -\int\rmd^dx\,\lambda\lambda^+~,~~
                &\inner{{\bar\lambda}}{{\bar\lambda}^+}\ &\coloneqq\ -\int\rmd^dx\,{\bar\lambda}{\bar\lambda}^+~,
                \\
                \inner{\Lambda}{\Lambda^+}\ &\coloneqq\ \int\rmd^dx\,\Lambda^\mu\Lambda^+_\mu~,~~
                &\inner{{\bar\Lambda}}{{\bar\Lambda}^+}\ &\coloneqq\ -\int\rmd^dx\,{\bar\Lambda}^\mu{\bar\Lambda}^+_\mu~,
                \\
                \inner{B}{B^+}\ &\coloneqq\ \int\rmd^dx\, B^{\mu\nu}B^+_{\mu\nu}~,
                \\
                \inner{\alpha}{\alpha^+}\ &\coloneqq\ \int\rmd^dx\,\alpha^\mu\alpha^+_\mu~,~~
                &\inner{\eps}{\eps^+}\ &\coloneqq\ \int\rmd^dx\,\eps\eps^+~,
                \\
                \inner{\gamma}{\gamma^+}\ &\coloneqq\ \int\rmd^dx\,\gamma\gamma^+~,~~
                &\inner{{\bar\gamma}}{{\bar\gamma}^+}\ &\coloneqq\ -\int\rmd^dx\,{\bar\gamma}{\bar\gamma}^+~.
            \end{aligned}
        \end{equation}
        Again, the superfield homotopy Maurer--Cartan action~\eqref{eq:CBVAction} of $\frL^{\rm KR}_{\rm BV}$ with higher products~\eqref{eq:translation_BV_L_infty} is the BV action~\eqref{eq:BVActionKR}.
        
        \paragraph{Gauge fixing.}
        Recall the general gauge-fixing procedure in the BV formalism from \cref{sec:BVMotivation}. The most general Lorentz covariant linear gauge choices are implemented by the gauge-fixing fermion
        \begin{equation}\label{eq:gaugeFixingFermionKR}
            \Psi\ \coloneqq\ -\int\rmd^dx\,\Big\{\bar\Lambda^\nu\big(\partial^\mu B_{\mu\nu}+\tfrac{\zeta_1}{2}\alpha_\nu\big)-\bar\lambda\big(\partial^\mu\Lambda_\mu+\zeta_2\gamma\big)+\eps\big(\partial^\mu\bar\Lambda_\mu+\zeta_3\bar\gamma\big)\Big\}
        \end{equation}
        for some real parameters $\zeta_{1,2,3}$. The resulting gauge-fixed action (after putting to zero the anti-fields) is
        \begin{equation}\label{eq:BRSTactionKR}
            \begin{aligned}
                S^{\rm KR}_{\rm BRST}\ &=\ \int\rmd^dx\,\Big\{\tfrac14B_{\mu\nu}\wave B^{\mu\nu}+\tfrac12(\partial^\mu B_{\mu\nu})(\partial_\kappa B^{\kappa\nu})-\bar\Lambda_\mu\wave\Lambda^\mu\,-
                \\
                &\kern2cm-(\partial^\mu\bar\Lambda_\mu)(\partial_\nu\Lambda^\nu)-\bar\lambda\wave\lambda+\tfrac{\zeta_1}{2}\alpha_\mu\alpha^\mu+\alpha^\nu\partial^\mu B_{\mu\nu}\,+
                \\
                &\kern4cm+\eps\partial_\mu\alpha^\mu-(\zeta_2+\zeta_3)\,\bar\gamma\gamma+\gamma\partial_\mu\bar\Lambda^\mu-\bar\gamma\partial_\mu\Lambda^\mu\Big\}\,.
            \end{aligned}
        \end{equation}
        
        \subsection{Einstein--Hilbert gravity}\label{ssec:ex:EH}
        
        The fourth relevant theory is Einstein--Hilbert gravity on a $d$-dimensional Lorentzian manifold $M^d$ with metric $g\in\Gamma(M^d,\odot^2T^*M^d)$. Let $\nabla$ be the Levi--Civita connection for $g$. Recall that infinitesimal gauge transformations of the metric are parametrised by a vector field $\chi$ and act as 
        \begin{equation}
            g_{\mu\nu}\ \mapsto\ \tilde g_{\mu\nu}\ \coloneqq\ g_{\mu\nu}+(\caL_\chi g)_{\mu\nu}~,
        \end{equation}
        where $\caL_\chi$ denotes the Lie derivative along $\chi$.
        
        \paragraph{BV action and BV operator.}
        The list of all the fields required in the BV formulation of Einstein--Hilbert gravity together with their properties is found in \cref{tab:fields:EH} and the BV action (cf.~e.g.~\cite{Fredenhagen:2011an} or~\cite{Nakanishi:1977gt} for the gauge-fixed version) is
        \begin{equation}
            S_{\rm BV}^{\rm EH}\ \coloneqq\ \int\rmd^dx\,\Big\{\!-\!\tfrac{1}{\kappa^2}\sqrt{-g}\,R+g^{+\mu\nu}(\caL_\chi g)_{\mu\nu}+\tfrac12\chi^+_{\mu}(\caL_\chi\chi)^\mu-\varrho^\mu\bar\chi^+_\mu\Big\}\,,
        \end{equation}
        where $R$ denotes the Ricci scalar and $2\kappa^2=16\pi G_{\rm N}^{(d)}$ Einstein's gravitational constant. Consequently, the BV operator~\eqref{eq:CQ} acts as
        \begin{equation}\label{eq:BVOperatorEH}
            \begin{aligned}
                \chi^\mu\ &\mapstomap{Q_{\rm BV}}\ -\tfrac12(\caL_\chi\chi)^\mu~,~~~&
                \chi^+_\mu\ &\mapstomap{Q_{\rm BV}}\ 
                -2\nabla^\nu g^+_{\nu\mu}+(\caL_\chi\chi^+)_\mu~,
                \\
                g_{\mu\nu}\ &\mapstomap{Q_{\rm BV}}\ (\caL_\chi g)_{\mu\nu}~,~~~&
                g^{+\mu\nu}\ &\mapstomap{Q_{\rm BV}}\ -\tfrac{1}{\kappa^2}\sqrt{-g}\big(R^{\mu\nu}-\tfrac12g^{\mu\nu}R\big)+(\caL_\chi g^+)^{\mu\nu}~,
                \\
                \varrho^\mu\ &\mapstomap{Q_{\rm BV}}\ 0~,~~~&
                \varrho^+_\mu\ &\mapstomap{Q_{\rm BV}}\ -\bar\chi_\mu~,
                \\
                \bar\chi^\mu\ &\mapstomap{Q_{\rm BV}}\ \varrho^\mu~,~~~&
                \bar\chi^+_\mu\ &\mapstomap{Q_{\rm BV}}\ 0~,
            \end{aligned}    
        \end{equation}
        where $R_{\mu\nu}$ is the Ricci tensor.
        
        \begin{table}[ht]
            \resizebox{\textwidth}{!}{
                \begin{tabular}{|c|l|c|c|c|c|c|c|c|}
                    \hline
                    \multicolumn{5}{|c|}{fields} & \multicolumn{4}{c|}{anti-fields}
                    \\
                    \hline
                    & role & $|-|_{\rm gh}$ & $|-|_\frL$ & dim & & $|-|_{\rm gh}$ & $|-|_\frL$ & dim
                    \\
                    \hline
                    $\chi^\mu$& ghost field & 1 & 0 & $-1$ & $\chi^+_\mu$ & $-2$ & 3 & $d+1$
                    \\
                    $g_{\mu\nu}$ & physical field & 0 & 1 & 0 & $g^{+\mu\nu}$ & $-1$ & 2 & $d$
                    \\
                    $\varrho^\mu$ & Nakanishi--Lautrup field & 0 & 1 & $\tfrac{d}{2}$ & $\varrho^+_\mu$ & $-1$ & $2$ & $\tfrac{d}{2}$
                    \\
                    $\bar \chi^\mu$ & anti-ghost field & $-1$ & $2$ & $\tfrac{d}{2}$ & $\bar \chi^+_\mu$ & 0 & 1 & $\tfrac{d}{2}$ 
                    \\
                    \hline
                \end{tabular}}
            \caption{The full set of BV fields for Einstein--Hilbert gravity, including their ghost numbers, their $L_\infty$-degrees, and their mass dimensions. The mass dimension of the coupling constant $\kappa$ is $1-\frac{d}{2}$. Note that all fields are tensors and all anti-fields are tensor densities.}\label{tab:fields:EH}
        \end{table} 
        
        \paragraph{Perturbation theory.}
        Let us now restrict to a Lorentzian manifold $M^d$ for which the metric can be seen as a fluctuation $h_{\mu\nu}$ about the Minkowski metric $\eta_{\mu\nu}$ on $\IM^d$, that is,
        \begin{subequations}\label{eq:expand_metric}
            \begin{equation}
                g_{\mu\nu}\ \eqqcolon\ \eta_{\mu\nu}+\kappa h_{\mu\nu}~.
            \end{equation}
            For future reference, we note that
            \begin{equation}
                g^{\mu\nu}\ =\ \eta^{\mu\nu}-\kappa h^{\mu\nu}+\kappa^2h^{\mu\rho}h_{\rho}{}^{\nu}-\kappa^3h^{\mu\rho}h_{\rho}{}^{\sigma}h_{\sigma}{}^{\nu}+\caO(\kappa^4)~,
            \end{equation}
            where ${h_\mu}^\nu\coloneqq \eta^{\nu\lambda}h_{\mu\lambda}$ and $h^{\mu\nu}\coloneqq\eta^{\mu\kappa}\eta^{\nu\lambda}h_{\kappa\lambda}$. Likewise,
            \begin{equation}
                \begin{aligned}
                    \sqrt{-g}\ &=\ 1+\tfrac{1}{2}\kappa\mathring{h}+\kappa^2\big(\tfrac18\mathring{h}^2-\tfrac14h_{\mu}{}^{\nu}h_{\nu}{}^\mu\big)\,+
                    \\
                    &\kern1cm+\,\kappa^3\big(\tfrac{1}{48}\mathring{h}^3-\tfrac18\mathring{h}h_{\mu}{}^{\nu}h_{\nu}{}^\mu+\tfrac16h_{\mu}{}^{\nu}h_{\nu}{}^{\rho} h_{\rho}{}^{\mu}\big)+\caO(\kappa^4)~,
                \end{aligned}
            \end{equation}
        \end{subequations}
        where $\mathring{h}\coloneqq\eta^{\mu\nu}h_{\mu\nu}$.
        
        We also introduce the following rescaled anti-fields and unphysical fields:
        \begin{equation}
            \begin{gathered}
                h^{+\mu\nu}\ \coloneqq\ \tfrac{\kappa}{\sqrt{-g}}g^{+\mu\nu}~,
                \\
                X^\mu\ \coloneqq\ \tfrac1\kappa\chi^\mu
                ~,~~~
                X^+_\mu\ \coloneqq\ \tfrac{\kappa}{\sqrt{-g}}\chi^+_\mu
                ~,~~~
                \bar X^\mu\ \coloneqq\ \bar\chi^\mu
                ~,~~~
                \bar X^+_\mu\ \coloneqq\ \tfrac{1}{\sqrt{-g}}\bar\chi^+_\mu~,
                \\
                \varpi^\mu\ \coloneqq\ \varrho^\mu
                ~,~~~
                \varpi^+_\mu\ \coloneqq\ \tfrac{1}{\sqrt{-g}}\varrho^+_\mu~.
            \end{gathered}
        \end{equation}
        In addition to these, we introduce two trivial pairs $(\beta,\delta)$ and $(\pi,\bar \beta)$, together with the corresponding anti-fields. These do not modify the physical observables; as we shall see later, however, they do arise rather naturally in the double copy and are crucial once the dilaton enters. We also note that precisely these trivial pairs were also introduced in~\cite{Baulieu:2020obv} in order to explain a unimodular gauge fixing in the BV formalism. The full list of fields and their properties is given in \cref{tab:fields:EH_lin}.
        
        \begin{table}[ht]
            \resizebox{\textwidth}{!}{
                \begin{tabular}{|c|l|c|c|c|c|c|c|c|}
                    \hline
                    \multicolumn{5}{|c|}{fields} & \multicolumn{4}{c|}{anti-fields}
                    \\
                    \hline
                    & role & $|-|_{\rm gh}$ & $|-|_\frL$ & dim & & $|-|_{\rm gh}$ & $|-|_\frL$ & dim
                    \\
                    \hline
                    $X^\mu$& ghost field & 1 & 0 & $\tfrac{d}{2}-2$ & $X^+_\mu$ & $-2$ & 3 & $\tfrac{d}{2}+2$
                    \\
                    $\beta$ & trivial pair partner of $\delta$ & 1 & 0 & $\tfrac{d}{2}-1$ & $\beta^+$ & $-2$ & $3$ & $\tfrac{d}{2}+1$
                    \\
                    $h_{\mu\nu}$ & physical field & 0 & 1 & $\tfrac{d}{2}-1$ & $h^{+\mu\nu}$ & $-1$ & 2 & $\tfrac{d}{2}+1$ 
                    \\
                    $\varpi^\mu$ & Nakanishi--Lautrup field & 0 & 1 & $\tfrac{d}{2}$ & $\varpi^+_\mu$ & $-1$ & $2$ & $\tfrac{d}{2}$
                    \\
                    $\pi$ & trivial pair partner of $\bar\beta$  & 0 & $1$ & $\tfrac{d}{2}+1$ & $\pi^+$ & $-1$ & 2 & $\tfrac{d}{2}-1$
                    \\
                    $\delta$ & trivial pair partner of $\beta$  & 0  & $1$ & $\tfrac{d}{2}-1$ & $\delta^+$ & $-1$ & 2 & $\tfrac{d}{2}+1$
                    \\
                    $\bar X^\mu$ & anti-ghost field & $-1$ & $2$ & $\tfrac{d}{2}$ & $\bar X^+_\mu$ & 0 & 1 & $\tfrac{d}{2}$
                    \\
                    $\bar\beta$ & trivial pair partner of $\pi$  & $-1$ & 2 & $\tfrac{d}{2}+1$ & $\bar\beta^+$ & 0 & 1 &  $\tfrac{d}{2}-1$
                    \\
                    \hline
                \end{tabular}
            }
            \caption{The full set of BV fields for perturbative Einstein--Hilbert gravity, including their ghost numbers, their $L_\infty$-degrees, and their mass dimension. All the fields are regarded as tensors on Minkowski space.}\label{tab:fields:EH_lin}
        \end{table}
        
        The action itself can now be expanded in orders of $\kappa$,
        \begin{equation}\label{eq:BVActionEHkappa}
            \begin{aligned}
                S_{\rm BV}^{\rm eEH}\ &=\ \int\rmd^dx\,\sqrt{-g}\,\Big\{\!-\!\tfrac{1}{\kappa^2}R+\tfrac{2}{\sqrt{-g}}g^{+\mu\nu}\nabla_\mu\chi_\nu\,+
                \\
                &\kern4cm+\tfrac{1}{2\sqrt{-g}}\chi^+_\mu(\caL_\chi\chi)^\mu-\tfrac{1}{\sqrt{-g}}\varpi^\mu\bar\chi^+_\mu+\beta\delta^++\pi\bar\beta^+\Big\}
                \\
                &=\ \int\rmd^dx\,\sqrt{-g}\,\Big\{\!-\!\tfrac{1}{\kappa^2}R+2h^{+\mu\nu}\nabla_\mu X_\nu+\tfrac\kappa2 X^+_\mu(\caL_XX)^\mu-\varpi^\mu\bar\chi^+_\mu+\beta\delta^++\pi\bar\beta^+\Big\}
                \\
                &\eqqcolon\ \int\rmd^dx\,\sum_{n=0}^\infty\kappa^n\scL^{\rm eEH}_n
            \end{aligned}
        \end{equation}
        with indices now raised and lowered with the Minkowski metric. The lowest-order Lagrangian $\scL_0$ is given by the \uline{Fierz--Pauli Lagrangian} plus the terms containing ghosts and other unphysical fields,
        \begin{equation}
            \begin{aligned} 
                \scL^{\rm eEH}_0\ &=\ -\tfrac14\partial^\mu h^{\nu\rho}\partial_\mu h_{\nu\rho}+\tfrac12\partial^\mu h^{\nu\rho}\partial_\nu h_{\mu\rho}-\tfrac12\partial^\mu\mathring{h}\partial^\nu h_{\mu\nu}+\tfrac14\partial^\mu\mathring{h}\partial_\mu\mathring{h}\,+
                \\
                &\kern2cm+2h^{+\mu\nu}\partial_\mu X_\nu-\varpi^\mu\bar X^+_\mu+\beta\delta^++\pi\bar\beta^+~,
            \end{aligned}
        \end{equation}
        cf.~e.g.~\cite{Ortin:2004ms}. To first order in $\kappa$, we have
        \begin{equation}
            \begin{aligned}
                \scL^{\rm eEH}_1\ &= \ -h^{\mu\nu}\Big\{\tfrac12\partial_\mu h^{\rho\sigma}\partial_\nu h_{\rho \sigma}-\tfrac14\eta_{\mu\nu}\partial_\sigma h_{\tau\rho}\partial^\sigma h^{\tau\rho}+\partial_\nu\mathring{h}\left(\partial_\rho h_\mu{}^\rho-\tfrac12\partial_\mu\mathring{h}\right)+
                \\
                &\kern1.5cm+\partial_\nu h_\mu{}^\rho\partial_\rho\mathring{h}-\partial_\rho\mathring{h}\partial^\rho h_{\mu\nu}-\tfrac12\eta_{\mu\nu}\partial^\rho\mathring{h}\left(\partial_\sigma h_\rho{}^{\sigma}-\tfrac12\partial_\rho\mathring{h}\right)+\partial^\rho h_{\mu\nu}\partial_\sigma h_\rho{}^{\sigma}\,-
                \\
                &\kern1.5cm-2\partial_\nu h_{\rho\sigma}\partial^\sigma h_\mu{}^\rho-\partial_\rho h_{\nu\sigma}\partial^\sigma h_\mu{}^\rho+\partial_\sigma h_{\nu\rho}\partial^\sigma h_\mu{}^\rho+\tfrac12\eta_{\mu\nu}\partial_\rho h_{\tau\sigma}\partial^\sigma h^{\tau\rho}\Big\}\,+
                \\
                &\kern1.5cm+2h^{+\mu\nu}\Big\{h_{\nu\lambda}\partial_\mu X^\lambda+\tfrac12(\partial_\mu h_{\lambda\nu}+\partial_\lambda h_{\mu\nu}-\partial_\nu h_{\mu\lambda})X^\lambda\Big\}\,+
                \\
                &\kern1.5cm+\tfrac12X^+_\mu(\caL_X X)^\mu+\tfrac12\mathring{h}(-\varpi^\mu\bar X^+_\mu+\beta\delta^++\pi\bar\beta^+)~.
            \end{aligned}
        \end{equation}
        
        \paragraph{\mathversion{bold}$L_\infty$-algebra.}
        The full $L_\infty$-algebra $\frL^{\rm eEH}_{\rm BV}$ of Einstein--Hilbert gravity has the underlying complex
        \begin{subequations}
            \begin{equation}
                \begin{tikzcd}
                    \stackrel{X_\mu}{\Omega^1(\IM^d)}\rar["2\partial_{(\nu}"]&\stackrel{h_{\mu\nu}}{\Omega^2(\IM^d)}\arrow{r}{\mu_1}
                    &\stackrel{h^+_{\mu\nu}}{\Omega^2(\IM^d)}\rar["2\partial^\nu"]&\stackrel{X_\mu^+}{\Omega^1(\IM^d)}
                    \\
                    &\stackrel{\bar X_\mu^+}{\Omega^1(\IM^d)}\arrow[start anchor=south east, end anchor= north west,dr,"~\sfid",pos=0.05]&\stackrel{\bar X_\mu}{\Omega^1(\IM^d)} 
                    \\
                    & \stackrel{\varpi^{\phantom{+}}_\mu}{\Omega^1(\IM^d)}\arrow[start anchor=north east, end anchor= south west,crossing over,ur,"-\sfid",pos=0.05,swap]& \stackrel{\varpi^+_\mu}{\Omega^1(\IM^d)} & &
                    \\
                    &\stackrel{\bar\beta^+}{\scC^\infty(\IM^d)}\arrow[start anchor=south east, end anchor= north west,dr,"-\sfid",pos=0.05]&\stackrel{\bar\beta}{\scC^\infty(\IM^d)} 
                    \\
                    & \stackrel{\pi^{\phantom{+}}}{\scC^\infty(\IM^d)}\arrow[start anchor=north east, end anchor= south west,crossing over,ur,"~\sfid",pos=0.05,swap]& \stackrel{\pi^+}{\scC^\infty(\IM^d)} &  &
                    \\
                    \underbrace{\spacer{2ex}\stackrel{\beta}{\scC^\infty(\IM^d)}}_{\eqqcolon\,\frL^{\rm eEH}_{{\rm BV},\,0}}\rar["\sfid"] & 
                    \underbrace{\spacer{2ex}\stackrel{\delta}{\scC^\infty(\IM^d)}}_{\eqqcolon\,\frL^{\rm eEH}_{{\rm BV},\,1}}
                    & \underbrace{\spacer{2ex}\stackrel{\delta^+}{\scC^\infty(\IM^d)}}_{\eqqcolon\,\frL^{\rm eEH}_{{\rm BV},\,2}}\rar["-\sfid"]
                    & \underbrace{\spacer{2ex}\stackrel{\beta^+}{\scC^\infty(\IM^d)}}_{\eqqcolon\,\frL^{\rm eEH}_{{\rm BV},\,3}}                   
                \end{tikzcd}
            \end{equation}
            with  
            \begin{equation}
                \begin{gathered}
                    \frL^{\rm eEH}_{{\rm BV},\,0}\ =\ \frL^{\rm eEH}_{{\rm BV},\,0,\,X}\oplus\frL^{\rm eEH}_{{\rm BV},\,0,\,\beta}~,~~~
                    \frL^{\rm eEH}_{{\rm BV},\,1}\ =\ \spacee{\frL^{\rm eEH}}{{\rm BV},\,1}{h,\,\bar X^+,\,\varpi,\,\bar\beta^+,\,\pi,\,\delta}~,
                    \\
                    \frL^{\rm eEH}_{{\rm BV},\,2}\ =\ \spacee{\frL^{\rm eEH}}{{\rm BV},\,2}{h^+,\,\bar X,\,\varpi^+,\,\bar\beta,\,\pi^+,\,\delta^+}~,~~~
                    \frL^{\rm eEH}_{{\rm BV},\,3}\ =\ \frL^{\rm eEH}_{{\rm BV},\,0,\,X^+}\oplus\frL^{\rm eEH}_{{\rm BV},\,0,\,\beta^+}~.
                \end{gathered}
            \end{equation}
            The $L_\infty$-algebra $\frL^{\rm eEH}_{\rm BV}$ comes with non-trivial higher products of arbitrarily high order, with $\mu_i$ encoding the Lagrangian $\scL^{\rm eEH}_{{\rm BV},\,i-1}$. Below, we merely list $\mu_1$ and $\mu_2$ to prepare our discussion of the double copy later on. The differentials are
            \begin{equation}
                \begin{aligned}
                    \colvec{
                        X_\mu,
                        \beta
                    }
                    \ &\mapstomap{\mu_1}\ 
                    \colvec{
                        -\partial_{(\mu} X_{\nu)},
                        \beta
                    }
                    \ \in\ \spacee{\frL^{\rm eEH}}{{\rm BV},\,1}{h,\,\delta}~,
                    \\[10pt]
                    \colvec{
                        h_{\mu\nu},
                        \bar X_{\mu}^+,
                        \varpi^\mu,
                        \bar\beta^+,
                        \pi
                    }
                    \ &\mapstomap{\mu_1}\ 
                    \colvec{
                        \left[\frac{1}{2}(\delta^{\rho}_{\mu}\delta^{\sigma}_{\nu}-\eta_{\mu\nu}\eta^{\rho\sigma})\wave-(\delta^{\sigma}_{\nu}\eta_{\mu\kappa}-\delta^{\sigma}_{\kappa}\eta_{\mu\nu})\partial^{\rho}\partial^{\kappa}\right]h_{\rho\sigma},
                        -\varpi^\mu,
                        \bar X^+_\mu,
                        \pi,
                        -\bar\beta^+
                    }
                    \\[5pt]
                    &\kern2cm\in\ \spacee{\frL^{\rm eEH}}{{\rm BV},\,2}{h^+,\,\bar{X},\varpi^+,\,\bar{\beta},\,\pi^+}~,
                    \\[10pt]
                    \colvec{
                        h_{\mu\nu}^+,
                        \delta^+
                    }
                    \ &\mapstomap{\mu_1}\ 
                    \colvec{
                        -\partial^\nu h_{\nu\mu},
                        -\delta^+
                    }
                    \ \in\ \spacee{\frL^{\rm eEH}}{{\rm BV},\,3}{X^+,\,\beta^+}~,
                \end{aligned}
            \end{equation}
            and the cubic interactions are encoded in the binary products
            \begin{equation*}
                \begin{gathered}
                    (X_{1\mu},X_{2\nu})\ \mapstomap{\mu_2}\ (\caL_{X_1}X_2)_\mu\ \in\ \frL^{\rm eEH}_{{\rm BV},\,0,\,X}~,
                    \\
                    (X_\mu,X_\nu^+)\ \mapstomap{\mu_2}\ (\partial_\mu X^\nu)X^+_\nu+\partial_\nu(X^\nu X_\mu^+)\in\ \frL^{\rm eEH}_{{\rm BV},\,3,\,X^+}~,
                    \\
                    (\varpi,\bar X_\mu^+)\ \mapstomap{\mu_2}\ \tfrac12\varpi^\rho\bar X^+_\rho\eta_{\mu\nu}\ \in\ \frL^{\rm eEH}_{{\rm BV},\,2,\,h^+}~,
                    \\
                    (h_{\mu\nu},\varpi)\ \mapstomap{\mu_2}\ -\tfrac12\mathring{h}\varpi_\mu\ \in\ \frL^{\rm eEH}_{{\rm BV},\,2,\,{\bar X}}~,
                    \\
                    (h_{\mu\nu},\bar X_\rho^+)\ \mapstomap{\mu_2}\ \tfrac12\mathring{h}\bar X^+_\mu\ \in\ \frL^{\rm eEH}_{{\rm BV},\,2,\,\varpi^+}~,
                    \\
                    (\beta,\delta^+)\ \mapstomap{\mu_2}\ \tfrac12\beta\delta^+\eta_{\mu\nu}\ \in\ \frL^{\rm eEH}_{{\rm BV},\,2,\,h^+}~,
                    \\
                    (h_{\mu\nu},\beta)\ \mapstomap{\mu_2}\ \tfrac12\mathring{h}\beta\ \in\ \frL^{\rm eEH}_{1,\delta}~,
                    \\
                    (h_{\mu\nu},\delta^+)\ \mapstomap{\mu_2}\ -\tfrac12\mathring{h} \delta^+\ \in\ \frL^{\rm eEH}_{{\rm BV},\,3,\,\beta^+}~,
                    \\
                    (\pi,\bar\beta^+)\ \mapstomap{\mu_2}\ \tfrac12\pi\bar\beta^+\eta_{\mu\nu}\ \in\ \frL^{\rm eEH}_{{\rm BV},\,2,\,h^+}~,~~(h_{\mu\nu},\pi)\ \mapstomap{\mu_2}\ -\tfrac12\mathring{h}\pi\ \in\ \frL^{\rm eEH}_{{\rm BV},\,2,\,{\bar\beta}}~,
                    \\
                    (h_{\mu\nu},\bar{\beta}^+)\ \mapstomap{\mu_2}\ \tfrac12\mathring{h}\bar \beta^+\ \in\ \frL^{\rm eEH}_{{\rm BV},\,2,\,\pi^+}~,
                    \\
                    (X_\mu,h_{\nu\rho})\ \mapstomap{\mu_2}\ -2h_{\nu\kappa}\partial_\mu X^\kappa-(\partial_\mu h_{\kappa\nu}+\partial_\kappa h_{\mu\nu}-\partial_\nu h_{\mu\kappa})X^\kappa\ \in\ \frL^{\rm eEH}_{{\rm BV},\,1,\,h}~,
                    \\
                \end{gathered}
            \end{equation*}
            \begin{equation}
                \begin{gathered}
                    (h_{\mu\nu}^+,h_{\rho\sigma})\ \mapstomap{\mu_2}\ -2\partial_\kappa (h^{+\kappa\nu}h_{\nu\mu})+h^{+\kappa\nu}(\partial_\kappa h_{\mu\nu}+\partial_\mu h_{\kappa\nu}-\partial_\nu h_{\kappa\mu})\ \in\ \frL^{\rm eEH}_{{\rm BV},\,3,\,X^+}~,
                    \\
                    (h_{\mu\nu}^+,X_\rho)\ \mapstomap{\mu_2}\ -2h^{+}_{\kappa\mu}\partial^\kappa X_\nu+\partial^\kappa(h^+_{\kappa\nu}X_\mu)+\partial_\kappa(h^{+\mu\nu}X^\kappa)-\partial^\kappa(h^+_{\mu\kappa}X_\nu)\ \in\ \frL^{\rm eEH}_{{\rm BV},\,2,\,h^+}~,
                    \\
                    \begin{aligned}
                        (h_{1\mu\nu},h_{2\rho\sigma})\ \mapstomap{\mu_2}\ 3\Big\{&\tfrac12\partial_\mu h_{1}^{\rho\sigma}\partial_\nu h_{2\rho \sigma}-\tfrac14\eta_{\mu\nu}\partial_\sigma h_{1\tau\rho}\partial^\sigma h_2^{\tau\rho}+\partial_\nu\mathring{h}_1\left(\partial_\rho h_{2\mu}{}^\rho-\tfrac12\partial_\mu\mathring{h}_2\right)+
                        \\
                        &+\partial_\nu h_{1\mu}{}^\rho\partial_\rho\mathring{h}_2-\partial_\rho\mathring{h}_1\partial^\rho h_{2\mu\nu}-\tfrac12\eta_{\mu\nu}\partial^\rho\mathring{h}_1\left(\partial_\sigma h_{2\rho}{}^{\sigma}-\tfrac12\partial_\rho\mathring{h}_2\right)\,+
                        \\
                        &+\partial^\rho h_{1\mu\nu}\partial_\sigma h_{2\rho}{}^{\sigma}-2\partial_\nu h_{1\rho\sigma}\partial^\sigma h_{2\mu}{}^\rho-\partial_\rho h_{1\nu\sigma}\partial^\sigma h_{2\mu}{}^\rho\,+
                        \\
                        &+\partial_\sigma h_{1\nu\rho}\partial^\sigma h_{2\mu}{}^\rho+\tfrac12\eta_{\mu\nu}\partial_\rho h_{1\tau\sigma}\partial^\sigma h_2^{\tau\rho}\Big\}+\,
                        \\
                        +&(1\leftrightarrow2)\ \in\ \frL^{\rm eEH}_{{\rm BV},\,2,\,h^+}~.
                    \end{aligned}
                \end{gathered}
            \end{equation}
        \end{subequations}
        The cyclic structure is given by the following integrals:
        \begin{equation}
            \begin{aligned}
                \inner{X}{X^+}\ &\coloneqq\ \int\rmd^dx\,X^\mu X^+_\mu~,~~
                &\inner{\bar X}{\bar X^+}\ &\coloneqq\ -\int\rmd^dx\,{\bar X}^\mu{\bar X}^+_\mu~,
                \\
                \inner{\beta}{\beta^+}\ &\coloneqq\ \int\rmd^dx\,\beta\beta^+~,~~
                &\inner{\bar\beta}{\bar\beta^+}\ &\coloneqq\ -\int\rmd^dx\,{\bar\beta}{\bar\beta}^+~,
                \\
                \inner{h}{h^+}\ &\coloneqq\ \int\rmd^dx\, h^{\mu\nu} h^+_{\mu\nu}~,
                ~~~
                &\inner{\varpi}{\varpi^+}\ &\coloneqq\ \int\rmd^dx\,\varpi^\mu\varpi^+_\mu~,~~
                &
                \\
                \inner{\pi}{\pi^+}\ &\coloneqq\ \int\rmd^dx\,\pi\pi^+~,~~
                &\inner{\delta}{\delta^+}\ &\coloneqq\ \int\rmd^dx\,\delta\delta^+~.
            \end{aligned}
        \end{equation}
        
        \paragraph{Gauge fixing.} Gauge fixing proceeds as usual in the BV formalism, but due to our two additional trivial pairs, we can now write down a much more general gauge-fixing fermion. We restrict ourselves to
        \begin{equation}\label{eq:gaugeFixingFermionEH}
            \begin{aligned}
                \Psi_0\ \coloneqq\ -\int \rmd^dx\,\Big\{ 
                &\bar X^\nu\Big(\zeta_4\partial^\mu h_{\mu\nu}-\tfrac{\zeta_5}{2}\varpi_\nu+\zeta_6\partial_\nu\mathring{h}-\zeta_7\partial_\nu\delta+\zeta_{8}\frac{\partial_\nu\pi}{\wave}\Big)\,+
                \\
                &+\bar\beta\Big(\zeta_9\mathring{h}-\zeta_{10}\delta+\zeta_{11}\frac{\partial^\mu\partial^\nu h_{\mu\nu}}{\wave}\Big)\Big\}~,
            \end{aligned}
        \end{equation}        
        and this is the freedom required for the discussion of the double copy. The resulting Lagrangian, to lowest order in $\kappa$, reads as
        \begin{equation}\label{eq:canonicallyRedefinedActionEHZerothOrder}
            \begin{aligned}
                \scL_0^{\rm eEH,\,gf}\ &=\ \tfrac14h^{\mu\nu}\wave h_{\mu\nu}+\tfrac12(\partial^\mu h_{\mu\nu})^2+\tfrac12\mathring{h}\partial^\mu\partial^\nu h_{\mu\nu}-\tfrac14\mathring{h}\wave\mathring{h}\,+
                \\
                &\kern1cm+\zeta_4\varpi^\nu\partial^\mu h_{\mu\nu}-\tfrac{\zeta_5}{2}\varpi^\mu\varpi_\mu+\zeta_6\varpi^\mu\partial_\mu\mathring{h}-\zeta_7\varpi^\mu\partial_\mu\delta+\zeta_{8}\varpi^\mu\frac{\partial_\mu\pi}{\wave}\,-
                \\
                &\kern1cm-\pi\zeta_9\mathring{h}+\zeta_{10}\pi\delta-\zeta_{11}\pi\frac{\partial^\mu\partial^\nu h_{\mu\nu}}{\wave}\,+
                \\
                &\kern1cm+\zeta_4(\partial^\mu \bar X^\nu+\partial^\nu \bar X^\mu)\partial_\mu X_\nu+\zeta_6(\partial_\mu\bar X^\mu)(\partial_\nu X^\nu)-\zeta_9\bar\beta\partial^\mu X_\mu\,-
                \\
                &\kern1cm-\zeta_{11}\frac{\partial^\mu\partial^\nu\bar \beta}{\wave}\partial_\mu X_\nu+\zeta_7\beta\partial_\nu \bar X^\nu-\zeta_{10}\bar \beta\beta~,
            \end{aligned}
        \end{equation}
        after putting to zero the anti-fields.
        
        \subsection{\texorpdfstring{$\caN=0$}{N=0} supergravity}\label{ssec:N=0SUGRA}
        
        The actions for the free Kalb--Ramond field and Einstein--Hilbert gravity are combined and coupled to an additional scalar field $\varphi$, the dilaton, in $\caN=0$ supergravity. This is the common, or universal, Neveu--Schwarz-Neveu--Schwarz sector of the $\alpha'\rightarrow 0$ limit of closed string theories, and the action reads as
        \begin{equation}\label{eq:action0_N0_SUGRA}
            S^{\caN=0}\ \coloneqq\ \int\rmd^dx\,\sqrt{-g}\,\Big\{\!-\!\tfrac{1}{\kappa^2}R-\tfrac{1}{d-2}\partial_\mu\varphi\partial^\mu\varphi-\tfrac{1}{12}\rme^{-\frac{4\kappa}{d-2}\varphi}H_{\mu\nu\kappa}H^{\mu\nu\kappa}\Big\}\,.
        \end{equation}
        The solutions of the associated equations of motions give backgrounds (with vanishing cosmological constant) around which strings can be quantised to lowest order in $\alpha'$ and string coupling. They also ensure conformal invariance of the string is non-anomalous in critical dimensions. 
        
        We note that the free part of $\caN=0$ supergravity is a sum of the free Kalb--Ramond two form, Einstein--Hilbert gravity and a free scalar field. Therefore, the free parts of the BV formalism as well as the $L_\infty$-algebra description just add in a straightforward manner. The interaction terms then consist of the interaction terms of Einstein--Hilbert gravity as presented in the previous section, as well as additional terms of arbitrary order involving the dilaton and the Kalb--Ramond two-form. These are readily read off~\eqref{eq:action0_N0_SUGRA}, but their explicit form will not be of relevance to us.
        
        \section{Factorisation of homotopy algebras and colour ordering}
        
        A key point of our discussion of the double copy is the factorisation of the $L_\infty$-algebras of Yang--Mills theory and $\caN=0$ supergravity into common factors. In this section, we discuss the basics of tensor products of homotopy algebras and introduce a twisted generalisation that will prove to be the key to understanding the double copy from a homotopy algebraic perspective.
        
        \subsection{Tensor products of homotopy algebras}\label{ssec:tensor_prod_homotopy_algebras}
        
        \paragraph{Tensor products of strict homotopy algebras.}
        Let $\CatAss$, $\CatCom$, and $\CatLie$ denote (the categories of) associative, commutative, and Lie algebras, respectively. Schematically, we have \uline{tensor products} of the form
        \begin{equation}\label{eq:ex_tensor_products}
            \begin{aligned}
                \otimes\,:\,\CatAss\times\CatAss\ \rightarrow\ \CatAss~,~~~
                \otimes\,:\,\CatCom\times\CatAss\ \rightarrow\ \CatAss~,~~~
                \otimes\,:\,\CatAss\times\CatCom\ \rightarrow\ \CatAss~,
                \\
                \otimes\,:\,\CatCom\times\CatCom\ \rightarrow\ \CatCom~,~~~
                \otimes\,:\,\CatCom\times\CatLie\ \rightarrow\ \CatLie~,~~~
                \otimes\,:\,\CatLie\times\CatCom\ \rightarrow\ \CatLie~.
            \end{aligned}
        \end{equation}
        In particular, let $\frA$ and $\frB$ be two algebras from this list for which there is a tensor product. The vector space underlying the tensor product algebra $\frA\otimes \frB$ is simply the ordinary tensor product of vector spaces and the product $\sfm_{2}^{\frA\otimes\frB}$ is given by
        \begin{equation}\label{eq:m2_untwisted}
            \sfm_{2}^{\frA\otimes\frB}(a_1\otimes b_1,a_2\otimes b_2)\ \coloneqq\ \sfm_{2}^{\frA}(a_1,a_2)\otimes\sfm_{2}^{\frB}(b_1,b_2)
        \end{equation}
        for $a_1,a_2\in\frA$ and $b_1,b_2\in\frB$.
        
        On the other hand, the tensor product of two cochain complexes $(\frA,\sfm^\frA_1)$ and $(\frB,\sfm^\frB_1)$ is defined as the tensor product of the underlying (graded) vector spaces $\frA$ and $\frB$,
        \begin{subequations}
            \begin{equation}
                \frA\otimes\frB\ =\ \bigoplus_{k\in\IZ}(\frA\otimes\frB)_k
                \ewith
                (\frA\otimes\frB)_k\ \coloneqq\ \bigoplus_{i+j=k}\frA_i\otimes\frB_j~,
            \end{equation}
            cf.~\eqref{eq:tensor_product_graded_vector_spaces}. The differential $\sfm_1^{\frA\otimes\frB}$ is defined as
            \begin{equation}\label{eq:m1_untwisted}
                \sfm_1^{\frA\otimes\frB}(a\otimes b)\ \coloneqq\ \sfm^\frA_1(a)\otimes b+(-1)^{|a|_\frA}a\otimes\sfm^\frB_1(b)
            \end{equation}
        \end{subequations}
        for $a\in\frA$ and $b\in\frB$.
        
        Strict $A_\infty$-, $C_\infty$-, and $L_\infty$-algebras are nothing but differential graded associative, commutative, and Lie algebras, respectively. For such algebras $\frA$ and $\frB$, the above formulas combine to 
        \begin{equation}
            \begin{aligned}
                \sfm^{\frA\otimes\frB}_1(a_1\otimes b_1)\ &\coloneqq\ \sfm^\frA_1(a_1)\otimes b_1+(-1)^{|a_1|_\frA}a_1\otimes\sfm^\frB_1(b_1)~,
                \\
                \sfm_2^{\frA\otimes\frB}(a_1\otimes b_1,a_2\otimes b_2)\ &\coloneqq\ (-1)^{|b_1|_\frB|a_2|_\frA}\sfm_2^\frA(a_1,a_2)\otimes\sfm_2^\frB(b_1,b_2)
            \end{aligned}
        \end{equation}
        for $a_1,a_2\in\frA$ and $b_1,b_2\in\frB$. If, in addition, the two differential graded algebras carry cyclic inner products $\inner{-}{-}_\frA$ and $\inner{-}{-}_\frB$, then the tensor product carries the cyclic inner product
        \begin{equation}\label{eq:cyclic_inner_product}
            \inner{a_1\otimes b_1}{a_2\otimes b_2}_{\frA\otimes\frB}\ \coloneqq\ (-1)^{|b_1|_\frB|a_2|_\frA+s(|a_1|_\frA+|a_2|_\frA)}\inner{a_1}{a_2}_\frA\,\inner{b_1}{b_2}_\frB
        \end{equation}
        for $a_1,a_2\in\frA$ and $b_1,b_2\in\frB$. Here, $s\coloneqq|\inner{-}{-}_\frB|_\frB$ is the degree of the inner product on~$\frB$.
        
        \paragraph{Tensor products of general homotopy algebras.} There is a simple argument that extends the above tensor product of strict homotopy algebras to general homotopy algebras, using not much more than the homological perturbation lemma. Let us therefore also briefly consider this case, even though we will only make use of it in passing when discussing colour-stripping of Yang--Mills amplitudes.
        
        An extension from the strict case to the general case can be performed as follows. Recall that the strictification theorem asserts that every homotopy algebra is quasi-isomorphic to a strict homotopy algebra, see \cref{app:structure_theorems} for details. Using this theorem, we first strictify each of the factors $\frA$ and $\frB$ in the tensor product $\frA\otimes\frB$ we wish to define. We then compute the tensor product $\frA^{\rm st}\otimes\frB^{\rm st}$ of the strictified factors. This is a homotopy algebra whose underlying cochain complex $\sfCh(\frA^{\rm st}\otimes\frB^{\rm st})$ is quasi-isomorphic to the tensor product $\sfCh(\frA)\otimes\sfCh(\frB)$ of the two differential complexes underlying the factors $\frA$ and $\frB$. We can then use the homological perturbation lemma, most readily in the form used e.g.~in~\cite{Jurco:2019yfd} for the coalgebra formulation of homotopy algebras, to transfer the full homotopy structure from $\sfCh(\frA^{\rm st}\otimes\frB^{\rm st})$ to $\sfCh(\frA)\otimes\sfCh(\frB)$ along the quasi-isomorphism between the cochain complexes. This yields a homotopy algebra structure on $\sfCh(\frA)\otimes \sfCh(\frB)$ together with a quasi-isomorphism to the tensor product of the strictified factors. We stress that this transfer is not unique and depends on the choice of contracting homotopy (essentially, a choice of gauge). 
        
        We stress that the fact that the tensor products~\eqref{eq:ex_tensor_products} lift to corresponding tensor products of homotopy algebras is found in the literature for special cases, see e.g.~\cite{Saneblidze:0011065,Amorim:1311.4073} for the case of $A_\infty$-algebras, as well as~\cite[Appendix B]{Cattaneo:2017tef} for the case of tensor products of $C_\infty$-algebras with Lie algebras. 
        
        \paragraph{Tensor products of matrix and Lie algebras with homotopy algebras.}
        To capture the colour decomposition of amplitudes in Yang--Mills theory, it suffices to consider the tensor product between homotopy algebras and matrix (Lie) algebras. In particular, given a matrix algebra (or, more generally, an associative algebra) $\fra$ and an $A_\infty$-algebra $(\frA,\sfm_i)$, then the tensor product $\fra\otimes\frA$ is equipped with the higher products
        \begin{equation}\label{eq:fac_higher_products_A}
            \sfm_i^{\fra\otimes\frA}(\cb_1\otimes a_1,\ldots,\cb_i\otimes a_i)\ \coloneqq\ \cb_1\cdots\cb_i\otimes\sfm_i(a_1,\ldots,a_i)
        \end{equation}
        for all $\cb_1,\ldots,\cb_i\in\fra$ and $a_1,\ldots,a_i\in\frA$ and $i\in\IN^+$. Evidently, these formulas can also be applied to the tensor product between a matrix algebra and a $C_\infty$-algebra, however, the result will, in general, be an $A_\infty$-algebra rather than a $C_\infty$-algebra as, for instance, the binary product on the tensor product will not necessarily be graded commutative.
        
        Next, we may consider the tensor product $\frg\otimes\frC$ between a Lie algebra $(\frg,[-,-])$ and a $C_\infty$-algebra $(\frC,\sfm_i)$. We obtain an $L_\infty$-algebra $(\frL,\mu_i)$ with $\frL\coloneqq\frg\otimes\frC$, however, the higher products $\mu_i$ are less straightforward than the ones in~\eqref{eq:fac_higher_products_A} for $A_\infty$-algebras. Nevertheless, they can be computed iteratively, and we obtain for the lowest products\footnote{As detailed in~\eqref{eq:A_infty_to_L_infty}, the graded anti-symmetrisation of any $A_\infty$-algebra yields an $L_\infty$-algebra, and so the form of the higher products can be gleaned from the graded anti-symmetrisation of~\eqref{eq:fac_higher_products_A}.}
        \begin{subequations}\label{eq:fac_higher_products_L}
            \begin{equation}
                \begin{aligned}
                    \mu_1(\cb_1\otimes c_1)\ &\coloneqq\ \cb_1\otimes\sfm_1(c_1)~,
                    \\
                    \mu_2(\cb_1\otimes c_1,\cb_2\otimes c_2)\ &\coloneqq\ [\cb_1,\cb_2]\otimes\sfm_2(c_1,c_2)~,
                \end{aligned}
            \end{equation}
            and
            \begin{equation}
                \begin{aligned}
                    \mu_3(\cb_1\otimes c_1,\cb_2\otimes c_2,\cb_3\otimes c_3)\ &\coloneqq\ [\cb_1,[\cb_2,\cb_3]]\otimes \sfm_3(c_1,c_2,c_3)\,-
                    \\
                    &\kern1cm-(-1)^{|c_1|_\frC|c_2|_\frC}[\cb_1,[\cb_2,\cb_3]]\otimes\sfm_3(c_2,c_1,c_3)\,+
                    \\
                    &\kern1cm+(-1)^{|c_1|_\frC|c_2|_\frC}[[\cb_1,\cb_2],\cb_3]\otimes\sfm_3(c_2,c_1,c_3)~,
                    \\
                    \mu_4(\cb_1\otimes c_1,\cb_2\otimes c_2,\cb_3\otimes c_3,\cb_4\otimes c_4)\ &\coloneqq\ [\cb_1,[\cb_2,[\cb_3,\cb_4]]]\otimes\sfm_4(c_1,c_2,c_3,c_4)\,-
                    \\
                    &\kern-3cm-(-1)^{|c_2|_\frC|c_3|_\frC}[\cb_1,[\cb_3,[\cb_2,\cb_4]]]\otimes\sfm_4(c_1,c_3,c_2,c_4)\,-
                    \\
                    &\kern-3cm-(-1)^{|c_1|_\frC|c_2|_\frC}[\cb_2,[\cb_1,[\cb_3,\cb_4]]]\otimes\sfm_4(c_2,c_1,c_3,c_4)\,+
                    \\
                    &\kern-3cm+(-1)^{|c_1|_\frC(|c_2|_\frC+|c_3|_\frC)}[[[\cb_1,\cb_4],\cb_3],\cb_2]\otimes\sfm_4(c_2,c_3,c_1,c_4)\,-
                    \\
                    &\kern-3cm-(-1)^{(|c_1|_\frC+|c_2|_\frC)|c_3|_\frC}[[\cb_1,[\cb_2,\cb_4]],\cb_3]\otimes\sfm_4(c_3,c_1,c_2,c_4)\,-
                    \\
                    &\kern-3cm-(-1)^{|c_1|_\frC(|c_2|_\frC+|c_3|_\frC)+|c_2|_\frC|c_3|_\frC}[[[\cb_1,\cb_4],\cb_2],\cb_3]\otimes\sfm_4(c_3,c_2,c_1,c_4)
                    \\
                    &\ \vdots
                \end{aligned}
            \end{equation}
        \end{subequations}
        for all $\cb_1,\ldots,\cb_4\in\frg$ and $c_1,\ldots,c_4\in\frC$. We list these formulas here as they are useful in colour-stripping in Yang--Mills theory, and we have not been able to find them in the literature.
        
        \subsection{Colour-stripping in Yang--Mills theory}\label{sec:colourDecompositionYM}
        
        As an example of the above factorisations, let us discuss colour-stripping in Yang--Mills theory and show that this is nothing but a factorisation of homotopy algebras. For concreteness, let us consider the gauge-fixed action~\eqref{eq:YMBVActionComponents} and the corresponding $L_\infty$-algebra $\frL^{\rm YM,\,gf}_{\rm BV}$.
        
        If the gauge Lie algebra $\frg$ is a matrix Lie algebra, then the $L_\infty$-algebra $\frL^{\rm YM,\,gf}_{\rm BV}$ is the total anti-symmetrisation via~\eqref{eq:A_infty_to_L_infty} of a family of $A_\infty$-algebras. One of these is special in that it is totally graded anti-symmetric~\cite{Jurco:2019yfd} and thus is also a $C_\infty$-algebra.
        
        More generally, we can factorise $\frL^{\rm YM,\,gf}_{\rm BV}$ into a gauge Lie algebra $\frg$ and a colour $C_\infty$-algebras $\frC^{\rm YM,\,gf}_{\rm BV}$ using formula~\eqref{eq:fac_higher_products_A},
        \begin{equation}\label{eq:colourDecompositionYM}
            \frL^{\rm YM,\,gf}_{\rm BV}\ =\ \frg\otimes\frC^{\rm YM,\,gf}_{\rm BV}
        \end{equation}
        Explicitly, the $C_\infty$-algebra $\frC^{\rm YM,\,gf}_{\rm BV}$ has the underlying cochain complex
        \begin{subequations}
            \begin{equation}\label{eq:BRSTComplexYM}
                \begin{tikzcd}[column sep=50pt]
                    & \stackrel{A_\mu}{\Omega^1(\IM^d)}\arrow[r,"-(\partial_\nu\partial^\mu-\delta^\mu_\nu\wave)"] \arrow[rd,"-\partial^\mu",pos=0.85] & \stackrel{A^+_\mu}{\Omega^1(\IM^d)}\arrow[rdd,"-\partial^\mu"] & 
                    \\
                    &\stackrel{b}{\scC^\infty(\IM^d)} \arrow[r,"\xi"] \arrow[ru,"\partial_\mu",crossing over,pos=0.15] & \stackrel{b^{+}}{\scC^\infty(\IM^d)} 
                    \\
                    \underbrace{\spacer{2ex}\stackrel{c}{\scC^\infty(\IM^d)}}_{\eqqcolon\,\frC^{\rm YM,\,gf}_{{\rm BV},\,0}}
                    \arrow[ruu,"-\partial_\mu"] \arrow[r,"-\wave"]& \underbrace{\spacer{2ex}\stackrel{\bar c^{+}}{\scC^\infty(\IM^d)}}_{\eqqcolon\,\frC^{\rm YM,\,gf}_{{\rm BV},\,1}} & \underbrace{\spacer{2ex}\stackrel{\bar c}{\scC^\infty(\IM^d)}}_{\eqqcolon\,\frC^{\rm YM,\,gf}_{{\rm BV},\,2}}\arrow[r,"-\wave"] & \underbrace{\spacer{2ex}\stackrel{c^{+}}{\scC^\infty(\IM^d)}}_{\eqqcolon\,\frC^{\rm YM,\,gf}_{{\rm BV},\,3}}
                \end{tikzcd}
            \end{equation}
            where we label subspaces again by the fields parametrising them
            \begin{equation}
                \begin{aligned}
                    \frC^{\rm YM,\,gf}_{{\rm BV},\,0}\ &=\ \frC^{\rm YM,\,gf}_{{\rm BV},\,0,\,c}
                    ~,~~~&
                    \frC^{\rm YM,\,gf}_{{\rm BV},\,1}\ &=\ \spacee{\frC^{\rm YM,\,gf}}{{\rm BV},\,1}{A,\,b,\,\bar c^+}~,
                    \\
                    \frC^{\rm YM,\,gf}_{{\rm BV},\,2}\ &=\ \spacee{\frC^{\rm YM,\,gf}}{{\rm BV},\,2}{A^+,\,b^+,\,\bar c}
                    ~,~~~&
                    \frC^{\rm YM,\,gf}_{{\rm BV},\,3}\ &=\ \frC^{\rm YM,\,gf}_{{\rm BV},\,3,\,c^+}~.
                \end{aligned}
            \end{equation}
            The non-trivial actions of the differential $\sfm_1$ are
            \begin{equation}
                \begin{aligned}
                    c\ &\mapstomap{\sfm_1}\ 
                        \colvec{
                            -\partial_\mu c,
                            0,
                            -\wave c
                        }
                    \ \in\ \spacee{\frC^{\rm YM,\,gf}}{{\rm BV},\,1}{A,\,b,\,\bar c^+}~,
                    \\
                    \colvec{
                        A_\mu,
                        b,
                        \bar c^+
                    }
                    \ &\mapstomap{\sfm_1}\
                        \colvec{
                            -(\partial_\mu\partial^\nu-\delta^\nu_\mu\wave)A_\nu-\partial_\mu b,
                            \partial^\mu A_\mu+\xi b,
                            0
                        }
                    \ \in\ \spacee{\frC^{\rm YM,\,gf}}{{\rm BV},\,2}{A^+,\,b^+,\,\bar c}~,
                    \\
                    \colvec{
                        A^+_\mu,
                        b^+,
                        \bar c
                    }
                    \ &\mapstomap{\sfm_1}\ -\partial^\mu(A^+_\mu+\partial_\mu\bar c)\ \in\ \frC^{\rm YM,\,gf}_{{\rm BV},\,3,\,c^+}~,
                \end{aligned}
            \end{equation}        
            the binary product $\sfm_2$ acts as 
            \begin{equation}
                \begin{aligned}
                    (c_1,c_2)\ &\mapstomap{\sfm_2}\ gc_1c_2\ \in\ \frC^{\rm YM,\,gf}_{{\rm BV},\,0,\,c}~,
                    \\
                    \left(c,
                    \colvec{
                        A_\mu,
                        b,
                        \bar c^+
                    }
                    \right)\ &\mapstomap{\sfm_2}\ -g 
                        \colvec{
                            cA_\mu,
                            0,
                            \partial^\mu(cA_\mu)
                        }
                    \ \in\ \spacee{\frC^{\rm YM,\,gf}}{{\rm BV},\,1}{A,\,b,\,\bar c^+}~,
                    \\
                    \left(c,
                    \colvec{
                        A^+_\mu,
                        \bar c,
                        b^+
                    }
                    \right)\ &\mapstomap{\sfm_2}\ g
                    \colvec{
                        c(A^+_\mu+\partial_\mu \bar c),
                        0,
                        0
                    }
                    \ \in\ \spacee{\frC^{\rm YM,\,gf}}{{\rm BV},\,2}{A^+,\,b^+,\,\bar c}~,
                    \\
                    (c,c^+)\ &\mapstomap{\sfm_2}\ gcc^+\ \in\ \frC^{\rm YM,\,gf}_{{\rm BV},\,3,\,c^+}~,
                    \\ 
                    \left(
                    \colvec{
                        A_\mu,
                        b,
                        \bar c^+
                    },
                    \colvec{
                        A^+_\nu,
                        \bar c,
                        b^+
                    }
                    \right)\ &\mapstomap{\sfm_2}\ -gA^\mu(A^+_\mu+\partial_\mu\bar c)\ \in\ \frC^{\rm YM,\,gf}_{{\rm BV},\,3,\,c^+}~,
                    \\
                    \left(
                    \colvec{
                        A_{1\mu},
                        b_1,
                        \bar c^+_1
                    },
                    \colvec{
                        A_{2\nu},
                        b_2,
                        \bar c^+_2
                    }
                    \right)\ &\mapstomap{\sfm_2}\ 2g
                        \colvec{
                            \partial^\nu (A_{1[\nu}A_{2\mu]})+A_{1}^{\nu}\partial_{[\nu}A_{2\mu]}-\partial_{[\nu}A_{1\mu]}A_{2}^{\nu},
                            0,
                            0
                        }
                    \\[5pt]
                    &\kern2cm\in\ \spacee{\frC^{\rm YM,\,gf}}{{\rm BV},\,2}{A^+,\,b^+,\,\bar c}~,
                \end{aligned}
            \end{equation}
            and the ternary product $\sfm_3$ acts as
            \begin{equation}
                \begin{aligned}
                    \left(
                    \colvec{
                        A_{1\mu},
                        b_1,
                        \bar c^+_1
                    },
                    \colvec{
                        A_{2\nu},
                        b_2,
                        \bar c^+_2
                    },
                    \colvec{
                        A_{3\kappa},
                        b_3,
                        \bar c^+_4
                    }
                    \right)
                    \ &\mapstomap{\sfm_3}\ -2g^2
                        \colvec{
                            A_1^{\nu}A_{2[\mu}A_{3\nu]} - A_{1[\mu}A_{2\nu]}A_3^{\nu},
                            0,
                            0
                        }
                    \\[5pt]
                    &\hspace{2cm}\in\ \spacee{\frC^{\rm YM,\,gf}}{2}{A^+,b^+,\bar c}~.
                \end{aligned}
            \end{equation}
        \end{subequations}
        It is a straightforward exercise to check that these higher products do indeed satisfy the $C_\infty$-algebra relations~\eqref{eq:hJI_associative} and~\eqref{eq:homotopy_commutativity}. 
        
        The factorisation~\eqref{eq:colourDecompositionYM} descends to the minimal model $\frL^{\rm YM,\,gf\,\circ}_{\rm BV}$,
        \begin{equation}
            \frL^{\rm YM,\,gf\,\circ}_{\rm BV}\ =\ \frg\otimes\frC^{\rm YM,\,gf\,\circ}_{\rm BV}~,
        \end{equation}
        and the higher products in the $C_\infty$-algebra $\frC^{\rm YM,\,gf\,\circ}$ describe the colour-ordered tree-level scattering amplitudes. We set
        \begin{equation}
            \scA_{n,0}(1,\ldots,n)\ \eqqcolon\ \rmi\sum_{\sigma\in S_n/\IZ_n}\tr(\cb_{a_{\sigma(1)}}\cdots\cb_{a_{\sigma(n)}})A_{n,0}(\sigma(1),\ldots,\sigma(n))~,
        \end{equation}
        and we have the formula 
        \begin{equation}
            A_{n,0}(1,\ldots,n)\ =\ \inner{n}{\sfm^\circ_{n-1}(1,\ldots,n-1)}~,
        \end{equation}
        where the numbers $1,\ldots,n$ represent the external free fields. The symmetry of the colour-stripped amplitude is reflected in the graded anti-symmetry of the higher products $\sfm^\circ_i$ in the $C_\infty$-algebra $\frC^{\rm YM,\,gf\,\circ}$, because all fields are of degree one.
        
        \subsection{Twisted tensor products of strict homotopy algebras}\label{ssec:twisted_tp_homotopy_algebras}
        
        The factorisation of the $L_\infty$-algebras corresponding to the field theories involved in the double copy is a twisted factorisation, and we define our notion of twisted tensor products in the following. 
        
        \paragraph{Cochain complexes.}
        A graded vector space is a particular example of a cochain complex with trivial differential. In our situation, we would like the vector space to act as an Abelian Lie algebra on the cochain complex. We therefore generalise the usual tensor product as follows. Given a graded vector space $\frV$ together with a cochain complex $(\frA,\sfm)$, we define a \uline{twist datum} $\tau_1$ to be a linear map
        \begin{equation}
            \begin{aligned}
                \tau_1\,:\,\frV\ &\rightarrow\ \frV\otimes\sfEnd(\frA)~,
                \\
                v\ &\mapsto\ \tau_1(v)\ \coloneqq\ \sum_\pi \tau_1^{\pi,1}(v)\otimes\tau_1^{\pi,2}(v)~,
            \end{aligned}
        \end{equation}
        where the index $\pi$ labels the summands in the twist element $\tau_1(v)$.\footnote{In Sweedler notation, popular e.g.~in the context of Hopf algebras, we would simply write
            \begin{equation}
                \tau_1(v):=\tau_1^{(1)}(v)\otimes \tau^{(2)}_1(v)~.
            \end{equation}
            } Given a twist datum $\tau_1$, the \uline{twisted differential} is defined by
        \begin{equation}\label{eq:twisted_differential}
            \sfm^{\tau_1}_1(v\otimes a)\ \coloneqq\ \sum_\pi(-1)^{|\tau^{\pi,1}_1(v)|_\frV}~\tau^{\pi,1}_1(v)\otimes\sfm_1(\tau^{\pi,2}_1(v)(a))
        \end{equation}
        for $v\otimes a\in\frV\otimes\frA$. This somewhat cumbersome formula describes a rather simple procedure, and it will become fully transparent in concrete examples. Evidently, there are constraints on admissible twist data. Firstly, $\sfm^{\tau_1}_1$ has to be a differential and satisfy
        \begin{equation}
            \sfm^{\tau_1}_1\circ\sfm^{\tau_1}_1\ =\ 0~,
        \end{equation}
        and secondly, $\sfm^{\tau_1}_1$ has to be cyclic with respect to the inner product~\eqref{eq:cyclic_inner_product} on the tensor product $\frV\otimes\frA$. We note that as it stands, the twisted tensor product is not necessarily compatible with quasi-isomorphisms because its cohomology is, in general, independent of those of the underlying factors. This is not an issue for our constructions, but explains why the above twist is not readily found in the mathematical literature.
        
        As we shall see momentarily, one of the key roles of the twist is the construction of a complex of differential forms from a complex of functions. The following toy example exemplifies what we have in mind. 
        
        \begin{example}\label{ex:factorisation_deRham}
            Consider the graded vector space $\frV$ and the cochain complex $(\frA,\sfm_1)$ defined by
            \begin{equation}
                \frV\ \coloneqq\ \underbrace{\IM^d\oplus\IR}_{\eqqcolon\,\frV_0}
                \eand
                \frA\ \coloneqq\ \Big(~\underbrace{\scC^\infty(\IM^d)}_{\eqqcolon\,\frA_1}\xrightarrow{~\sfid~}\underbrace{\scC^\infty(\IM^d)}_{\eqqcolon\,\frA_2}~\Big)~.
            \end{equation}
            For a basis $(\ttv^\mu,\ttn)$ of $\IM^d\oplus\IR$, a choice of twist datum is given by
            \begin{equation}
                \tau_1(\ttv^\mu)\ \coloneqq\ 0\otimes0
                \eand
                \tau_1(\ttn)\ \coloneqq\ \ttv^\mu\otimes\parder{x^\mu}~.
            \end{equation}
            The complex $\frV\otimes_\tau\frA$ with the twisted differential $\sfm^\tau_1$ is then
            \begin{equation}
                \frV\otimes_\tau\frA\ =\
                \left(
                \begin{tikzcd}[row sep=0cm,column sep=2cm]
                    \Omega^1(\IM^d)\ \cong\ \IM^d\otimes \scC^\infty(\IM^d) & \Omega^1(\IM^d)\ =\ \IM^d\otimes\scC^\infty(\IM^d)
                    \\
                    {} \oplus & \oplus 
                    \\
                    \scC^\infty(\IM^d)\ \cong\ \IR\otimes\scC^\infty(\IM^d) \arrow[uur,"\rmd"] & \scC^\infty(\IM^d)\ =\ \IR\otimes\scC^\infty(\IM^d)
                \end{tikzcd}
                \right)
            \end{equation}
            Hence, we obtain a description of the cochain complex $(\scC^\infty(\IM^d)\oplus\Omega^1(\IM^d),\rmd)$, albeit with some amount of redundancy.
        \end{example}
        
        \paragraph{Differential graded algebras.}
        Twisted tensor products for unital algebras were discussed in various places in the literature, e.g., in~\cite{Cap:1995aa}. We would like to twist the ordinary tensor product of differential graded algebras introduced in \cref{ssec:tensor_prod_homotopy_algebras}, by extending the notion of twist datum from cochain complexes as follows. Given a graded vector space $\frV$ and a differential graded algebra $(\frA,\sfm_1,\sfm_2)$, a \uline{twist datum} is a pair of maps, one linear and the other one bilinear,
        \begin{subequations}\label{eq:twistDatumDGA}
            \begin{equation}
                \begin{aligned}
                    \tau_1\,:\,\frV\ &\rightarrow\ \frV\otimes\sfEnd(\frA)~,
                    \\
                    v\ &\mapsto\ \tau_1(v)\ \coloneqq\ \sum_\pi \tau_1^{\pi,1}(v)\otimes\tau_1^{\pi,2}(v)~,
                \end{aligned}
            \end{equation}
            and
            \begin{equation}
                \begin{aligned}
                    \tau_2\,:\,\frV\otimes\frV\ &\rightarrow\ \frV\otimes\sfEnd(\frA)\otimes\sfEnd(\frA)~,
                    \\
                    v_1\otimes v_2\ &\mapsto\ \tau_2(v_1,v_2)\ \coloneqq\ \sum_\pi\tau_2^{\pi,1}(v_1,v_2)\otimes\tau_2^{\pi,2}(v_1,v_2)\otimes\tau_2^{\pi,3}(v_1,v_2)~,
                \end{aligned}
            \end{equation}
        \end{subequations}
        where we again label summands in the tensor product by $\pi$. The twisted tensor product has then higher maps
        \begin{equation}\label{eq:twisted_products}
            \begin{aligned}
                \sfm^{\tau_1}_1(v\otimes a)\ &\coloneqq\ \sum_\pi(-1)^{|\tau^{\pi,1}_1(v)|_\frV}\tau^{\pi,1}_1(v)\otimes \sfm_1(\tau^{\pi,2}_1(v)(a))~,
                \\
                \sfm^{\tau_2}_2(v_1\otimes a_1,v_2\otimes a_2)\ &\coloneqq\
                \\
                &\kern-2cm\coloneqq\ (-1)^{|v_2|_\frV\,|a_1|_\frA}\sum_\pi\tau^{\pi,1}_2(v_1,v_2)\otimes\sfm_2(\tau^{\pi,2}_2(v_1,v_2)(a_1),\tau^{\pi,3}_2(v_1,v_2)(a_2))~.
            \end{aligned}
        \end{equation}
        Note that in general, one may want to insert an additional sign $(-1)^{|\tau_2^{\pi,3}(v_1,v_2)|_\frV\,|a_1|_\frA}$ into this equation; all our twist, however, satisfy $|\tau_2^{\pi,3}(v_1,v_2)|_\frV=0$.
        
        Clearly, not every twist datum leads to a valid homotopy algebra, and just as in the case of cochain complexes, one has to check that this works for a given twist by hand. We also note that the twist datum relevant for the double copy will be able to mix types of homotopy algebras, that is, for $\frA$ an $L_\infty$-algebra, we obtain a $C_\infty$-algebra and for $\frA$ a $C_\infty$-algebra, we obtain again an $L_\infty$-algebra.
        
        Altogether, our twisted tensor products are a way of factorising strict homotopy algebras in a unique fashion as necessary for the double copy. However, it remains to be seen if our construction in its present form is mathematically interesting in a wider context.

        \section{Factorisation of free field theories and free double copy}\label{sec:factorisation_free_field_theories}
        
        In this section, we factorise the cochain complexes of the $L_\infty$-algebras of biadjoint scalar field theory, Yang--Mills theory, and $\caN=0$ supergravity into common factors. This exposes the factorisation of field theories underlying the double copy at the linearised level.
        
        \paragraph{Summary.}
        Recall that the unary product $\mu_1$ in any $L_\infty$-algebra is a differential. Consequently, any $L_\infty$-algebra $(\frL,\mu_i)$ naturally comes with an underlying cochain complex
        \begin{equation}
            \begin{tikzcd}
                \sfCh(\frL)\ \coloneqq\ \big(~~\tikzcdset{arrow style=tikz, diagrams={>=stealth'}}
                \cdots\arrow[r,"\mu_1"] & \frL_0\arrow[r,"\mu_1"] & \frL_1\arrow[r,"\mu_1"] & \frL_2\arrow[r,"\mu_1"] & \frL_3\arrow[r,"\mu_1"] & \cdots~~\big)~.
            \end{tikzcd}
        \end{equation}
        In an $L_\infty$-algebra corresponding to a field theory, the cochain complex $\sfCh(\frL)$ is the $L_\infty$-algebra of the free theory with all coupling constants put to zero. In each factorisation, we thus expose the field content as well as the free fields that parametrise the theory's scattering amplitudes.
        
        We will obtain the following factorisations of cochain complexes isomorphic to the cochain complexes underlying the $L_\infty$-algebras of biadjoint scalar field theory, Yang--Mills theory in $R_\xi$-gauge, and gauge-fixed $\caN=0$ supergravity:
        \begin{equation}\label{eq:summary_factorisations}
            \begin{gathered}
                \sfCh(\frL^{\rm biadj}_{\rm BRST})\ =\ \sfCh(\tilde\frL^{\rm biadj}_{\rm BRST})\ =\ \frg\otimes(\bar\frg\otimes\sfCh(\frScal))~,
                \\
                \sfCh(\frL^{\rm YM}_{\rm BRST})\ \cong\ \sfCh(\tilde\frL^{\rm YM}_{\rm BRST})\ =\ \frg\otimes(\frKin\otimes_{\tau_1}\sfCh(\frScal))~,
                \\
                \sfCh(\frL^{\caN=0}_{\rm BRST})\ \cong\ \sfCh(\tilde\frL^{\caN=0}_{\rm BRST})\ =\ \frKin\otimes_{\tau_1}(\frKin\otimes_{\tau_1}\sfCh(\frScal))~,
            \end{gathered}
        \end{equation}
        where $\frg$ and $\bar\frg$ are semi-simple compact matrix Lie algebras corresponding to the colour factors, $\frKin$ is a graded vector space and $\frScal$ is the $L_\infty$-algebra of a scalar field theory. We see that the cochain complex $\sfCh(\tilde\frL^{\caN=0}_{\rm BRST})$ is fully determined by the factorisation of $\sfCh(\tilde\frL^{\rm YM}_{\rm BRST})$, which is nothing but the double copy at the linearised level.
        
        There are two points to note concerning the factorisations of all those field theories but that of biadjoint scalar field theory. Firstly, these factorisations are most conveniently performed in particular field bases. We explain the required changes of basis, which are canonical transformations on the relevant BV field spaces. Secondly, these factorisations are twisted factorisation of cochain complexes of the type introduced in \cref{ssec:twisted_tp_homotopy_algebras}, with common twist datum $\tau_1$, as indicated in~\eqref{eq:summary_factorisations}.
        
        \subsection{Factorisation of the cochain complex of biadjoint scalar field theory}\label{sec:factorisationBiadjointyScalars}
        
        We start with biadjoint scalar field theory as introduced in \cref{sec:biadjoint}. This case is particularly simple as its cochain complex $\sfCh(\frL^{\rm biadj}_{\rm BRST})$ factorises as an ordinary tensor product.
        
        \paragraph{Factorisation of the cochain complex.}
        We can factor out the colour Lie algebras $\frg$ and $\bar\frg$ leaving us with the $L_\infty$-algebra $\frScal$ of a plain scalar theory,
        \begin{equation}\label{eq:fac:diff_comp:biadjoint}
            \sfCh(\frL^{\rm biadj}_{\rm BRST})\ =\ \frg\otimes(\bar\frg\otimes\sfCh(\frScal))~,
        \end{equation}
        where $\frScal$ is a homotopy algebra of cubic scalar field theory which we will fully identify later in~\eqref{eq:L_infty_scalar}. The natural cochain complex is\footnote{See~\eqref{eq:shift_notation} for the notation $\frF[k]$.}
        \begin{equation}\label{eq:Scal}
            \sfCh(\frScal)\ \coloneqq\ \left(~\underbrace{\stackrel{\tts_x}{\frF[-1]}}_{\frScal_1}~\xrightarrow{~\wave~}~\underbrace{\stackrel{\tts^+_x}{\frF[-2]}}_{\frScal_2}~\right),
        \end{equation}
        concentrated in degrees one and two, cf.~\cite{Jurco:2018sby,Macrelli:2019afx}. Here, $\tts_x$ and $\tts_x^+$ are basis vectors for the function spaces $\frF[-1]$ and $\frF[-2]$ with $\frF$ given in~\eqref{eq:properFunctionSpace}. Their inner product reads as
        \begin{equation}
            \inner{\tts_{x_1}}{\tts^+_{x_2}}\ \coloneqq\ \delta^{(d)}(x_1-x_2)~.
        \end{equation}
        
        \begin{table}[ht]
            \vspace{10pt}
            \begin{center}
                \begin{tabular}{|c|c|c|c|c|c|c|c|c|}
                    \hline
                    \multicolumn{4}{|c|}{fields} & \multicolumn{4}{c|}{anti-fields}
                    \\
                    \hline
                    & $|-|_{\rm gh}$ & $|-|_\frL$ & dim & & $|-|_{\rm gh}$ & $|-|_\frL$ & dim 
                    \\
                    \hline
                    $\tts_x$ & $0$ & $1$ & $\frac{d}{2}-1$ & $\tts^+_x$ & $-1$ & $2$ & $\frac{d}{2}+1$
                    \\
                    \hline
                \end{tabular}
            \end{center}
            \caption{The basis vectors of $\frScal$ with their $L_\infty$-degrees, their ghost numbers, and their mass dimensions.}\label{tab:fields:scal}
        \end{table}
        \begin{table}[ht]
            \begin{center}
                \resizebox{\textwidth}{!}{
                    \begin{tabular}{|c|c|c|c|c|c|c|c|}
                        \hline
                        \multicolumn{4}{|c|}{fields} & \multicolumn{4}{c|}{anti-fields}
                        \\
                        \hline
                        factorisation & $|-|_{\rm gh}$ & $|-|_\frL$ & dim & factorisation & $|-|_{\rm gh}$ & $|-|_\frL$ & dim \\
                        \hline
                        $\varphi=\cb_a\bar\cb_{\bar a}\tts_x\varphi^{a\bar a}(x)$ & $0$ & $1$ & $\tfrac{d}{2}-1$ & $\varphi^+=\cb_a\bar\cb_{\bar a}\tts^+_x\varphi^{+a\bar a}(x)$ & $-1$ & $2$ & $\tfrac{d}{2}+1$ 
                        \\
                        \hline
                    \end{tabular}
                }
            \end{center}
            \caption{Factorisation of the BV fields in the theory of biadjoint scalars. Note that we suppressed the integrals over $x$ and the tensor products for simplicity.}\label{tab:fac_vec_biadj}
        \end{table}
        The $L_\infty$-degrees correspond to the evident ghost numbers and the differential induces mass dimensions, and both are summarised in \cref{tab:fields:scal}. The factorisation of the BV fields is listed in \cref{tab:fac_vec_biadj}. The differential $\mu_1:\frL^{\rm biadj}_{{\rm BRST},\,1}\rightarrow\frL^{\rm biadj}_{{\rm BRST},\,2}$ is given by~\eqref{eq:m1_untwisted} for the untwisted tensor product,
        \begin{equation}
            \begin{aligned}
                \mu_1(\varphi)\ &= \ \mu_1\left(\cb_a\otimes\bar\cb_{\bar a}\otimes\int\rmd^dx\,\tts_x\varphi^{a\bar a}(x)\right)
                \\
                &=\ \cb_a\otimes\bar\cb_{\bar a}\otimes\mu_1^\frScal\left(\int\rmd^dx\,\tts_x\varphi^{a\bar a}(x)\right)\ =\ \wave\varphi~,
            \end{aligned}
        \end{equation}
        where $\mu_1^\frScal$ is the product appearing in~\eqref{eq:Scal}. Furthermore, the inner product is
        \begin{equation}
            \begin{aligned}
                \inner{\varphi}{\varphi^+}\ &=\ \tr_\frg(\cb_a\cb_b)\,\tr_{\bar\frg}(\bar\cb_{\bar a}\bar\cb_{\bar b})\int\rmd^dx_1\int\rmd^dx_2\,\inner{\tts_{x_1}}{\tts^+_{x_2}}\,\varphi^{a\bar a}(x_1)\varphi^{+b\bar b}(x_2)
                \\
                &=\ \int\rmd^dx\,\varphi^{a\bar a}(x)\varphi^+_{a\bar a}(x)~.
            \end{aligned}
        \end{equation}
        In conclusion, we have thus verified the factorisation of the cochain complex~\eqref{eq:fac:diff_comp:biadjoint}.
        
        \subsection{Factorisation of the cochain complex of Yang--Mills theory}\label{sec:canonicalTransformationYM}
        
        The case of Yang--Mills theory is more involved than the previous one. We start with the gauge fixed BV action~\eqref{eq:CGFYMA} and perform a canonical transformation on BV field space, which then allows for a convenient factorisation of the resulting cochain complex $\sfCh(\tilde\frL^{\rm YM}_{\rm BRST})$. For the following discussion, recall the gauge-fixing procedure and the gauge-fixed action from \cref{ssec:ex:YM}.
        
        \paragraph{Canonical transformation.}
        We note that the term $\partial^\mu A^a_\mu$ will vanish for physical states due to the polarisation condition $p\cdot\eps=0$ where $p_\mu$ is the momentum and $\eps_\mu$ is the polarisation vector for $A^a_\mu$. Off-shell, and at the level of the action, our gauge-fixing terms allow us to absorb quadratic terms in $\partial^\mu A^a_\mu$ in a field redefinition\footnote{The redefinition of the anti-fields preserves the cyclic structure of the $L_\infty$-algebra; it is mostly irrelevant for our discussion.} of the Nakanishi--Lautrup field $b^a$. We further rescale the field $b^a$ in order to homogenise its mass dimension with that of $A^a_\mu$, which will prove useful in our later discussion. Explicitly, we perform the field redefinitions
        \begin{equation}\label{eq:canonicalFieldRedefinitionYM}
            \begin{aligned}
                \tilde c^a\ &\coloneqq\ c^a~,
                &
                \tilde c^{+a}\ &\coloneqq\ c^{+a}~,
                \\
                \tilde A^a_\mu\ &\coloneqq\ A^a_\mu~,
                &
                \tilde A^{+a}_\mu\ &\coloneqq\ A_\mu^{+a}+\frac{1-\sqrt{1-\xi}}{\xi}\,\partial_\mu b^{+a}~,
                \\
                \tilde b^a\ &\coloneqq\ \sqrt{\frac{\xi}{\wave}}\left(b^a+\frac{1-\sqrt{1-\xi}}{\xi}\,\partial^\mu A^a_\mu\right),
                &
                \tilde b^{+a}\ &\coloneqq\ \sqrt{\frac{\wave}{\xi}}b^{+a}~,
                \\
                \tilde{\bar c}^a\ &\coloneqq\ \bar c^a~,
                &
                \tilde{\bar c}^{+a}\ &\coloneqq\ \bar c^{+a}~.
            \end{aligned}
        \end{equation}
        Under these field redefinitions, the action~\eqref{eq:CGFYMA}
        \begin{equation}
            S^{\rm YM}_{\rm BRST}\ =\ \int\rmd^dx\,\Big\{\tfrac12 A_{a\mu}\wave A^{a\mu}+\tfrac12(\partial^\mu A^a_\mu)^2-\bar c_a\,\wave c^a+\tfrac\xi2b_ab^a+b_a\partial^\mu A^a_\mu\Big\}+S^{\rm YM,\,int}_{\rm BRST}~,
        \end{equation}
        where $S^{\rm YM,\,int}_{\rm BRST}$ represents the interaction terms, turns into
        \begin{equation}\label{eq:YM_field_redefined_action}
            \tilde S^{\rm YM}_{\rm BRST}\ \coloneqq\ \int \rmd^dx\,\Big\{\tfrac12\tilde A_{a\mu}\wave\tilde A^{a\mu}-\tilde{\bar c}_a\wave\tilde c^a+\tfrac12\tilde b_a\wave\tilde b^a+\tilde\xi\,\tilde b_a\,\sqrt{\wave}\,\partial^\mu\tilde A^a_\mu\Big\}+\tilde S^{\rm YM,\,int}_{\rm BRST}~,
        \end{equation}
        where we rewrote the gauge-fixing parameter as
        \begin{equation}\label{eq:xi-tilde}
            \tilde\xi\ \coloneqq\ \sqrt{\frac{1-\xi}{\xi}}~.
        \end{equation}
        Note that at the level of the BV field space, the redefinitions~\eqref{eq:canonicalFieldRedefinitionYM} constitute a canonical transformation. For a more detailed analytical discussion, including the precise meaning of the inverses of the $\wave$ operator, see \cref{app:analytical_details}.
        
        \paragraph{\mathversion{bold}$L_\infty$-algebra.}
        The action~\eqref{eq:YM_field_redefined_action} is now the superfield homotopy Maurer--Cartan action~\eqref{eq:superfield_hMC_action} for an $L_\infty$-algebra $\tilde\frL^{\rm YM}_{\rm BRST}$. The complex underlying $\tilde\frL^{\rm YM}_{\rm BRST}$ is given as 
        \begin{subequations}\label{eq:L_infty_BRST_YM}
            \begin{equation}\label{eq:YM_diff_complex}
                \begin{tikzcd}
                    & \stackrel{\tilde A_\mu^a}{\Omega^1(\IM^d)\otimes\frg}\arrow[r,"\wave"]\arrow[start anchor=south east, end anchor= north west,rd, "-\tilde\xi\sqrt{\wave}\,\partial^\mu",pos=0.02, swap] &[1cm] \stackrel{\tilde A_\mu^{+a}}{\Omega^1(\IM^d)\otimes\frg} & 
                    \\[0.8cm]
                    &\stackrel{\tilde b^a}{\scC^\infty(\IM^d)\otimes\frg} \arrow[r,"\wave",swap]\arrow[start anchor=north east, end anchor= south west,ur, "\tilde\xi\sqrt{\wave}\,\partial_\mu",pos=0.02, crossing over] & \stackrel{\tilde b^{+a}}{\scC^\infty(\IM^d)\otimes\frg} 
                    \\
                    \underbrace{\spacer{2ex}\stackrel{\tilde c^a}{\scC^\infty(\IM^d)\otimes\frg}}_{\eqqcolon\,\tilde\frL^{\rm YM}_{{\rm BRST},\,0}}
                    \arrow[r,"-\wave"] & \underbrace{\spacer{2ex}\stackrel{\tilde{\bar c}^{+a}}{\scC^\infty(\IM^d)\otimes\frg}}_{\eqqcolon\,\tilde\frL^{\rm YM}_{{\rm BRST},\,1}} & \underbrace{\spacer{2ex}\stackrel{\tilde{\bar c}^a}{\scC^\infty(\IM^d)\otimes\frg}\arrow[r,"-\wave"]}_{\eqqcolon\,\tilde\frL^{\rm YM}_{{\rm BRST},\,2}} & \underbrace{\spacer{2ex}\stackrel{\tilde c^{+a}}{\scC^\infty(\IM^d)\otimes\frg}}_{\eqqcolon\,\tilde\frL^{\rm YM}_{{\rm BRST},\,3}}
                \end{tikzcd}
            \end{equation}
            with
            \begin{equation}
                \begin{aligned}
                    \tilde\frL^{\rm YM}_{{\rm BRST},\,0}\ &=\ \tilde\frL^{\rm YM}_{{\rm BRST},\,0,\,\tilde c}
                    ~,~~~&
                    \tilde\frL^{\rm YM}_{{\rm BRST},\,1}\ &=\ \spacee{\tilde\frL^{\rm YM}}{{\rm BRST},\,1}{\tilde A,\,\tilde b,\,\tilde{\bar c}^+}~,
                    \\
                    \tilde\frL^{\rm YM}_{{\rm BRST},\,2}\ &=\ \spacee{\tilde\frL^{\rm YM}}{{\rm BRST},\,1}{\tilde A^+,\,\tilde b^+,\,\tilde{\bar c}}
                    ~,~~~&
                    \tilde\frL^{\rm YM}_{{\rm BRST},\,3}\ &=\ \tilde\frL^{\rm YM}_{{\rm BRST},\,3,\,\tilde c^+}~.
                \end{aligned}
            \end{equation}
            The differential $\mu_1$ acts on the various fields as follows
            \begin{equation}\label{eq:differentialsYMRedefined}
                \begin{aligned}
                    (\tilde c^a)\ &\mapstomap{\mu_1}\ -\wave\tilde c^a\ \in\ \tilde\frL^{\rm YM}_{{\rm BRST},\,1,\,\tilde{\bar c}^+}~,
                    \\
                    \colvec{
                        \tilde A_{\mu}^a,
                        \tilde b^a
                    }
                    \ &\mapstomap{\mu_1}\ 
                    \colvec{
                        \wave\tilde A^a_\mu-\tilde\xi\sqrt{\wave}\,\partial_\mu\tilde b^a,
                        \wave\tilde b^a+\tilde\xi\sqrt{\wave}\,\partial^\mu\tilde A^a_\mu
                    }
                    \ \in\ \spacee{\tilde\frL^{\rm YM}}{{\rm BRST},\,2}{\tilde A^+,\,\tilde b^+}~,
                    \\
                    (\tilde{\bar c}^a)\ &\mapstomap{\mu_1}\ -\wave\tilde{\bar c}^a\ \in\ \tilde\frL^{\rm YM}_{{\rm BRST},\,3,\,\tilde c^+}
                \end{aligned}
            \end{equation}
            with all other actions trivial. The non-vanishing images of the higher products $\mu_2$ and $\mu_3$ are
            \begin{equation}
                \begin{aligned}
                    (\tilde A_\mu^a,\tilde c^b)\ &\mapstomap{\mu_2}\ -g{f_{bc}}^a\partial^\mu(\tilde A_\mu^b\tilde c^c)\ \in\ \tilde\frL^{\rm YM}_{{\rm BRST},\,1,\,\tilde{\bar c}^+}~,
                    \\
                    (\tilde c^a,\tilde{\bar c}^b)\ &\mapstomap{\mu_2}\ -g{f_{bc}}^a\tilde c^b\partial_\mu\tilde{\bar c}^c\ \in\ \tilde\frL^{\rm YM}_{{\rm BRST},\,2,\,\tilde A^+}~,
                    \\
                    (\tilde A_\mu^a,\tilde A_\nu^b)\ &\mapstomap{\mu_2}\ 3!g{f_{bc}}^a\partial^\nu(\tilde A_\nu^b\tilde A_\mu^c)\ \in\ \tilde\frL^{\rm YM}_{{\rm BRST},\,2,\,\tilde A^+}~,
                    \\
                    (\tilde A_\mu^a,\tilde{\bar c}^b)\ &\mapstomap{\mu_2}\ -g{f_{bc}}^a\tilde A_\mu^b\partial^\mu\tilde{\bar c}^c\ \in\ \tilde\frL^{\rm YM}_{{\rm BRST},\,3,\,\tilde c^+}~,
                    \\
                    (\tilde A_\mu^a,\tilde A_\nu^b,\tilde A_\kappa^c)\ &\mapstomap{\mu_3}\ -3!g^2 {f_{bc}}^a{f_{de}}^b\tilde A^{\nu c}\tilde A_\nu^d\tilde A_\mu^e \ \in\ \tilde\frL^{\rm YM}_{{\rm BRST},\,2,\,\tilde A^+}~,
                \end{aligned}
            \end{equation}
        \end{subequations}
        and the general expressions follow from anti-symmetrisation and polarisation. We note that the formulas~\eqref{eq:translation_BV_L_infty} are useful in the derivation of the explicit form of these higher products. 
        
        By construction, $(\tilde\frL^{\rm YM}_{\rm BRST},\mu_i)$ forms an $L_\infty$-algebra, and with the inner products
        \begin{equation}\label{eq:innerProductYMRedefined}
            \begin{aligned}
                \inner{\tilde A}{\tilde A^+}\ &\coloneqq\ \int \rmd^dx\,\tilde A^a_\mu\tilde A^{+\mu}_a~,~~
                &
                \inner{\tilde b}{\tilde b^+}\ &\coloneqq\ \int \rmd^dx\,\tilde b^a\tilde b^+_a~,
                \\
                \inner{\tilde c}{\tilde c^+}\ &\coloneqq\ \int \rmd^dx\,\tilde c^a\tilde c^+_a~,~~
                &
                \inner{\tilde{\bar c}}{\tilde{\bar c}^+}\ &\coloneqq\ -\int \rmd^dx\,\tilde{\bar c}^a\tilde{\bar c}^+_a~,
            \end{aligned}
        \end{equation}
        it is cyclic. 
        
        We stress that the Chevalley--Eilenberg differential of the $L_\infty$-algebra $\tilde\frL^{\rm YM}_{\rm BRST}$ is {\em not} the usual gauge-fixed BV operator\footnote{Here, $|_{\tilde\Phi^+_I=0}$ is again the restriction to the subspace of the BV field space where all anti-fields are zero.}
        \begin{equation}
            \tilde Q_{\rm BV}^{\rm YM,\,gf}\ \coloneqq\ \big\{\tilde S_{\rm BV}^{\rm YM,\,gf},-\big\}\big|_{\tilde\Phi^+_I=0}~,
        \end{equation}
        where $\tilde S_{\rm BV}^{\rm YM,\,gf}$ is the gauge-fixed BV action that is obtained from~\eqref{eq:gaugeFixingCanonicalTrafo} by the canonical transformation determined by the gauge-fixing fermion~\eqref{eq:gaugeFixingFermionYM}. Instead, we are merely using the general correspondence between Lagrangians and $L_\infty$-algebras as pointed out in \cref{sec:BVFormalismLinfty}. This is reflected in the images of all higher products of~\eqref{eq:YM_diff_complex} lying in spaces parametrised by anti-fields.
        
        \paragraph{Factorisation of the cochain complex.}
        As explained in \cref{sec:colourDecompositionYM}, we may factor out the gauge Lie algebra $\frg$, and we are left with a $C_\infty$-algebra. This $C_\infty$-algebra can be further factorised into a twisted tensor product, extending \cref{ex:factorisation_deRham}, and we obtain
        \begin{equation}\label{eq:factorisation_YM_diff_complex}
            \sfCh(\tilde\frL^{\rm YM}_{\rm BRST})\ =\ \frg\otimes(\frKin\otimes_{\tau_1}\sfCh(\frScal))~.
        \end{equation}
        Here, $\frg$ is the colour Lie algebra, $\sfCh(\frScal)$ is the cochain complex~\eqref{eq:Scal}, and $\frKin$ is the graded vector space\footnote{See~\eqref{eq:shift_notation} for the notation $\IR[k]$, etc.}
        \begin{equation}\label{eq:kinYM}
            \frKin\ \coloneqq\ \underbrace{\overset{\ttg}{\IR[1]}}_{\eqqcolon\,\frKin_{-1}}~\oplus~\underbrace{\big(\overset{\ttv^\mu}{\IM^d}\oplus\overset{\ttn}{\IR}\big)}_{\eqqcolon\,\frKin_0}~\oplus~\underbrace{\overset{\tta}{\IR[-1]}}_{\eqqcolon\,\frKin_1}~,
        \end{equation}
        where the typewriter letters label basis elements of the corresponding vector spaces. The natural degree-zero inner product on $\frKin$ is given by
        \begin{equation}\label{eq:inner_product_YM_1}
            \inner{\ttg}{\tta}\ \coloneqq\ -1~,~~~
            \inner{\ttv^\mu}{\ttv^\nu}\ \coloneqq\ \eta^{\mu\nu}~,~~~
            \inner{\ttn}{\ttn}\ \coloneqq\ 1~.
        \end{equation}
        The elements of $\frKin$ also carry mass dimensions, which are listed in \cref{tab:fields:kin}.
        \begin{table}[ht]
            \begin{center}
                \begin{tabular}{|c|c|c|c|}
                    \hline
                    & $|-|_{\rm gh}$ & $|-|_{\frL}$ & dim
                    \\
                    \hline
                    $\ttg$ & $1$ & $-1$ & $-1$
                    \\
                    $\ttv^\mu$ & $0$ & $0$ & $0$
                    \\
                    $\ttn$ & $0$ & $0$ & $0$
                    \\
                    $\tta$ & $-1$ & $1$ & $1$
                    \\
                    \hline
                \end{tabular}
            \end{center}
            \caption{The elements of $\frKin$ with their $L_\infty$-degrees, their ghost numbers, and their mass dimensions.}\label{tab:fields:kin}
        \end{table}
        
        We summarise the factorisation of individual Yang--Mills fields in  \cref{tab:fac_vec_YM}. A few remarks about the structure of the factorisation are in order. Whilst fields always have a factor of $\sfs_x$, anti-fields always have a factor of $\sfs^+_x$. This guarantees that the inner product is indeed that of the factorisation:~\eqref{eq:innerProductYMRedefined} is reproduced correctly using the factorisations given in \cref{tab:fac_vec_YM} and~\eqref{eq:inner_product_YM_1} complemented by the inner product $\inner{\cb_a}{\cb_b}=-\tr(\cb_a\cb_b)=\delta_{ab}$ on $\frg$:
        \begin{equation}
            \begin{aligned}
                \inner{\tilde c}{\tilde c^+}\ &=\ \innerLarge{\cb_a\otimes\ttg\otimes\int \rmd^dx_1\,\tts_{x_1}\tilde c^a(x_1)}{\cb_b\otimes\tta\otimes\int\rmd^dx_2\,\tts^+_{x_2}\tilde c^{+b}(x_2)}
                \\
                &=\ -\inner{\cb_a}{\cb_b}\,\inner{\ttg}{\tta}\int\rmd^dx_1\,\int\rmd^dx_2\,\delta^{(d)}(x_1-x_2)\tilde c^a(x_1)\,\tilde c^{+b}(x_2)
                \\
                &=\ \int\rmd^dx\,\tilde c^a(x)\,\tilde c^+_a(x)~,
                \\
                \inner{\tilde A}{\tilde A^+}\ &=\ \innerLarge{\cb_a\otimes\ttv^\mu\otimes\int\rmd^dx_1\,\tts_{x_1}\tilde A^a_\mu(x_1)}{\cb_b\otimes\ttv^\nu\otimes\int\rmd^dx_2\,\tts_{x_2}^+\tilde A^{+b}_\nu(x_2)}
                \\
                &=\ \inner{\cb_a}{\cb_b}\,\inner{\ttv^\mu}{\ttv^\nu}\int\rmd^dx_1\,\int\rmd^dx_2\,\delta^{(d)}(x_1-x_2)\tilde A^a_\mu(x_1)\,\tilde A^{+b}_\nu(x_2)
                \\
                &=\ \int\rmd^dx\,\tilde A_\mu^a(x)\,\tilde A^{+\mu}_a(x)~,
                \\
                \inner{\tilde b}{\tilde b^+}\ &=\ \innerLarge{\cb_a\otimes\ttn\otimes\int\rmd^dx_1\,\tts_{x_1}\tilde b^a(x_1)}{\cb_b\otimes\ttn\otimes\int\rmd^dx_2\,\tts^+_{x_2}\tilde b^{+b}(x_2)}
                \\
                &=\ \inner{\cb_a}{\cb_b}\,\inner{\ttn}{\ttn}\int\rmd^dx_1\,\int\rmd^dx_2\,\delta^{(d)}(x_1-x_2)\tilde c^a(x_1)\,\tilde c^{+b}(x_2)
                \\
                &=\ \int\rmd^dx\,\tilde b^a(x)\,\tilde b^+_a(x)~,
                \\
                \inner{\tilde{\bar c}}{\tilde{\bar c}^+}\ &=\ \innerLarge{\cb_a\otimes\tta\otimes\int \rmd^dx_1\,\tts_{x_1}\tilde{\bar c}^a(x_1)}{\cb_b\otimes\ttg\otimes\int\rmd^dx_2\,\tts^+_{x_2}\tilde{\bar c}^{+b}(x_2)}
                \\
                &=\ -\inner{\cb_a}{\cb_b}\,\inner{\tta}{\ttg}\int\rmd^dx_1\,\int\rmd^dx_2\,\delta^{(d)}(x_1-x_2)\tilde{\bar c}^a(x_1)\tilde{\bar c}^{+b}(x_2)
                \\
                &=\ -\int\rmd^dx\,\tilde{\bar c}^a(x)\,\tilde{\bar c}^+_a(x)~.
            \end{aligned}
        \end{equation}
        Note that the kinematic factor $\frKin$ essentially arranges the fields in a quartet: the physical field has a ghost, a Nakanishi--Lautrup field, and an anti-ghost. These patterns reoccur in the double copy. 
        
        \begin{table}[ht]
            \begin{center}
                \resizebox{\textwidth}{!}{
                    \begin{tabular}{|c|c|c|c|c|c|c|c|}
                        \hline
                        \multicolumn{4}{|c|}{fields} & \multicolumn{4}{c|}{anti-fields}
                        \\
                        \hline
                        factorisation & $|-|_{\rm gh}$ & $|-|_\frL$ & dim & factorisation & $|-|_{\rm gh}$ & $|-|_\frL$ & dim
                        \\
                        \hline
                        $\tilde c=\cb_a{\ttg}\tts_x\tilde c^a(x)$ & $1$ & $0$ & $\tfrac{d}{2}-2$ & $\tilde c^+=\cb_a{\tta}\tts^+_x\tilde c^{+a}(x)$ & $-2$ & $3$ & $\tfrac{d}{2}+2$ 
                        \\
                        $\tilde A=\cb_a{\ttv^\mu}\tts_x\tilde A^a_\mu(x)$ & $0$ & $1$ & $\tfrac{d}{2}-1$& $\tilde A^+=\cb_a{\ttv^\mu}\tts^+_x\tilde A^{+a}_\mu(x)$ & $-1$ & $2$ & $\tfrac{d}{2}+1$
                        \\
                        $\tilde b=\cb_a{\ttn}\tts_x \tilde b^a(x)$ & $0$ & $1$ & $\tfrac{d}{2}-1$& $\tilde b^+=\cb_a{\ttn}\tts^+_x \tilde b^{+a}(x)$ & $-1$ & $2$ & $\tfrac{d}{2}+1$
                        \\
                        $\tilde{\bar c}=\cb_a{\tta}\tts_x\tilde{\bar c}^a(x)$ & $-1$ & $2$ & $\tfrac{d}{2}$ & $\tilde{\bar c}^+=\cb_a{\ttg}\tts^+_x\tilde{\bar c}^{+a}(x)$ & $0$ & $1$ & $\tfrac{d}{2}$
                        \\
                        \hline
                    \end{tabular}
                }
            \end{center}
            \caption{Factorisation of the redefined BV fields for Yang--Mills theory from \cref{tab:fields:YM} after the field redefinitions~\eqref{eq:canonicalFieldRedefinitionYM}. Here, $\cb_a$ denote the basis vectors of $\frg$. Likewise, $\ttg$, $\ttn$, $\ttv^\mu$, and $\tta$ denote the basis vectors of $\frKin$ defined in~\eqref{eq:kinYM}. Furthermore, $\tts_x$ and $\tts_x^+$ are the basis vectors of $\frScal$ from \cref{tab:fields:scal}. Note that we suppressed the integrals over $x$ and the tensor products for simplicity.\label{tab:fac_vec_YM}}
        \end{table}
        
        To extend this factorisation of graded vector spaces to a factorisation of cochain complexes, we introduce the twist datum $\tau_1$ given by
        \begin{equation}\label{eq:twistYM}
            \tau_1(\ttg)\ \coloneqq\ \ttg\otimes\sfid~,~~~
            \begin{aligned}
                \tau_1(\ttv^\mu)\ &\coloneqq\ \ttv^\mu\otimes\sfid+\tilde\xi\ttn\otimes\frac{1}{\sqrt{\wave}}\partial^\mu~,
                \\
                \tau_1(\ttn)\ &\coloneqq\ \ttn\otimes\sfid-\tilde\xi\ttv^\mu\otimes\frac{1}{\sqrt{\wave}}\partial_\mu~,
            \end{aligned}
            ~~~
            \tau_1(\tta)\ \coloneqq\ \tta\otimes\sfid~,
        \end{equation}
        and we shall use the convenient shorthand notation
        \begin{equation}
            \tau_1(\ttv^\mu,\ttn)
            \colvec{
                \int\rmd^dx\,\tts_x\tilde A_\mu^a(x),
                \int\rmd^dx\,\tts_x\tilde b^a(x)
            }
            \ =\ (\ttv^\mu,\ttn)\otimes
            \colvec{
                \sfid & -\frac{\tilde\xi}{\sqrt{\wave}}\partial_\mu,
                \frac{\tilde\xi}{\sqrt{\wave}}\partial^\mu & \sfid
            }
            \colvec{
                \int\rmd^dx\,\tts_x\tilde A_\mu^a(x),
                \int\rmd^dx\,\tts_x\tilde b^a(x)
            }.
        \end{equation}
        
        The twisted differentials on $\frg\otimes(\frKin\otimes_{\tau_1}\frScal)$ are now indeed those of~\eqref{eq:differentialsYMRedefined}:
        \begingroup
        \allowdisplaybreaks
        \begin{subequations}
            \begin{align*}
                \mu_1(\tilde c)\ &=\ \mu_1\left(\cb_a\otimes\ttg\otimes\int\rmd^dx\,\tts_x\tilde c^a(x)\right)
                \\ 
                &=\ -\cb_a\otimes\ttg\otimes\mu_1^\frScal\left(\int\rmd^dx\,\tts_x\tilde c^a(x)\right)
                \\
                &=\ \cb_a\otimes\ttg\otimes\int\rmd^dx\,\tts^+_x\big\{-\wave\tilde c^a(x)\big\}~,\numberthis
            \end{align*}
            \pagebreak
            \begin{align*}
                \mu_1
                \begin{pmatrix}
                    \tilde A
                    \\
                    \tilde b 
                \end{pmatrix}
                \ &=\ \mu_1
                \left(
                \cb_a\otimes(\ttv^\mu,\ttn)\otimes
                \colvec{
                    \int\rmd^dx\,\tts_x\tilde A_\mu^a(x),
                    \int\rmd^dx\,\tts_x\tilde b^a(x)
                }
                \right)
                \\ 
                &=\ \cb_a\otimes(\ttv^\mu,\ttn)\otimes\mu_1^\frScal
                \left(
                {\colvec{
                        \sfid & -\frac{\tilde\xi}{\sqrt{\wave}}\partial_\mu,
                        \frac{\tilde\xi}{\sqrt{\wave}}\partial^\mu & \sfid
                    }}
                \colvec{
                    \int\rmd^dx\,\tts_x\tilde A_\mu^a(x),
                    \int\rmd^dx\,\tts_x\tilde b^a(x)
                }
                \right)
                \\
                &=\ \cb_a\otimes(\ttv^\mu,\ttn)\otimes
                \colvec{
                    \int\rmd^dx\,\tts^+_x\big\{\wave\tilde A_\mu^a(x)-\tilde\xi\sqrt{\wave}\,\partial_\mu\tilde b^a(x)\big\},
                    \int\rmd^dx\,\tts^+_x\big\{\wave\tilde b^a(x)+\tilde\xi\sqrt{\wave}\,\partial^\mu\tilde A_\mu^a(x)\big\}
                }
                \\
                &=\ \cb_a\otimes
                \colvec{
                    \ttv^\mu\otimes\int\rmd^dx\,\tts^+_x\big\{\wave\tilde A_\mu^a(x)-\tilde\xi\sqrt{\wave}\,\partial_\mu\tilde b^a(x)\big\},
                    \ttn\otimes\int\rmd^dx\,\tts^+_x\big\{\wave\tilde b^a(x)+\tilde\xi\sqrt{\wave}\,\partial^\mu\tilde A_\mu^a(x)\big\}
                },\numberthis
                \\
                \mu_1(\tilde{\bar c})\ &=\ \mu_1\left(\cb_a\otimes\tta\otimes\int\rmd^dx\,\tts_x\tilde{\bar c}^a(x)\right)
                \\ 
                &=\ -\cb_a\otimes\tta\otimes\mu_1^\frScal\left(\int\rmd^dx\,\tts_x\tilde{\bar c}^a(x)\right)
                \\
                &=\ \cb_a\otimes\tta\otimes\int\rmd^dx\,\tts^+_x\big\{-\wave\tilde{\bar c}^a(x)\big\}~.\numberthis
            \end{align*}
        \end{subequations}
        \endgroup
        Altogether, we saw that the factorisation~\eqref{eq:factorisation_YM_diff_complex} is valid for twist datum $\tau_1$.
        
        \subsection{Canonical transformation for the free Kalb--Ramond two-form}
        
        To keep our discussion manageable, we shall discuss the canonical transformations for the free Kalb--Ramond two-form and Einstein--Hilbert gravity separately. For the following discussion, recall the gauge-fixing procedure and the gauge-fixed action from \cref{ssec:ex:KR}.
        
        \paragraph{Canonical transformation.}
        Analogously to the case of Yang--Mills theory, we can now perform a field redefinition in order to eliminate the quadratic terms that would vanish on-shell in Lorenz gauge due to contractions between momenta and polarisation tensors. We also insert inverses of the wave operator to match the mass dimensions of fields of $L_\infty$-degree one. The field redefinitions are
        \begin{subequations}\label{eq:canonicalFieldRedefinitionKR}
            \begin{equation}
                \begin{aligned}
                    \tilde\lambda\ &\coloneqq\ \lambda~,
                    &
                    \tilde\lambda^+\ &\coloneqq\ \lambda^+~,
                    \\
                    \tilde\Lambda_\mu\ &\coloneqq\ \Lambda_\mu~,
                    &
                    \tilde\Lambda^+_\mu\ &\coloneqq\ \Lambda^+_\mu+\frac{1-\sqrt{1-\xi}}{\xi}\partial_\mu\gamma^+~,
                    \\
                    \tilde\gamma\ &\coloneqq\ \sqrt{\frac{\xi}{\wave}}\left(\gamma+\frac{1-\sqrt{1-\xi}}{\xi}\partial^\mu\Lambda_\mu\right),
                    &
                    \tilde\gamma^+\ &\coloneqq\ \sqrt{\frac{\wave}{\xi}}\gamma^+~,
                    \\
                    \tilde B_{\mu\nu}\ &\coloneqq\ B_{\mu\nu}~,
                    &
                    \tilde B^+_{\mu\nu}\ &\coloneqq\ B^+_{\mu\nu}+\frac{1-\sqrt{1-\xi}}{\xi}\partial_{[\mu}\alpha^+_{\nu]}~,
                    \\
                    \tilde\alpha_\mu\ &\coloneqq\ \sqrt{\frac{\xi}{\wave}}\left(\alpha_\mu-\partial_\mu\eps\,-\,\phantom{\frac12}\right.
                    &
                    \tilde\alpha^+_\mu\ &\coloneqq\ \sqrt{\frac{\wave}{\xi}}\left(\alpha_\mu^++\frac{1-\xi}{2\wave}\partial_\mu\eps^+\right),
                    \\
                    &\kern2cm-\frac{1-\xi}{2\wave}\partial_\mu\partial^\nu\alpha_\nu\,+
                    \\
                    &\kern2cm\left.+\frac{1-\sqrt{1-\xi}}{\xi}\partial^\nu B_{\nu\mu}\right),
                    \\
                    \tilde\eps\ &\coloneqq\ \eps+\frac{1-\xi}{2\wave}\partial^\mu\alpha_\mu~,
                    &
                    \tilde\eps^+\ &\coloneqq\ \frac{1+\xi}{2}\eps^+-\partial^\mu\alpha_\mu^+~,
                    \\
                    \tilde{\bar\Lambda}_\mu\ &\coloneqq\ \bar\Lambda_\mu~,
                    &
                    \tilde{\bar\Lambda}^+_\mu\ &\coloneqq\ \bar\Lambda^+_\mu+\frac{1-\sqrt{1-\xi}}{\xi}\partial_\mu\bar\gamma^+~,
                    \\
                    \tilde{\bar\gamma}\ &\coloneqq\ \sqrt{\frac{\xi}{\wave}}\left(\bar\gamma+\frac{1-\sqrt{1-\xi}}{\xi}\partial^\mu\bar\Lambda_\mu\right),
                    &
                    \tilde{\bar\gamma}^+\ &\coloneqq\ \sqrt{\frac{\wave}{\xi}}\bar\gamma^+~,
                    \\
                    \tilde{\bar\lambda}\ &\coloneqq\ \bar\lambda~,
                    &
                    \tilde{\bar\lambda}^+\ &\coloneqq\ \bar\lambda^+~,
                \end{aligned}
            \end{equation}
            with
            \begin{equation}
                \xi\ \coloneqq\ \xi_1\ =\ \xi_3-\xi_2
            \end{equation}
        \end{subequations}
        from~\eqref{eq:gaugeFixingFermionKR}. These redefinitions constitute canonical transformations on the BV field space. Upon applying these transformations to the action~\eqref{eq:BRSTactionKR}, we obtain 
        \begin{equation}\label{eq:canonicallyRedefinedActionKR}
            \begin{aligned}
                \tilde S^{\rm KR}_{\rm BRST}\ &\coloneqq\ \int\rmd^dx\,\Big\{\tfrac14\tilde B_{\mu\nu}\wave\tilde B^{\mu\nu}-\tilde{\bar\Lambda}_\mu\wave \tilde\Lambda^\mu+\tfrac12\tilde\alpha_\mu\wave\tilde\alpha^\mu-\tfrac{\tilde\xi^2}{2}(\partial^\mu\tilde\alpha_\mu)^2+\tfrac12\tilde\eps\wave\tilde\eps-\tilde{\bar\lambda}\wave\tilde\lambda\,-
                \\
                &\kern2cm
                -\tilde{\bar\gamma}\wave\tilde\gamma+\tilde\xi\tilde\alpha^\nu\sqrt{\wave}\,\partial^\mu\tilde B_{\mu\nu}+\tilde\xi\tilde\gamma\sqrt{\wave}\,\partial_\mu\tilde{\bar\Lambda}^\mu-\tilde\xi\tilde{\bar\gamma}\sqrt{\wave}\,\partial_\mu\tilde\Lambda^\mu\Big\}\,,
            \end{aligned}
        \end{equation}
        where we have again used the shorthand $\tilde\xi\ \coloneqq\ \sqrt{\frac{1-\xi}{\xi}}$, cf.~\eqref{eq:xi-tilde}.
        
        \paragraph{\mathversion{bold}$L_\infty$-algebra.} 
        The action~\eqref{eq:canonicallyRedefinedActionKR} is the superfield homotopy Maurer--Cartan action~\eqref{eq:superfield_hMC_action} of an $L_\infty$-algebra, denoted by $\tilde\frL^{\rm KR}_{\rm BRST}$, that is given by
        \begin{subequations}
            \begin{equation}
                \begin{tikzcd}[
                    every label/.append style={scale=.95},
                    cells={nodes={scale=.95}}
                    ]
                    &&\stackrel{\tilde\eps}{\scC^\infty(\IM^d)}\rar["\wave"] & \stackrel{\tilde\eps^+}{\scC^\infty(\IM^d)}
                    \\
                    &\stackrel{\tilde\Lambda_\mu}{\Omega^1(\IM^d)}\rar["-\wave"]\drar["-\tilde\xi\sqrt{\wave}\,\partial^\mu" very near start, swap] & \stackrel{\tilde{\bar\Lambda}_\mu^+}{\Omega^1(\IM^d)} & \stackrel{\tilde{\bar\Lambda}_\mu}{\Omega^1(\IM^d)}\rar["-\wave"]\drar["-\tilde\xi\sqrt{\wave}\,\partial^\mu" very near start, swap]&\stackrel{\tilde\Lambda^+_\mu}{\Omega^1(\IM^d)} 
                    \\
                    & \stackrel{\tilde\gamma}{\scC^\infty(\IM^d)}\rar["-\wave"]\urar["\tilde\xi\sqrt{\wave}\,\partial_\mu" very near end, swap, crossing over]& \stackrel{\tilde{\bar\gamma}^+}{\scC^\infty(\IM^d)}& \stackrel{\tilde{\bar\gamma}}{\scC^\infty(\IM^d)}\rar["-\wave"]\urar["\tilde\xi\sqrt{\wave}\,\partial^\mu" very near end, swap, crossing over]& \stackrel{\tilde\gamma^+}{\scC^\infty(\IM^d)}
                    \\
                    \stackrel{\tilde\lambda}{\scC^\infty(\IM^d)}\rar["\wave"] &\stackrel{\tilde{\bar\lambda}^+}{\scC^\infty(\IM^d)}&\stackrel{\tilde B_{\mu\nu}}{\Omega^2(\IM^d)}\rar["\wave"]\drar["-\tilde\xi\sqrt{\wave}\,\partial^\nu" very near start, swap]&\stackrel{\tilde B^+_{\mu\nu}}{\Omega^2(\IM^d)} &\stackrel{\tilde{\bar\lambda}}{\scC^\infty(\IM^d)}\rar["\wave"] &\stackrel{\tilde\lambda^+}{\scC^\infty(\IM^d)}
                    \\
                    \underbrace{\spacer{2ex}~~~~~~~~~~}_{\eqqcolon\,\tilde\frL^{\rm KR}_{{\rm BRST},\,-1}} & \underbrace{\spacer{2ex}~~~~~~~~~~}_{\eqqcolon\,\tilde\frL^{\rm KR}_{{\rm BRST},\,0}} & \underbrace{\spacer{2ex}\stackrel{\tilde\alpha_\mu}{\Omega^1(\IM^d)}}_{\eqqcolon\,\tilde\frL^{\rm KR}_{{\rm BRST},\,1}}
                    \arrow{r}{\wave}[below]{\tilde\xi^2\partial_{\nu}\partial^{\mu}}
                    \urar["\tilde\xi\sqrt{\wave}\,\partial_{[\nu}" very near end, swap, crossing over]& \underbrace{\spacer{2ex}\stackrel{\tilde\alpha^+_\mu}{\Omega^1(\IM^d)}}_{\eqqcolon\,\tilde\frL^{\rm KR}_{{\rm BRST},\,2}} & \underbrace{\spacer{2ex}~~~~~~~~~~}_{\eqqcolon\,\tilde\frL^{\rm KR}_{{\rm BRST},\,3}} & \underbrace{\spacer{2ex}~~~~~~~~~~}_{\eqqcolon\,\tilde\frL^{\rm KR}_{{\rm BRST},\,4}}
                \end{tikzcd}
            \end{equation}
            with 
            \begin{equation}
                \begin{aligned}
                    \tilde\frL^{\rm KR}_{{\rm BRST},\,-1}\ &=\ \tilde\frL^{\rm KR}_{{\rm BRST},\,-1,\,\tilde\lambda}
                    ~,~~~&
                    \tilde\frL^{\rm KR}_{{\rm BRST},\,0}\ &=\ \spacee{\tilde\frL^{\rm KR}}{{\rm BRST},\,0}{\tilde\Lambda,\,\tilde\gamma,\,\tilde{\bar\lambda}^+}~,
                    \\
                    \tilde\frL^{\rm KR}_{{\rm BRST},\,1}\ &=\ \spacee{\tilde\frL^{\rm KR}}{{\rm BRST},\,1}{\tilde\eps,\tilde{\bar\Lambda}^+,\,\tilde{\bar\gamma}^+,\,\tilde B,\,\tilde\alpha}
                    ~,~~~&
                    \tilde\frL^{\rm KR}_{{\rm BRST},\,2}\ &=\ \spacee{\tilde\frL^{\rm KR}}{{\rm BRST},\,2}{\tilde\eps^+,\tilde{\bar\Lambda},\tilde{\bar\gamma},\,\tilde B^+,\,\tilde\alpha^+}~,
                    \\
                    \tilde\frL^{\rm KR}_{{\rm BRST},\,3}\ &=\ \spacee{\tilde\frL^{\rm KR}}{{\rm BRST},\,3}{\tilde\Lambda^+,\,\tilde\gamma^+,\,\tilde{\bar\lambda}}
                    ~,~~~&
                    \tilde\frL^{\rm KR}_{{\rm BRST},\,4}\ &=\ \tilde\frL^{\rm YM}_{{\rm BRST},\,4,\,\tilde\lambda^+}~,
                \end{aligned}
            \end{equation}
            and the non-vanishing differential
            \begin{equation}
                \begin{aligned}
                    (\tilde\lambda)\ &\mapstomap{\mu_1}\ \wave\tilde\lambda\ \in\ \tilde\frL^{\rm KR}_{{\rm BRST},\,0,\,\tilde{\bar\lambda}^+}~,
                    \\
                    \colvec{
                        \tilde\Lambda_\mu,
                        \tilde\gamma
                    }\ &\mapstomap{\mu_1}\ 
                    -\colvec{
                        \wave\tilde\Lambda_\mu-\tilde\xi\sqrt{\wave}\,\partial_\mu\tilde\gamma,
                        \wave\tilde\gamma+\tilde\xi\sqrt{\wave}\,\partial^\mu\tilde\Lambda_\mu
                    }\ \in\ \spacee{\tilde\frL^{\rm KR}}{{\rm BRST},\,1}{\tilde{\bar{\Lambda}}^+,\,\tilde{\bar{\gamma}}^+}~,
                    \\
                    \colvec{
                        \tilde B_{\mu\nu},
                        \tilde\alpha_\mu
                    }\ &\mapstomap{\mu_1}\ 
                    \colvec{
                        \wave\tilde B_{\mu\nu}-2\tilde\xi\sqrt{\wave}\,\partial_{[\mu}\tilde\alpha_{\nu]},
                        \wave\tilde\alpha_\mu+\tilde\xi\sqrt{\wave}\,\partial^\nu\tilde B_{\nu\mu}+\tilde\xi^2\partial_\mu\partial^\nu\tilde\alpha_\nu
                    }\ \in\ \spacee{\tilde\frL^{\rm KR}}{{\rm BRST},\,2}{\tilde{B}^+,\,\tilde{\alpha}^+}~,
                    \\
                    \colvec{
                        \tilde{\bar\Lambda}_\mu,
                        \tilde{\bar\gamma}
                    }
                    \ &\mapstomap{\mu_1}\ 
                    -\colvec{
                        \wave\tilde{\bar\Lambda}_\mu-\tilde\xi\sqrt{\wave}\,\partial_\mu\tilde{\bar\gamma},
                        \wave\tilde{\bar\gamma}+\tilde\xi\sqrt{\wave}\,\partial^\mu\tilde{\bar\Lambda}_\mu
                    }
                    \ \in\ \spacee{\tilde\frL^{\rm KR}}{{\rm BRST},\,3}{\tilde\Lambda^+,\,\tilde\gamma^+}~,
                    \\
                    (\tilde{\bar\lambda})\ &\mapstomap{\mu_1} \ \wave\tilde{\bar\lambda}\ \in\ \tilde\frL^{\rm KR}_{{\rm BRST},\,4,\,\tilde{\lambda}^+}~.
                \end{aligned}
            \end{equation}
        \end{subequations}
        There are no additional higher products because the theory is free. The expressions
        \begin{equation}
            \begin{aligned}
                \inner{\tilde\lambda}{\tilde\lambda^+}\ &\coloneqq\ -\int\rmd^dx\,\tilde\lambda\tilde\lambda^+
                ~,~~~
                &\inner{\tilde{\bar\lambda}}{\tilde{\bar\lambda}^+}\ &\coloneqq\ -\int\rmd^dx\,\tilde{\bar\lambda}\tilde{\bar\lambda}^+~,
                \\
                \inner{\tilde\Lambda}{\tilde\Lambda^+}\ &\coloneqq\ \int \rmd^dx\,\tilde\Lambda^\mu\tilde\Lambda^+_\mu
                ~,~~~
                &\inner{\tilde{\bar\Lambda}}{\tilde{\bar\Lambda}^+}\ &\coloneqq\ -\int\rmd^dx\,\tilde{\bar\Lambda}^\mu\tilde{\bar\Lambda}^+_\mu~,
                \\
                \inner{\tilde B}{\tilde B^+}\ &\coloneqq\ \tfrac12\int\rmd^dx\,\tilde B^{\mu\nu}\tilde B^+_{\mu\nu}~,
                \\
                \inner{\tilde\alpha}{\tilde\alpha^+}\ &\coloneqq\ \int\rmd^dx\,\tilde\alpha^\mu\tilde\alpha^+_\mu
                ~,~~~
                &\inner{\tilde\eps}{\tilde\eps^+}\ &\coloneqq\ \int\rmd^dx\,\tilde\eps\tilde\eps^+~,
                \\
                \inner{\tilde\gamma}{\tilde\gamma^+}\ &\coloneqq\ \int\rmd^dx\,\tilde\gamma\tilde\gamma^+
                ~,~~~
                &\inner{\tilde{\bar\gamma}}{\tilde{\bar\gamma}^+}\ &\coloneqq\ -\int\rmd^dx\,\tilde{\bar\gamma}\tilde{\bar\gamma}^+
            \end{aligned}
        \end{equation}
        define a cyclic inner product on $(\tilde\frL^{\rm YM}_{\rm BRST},\mu_1)$.
        
        \subsection{Canonical transformation for Einstein--Hilbert gravity with dilaton}\label{ssec:canonicalTransformationEHD}
        
        The case of Einstein--Hilbert gravity with dilaton is now more involved that of the free Kalb--Ramond field. For the following discussion, recall the gauge-fixing procedure and the gauge-fixed action from \cref{ssec:ex:EH}.
        
        \paragraph{Canonical transformations.} We start from the Lagrangian~\eqref{eq:canonicallyRedefinedActionEHZerothOrder} but add a scalar kinetic term for the dilaton $\varphi$,
        \begin{equation}
                \scL_0^{\rm eEHD,\,gf}\ \coloneqq\ \scL_0^{\rm eEH,\,gf}+\tfrac{1}{2}\varphi\wave\varphi~.
        \end{equation}
        We perform a field redefinition analogous to the case of Yang--Mills theory and the Kalb--Ramond field, absorbing various terms that vanish on-shell, as well as the trace of $h_{\mu\nu}$ in $\delta$ and ensuring that all fields come with the right propagators. For the fields of non-vanishing ghost number, the transformation reads as 
        \begin{subequations}\label{eq:canonicalFieldRedefinitionEH}
            \begin{equation}
                \begin{aligned}
                    \tilde X_\mu\ &\coloneqq\ X_\mu~,
                    &
                    \tilde X^+_\mu\ &\coloneqq\ X^+_\mu~,
                    \\
                    \tilde\beta\ &\coloneqq\ \frac{1}{\sqrt{\wave}}\,\beta~,
                    &
                    \tilde\beta^+\ &\coloneqq\ \sqrt{\wave}\,\beta^+~,
                    \\
                    \tilde{\bar X}_\mu\ &\coloneqq\ \bar X_\mu~,
                    &
                    \tilde{\bar X}^+_\mu\ &\coloneqq\ \bar X^+_\mu-\frac{1-\sqrt{1-\xi}}{\sqrt{\xi}}\partial_\mu\bar \beta^+~,
                    \\
                    \tilde{\bar{\beta}}\ &\coloneqq\ \frac{1}{\sqrt{\wave}}\left(\bar\beta-\frac{1-\sqrt{1-\xi}}{\sqrt{\xi}}\partial^\mu \bar X_\mu\right),
                    &
                    \tilde{\bar\beta}^+\ &\coloneqq\ \sqrt{\wave}\,\bar\beta^+~,
                \end{aligned}
            \end{equation}
            where we worked in the special gauge 
            \begin{equation}
                \begin{aligned}
                    &\zeta_4\ =\ 1~,~~~\zeta_5\ =\ \frac{1-\sqrt{1-\xi}}{\sqrt{\xi}}~,~~~\zeta_6\ =\ -\tfrac12~,~~~\zeta_7\ =\ -\frac{4(\xi+2\sqrt{1-\xi}\xi-\sqrt{1-\xi}-1)}{\sqrt{\xi}(4\xi-3)}~,
                    \\
                    &\zeta_8\ =\ \frac{1}{4(3-4\xi)^2\sqrt{\xi}}\Big\{50\big(1+\sqrt{1-\xi}\big)-\xi\Big[5\big(34+29\sqrt{1-\xi}\big)\,+
                    \\
                    &\kern4cm+8\xi\Big(-23-15\sqrt{1-\xi}+2\big(4+\sqrt{1-\xi}\big)\xi\Big)\Big]\Big\}\,,
                    \\
                    &\zeta_9\ =\ 0~,~~~\zeta_{10}\ =\ \frac{1}{2\xi+\sqrt{1-\xi}-1}~,~~~\zeta_{11}\ =\ 0
                \end{aligned}
            \end{equation}
            from~\eqref{eq:gaugeFixingFermionEH}. The expressions for $\zeta_7$ and $\zeta_8$ make it apparent that the field redefinitions we would like to perform here are much more involved than in the case of the Kalb--Ramond field.\footnote{We suspect that there is a simpler field redefinition in a simpler gauge which we have not been able to identify yet.} Because the resulting expressions for the fields of ghost number zero are too involved and not very illuminating, we restrict ourselves to the case $\xi=1$ corresponding to Feynman gauge in Yang--Mills theory. Here, we have the inverse field transformations
            \begin{equation}
                \begin{aligned}
                    h_{\mu\nu}\ &=\ \tilde h_{\mu\nu}-\frac{\partial_\mu}{\wave}\Big(\partial^\kappa\tilde h_{\kappa\nu}-\frac12\partial_\nu\mathring{\tilde h}\Big)-\frac{\partial_\nu}{\wave}\Big(\partial^\kappa\tilde h_{\kappa\mu}-\frac12\partial_\mu\mathring{\tilde h}\Big)-\frac{1}{\sqrt{\wave}}\Big(\partial_\mu\tilde\varpi_\nu+\partial_\nu\tilde\varpi_\mu\Big)\,,
                    \\
                    \varpi_\mu\ &=\ -\partial_\mu\tilde\delta-\partial^\nu\tilde h_{\mu\nu}-\sqrt{\wave}\,\tilde\varpi_\mu~,
                    \\
                    \pi\ &=\ -2\wave\tilde\delta+\wave\tilde\pi-\partial^\mu\partial^\nu\tilde h_{\mu\nu}~,
                    \\
                    \delta\ &=\ \frac{\tilde\delta}{2}+\frac{\tilde\pi}{4}+\frac{\mathring{\tilde h}}{8}+\frac{\partial^\mu}{4\wave}\Big(\partial^\nu\tilde h_{\mu\nu}-\frac12\partial_\mu\mathring{\tilde h}\Big)\,,
                    \\
                    \varphi\ &=\ \frac{\mathring{\tilde h}}{2\sqrt{2}}-\frac{\partial^\mu}{\sqrt{2}\wave}\Big(\partial^\nu\tilde h_{\mu\nu}-\frac12\partial_\mu\mathring{\tilde h}\Big)
                \end{aligned}
            \end{equation}
        \end{subequations}
        with readily computed antifield transformations. Jumping ahead of our story a bit, we note that the field redefinition for $\varphi$ agrees precisely with the expectation of how the dilaton should be extracted from the double copied metric perturbation $\tilde h_{\mu\nu}$.
        
        For general $\xi$, the total Lagrangian, to lowest order in $\kappa$, reads as
        \begin{equation}\label{eq:LeEHD}
            \begin{aligned}
                \tilde\scL^{\rm eEHD}_{\rm BRST,\,0}\ &=\ \tfrac14\tilde h_{\mu\nu}\wave\tilde h^{\mu\nu}
                +\tfrac12\tilde\varpi_\mu\wave\tilde\varpi^\mu
                +\tfrac12\tilde\xi^2(\partial^\mu\tilde\varpi_\mu)^2+\tilde\xi\tilde\varpi^\nu\sqrt{\wave}\partial^\mu\tilde h_{\mu\nu}\ -
                \\
                &\kern0.5cm
                -\tfrac12\tilde\delta\wave\tilde\delta+\tfrac14\tilde\pi\wave\tilde\pi
                +\tilde\xi\tilde\pi\sqrt{\wave}\partial_\mu\tilde\varpi^\mu\,+\tfrac12\tilde\xi^2\tilde\pi\partial_\mu\partial_\nu\tilde h^{\mu\nu}\ -
                \\
                &\kern0.5cm
                -\tilde{\bar X}_\mu\wave\tilde X^\mu-\tilde{\bar\beta}\wave\tilde\beta
                +\tilde\xi\tilde\beta\sqrt{\wave}\partial_\mu\tilde{\bar X}^\mu
                -\tilde\xi\tilde{\bar\beta}\sqrt{\wave}\partial_\mu\tilde X^\mu~.
            \end{aligned}
        \end{equation}
        This is the quadratic part of the Lagrangian of the superfield homotopy Maurer--Cartan action~\eqref{eq:superfield_hMC_action} for an $L_\infty$-algebra $\tilde\frL^{\rm eEHD}_{\rm BRST}$. The latter has underlying complex
        \begin{subequations}
            \begin{equation}
                \begin{tikzcd}
                    &&\stackrel{\tilde\varphi}{\scC^\infty(\IM^d)}\rar["0"]  &\stackrel{\tilde\varphi^+}{\scC^\infty(\IM^d)}
                    \\
                    &&\stackrel{\tilde\delta}{\scC^\infty(\IM^d)}\rar["\wave"]  &\stackrel{\tilde\delta^+}{\scC^\infty(\IM^d)}
                    \\
                    &\stackrel{\tilde X^\mu}{\Omega^1(\IM^d)}\rar["-\wave"]\drar[]  &\stackrel{\tilde{\bar X}^{+\mu}}{\Omega^1(\IM^d)} &\stackrel{\tilde{\bar X}^\mu}{\Omega^1(\IM^d)}\rar["-\wave"]\drar[]&\stackrel{\tilde X^{+\mu}}{\Omega^1(\IM^d)} 
                    \\
                    & \stackrel{\tilde\beta}{\scC^\infty(\IM^d)}\rar["-\wave"] \urar[crossing over]& \stackrel{\tilde{\bar\beta}^+}{\scC^\infty(\IM^d)}& \stackrel{\tilde{\bar\beta}}{\scC^\infty(\IM^d)}\rar["-\wave"]\urar[crossing over]& \stackrel{\tilde\beta^+}{\scC^\infty(\IM^d)}
                    \\
                    &&\stackrel{\tilde h_{\mu\nu}}{\Omega^2(\IM^d)}\rar["\wave"]\drar[]\arrow[ddr,crossing over]&\stackrel{\tilde h^+_{\mu\nu}}{\Omega^2(\IM^d)} & &
                    \\
                    && \stackrel{\tilde\varpi^\mu}{\Omega^1(\IM^d)}\rar["\wave-\tilde \xi^2\partial_\mu\partial^\nu", crossing over]\urar[crossing over]\drar[]& \stackrel{\tilde\varpi^{+\mu}}{\Omega^1(\IM^d)}
                    \\
                    & \underbrace{\spacer{2ex}~~~~~~~~~~}_{\eqqcolon\,\tilde\frL^{\rm eEHD}_{{\rm BRST},\,0}} & \underbrace{\spacer{2ex}\stackrel{\tilde\pi}{\scC^\infty(\IM^d)}}_{\eqqcolon\,\tilde\frL^{\rm eEHD}_{{\rm BRST},\,1}} \rar["\wave"]\urar[crossing over]\arrow[uur] & \underbrace{\spacer{2ex}\stackrel{\tilde\pi^+}{\scC^\infty(\IM^d)}}_{\eqqcolon\,\tilde\frL^{\rm eEHD}_{{\rm BRST},\,2}} & \underbrace{\spacer{2ex}~~~~~~~~~~}_{\eqqcolon\,\tilde\frL^{\rm eEHD}_{{\rm BRST},\,3}}
                \end{tikzcd}
            \end{equation}
            with 
            \begin{equation}
                \begin{aligned}
                    \tilde\frL^{\rm eEHD}_{{\rm BRST},\,0}\ &=\ \spacee{\tilde\frL^{\rm eEHD}}{{\rm BRST},\,0}{\tilde\beta,\,\tilde X}
                    ~,~~~&
                    \tilde\frL^{\rm eEHD}_{{\rm BRST},\,1}\ &=\ \spacee{\tilde\frL^{\rm eEHD}}{{\rm BRST},\,1}{\tilde\delta,\,\tilde{\bar X}^+,\,\tilde{\bar\beta}^+,\,\tilde h,\,\tilde\varpi,\,\tilde\pi}~,
                    \\
                    \tilde\frL^{\rm eEHD}_{{\rm BRST},\,3}\ &=\ \spacee{\tilde\frL^{\rm eEHD}}{{\rm BRST},\,3}{\tilde\beta^+,\,\tilde X^+}
                    ~,~~~&
                    \tilde\frL^{\rm eEHD}_{{\rm BRST},\,2}\ &=\ \spacee{\tilde\frL^{\rm eEHD}}{{\rm BRST},\,2}{\tilde\delta^+,\,\tilde{\bar X},\,\tilde{\bar\beta},\,\tilde h^+,\,\tilde\varpi^+,\,\tilde\pi^+}~,
                \end{aligned}
            \end{equation}
            and the lowest non-vanishing products
            \begin{equation}
                \begin{aligned}
                    \colvec{
                        \tilde X_{\mu},
                        \tilde\beta
                    }
                    \ &\mapstomap{\mu_1}\ 
                    -\colvec{
                      \wave\tilde X_\mu-\tilde\xi\sqrt{\wave}\partial_\mu\tilde\beta,
                       \wave\tilde\beta+\tilde\xi\sqrt{\wave}\partial_\mu\tilde X^\mu
                    }
                    \ \in\ \spacee{\tilde\frL^{\rm eEHD}}{{\rm BRST},\,1}{\tilde{\bar{X}}^+,\,\tilde{\bar{\beta}}^+}~,
                    \\
                    \colvec{
                        \tilde h_{\mu\nu},
                        \tilde\varpi_\mu,
                        \tilde\pi
                    }
                    \ &\mapstomap{\mu_1}\ 
                    \colvec{
                        \wave\tilde h_{\mu\nu}-2\tilde\xi\sqrt{\wave}\partial_\mu\tilde\varpi_\nu+\tilde\xi^2\partial_\mu\partial_\nu\tilde\pi,
                        \wave\tilde\varpi_\mu+\tilde\xi\sqrt{\wave}\partial^\mu\tilde h_{\mu\nu}-\tilde\xi\sqrt{\wave}\partial_\mu\tilde\pi-\tilde \xi^2\partial_\mu\partial^\nu\tilde\varpi_\nu,
                        \wave\tilde \pi_\mu(x)+2\tilde\xi\sqrt{\wave}\partial^\mu\tilde\varpi_\mu(x)+\tilde\xi^2\partial^\mu\partial^\nu\tilde h_{\mu\nu}
                    }
                    \ \in\ \spacee{\tilde\frL^{\rm eEHD}}{{\rm BRST},\,2}{\tilde{h}^+,\,\tilde{\varpi}^+,\,\tilde{\pi}^+}~,
                    \\
                    \colvec{
                        \tilde{\bar X}_\mu,
                        \tilde{\bar\beta}
                    }
                    \ &\mapstomap{\mu_1}\ 
                    -\colvec{
                        \wave\tilde{\bar X}_\mu-\tilde\xi\sqrt{\wave}\partial_\mu\tilde{\bar\beta},
                        \wave\tilde{\bar\beta}+\tilde\xi\sqrt{\wave}\partial_\mu\tilde{\bar X}^\mu
                    }
                    \in\spacee{\tilde\frL^{\rm eEHD}}{{\rm BRST},\,3}{\tilde X^+,\,\tilde\beta^+}~.
                \end{aligned}
            \end{equation}
        \end{subequations}
        The $\tilde\frL^{\rm eEHD}_{\rm BRST}$ algebra is endowed with the following cyclic structure:
        \begin{equation}
            \begin{aligned}
                \inner{\tilde X}{\tilde X^+}\ &\coloneqq\ \int\rmd^dx\,\tilde X^\mu\tilde X^+_\mu~,~~
                &\inner{\tilde{\bar X}}{\tilde{\bar X}^+}\ &\coloneqq\ -\int\rmd^dx\,\tilde{\bar X}^\mu\tilde{\bar X}^+_\mu~,
                \\
                \inner{\tilde\beta}{\tilde\beta^+}\ &\coloneqq\ \int\rmd^dx\,\tilde\beta\tilde\beta^+~,~~
                &\inner{\tilde{\bar\beta}}{\tilde{\bar\beta}^+}\ &\coloneqq\ -\int \rmd^dx\,\tilde{\bar\beta}\tilde{\bar\beta}^+~,
                \\
                \inner{\tilde h}{\tilde h^+}\ &\coloneqq\ \tfrac12\int\rmd^dx\,\tilde h^{\mu\nu}\tilde h^+_{\mu\nu}~,
                \\
                \inner{\tilde\varpi}{\tilde\varpi^+}\ &\coloneqq\ \int\rmd^dx\,\tilde\varpi^\mu\tilde\varpi^+_\mu~,~~
                &
                \\
                \inner{\tilde\pi}{\tilde\pi^+}\ &\coloneqq\ \tfrac12\int\rmd^dx\,\tilde\pi\tilde\pi^+~,~~
                &\inner{\tilde\delta}{\tilde\delta^+}\ &\coloneqq\ -\int\rmd^dx\,\tilde\delta\tilde\delta^+~.
            \end{aligned}
        \end{equation}

        \subsection{Factorisation of the cochain complex of \texorpdfstring{$\caN=0$}{N=0} supergravity}\label{ssec:Factorisation_CC_N=0}
        
        The factorisation of the cochain complex of the $L_\infty$-algebra for Yang--Mills theory now fixes completely the factorisation of the cochain complex of the $L_\infty$-algebra of $\caN=0$ supergravity. In view of~\eqref{eq:factorisation_YM_diff_complex}, it thus merely remains to verify that
        \begin{equation}\label{eq:fac_N0_dgc}
            \sfCh(\tilde\frL^{\caN=0}_{\rm BRST})\ =\ \frKin\otimes_{\tau_1}(\frKin\otimes_{\tau_1}\sfCh(\frScal))
        \end{equation}
        at the level of cochain complexes, where $\frKin$ is given in~\eqref{eq:kinYM} and $\sfCh(\frScal)$ in~\eqref{eq:Scal}. Furthermore, the twist in the outer tensor product of~\eqref{eq:fac_N0_dgc} will only affect $\sfCh(\frScal)$ and commute with the other factor of $\frKin$. Let us stress that we could have allowed for two different twist parameters for each of the tensor products. This, however, would make our discussion unnecessarily involved.
        
        \paragraph{Factorisation of fields.} It is not surprising that the identification works at the level of graded vector spaces for the physical fields. This is merely the statement that a rank-two (covariant) tensor decomposes into its symmetric part and its anti-symmetric part. The symmetric part splits further into the trace, which can be identified with the dilaton, and the remaining components, which describe gravitational modes. More interesting is the sector of unphysical fields, and the complete factorisation of all fields is given in \cref{tab:fac_vec_N0}.
        \begin{table}
            \begin{center}
                \resizebox{\textwidth}{!}{
                    \begin{tabular}{|c|l|c|c|c|c|c|c|c|}
                        \hline
                        \multicolumn{4}{|c|}{fields} & \multicolumn{3}{c|}{anti-fields}
                        \\
                        \hline
                        factorisation & $|-|_{\rm gh}$ & $|-|_{\frL}$ & dim  & factorisation & $|-|_{\frL}$ & dim \\
                        \hline
                        $\tilde\lambda=-[\ttg,\ttg]\tts_x\tfrac12\tilde\lambda(x)$ & $2$ & $-1$ & $\tfrac{d}{2}-3$ & $\tilde\lambda^+=-[\tta, \tta]\tts_x^+\tfrac12\tilde\lambda^+(x)$ & 4 & $\tfrac{d}{2}+3$ 
                        \\
                        $\tilde\Lambda=[\ttg,\ttv^\mu]\tts_x\tfrac1{\sqrt{2}}\tilde\Lambda_\mu(x)$ &  $1$ & 0 & $\tfrac{d}{2}-2$ & $\tilde\Lambda^+=[\tta, \ttv^\mu]\tts^+_x\tfrac1{\sqrt{2}}\tilde\Lambda_\mu^+$ & 3 & $\tfrac{d}{2}+2$ 
                        \\
                        $\tilde\gamma=[\ttg,\ttn]\tts_x\tfrac1{\sqrt{2}}\tilde\gamma(x)$ & $1$ & 0 & $\tfrac{d}{2}-2$ & $\tilde\gamma^+=[\tta,\ttn]\tts_x^+\tfrac1{\sqrt{2}}\tilde\gamma^+(x)$ & $3$ & $\tfrac{d}{2}+2$ 
                        \\
                        $\tilde B=[\ttv^{\mu},\ttv^\nu] \tts_x\tfrac1{2\sqrt{2}}\tilde B_{\mu\nu}(x)$ & $0$  & 1 & $\tfrac{d}{2}-1$ & $\tilde B^+=[\ttv^{\mu},\ttv^{\nu}] \tts_x^+\tfrac1{2\sqrt{2}} \tilde B^+_{\mu\nu}(x)$ & 2 & $\tfrac{d}{2}+1$ 
                        \\
                        $\tilde{\alpha}=[\ttn,\ttv^\mu]\tts_x\tfrac1{\sqrt{2}}\tilde{\alpha}_\mu(x)$  & $0$ & 1 & $\tfrac{d}{2}-1$ & $\tilde{\alpha}^+=[\ttn,\ttv^\mu]\tts_x^+\tfrac1{\sqrt{2}}\tilde{\alpha}^+_\mu(x)$ & $2$ & $\tfrac{d}{2}+1$ 
                        \\
                        $\tilde\eps=-[\ttg,\tta]\tts_x\tfrac1{\sqrt{2}}\tilde\eps(x)$ & $0$  & $1$ & $\tfrac{d}{2}-1$ & $\tilde\eps^+=-[\ttg,\tta]\tts_x^+\tfrac1{\sqrt{2}}\tilde\eps^+(x)$ & 2 & $\tfrac{d}{2}+1$
                        \\
                        $\tilde{\bar\Lambda}=[\tta, \ttv^\mu]\tts_x\tfrac1{\sqrt{2}}\tilde{\bar\Lambda}_\mu(x)$ & $-1$ & $2$ & $\tfrac{d}{2}$ & $\tilde{\bar \Lambda}^+=[\ttg,\ttv^\mu]\tts_x^+\tfrac1{\sqrt{2}}\tilde{\bar \Lambda}_\mu^+(x)$ & 1 & $\tfrac{d}{2}$
                        \\
                        $\tilde{\bar\gamma}=[\tta,\ttn]\tts_x\tfrac1{\sqrt{2}}\tilde{\bar\gamma}(x)$  & $-1$ & 2 & $\tfrac{d}{2}$ & $\tilde{\bar\gamma}^+=[\ttg,\ttn]\tts_x^+\tfrac1{\sqrt{2}}\tilde{\bar \gamma}^+(x)$ & $1$ & $\tfrac{d}{2}$ 
                        \\
                        $\tilde{\bar\lambda}=-[\tta,\tta]\tts_x\tfrac12\tilde{\bar\lambda}(x)$ & $-2$ & $3$ & $\tfrac{d}{2}+1$ & $\tilde{\bar\lambda}^+=-[\ttg,\ttg]\tts_x^+\tfrac12\tilde{\bar\lambda}^+(x)$ & 0 & $\frac{d}{2}-1$
                        \\
                        \hline
                        $\tilde X=(\ttg,\ttv^\mu)\tts_x\tfrac1{\sqrt{2}}\tilde X_\mu(x)$ & $1$ & 0 & $\tfrac{d}{2}-2$ & $\tilde X^+=(\tta, \ttv^\mu)\tts_x^+\tfrac1{\sqrt{2}}\tilde X^+_\mu(x)$ & 3 & $\tfrac{d}{2}+2$ 
                        \\
                        $\tilde\beta=(\ttg,\ttn)\tts_x\tfrac1{\sqrt{2}}\tilde\beta(x)$ & $1$ & 0 & $\tfrac{d}{2}-2$ & $\tilde\beta^+=(\tta,\ttn)\tts_x^+\tfrac1{\sqrt{2}}\tilde\beta^+(x)$ & $3$ & $\tfrac{d}{2}+2$ 
                        \\
                        $\tilde h=(\ttv^{\mu},\ttv^{\nu})\tts_x\tfrac1{2\sqrt{2}}\tilde h_{\mu\nu}(x)$ & $0$ & 1 & $\tfrac{d}{2}-1$ & $\tilde h^+=(\ttv^{\mu},\ttv^{\nu})\tts_x^+\tfrac1{2\sqrt{2}}\tilde h^+_{\mu\nu}(x)$ & 2 & $\tfrac{d}{2}+1$ 
                        \\
                        $\tilde\varpi=-(\ttn,\ttv^\mu)\tts_x\tfrac1{\sqrt{2}}\tilde\varpi_\mu(x)$ & $0$ & 1 & $\tfrac{d}{2}-1$ & $\tilde\varpi^+=-(\ttn,\ttv^\mu)\tts_x^+\tfrac1{\sqrt{2}}\tilde\varpi^+_\mu(x)$ & $2$ & $\tfrac{d}{2}+1$
                        \\
                        $\tilde\pi=(\ttn,\ttn)\tts_x\tfrac1{2\sqrt{2}}\tilde\pi(x)$ &  $0$ & $1$ & $\tfrac{d}{2}-1$ & $\tilde\pi^+=(\ttn,\ttn)\tts_x^+\tfrac1{2\sqrt2}\tilde\pi^+(x)$ & 2 & $\tfrac{d}{2}+1$ 
                        \\
                        $\tilde\delta=-(\ttg,  \tta)\tts_x\tfrac1{\sqrt{2}}\tilde \delta(x)$ & $0$  & $1$ & $\tfrac{d}{2}-1$ & $\tilde\delta^+=-(\ttg,\tta)\tts_x^+\tfrac1{\sqrt{2}}\tilde \delta^+(x)$ & 2 & $\tfrac{d}{2}+1$ 
                        \\
                        $\tilde{\bar X}=(\tta,\ttv^\mu)\tts_x\tfrac1{\sqrt{2}}\tilde{\bar X}_\mu(x)$ & $-1$ & $2$ & $\tfrac{d}{2}$ & $\tilde{\bar X}^+=(\ttg,\ttv^\mu)\tts_x^+\tfrac1{\sqrt{2}}\tilde{\bar X}_\mu(x)$ & 1 & $\tfrac{d}{2}$
                        \\
                        $\tilde{\bar\beta}=(\tta,\ttn)\tts_x\tfrac1{\sqrt{2}}\tilde{\bar\beta}(x)$ & $-1$ & 2 & $\tfrac{d}{2}$ & $\tilde{\bar\beta}^+=(\ttg,\ttn)\tts_x^+\tfrac1{\sqrt{2}}\tilde{\bar \beta}^+(x)$ & $1$ & $\tfrac{d}{2}$ 
                        \\
                        \hline
                    \end{tabular}
                }
            \end{center}
            \caption{Factorisation of the redefined BV fields for $\caN=0$ supergravity. Just as in the case of Yang--Mills theory, all fields have a factor of $\tts_x$, while all anti-fields have a factor of $\tts_x^+$. Here, we again suppressed the integrals over $x$, and we used the notation $[\ttx,\tty]\coloneqq\ttx\otimes\tty-(-1)^{|\ttx|\,|\tty|}\tty\otimes\ttx$ and $(\ttx,\tty)\coloneqq\ttx\otimes \tty+(-1)^{|\ttx|\,|\tty|}\tty\otimes\ttx$ for $\ttx,\tty\in\frKin$.}\label{tab:fac_vec_N0}
        \end{table}
        
        The elements of $\frKin$ form a quartet, which is reflected in the well-known quartet of fields in the gauge-fixed Yang--Mills action:
        \begin{equation}
            \begin{tikzpicture}[
                scale=1,
                every node/.style={scale=1},
                baseline={([yshift=-.5ex]current bounding box.center)}
                ]
                \matrix (m) [
                matrix of nodes,
                ampersand replacement=\&,
                column sep=0.3cm,
                row sep=0.4cm
                ]{
                    {} \& $\ttn$ \& {}
                    \\
                    {} \& $\ttv^\mu$ \& {}
                    \\
                    $\ttg$ \& {} \& $\tta$
                    \\
                };
                \draw [->] (m-2-2) -- (m-1-2);
                \draw [->] (m-2-2) -- (m-3-1);
                \draw [->] (m-2-2) -- (m-3-3);
            \end{tikzpicture}~~~\longrightarrow~~~
            \begin{tikzpicture}[
                scale=1,
                every node/.style={scale=1},
                baseline={([yshift=-.5ex]current bounding box.center)}
                ]
                \matrix (m) [
                matrix of nodes,
                ampersand replacement=\&,
                column sep=0.3cm,
                row sep=0.4cm
                ]{
                    {} \& $b^a$ \& {}
                    \\
                    {} \& $A_\mu^a$ \& {}
                    \\
                    $c^a$ \& {} \& $\bar c^a$
                    \\
                };
                \draw [->] (m-2-2) -- (m-1-2);
                \draw [->] (m-2-2) -- (m-3-1);
                \draw [->] (m-2-2) -- (m-3-3);
            \end{tikzpicture}
        \end{equation}    
        Each field in $\sfCh(\tilde\frL^{\caN=0}_{\rm BRST})$ thus lives in the tensor product of two such quartets. This tensor product further splits into (graded) symmetric, anti-symmetric, and trace parts, which belong to the two-form $B_{\mu\nu}$, the graviton modes $h_{\mu\nu}$, and the dilaton $\varphi$. Because the product of two ghosts $\ttg\tilde\ttg$ is automatically anti-symmetric, only the $B$-field has a ghost for ghost $\lambda$. On the graviton/dilaton side, we do not have the higher gauge transformations, but contrary to Yang--Mills theory, the ghost is a vector. We can summarise the relations between the fields in the following two diagrams:
        \begin{equation}
            ~~~~
            \begin{tikzpicture}[
                scale=1,
                every node/.style={scale=1},
                baseline={([yshift=-.5ex]current bounding box.center)}
                ]
                \matrix (m) [
                matrix of nodes,
                ampersand replacement=\&,
                column sep=0.3cm,
                row sep=0.4cm
                ]{
                    {} \& {} \& $\alpha_\mu$ \& {} \& {}
                    \\
                    {} \& $\gamma$ \& $B_{\mu\nu}$ \& $\bar\gamma$ \& {}
                    \\
                    {} \& $\Lambda_\mu$ \& {} \& $\bar \Lambda_\mu$ \& {}
                    \\
                    $\lambda$ \& {} \& $\eps$ \& {} \& $\bar \lambda$
                    \\
                };
                \draw [->] (m-1-3) -- (m-2-2);
                \draw [->] (m-1-3) -- (m-2-4);
                \draw [->] (m-2-3) -- (m-1-3);
                \draw [->] (m-2-3) -- (m-3-2);
                \draw [->] (m-2-3) -- (m-3-4);
                \draw [->] (m-3-2) -- (m-2-2);
                \draw [->] (m-3-2) -- (m-4-1);
                \draw [->] (m-3-2) -- (m-4-3);
                \draw [->] (m-3-4) -- (m-2-4);
                \draw [->] (m-3-4) -- (m-4-3);
                \draw [->] (m-3-4) -- (m-4-5);
            \end{tikzpicture}
            ~~~~
            \begin{tikzpicture}[
                scale=1,
                every node/.style={scale=1},
                baseline={([yshift=-.5ex]current bounding box.center)}
                ]
                \matrix (m) [
                matrix of nodes,
                ampersand replacement=\&,
                column sep=0.3cm,
                row sep=0.4cm
                ]{
                    {} \& $\pi$ \& {}
                    \\
                    {} \& $\varpi_\mu$ \& {}
                    \\
                    $\beta$ \& $h_{\mu\nu}$ \& $\bar\beta$
                    \\
                    $X_\mu$ \& {} \& $\bar X_\mu$
                    \\
                    {} \& $\delta$ \& {}
                    \\
                };
                \draw [->] (m-2-2) -- (m-1-2);
                \draw [->] (m-2-2) -- (m-3-1);
                \draw [->] (m-2-2) -- (m-3-3);
                \draw [->] (m-3-2) -- (m-2-2);
                \draw [->] (m-3-2) -- (m-4-1);
                \draw [->] (m-3-2) -- (m-4-3);
                \draw [->] (m-4-1) -- (m-3-1);
                \draw [->] (m-4-3) -- (m-3-3);
                \draw [->] (m-4-1) -- (m-5-2);
                \draw [->] (m-4-3) -- (m-5-2);
            \end{tikzpicture}
        \end{equation}
        where upper, lower left, and lower right arrows point to fields where a vector factor $\ttv^\mu$ has been replaced by a factor $\ttn$, $\ttg$, and $\tta$, respectively. The $L_\infty$-degrees of the fields are the same in each column, increasing from left to right by one.

        \paragraph{Factorisation as cyclic complex.}
        From \cref{tab:fac_vec_N0}, it is clear that the tensor product~\eqref{eq:fac_N0_dgc} is indeed correct at the level of graded vector spaces. The inner product structure on the anti-symmetric part is given by
        \begin{subequations}
            \begin{equation}
                \begin{aligned}
                    \inner{\tilde\lambda}{\tilde\lambda^+}\ &=\ \innerLarge{-\ttg\otimes\ttg\otimes\int\rmd^d x_1\,\tts_{x_1}\tilde\lambda(x_1)}{-\tta\otimes\tta\otimes\int\rmd^dx_2\,\tts_{x_2}^+\tilde\lambda^+(x_2)}
                    \\
                    &=\ -\inner{\ttg}{\tta}\inner{\ttg}{\tta}\int\rmd^dx_1\,\int\rmd^dx_2\,\delta^{(d)}(x_1-x_2)\tilde\lambda(x_1)\tilde\lambda^+(x_2)
                    \\
                    &=\ -\int\rmd^dx~\tilde\lambda(x)\tilde\lambda^+(x)~,
                \end{aligned}
            \end{equation}
            Similarly,
            \begin{equation}
                \begin{aligned}
                    \inner{\tilde\Lambda}{\tilde\Lambda^+}\ &=\ \int\rmd^dx\,\tilde\Lambda^\mu(x)\tilde\Lambda^+_\mu(x)~,~~~
                    &\inner{\tilde{\bar\Lambda}}{\tilde{\bar\Lambda}^+}\ &=\ -\int\rmd^dx\,\tilde{\bar\Lambda}^\mu(x)\tilde{\bar\Lambda}^+_\mu(x)~,
                    \\
                    \inner{\tilde\gamma}{\tilde\gamma^+}\ &=\ \int\rmd^dx\,\tilde\gamma(x)\tilde\gamma^+(x)~,~~~
                    &\inner{\tilde{\bar\gamma}}{\tilde{\bar\gamma}^+}\ &=\ -\int\rmd^dx\,\tilde{\bar\gamma}(x)\tilde{\bar\gamma}^+(x)~,
                    \\
                    \inner{\tilde B}{\tilde B^+}\ &=\ \tfrac12\int\rmd^dx\,\tilde B^{\mu\nu}(x)\tilde B^+_{\mu\nu}(x)~,~~~
                    &\inner{\tilde\eps}{\tilde\eps^+}\ &=\  \int\rmd^dx\,\tilde\eps(x)\tilde\eps^+(x)~,
                    \\
                    \inner{\tilde\alpha}{\tilde\alpha^+}\ &=\ \int\rmd^dx\,\tilde\alpha^\mu(x)\tilde\alpha^+_\mu(x)~,
                    ~~~
                    &\inner{\tilde{\bar\lambda}}{\tilde{\bar\lambda}^+}\ &=\ -\int\rmd^dx\,\tilde{\bar\lambda}(x)\tilde{\bar\lambda}^+(x)~.
                \end{aligned}
            \end{equation}
            On the symmetric part, we have analogously 
            \begin{equation}
                \begin{aligned}
                    \inner{\tilde X}{\tilde X^+}\ &=\ \int\rmd^dx\,\tilde X^\mu(x)\tilde X^+_\mu(x)~,~~~
                    &\inner{\tilde\pi}{\tilde\pi^+}\ &=\ \tfrac12\int\rmd^dx\,\tilde\pi(x)\tilde\pi^+(x)~,
                    \\
                    \inner{\tilde\beta}{\tilde\beta^+}\ &=\ \int\rmd^dx\,\tilde\beta(x)\tilde\beta^+(x)~,~~~
                    &\inner{\tilde\delta}{\tilde\delta^+}\ &=\ -\int\rmd^dx\,\tilde\delta(x)\tilde\delta^+(x)~,
                    \\
                    \inner{\tilde h}{\tilde h^+}\ &=\ \tfrac12\int\rmd^dx\,\tilde h^{\mu\nu}(x)\tilde h^+_{\mu\nu}(x)
                    ~,~~~
                    &\inner{\tilde{\bar X}}{\tilde{\bar X}^+}\ &=\ -\int\rmd^dx\,\tilde{\bar X}^\mu(x)\tilde{\bar X}^+_\mu(x)~,
                    \\
                    \inner{\tilde\varpi}{\tilde\varpi^+}\ &=\ \int\rmd^dx\,\tilde\varpi^\mu(x)\tilde\varpi^+_\mu(x)~,~~~
                    &\inner{\tilde{\bar\beta}}{\tilde{\bar\beta}^+}\ &=\ -\int\rmd^dx\,\tilde{\bar\beta}(x)\tilde{\bar\beta}^+(x)~.
                \end{aligned}
            \end{equation}
        \end{subequations}
        
        Next, we compute the action of the differential $\mu_1$, which is completely fixed by the tensor product $\frKin\otimes_{\tau_1}(\frKin\otimes_{\tau_1}\frScal)$, cf.~definition~\eqref{eq:twisted_differential}. We have, for example,
        \begin{subequations}
            \begin{equation}            
                \begin{aligned}
                    \mu_1(\tilde\lambda)\ &=\ \mu_1\left(-[\ttg,\ttg]\otimes\tfrac12\int\rmd^dx\,\tts_x\tilde\lambda(x)\right) =\ -[\ttg,\ttg]\otimes\tfrac12\mu_1\left(\int\rmd^dx\,\tts_x\tilde\lambda(x)\right) =\ \wave\tilde\lambda~,
                    \\
                    \mu_1
                    \colvec{
                        \tilde\Lambda,
                        \tilde\gamma
                    }
                    \ &=\ \mu_1
                    \left(
                    ([\ttg,\ttv^\mu],[\ttg,\ttn])\otimes
                    \colvec{
                        \int\rmd^dx\,\tts_x\tfrac1{\sqrt{2}}\tilde\Lambda_\mu(x),
                        \int\rmd^dx\,\tts_x\tfrac1{\sqrt{2}}\tilde\gamma(x)
                    }
                    \right)
                    \\
                    &=\ -([\ttg,\ttv^\mu],[\ttg,\ttn])\otimes\mu_1
                    \left(
                    \begin{pmatrix}
                        \sfid & -\tilde\xi\wave{}^{-\frac12}\partial_\mu
                        \\
                        \tilde\xi\wave{}^{-\frac12}\partial^\mu & \sfid
                    \end{pmatrix}
                    \colvec{
                        \int\rmd^dx\,\tts_x\tfrac1{\sqrt{2}}\tilde\Lambda_\mu(x),
                        \int\rmd^dx\,\tts_x\tfrac1{\sqrt{2}}\tilde \gamma(x)
                    }
                    \right)
                    \\
                    &=\ -([\ttg,\ttv^\mu],[\ttg,\ttn])\otimes
                    \colvec{
                        \int\rmd^dx\,\tts^+_x\tfrac1{\sqrt{2}}\{\wave\tilde\Lambda_\mu(x)-\tilde\xi\sqrt{\wave}\partial_\mu \tilde\gamma(x)\},
                        \int\rmd^dx\,\tts^+_x\tfrac1{\sqrt{2}}\{\wave\tilde\gamma(x)+\tilde\xi\sqrt{\wave}\partial_\mu\tilde\Lambda^\mu(x)\}
                    },
                    \\
                    \mu_1
                    \colvec{
                        \tilde B,
                        \tilde\alpha
                    }
                    \ &=\ \mu_1
                    \left(
                    ([\ttv^\mu,\ttv^\nu],[\ttn,\ttv^\mu])\otimes
                    \colvec{
                        \int\rmd^dx\,\tts_x\tfrac1{2\sqrt{2}}\tilde B_{\mu\nu}(x),
                        \int\rmd^dx\,\tts_x\tfrac1{\sqrt{2}}\tilde \alpha_\mu(x)
                    }
                    \right)
                    \\
                    &=\ ([\ttv^\mu,\ttv^\nu],[\ttn,\ttv^\mu])\otimes
                    \colvec{
                        \int\rmd^dx\,\tts^+_x\tfrac1{\sqrt{2}}\{\tfrac12\wave\tilde B_{\mu\nu}(x)-\tilde\xi\sqrt{\wave}\partial_\mu\tilde\alpha_\nu(x)\},
                        \int\rmd^dx\,\tts^+_x\tfrac1{\sqrt{2}}\{\wave\tilde \alpha_\mu(x)+\tilde \xi\sqrt{\wave}\partial^\nu\tilde B_{\nu\mu}(x)+\tilde \xi^2\partial_\mu\partial^\nu\tilde\alpha_\nu(x)\}
                    },
                    \\
                    \mu_1
                    \colvec{
                        \tilde h,
                        \tilde\varpi,
                        \tilde \pi
                    }
                    \ &=\ \mu_1
                    \left(
                    ((\ttv^\mu,\ttv^\nu),(\ttn,\ttv^\mu),(\ttn,\ttn))\otimes
                    \colvec{
                        \int\rmd^dx\,\tts_x\tfrac1{2\sqrt{2}}\tilde h_{\mu\nu}(x),
                        \int\rmd^dx\,\tts_x\left(-\tfrac1{\sqrt{2}}\tilde \varpi_\mu(x)\right),
                        \int\rmd^dx\,\tts_x\tfrac1{2\sqrt{2}}\tilde\pi_\mu(x)
                    }
                    \right)
                    \\
                    &=\ ((\ttv^\mu,\ttv^\nu),(\ttn,\ttv^\mu),(\ttn,\ttn))\otimes M
                \end{aligned}
            \end{equation}
            with
            \begin{equation}
                M\ \coloneqq\
                \colvec{
                    \int\rmd^dx\,\tts^+_x\{\tfrac1{2\sqrt{2}}\wave\tilde h_{\mu\nu}(x)-\tfrac1{\sqrt{2}}\tilde\xi\sqrt{\wave}\partial_\mu\tilde\varpi_\nu(x)+\tfrac1{2\sqrt{2}}\tilde\xi^2\partial_\mu\partial_\nu\tilde\pi(x)\},
                    \int\rmd^dx\,\tts^+_x\{-\tfrac1{\sqrt2}\wave\tilde\varpi_\mu(x)-\tfrac1{\sqrt{2}}\tilde\xi\sqrt{\wave}\partial^\mu\tilde h_{\mu\nu}(x)+\tfrac1{\sqrt{2}}\tilde\xi\sqrt{\wave}\partial_\mu\tilde\pi(x)+\tfrac1{\sqrt{2}}\tilde \xi^2\partial_\mu\partial^\nu\tilde\varpi_\nu(x)\},
                    \int\rmd^dx\,\tts^+_x\{\tfrac1{2\sqrt{2}}\wave\tilde \pi_\mu(x)+\tfrac1{\sqrt{2}}\tilde\xi\sqrt{\wave}\partial^\mu\tilde\varpi_\mu(x)+\tfrac1{2\sqrt{2}}\tilde\xi^2\partial^\mu\partial^\nu\tilde h_{\mu\nu}\}
                }.
            \end{equation}
            Furthermore, we have
            \begin{equation}
                \begin{aligned}
                    \mu_1
                    \colvec{
                        \tilde{\bar\Lambda},
                        \tilde{\bar \gamma}
                    }
                    \ &=\ -([\tta,\ttv^\mu],[\tta,\ttn])\otimes
                    \colvec{
                        \int\rmd^dx\,\tts^+_x\tfrac1{\sqrt{2}}\{\wave\tilde{\bar\Lambda}_\mu(x)-\tilde\xi\sqrt{\wave}\partial_\mu\tilde{\bar\gamma}(x)\},
                        \int\rmd^dx\,\tts^+_x\tfrac1{\sqrt{2}}\{\wave\tilde{\bar\gamma}(x)+\tilde\xi\sqrt{\wave}\partial_\mu\tilde{\bar\Lambda}^\mu(x)\}
                    }~,
                    \\
                    \mu_1(\tilde\eps)\ &=\ \wave\tilde\eps~,
                    \\
                    \mu_1(\tilde{\bar\lambda})\ &=\ \wave\tilde{\bar\lambda}~,
                    \\
                    \mu_1
                    \colvec{
                        \tilde X,
                        \tilde\beta
                    }
                    \ &=\ -((\ttg,\ttv^\mu),(\ttg,\ttn))\otimes
                    \colvec{
                        \int\rmd^dx\,\tts^+_x\tfrac1{\sqrt{2}}\{\wave\tilde X_\mu(x)-\tilde\xi\sqrt{\wave}\partial_\mu\tilde\beta(x)\},
                        \int\rmd^dx\,\tts^+_x\tfrac1{\sqrt{2}}\{\wave\tilde\beta(x)+\tilde\xi\sqrt{\wave}\partial_\mu\tilde X^\mu(x)\}
                    }~,
                    \\
                    \mu_1
                    \colvec{
                        \tilde{\bar X},
                        \tilde{\bar\beta}
                    }
                    &=\ -([\tta,\ttv^\mu],[\tta,\ttn])\otimes
                    \colvec{
                        \int\rmd^dx\,\tts^+_x\tfrac1{\sqrt{2}}\{\wave\tilde{\bar X}_\mu(x)-\tilde\xi\sqrt{\wave}\partial_\mu\tilde{\bar\beta}(x)\},
                        \int\rmd^dx\,\tts^+_x\tfrac1{\sqrt{2}}\{\wave\tilde{\bar\beta}(x)+\tilde\xi\sqrt{\wave}\partial_\mu\tilde{\bar X}^\mu(x)\}
                    }~,
                    \\
                    \mu_1(\tilde\delta)\ &=\ \wave\tilde\delta~.
                \end{aligned}
            \end{equation}
        \end{subequations}
        The resulting superfield homotopy Maurer--Cartan action~\eqref{eq:CBVAction} for the superfield $\sfa=\tilde\lambda+\tilde\Lambda+\cdots+\tilde B+\tilde h$ is
        \begin{equation}
            \begin{aligned}
                \dbl{S}^{\rm DC}_0\ &\coloneqq\ \int\rmd^dx\,\Big\{\tfrac14\tilde B_{\mu\nu}\wave\tilde B^{\mu\nu}-\tilde{\bar\Lambda}_\mu\wave \tilde\Lambda^\mu+\tfrac12\tilde\alpha_\mu\wave\tilde\alpha^\mu-\tfrac{\tilde\xi^2}{2}(\partial^\mu\tilde\alpha_\mu)^2+\tfrac12\tilde\eps\wave\tilde\eps-\tilde{\bar\lambda}\wave\tilde\lambda\,-
                \\
                &\kern2cm-\tilde{\bar\gamma}\wave\tilde\gamma+\tilde\xi\tilde\alpha^\nu\sqrt{\wave}\partial^\mu\tilde B_{\mu\nu}+\tilde\xi\tilde\gamma\sqrt{\wave}\partial_\mu\tilde{\bar\Lambda}^\mu-\tilde\xi\tilde{\bar\gamma}\sqrt{\wave}\partial_\mu\tilde \Lambda^\mu\,+
                \\
                &\kern2cm+\tfrac14\tilde h_{\mu\nu}\wave\tilde h^{\mu\nu}-\tilde{\bar X}_\mu\wave\tilde X^\mu+\tfrac12\tilde\varpi_\mu\wave\tilde\varpi^\mu+\tfrac{\tilde\xi^2}2(\partial^\mu\tilde\varpi_\mu)^2\,-
                \\
                &\kern2cm-\tfrac12\tilde\delta\wave\tilde\delta+\tfrac14\tilde\pi\wave\tilde\pi-\tilde{\bar\beta}\wave\tilde\beta+\tilde\xi\tilde\varpi^\nu\sqrt{\wave}\partial^\mu\tilde h_{\mu\nu}+\tilde\xi\tilde\pi\sqrt{\wave}\partial_\mu\tilde\varpi^\mu\,+
                \\
                &\kern2cm+\tfrac12\tilde\xi^2\tilde\pi\partial_\mu\partial_\nu\tilde h^{\mu\nu}+\tilde\xi\tilde\beta\sqrt{\wave}\partial_\mu\tilde{\bar X}^\mu-\tilde\xi\tilde{\bar\beta}\sqrt{\wave}\partial_\mu\tilde X^\mu\Big\}\,.
            \end{aligned}
        \end{equation}
        This action is precisely the sum of the transformed Kalb--Ramond action~\eqref{eq:canonicallyRedefinedActionKR} and of the transformed zeroth-order gravity action augmented by a dilaton kinetic term~\eqref{eq:LeEHD}. Consequently, we see that our double copy prescription, arising from the factorisation of the $L_\infty$-algebras of Yang--Mills theory and $\caN=0$ supergravity into three factors, works at the level of cochain complexes. 
        
        \section{Quantum-field-theoretic preliminaries}\label{sec:QFT_preliminaries}
        
        Having completed the discussion at the free, linear level, we are almost ready to turn to the factorisation in the full, interacting picture, which is, perhaps not surprisingly, very involved, cf.~\cite{Borsten:2020zgj}.
        
        Firstly, as explained in \cref{sec:review}, the double copy of amplitudes is based on a reformulation of the underlying Feynman diagrams in terms of diagrams with exclusively trivalent vertices. At the level of the action, this means that we need to strictify the field theory, i.e.~to replace it by a physically equivalent one with exclusively cubic interaction terms. In this section, we will be relatively explicit, at least to lowest orders in the amplitude legs and coupling constants.
        
        Secondly, it is clear that the double copy of the factorisation of interacting Yang--Mills theory will be some form of strictified $\caN=0$ supergravity. We will not work out detailed expressions for this action but merely show that the produced action is quantum equivalent to $\caN=0$ supergravity. 
        
        To this end, we shall need a number of quantum field theoretic observations already made in~\cite{Borsten:2020zgj}. This section contains both a review and a much more detailed explanation of these observations than~\cite{Borsten:2020zgj}. 
        
        In the following, we shall always clearly distinguish between scattering amplitudes $\scA(\cdots)$ and correlation functions $\langle\cdots\rangle$. Correlation functions contain operators that create and annihilate arbitrary fields without any constraints. Scattering amplitudes, on the other hand, are labelled by \uline{external fields}, which usually are physical fields with on-shell momenta and physical polarisations. For our arguments, it is convenient to lift the restriction to physical polarisations and work with the BRST-extended Hilbert space of external fields which, in the case of Yang--Mills theory, includes gluons of arbitrary polarisations as well as the ghosts and anti-ghosts as explained next.
        
        \subsection{BRST-extended Hilbert space and Ward identities}\label{sec:BRSTExtension}
        
        The tree-level scattering amplitudes of Yang--Mills theory are parametrised by degree one elements of the minimal model of the $L_\infty$-algebra~\eqref{eq:L_infty_BRST_YM}. These are the physical, on-shell states. A convenient set of coordinates for these are the gluon's momentum $p_\mu$ as well as a discrete label indicating the gluon's helicity. More conveniently, we can replace the discrete labels by a linearly independent set of polarisation vectors $\eps_\mu$ that satisfy
        \begin{equation}
            (\eps_\mu)\ =\ \colvec{0,\vec\eps},~~~
            \vec p\cdot\vec\eps\ =\ 0~,
            \eand
            |\vec\eps\,|\ =\ 1~.
        \end{equation}
        
        \paragraph{BRST-extended Hilbert space.} We can extend this conventional Hilbert space of external fields to the full BRST field space $\frH^{\rm YM}_{\rm BRST}$ as done, e.g., in~\cite{Kugo:1977yx}. We thus have two additional, unphysical polarisations of the gluon, called \uline{forward} and \uline{backward} and denoted by $A^{\forw\,a}_\mu$ and $A^{\backw\,a}_\mu$, respectively. We can be a bit more explicit for general gluons with light-like momenta. Here, the polarisation vector $\eps^\forw_\mu$ is proportional to the momentum $p_\mu$ and the backwards polarisation vector $\eps^\backw_\mu$ is obtained by reversing the spatial part,
        \begin{subequations}\label{eq:on-shell_polarisation_vectors}
            \begin{equation}
                (\eps_\mu^\forw)\ =\ \frac{1}{\sqrt{2}|\vec p\,|}\colvec{p_0,\vec{p}}
                \eand
                (\eps_\mu^\backw)\ =\ \frac{1}{\sqrt{2}|\vec p\,|}\colvec{p_0,-\vec{p}},
            \end{equation}
            so that
            \begin{equation}
                \eps^\forw\cdot\eps^\forw\ =\ 0~,~~~
                \eps^\backw\cdot\eps^\backw\ =\ 0~,
                \eand
                \eps^\forw\cdot\eps^\backw\ =\ -1~.
            \end{equation}
        \end{subequations}
        We also have ghost and anti-ghost states. All scattering amplitudes we shall consider will be built from the Hilbert space $\frH^{\rm YM}_{\rm BRST}$. We note that the S-matrix of the physical Hilbert space $\frH^{\rm YM}_{\rm phys}$ is then the restriction of the S-matrix for the BRST extended Hilbert space $\frH^{\rm YM}_{\rm BRST}$. Although there are scattering amplitudes producing unphysical particles in $\frH^{\rm YM}_{\rm BRST}$ from physical gluons in $\frH^{\rm YM}_{\rm phys}$, this is consistent, because the restricted S-matrix is unitary. This is a consequence of the full S-matrix on $\frH^{\rm YM}_{\rm BRST}$ being unitary and BRST symmetry, cf.~\cite[Section~16.4]{Peskin:1995ev}.
        
        Evidently, $\frH^{\rm YM}_{\rm BRST}$ carries an action of the linearisation of the BRST operator, denoted by $Q^{\rm lin}_{\rm BRST}$, cf.~again~\cite{Kugo:1977yx} or the discussion in~\cite[Section~16.4]{Peskin:1995ev}. Note that after gauge-fixing, the full BRST transformations are given by the restriction of the BV transformations~\eqref{eq:BVOperatorYM} since the gauge-fixing fermion is assumed to be independent of the anti-fields. We have
        \begin{equation}\label{eq:BRSTOperatorYM}
            \begin{aligned}
                c^a\ &\mapstomap{Q^{\rm YM}_{\rm BRST}}\ -\tfrac12g{f_{bc}}^ac^bc^c~,~~~
                &\bar c^a\ &\mapstomap{Q^{\rm YM}_{\rm BRST}}\ b^a~,
                \\
                A^a_\mu\ &\mapstomap{Q^{\rm YM}_{\rm BRST}}\ (\nabla_\mu c)^a~,~~~
                &b^a\ &\mapstomap{Q^{\rm YM}_{\rm BRST}}\ 0~,
            \end{aligned}
        \end{equation}
        and $(Q^{\rm YM}_{\rm BRST})^2=0$ off-shell. In momentum space, it is then easy to see that the \uline{transversely-polarised} or \uline{physical} gluon states $A^{\trans\,a}_\mu$ are singlets under the action of the linearised BRST operator, $Q^{\rm YM,\,lin}_{\rm BRST}A^{\trans\,a}_\mu=0$. The remaining four states arrange into two doublets,
        \begin{equation}
            A^{\forw\,a}_\mu\ \mapstomap{Q^{\rm YM,\,lin}_{\rm BRST}}\ \partial_\mu c^a
            \eand
            \bar c^a \mapstomap{Q^{\rm YM,\,lin}_{\rm BRST}}\ b^a\ =\ \frac{1}{\xi}\partial^\mu A^{\backw\,a}_\mu~.
        \end{equation}
        
        \paragraph{Connected correlation functions.} 
        In our later analysis of the double copy, we shall compare correlation functions at the tree level. Recall that the partition function $Z$ and the free energy $W\coloneqq\log(Z)$ are the generating functionals for the correlation functions and the connected correlation functions, respectively. Evidently,  this implies that the connected correlation functions can be written as linear combinations of products of correlation functions. This simplifies our analysis as we can restrict ourselves to the contributions of connected Feynman diagrams to correlation functions.
        
        \begin{observation}\label{ob:BRST_connected}
            The set of connected correlation functions is BRST-invariant because the connected correlation functions can be written as linear combinations of products of correlation functions.
        \end{observation}
        
        \paragraph{Ward identities for scattering amplitudes.} 
        In order to translate colour--kinematics duality for scattering amplitudes from gluons to ghosts, we shall use supersymmetric on-shell Ward identities, cf.~\cite{Elvang:2015rqa,Elvang:2013cua}, and we focus on the supersymmetry generated by the linearised BRST operator $Q^{\rm YM,\,lin}_{\rm BRST}$ acting on the BRST-extended Hilbert space $\frH^{\rm YM}_{\rm BRST}$, whose elements label our scattering amplitudes. 
        
        The free vacuum is certainly invariant under the action of $Q^{\rm YM,\,lin}_{\rm BRST}$, cf.~again~\cite{Kugo:1977yx} or~\cite[Section~16.4]{Peskin:1995ev}. We therefore have the on-shell Ward identity 
        \begin{equation}\label{eq:on_shell_Ward}
            0\ =\ \langle 0|[Q^{\rm YM,\,lin}_{\rm BRST},{\scO_1\cdots\scO_n}]|0\rangle~.
        \end{equation}
        
        In order to use this Ward identity to link scattering amplitudes with $k$ ghost--anti-ghost pairs to amplitudes with $k+1$ such pairs, we consider the special case 
        \begin{equation}
            \scO_1\cdots\scO_n\ =\ A^\forw\bar c(c\bar c)^k A^\trans_1\cdots A^\trans_{n-2k-2}~,
        \end{equation}
        where the gluon $A^{\forw\,a}_\mu$ is forward polarised while all other gluons have physical polarisation. In this special case, the on-shell Ward identity~\eqref{eq:on_shell_Ward} becomes
        \begin{equation}
            \langle 0|(c\bar c)^{k+1}A_1^\trans\cdots A_{n-2k-2}^\trans|0\rangle\ \sim\ \langle 0|A^\forw(c\bar c)^kbA_1^\trans\cdots A_{n-2k-2}^\trans|0\rangle~.
        \end{equation}
        \begin{observation}\label{ob:ghost_pair_reduction}
            Any amplitude with $k+1$ ghost--anti-ghost pairs and all gluons transversely polarised is given by a sum of amplitudes with $k$ ghost pairs.
        \end{observation}
        The simplest non-trivial concrete example to illustrate \cref{ob:ghost_pair_reduction} is the case $n=4$, $k=0$ in Yang--Mills theory (the three-point scattering amplitudes vanish). We may then identify
        \begin{subequations}
            \begin{equation}
                \begin{aligned}
                    &\langle 0|\hat A^{\forw\,a}(p_1)\hat b^b(p_2)\hat A^{\trans\,c}_1(p_3)\hat A^{\trans\,d}_2(p_4)|0\rangle\ =
                    \\
                    &\kern2cm=\ p_2^0\scA_{AAAA}\big(\eps^\forw(p_1),p_1,a;\eps^\backw(p_2),p_2,b;\eps^\trans_1(p_3),p_3,c;\eps_2^\trans(p_4),p_4,d\big)
                \end{aligned}
            \end{equation}
            and
            \begin{equation}
                \begin{aligned}
                    &\langle0|\hat c^a(p_1)\hat{\bar c}^b(p_2)\hat A^{\trans\,c}_1(p_3)\hat A^{\trans\,d}_2(p_4)|0\rangle\ =
                    \\
                    &\kern2cm=\ p_1^0\scA_{c\bar cAA}\big(p_1,a;p_2,b;\eps^\trans_1(p_3),p_3,c;\eps^\trans_2(p_4),p_4,d\big)\,,
                \end{aligned}
            \end{equation}
        \end{subequations}
        where $\scA_{AAAA}$ and $\scA_{c\bar cAA}$ denote the four-gluon and two-ghost--two-gluon scattering amplitudes, respectively, with external particles labelled by polarisation vectors, momenta, and colour indices. The hat indicates the Fourier transform. A standard Feynman diagram computation then shows that
        \begin{subequations} 
            \begin{equation}
                \begin{aligned}
                    p_2^0\scA_{AAAA}\ &=\ \frac{f^{ade}f_e{}^{bc}}{\sqrt{2}}~\Big\{(\eps_2\cdot\eps_4)\big[(p_1\cdot\eps_3)+2(p_2\cdot\eps_3)\big]-(\eps_3\cdot\eps_4)\big[(p_1\cdot \eps_2)+2(p_3\cdot\eps_2)\big]\,-
                    \\
                    &\kern2.7cm-\frac{p_2^0(p_2\cdot\eps_3)(p_1\cdot\eps_4)}{\sqrt{2}\big((p_1\cdot p_2)+(p_1\cdot p_3)\big)}-(\eps_2\cdot \eps_3)(p_1\cdot\eps_4)
                    \\
                    &\kern2.7cm-2(\eps_2\cdot\eps_3)(p_2\cdot\eps_4)-\sqrt{2}p_2^0(\eps_3\cdot \eps_4)\Big\}\,+
                    \\
                    &\kern0.5cm+\frac{f^{abe}f_e{}^{cd}}{\sqrt{2}}~\Big\{-\frac{p_2^0}{\sqrt{2}(p_1\cdot p_2)}\big[2(p_1\cdot\eps_4)(p_2\cdot\eps_3)-2(p_1\cdot \eps_3)(p_2\cdot\eps_4)\big]\,-
                    \\
                    &\kern2.7cm-\frac{p_2^0}{\sqrt{2}(p_1\cdot p_2)}\big[(p_1\cdot p_2)-2(p_1\cdot p_3)\big](\eps_3\cdot \eps_4)\,-
                    \\
                    &\kern2.7cm-(\eps_2\cdot\eps_3)\big[(p_1\cdot\eps_4)+2(p_2\cdot\eps_4)\big]\,+
                    \\
                    &\kern2.7cm+(\eps_2\cdot\eps_4)\big[(p_1\cdot\eps_3)+2(p_2\cdot\eps_3)\big]\,-
                    \\
                    &\kern2.7cm-(\eps_3\cdot\eps_4)\big[(p_1\cdot\eps_2)+2(p_3\cdot\eps_2)\big]\Big\}\,+
                    \\
                    &\kern0.5cm+\frac{f^{ace}f_e{}^{bd}}{\sqrt{2}}~\Big\{\frac{p_2^0(p_1\cdot\eps_3)(p_2\cdot\eps_4)}{\sqrt{2}(p_1\cdot p_3)}+(\eps_2\cdot \eps_3)\big[(p_1\cdot\eps_4)+2(p_2\cdot\eps_4)\big]\,-
                    \\
                    &\kern2.7cm-(\eps_2\cdot\eps_4)\big[(p_1\cdot\eps_3)+2(p_2\cdot\eps_3)\big]\,+
                    \\
                    &\kern2.7cm+(\eps_3\cdot\eps_4)\big((p_1\cdot\eps_2)+2(p_3\cdot\eps_2)\big]
                    \\
                    &\kern2.7cm+\sqrt{2}p_2^0(\eps_3\cdot\eps_4)\Big\}
                \end{aligned}
            \end{equation} 
            and
            \begin{equation}
                \begin{aligned}
                    p_1^0\scA_{c\bar cAA}\ &=\ f^{ace}f_e{}^{bd}~\frac{p_2^0(p_1\cdot \eps_3)(p_2\cdot \eps_4)}{2 (p_1\cdot p_3)}
                    \\
                    &\kern0.5cm+f^{abe}f_e{}^{cd}~\frac{p_2^0}{(p_1\cdot p_2)}\Big\{(p_1\cdot\eps_3)(p_2\cdot\eps_4)-(p_1\cdot\eps_4)(p_2\cdot\eps_3)\,+
                    \\
                    &\kern5cm+\big[\tfrac12(p_1\cdot p_2)+(p_1\cdot p_3)\big](\eps_3\cdot\eps_4)\Big\}\,-
                    \\
                    &\kern0.5cm-f^{ade}f_e{}^{bc}~\frac{p_2^0(p_1\cdot\eps_4)(p_2\cdot\eps_3)}{2\big[(p_1\cdot p_2)+(p_1\cdot p_3)\big]}~.
                \end{aligned}
            \end{equation}
        \end{subequations}
        The sum of both terms vanishes,
        \begin{equation}
            p_2^0\scA_{AAAA}+p_1^0\scA_{c\bar cAA}\ =\ 0~,
        \end{equation}
        upon using momentum conservation, transversality of the physically polarised gluons, the explicit form of the on-shell polarisation vectors~\eqref{eq:on-shell_polarisation_vectors}, and the Jacobi identity. That is, the $s$-, $t$-, and $u$-channels are not related separately. This is not very surprising: as indicated in \cref{ssec:scattering_amplitudes_generalities}, the four-point gluon vertex can be distributed in different ways to the various channels and each distribution would imply a different relation between the channels of the two amplitudes. If we ensured colour--kinematics duality for the four-point vertex, however, then the relation between the two amplitudes would hold for each individual channel.
        
        When we come to discussing the double copy theory, we will be able to ensure BRST invariance of the action only on-shell. However, from the construction of correlators from Feynman diagrams it is clear that the action of $Q_{\rm BRST}^{\rm YM,\,lin}$ on the on-shell BRST-extended Hilbert space will still be preserved, and we again have~\eqref{eq:on_shell_Ward} with the corresponding link between scattering amplitudes with different number of ghost--anti-ghost pairs:
        \begin{observation}\label{ob:approximate_Ward}
            Suppose that $Q^{\rm YM}_{\rm BRST}S^{\rm YM}_{\rm BRST}=0$ and $(Q^{\rm YM}_{\rm BRST})^2=0$ only on-shell. Then, we still have an identification of scattering amplitudes with $k+1$ ghost--anti-ghost pairs and all gluons transversely polarised and a sum of amplitudes with $k$ ghost--anti-ghost pairs.
        \end{observation}
        
        \paragraph{Off-shell Ward identities.}
        BRST invariance of the action, being a global symmetry, induces an off-shell Ward identity for correlation functions, 
        \begin{equation}\label{eq:GeneralWardIdentity}
            \langle(\partial^\mu j_\mu(x))\scO_1(x_1)\cdots\scO_n(x_n)\rangle\ =\ \sum_{i=1}^n\mp\delta^{(d)}(x-x_i)\left\langle(Q_{\rm BRST}\scO_i(x_i))\prod_{j\neq i}\scO_j(x_j)\right\rangle,
        \end{equation}
        where $j_\mu$ is the BRST current and the sign is the Koszul sign arising from permuting operators of non-vanishing ghost number. Note that in general, $Q^{\rm YM}_{\rm BRST}$ is the renormalised BRST operator of the full quantum theory, cf.~\cite[Chapter 17.2]{Weinberg:1996kr}. As we will always discuss tree-level correlators, however, we can restrict ourselves to the classical BRST operator with action~\eqref{eq:BRSTOperatorYM}. We note that the left-hand side of~\eqref{eq:GeneralWardIdentity} vanishes after integration over $x$ and the Ward identity simplifies to 
        \begin{equation}\label{eq:simplified_off_shell_Ward}
            0\ =\ \sum_{i=1}^n\pm\left\langle(Q^{\rm YM}_{\rm BRST}\scO_i(x_i))\prod_{j\neq i}\scO_j(x_j)\right\rangle.
        \end{equation}
        
        When applying Ward identities to correlation functions, we can use \cref{ob:BRST_connected} to restrict the correlation functions to purely connected correlators, i.e.~the contribution arising from connected Feynman diagrams. Moreover, we can restrict the correlation functions to a particular order in the coupling constant $g$. This implies that for operators linear in the fields we can truncate the action of the BRST operator $Q^{\rm YM}_{\rm BRST}$ to the Abelian action. 
        
        As a short explicit example, let us consider~\eqref{eq:simplified_off_shell_Ward} for the special case $n=3$ with 
        \begin{equation}
            \hat\scO_1\ =\ \hat A_\mu^{a\forw}(p_1)~,~~~
            \hat\scO_2\ =\ \hat{\bar c}^b(p_2)~,~~~
            \hat\scO_3\ =\ \hat A_\mu^{c\forw}(p_3)~,
        \end{equation}
        and we switched to momentum space for simplicity. We obtain the identity
        \begin{equation}
            \begin{aligned}
                &\sfP^\forw_\mu{}^{\mu'}(p_1)\sfP^\forw_\nu{}^{\nu'}(p_3)\left(\langle\hat A^{a\forw}_{\mu'}(p_1)\hat b^b(p_2)\hat A^{c\forw}_{\nu'}(p_3)\rangle+\langle p_{1\,\mu'}\hat c^a(p_1)\hat{\bar c}^b(p_2)\hat A^{c\forw}_{\nu'}(p_3)\rangle\,-\right.
                \\
                &\kern4cm-\left.\langle\hat A^{a\forw}_{\mu'}(p_1)\hat{\bar c}^b(p_2)p_{3\,\nu'}\hat c^c(p_3)\rangle\right)\ =\ 0~,
            \end{aligned}
        \end{equation}
        where $\sfP^\forw_\mu{}^{\mu'}(p)$ is the projector onto (off-shell) forward polarised gluons. Explicitly,
        \begin{equation}\label{eq:offshell_projectors}
            \begin{aligned}
                \sfP^\forw_\mu{}^{\nu}(p)\ &\coloneqq\ p_\mu\frac{(p\cdot\tilde p)}{(p\cdot\tilde p)^2-(p\cdot p)^2}\left[\tilde p^\nu-\frac{(p\cdot p)}{(p\cdot\tilde p)}p^\nu\right],
                \\
                \sfP^\backw_\mu{}^{\nu}(p)\ &\coloneqq\ \tilde p_\mu\frac{(p\cdot\tilde p)}{(p\cdot\tilde p)^2-(p\cdot p)^2}\left[p^\nu-\frac{(p\cdot p)}{(p\cdot\tilde p)}\tilde p^\nu\right],
            \end{aligned}
        \end{equation}
        where $\tilde p_\mu$ is $p_\mu$ with spatial components reverted.
        
        The relevant vertices are clearly the cubic gluon vertex to which $\hat b^b(p_2)$ is linked by a propagator, as well as the ghost--anti-ghost--gluon vertex. At the tree-level, we thus obtain
        \begin{equation}
            \begin{gathered}
                \sfP^\forw_\mu{}^{\mu'}(p_1)\sfP^\forw_\nu{}^{\nu'}(p_3)\,\langle\hat A^{a\forw}_{\mu'}(p_1)\hat b^b(p_2)\hat A^{c\forw}_{\nu'}(p_3)\rangle\ =
                \\
                =\ f^{abc}\sfP^\forw_{\mu}{}^{\mu'}(p_1)\sfP^\forw_\nu{}^{\nu'}(p_3)\big[p_{2\mu'}p_{1\nu'}-p_{3\mu'}p_{2\nu'}+\eta_{\mu'\nu'}(p_3-p_1)\cdot(\sfP^\backw(p_2)\cdot p_2)\big]~,
                \\
                \sfP^\forw_\mu{}^{\mu'}(p_1)\sfP^\forw_\nu{}^{\nu'}(p_3)\,\langle p_{1\,\mu'}\hat c^a(p_1)\hat{\bar c}^b(p_2)\hat A^{c\forw}_{\nu'}(p_3)\rangle\ =\ f^{abc}\sfP^\forw_\mu{}^{\mu'}(p_1)\sfP^\forw_\nu{}^{\nu'}(p_3)p_{1\mu'}p_{2\nu'}~,
                \\
                \sfP^\forw_\mu{}^{\mu'}(p_1)\sfP^\forw_\nu{}^{\nu'}(p_3)\,\langle\hat A^{a\forw}_{\mu'}(p_1)\hat{\bar c}^b(p_2)p_{3\,\nu'}\hat c^c(p_3)\rangle\ =\ f^{cba}\sfP^\forw_\mu{}^{\mu'}(p_1)\sfP^\forw_\nu{}^{\nu'}(p_3)p_{3\nu'}p_{2\mu'}~.
            \end{gathered}
        \end{equation}
        The sum of these three terms is
        \begin{equation}
            f^{abc}\sfP^\forw_{\mu}{}^{\mu'}(p_1)\sfP^\forw_{\nu}{}^{\nu'}(p_3)\eta_{\mu'\nu'}\,\big[(p_3-p_1)\cdot (\sfP^\backw(p_2)\cdot p_2)\big]\,,
        \end{equation}
        which vanishes after inserting the explicit expressions~\eqref{eq:offshell_projectors}.
        
        We conclude with the following observation.
        \begin{observation}\label{ob:off-shell-Ward}
            We have Ward identities between tree-level correlation functions.
        \end{observation}

        \subsection{Quantum equivalence, correlation functions, and field redefinitions}
        
        Let us now leave the special case of Yang--Mills theory for a moment and reconsider notions of equivalence between field theories in general. As discussed in \cref{ssec:scatterin_amplitudes}, two field theories are classically equivalent if they are quasi-isomorphic and thus have a common minimal model. In the same section, it was explained how the minimal model of a field theory is constructed using the homological perturbation lemma.
        
        \paragraph{Perturbative quantum equivalence.} For the full quantum equivalence at the perturbative level, we have the following evident statement. 
        \begin{observation}\label{ob:quantum_equivalence}
            Two field theories are perturbatively quantum equivalent if all correlators built from polynomials of fields and their derivatives agree to any finite order in coupling constant  and loop level. Since correlators can be glued together from tree-level correlators (up to regularisation issues), it suffices if the tree level correlators agree. 
        \end{observation}
        \noindent We stress that we are only interested in the integrands of scattering amplitudes, which allows us to ignore all issues related to regularisation.
        
        To provide a link between the double-copied action and the action of $\caN=0$ supergravity, we will need to perform a sequence of field redefinitions. The field content of the theories will be the same from the outset, and we choose to work with the same path integral measure in both cases. We are therefore interested in field redefinitions that leave the path integral measure invariant. 
        
        There are large classes of such field redefinitions. The most evident such class of field redefinitions is
        \begin{equation}
            \phi\ \mapsto\ \tilde\phi\ \coloneqq\ \phi+f(\phi'_1,\ldots,\phi'_n)~,
        \end{equation}
        where $f$ is a polynomial function of a set of fields $\{\phi'_1,\ldots,\phi'_n\}$ and their derivatives with $\phi\notin\{\phi'_1,\ldots,\phi'_n\}$. Under such a field redefinition, the path integral measure remains unchanged; this becomes evident when imagining the finite-dimensional analogue of volume forms and a coordinate shifted by a function of different coordinates. 
        
        More subtle is the fact that field redefinitions of the form
        \begin{equation}\label{eq:polytrans}
            \phi\ \mapsto\ \tilde\phi\ \coloneqq\ \phi+\caO(\phi^2)~,
        \end{equation}
        where $\caO(\phi^2)$ denotes local polynomial functions in arbitrary fields and their derivatives which are at least of quadratic order in $\phi$ can also be considered as leaving the path integral measure invariant. 
        
        Invariance of the S-matrix under~\eqref{eq:polytrans}  without derivatives is captured by the Chisholm--Kamefuchi--O'Raifeartaigh--Salam equivalence theorem~\cite{Chisholm:1961tha,Kamefuchi:1961sb}. A proof using the BV formalism of perturbative quantum equivalence for local field redefinitions of the form~\eqref{eq:polytrans} allowing for derivatives  was given in~\cite{Tyutin:2000ht}. This is sufficient for our purposes as we are only concerned with the integrands of scattering amplitudes. Note, however, the well-known need to choose the counter-terms consistently, as emphasised in~\cite{Tyutin:2000ht}. With this in mind, the simplest approach is to use dimensional regularisation, since~\eqref{eq:polytrans} produces a Jacobian which is then regulated to unity, see~\cite{tHooft:1973wag,Leibbrandt:1975dj} as well as~\cite[Sections~18.2.3--4]{Henneaux:1992}.
        
        We sum up the above discussion as follows.
        \begin{observation}\label{ob:field_shifts}
            A shift of a field by products of fields and their derivatives which do not involve the field itself does not change the path integral measure. Local field redefinitions that are trivial at linear order are quantum mechanically safe as they produce a Jacobian that can be regulated to unity in dimensional regularisation.
        \end{observation}
        
        \paragraph{Nakanishi--Lautrup field shifts and changes of gauge.} Besides field redefinitions, we also adjust our choice of gauge to link equivalent field theories. In particular, we can shift the usual choice~\eqref{eq:gaugeFixingFermionYM} for $R_\xi$-gauge to 
        \begin{equation}\label{eq:Psi_shift}
            \Psi\ \mapsto\ \Psi+\Xi
            \ewith
            \Xi\ \coloneqq\ \int\rmd^dx\,\bar c^a Y_a~.
        \end{equation}
        Here, $Y^a$ is of ghost number zero, and we limit ourselves to terms $Y^a$ that are independent of the Nakanishi--Lautrup field. The shift~\eqref{eq:Psi_shift} leads to a shift of the gauge-fixed Lagrangian~\eqref{eq:CGFYMA} given by
        \begin{equation}\label{eq:shift_YM_Lagrangian}
            \scL^{\rm YM}_{\rm BRST}\ \mapsto\ \scL^{\rm YM}_{\rm BRST}+\delder[\Xi]{A^a_\mu}(\nabla_\mu c)^a+\frac g2{f_{bc}}^a\delder[\Xi]{c^a}c^bc^c-b^a\delder[\Xi]{\bar c^a}~.
        \end{equation}
        Evidently, this new Lagrangian is quantum-equivalent to the one with $Y^a=0$, as we merely chose to work in a different gauge.
        
        Subsequently, we may perform the shift
        \begin{equation}\label{eq:b-shifts}
            b^a\ \mapsto\ b^a+Z^a
        \end{equation}
        in the Nakanishi--Lautrup field with $Z^a$ polynomials in the fields and their derivatives. The combination of this shift and~\eqref{eq:Psi_shift} results in
        \begin{equation}\label{eq:add_terms}
            \begin{aligned}
                \scL^{\rm YM}_{\rm BRST}\ &\mapsto\ \scL^{\rm YM}_{\rm BRST}+\delder[\Xi]{A^a_\mu}(\nabla_\mu c)^a+\frac g2{f_{bc}}^a\delder[\Xi]{c^a}c^bc^c\,+
                \\
                &\kern1cm+\frac{\xi}2Z_aZ^a+Z_a(\xi b^a+\partial^\mu A_\mu^a)-(b^a+Z^a)\delder[\Xi]{\bar c^a}~.
            \end{aligned}
        \end{equation}
        We shall assume that $Z^a$ is independent of the Nakanishi--Lautrup field as this will yield a quantum-equivalent Lagrangian by \cref{ob:field_shifts}. We shall also assume that $Z^a$ depends at least quadratically on the other fields and their derivatives to preserve the linearised BRST action on the BRST-extended Hilbert space introduced in~\cref{sec:BRSTExtension}.
        
        \paragraph{Interaction terms linear in the Nakanishi--Lautrup fields.}
        Let us now consider the following special case: suppose that we are in $R_\xi$-gauge and that our Lagrangian contains a term $Z_a\partial^\mu A_\mu^a$ with $Z^a$ independent of the Nakanishi--Lautrup field and at least quadratic in the fields and their derivatives. On the physical Hilbert space with transversely polarised gluons, such expressions vanish. Off-shell, we can still remove such terms by the shifts~\eqref{eq:b-shifts}. Given the need to shift by $Z^a$, we can then iteratively  construct a $Y^a$ which cancels any new terms linear in $b^a$, as is clear from~\eqref{eq:add_terms}. Explicitly, we solve the equation
        \begin{equation}
            0\ =\ \xi Z_a-\delder[\Xi]{\bar c^a}\ =\ \xi Z_a-Y_a+\bar c^b\parder[Y_b]{\bar c^a}+\cdots~,
        \end{equation}
        where the ellipsis denotes terms containing partial derivatives with respect to derivatives of the anti-ghost field $\bar c^b$. Clearly, for consistency, $Y^a$ needs to be at least quadratic in the fields and their derivatives because $Z^a$ is. We are left with the terms 
        \begin{equation}\label{eq:add_terms2}
            -\frac{\xi}2Z_aZ^a+\delder[\Xi]{A^a_\mu}(\nabla_\mu c)^a+\frac g2{f_{bc}}^a\delder[\Xi]{c^a}c^bc^c~,
        \end{equation}
        which are either at least quartic in the fields or at least cubic in the fields, containing ghost fields. The ability to remove any terms of the form $Z_a(\partial^\mu A_\mu)^a$ through local shifts of the Nakanishi--Lautrup field,  absorbing them into $b^a$, and a compensating gauge choice is the `off-shell' Lagrangian analogue of being able to impose that the on-shell external gluons in an amplitude are transverse. We summarise as follows.
        \begin{observation}\label{ob:b-field-shifts}
            Interaction terms in the Lagrangian of degree $n\geq 3$ of the form $Z_a(\partial^\mu A_\mu)^a$ with $Z^a$ independent of the Nakanishi--Lautrup field can be removed in $R_\xi$-gauge by shifting the Nakanishi--Lautrup field according to~\eqref{eq:b-shifts}. This creates the additional terms~\eqref{eq:add_terms2}, which do not modify the scattering amplitudes by \cref{ob:field_shifts} and, in addition, contribute only to interaction vertices of degree $n$ with more ghost--anti-ghost pairs or to interaction vertices of degree greater than $n$.
        \end{observation}
        We also note that a shift of the gauge-fixing fermion by itself~\eqref{eq:Psi_shift} allows us to absorb physical terms proportional to the Nakanishi--Lautrup field without further affecting the physical sector.
        \begin{observation}\label{ob:match_Nakanishi}
            Terms in the action that are proportional to the Nakanishi--Lautrup field can be absorbed by choosing a suitable term $Y^a$. This leaves the physical sector invariant but it may modify the ghost sector. Because Nakanishi--Lautrup fields appear via trivial pairs in the BV action, this extends to general gauge theories, e.g.~with several Nakanishi--Lautrup fields and ghosts--for--ghosts.
        \end{observation}

        \paragraph{Actions related by field redefinitions.} Let us return to a general setting. Suppose that we are given two classical field theories which are specified by local actions $S$ and $\tilde S$, as power series in the fields and their derivatives, whose corresponding $L_\infty$-algebras have the same minimal model, the same field content and the same kinetic parts. 
        
        Consider the cubic interaction terms $\scL_3$ and $\tilde\scL_3$ in $S$ and $\tilde S$. Since the three-point amplitudes agree, these interaction terms can differ at most in terms that vanish on external fields. Therefore, these terms have to be proportional to either the on-shell equation for an external field or to a field with unphysical polarisation which is not contained in the external fields. Both types of terms can be cancelled by a local field redefinition which shifts the discrepancy into the quartic and higher interaction terms. Such field redefinitions constitute quasi-isomorphisms of $L_\infty$-algebras and leave the tree-level scattering amplitudes unmodified. We are left with two theories with the same tree-level scattering amplitudes and with the same interaction terms to cubic order.
        
        The discrepancy between the total quartic terms of both field theories after the above field redefinition is again invisible at the level of external fields, because the tree-level scattering amplitudes still agree. We then compensate again by further field redefinitions, shifting the discrepancy into quintic and higher interaction terms. In this way, we can remove the differences between the Lagrangians order by order in the interaction vertices, field-redefining the difference away to higher order interaction vertices. Since we are merely interested in perturbation theory, agreements to arbitrary finite orders are completely sufficient. 
        
        Altogether, we can conclude that for the purpose of perturbative quantum field theory, we can regard the actions $S$ and $\tilde S$ to be related by local field redefinitions. In certain cases it is even possible to give closed all order expression for (part of) the field redefinitions,  providing a formal non-perturbative equivalence.
        \begin{observation}\label{ob:classical_equivalence}
            If two field theories have the same tree-level scattering amplitudes, then the minimal models of the corresponding $L_\infty$-algebras coincide, cf.~\cite{Jurco:2018sby,Macrelli:2019afx}. If also the associated actions are local and given by power series of the fields and their derivatives, and have the same field content and kinetic parts, then they are related by local (invertible) field redefinitions.
        \end{observation} 

        The explicit example of Yang--Mills theory may be instructive. Consider the action~\eqref{eq:YM_field_redefined_action} of Yang--Mills theory in $R_\xi$-gauge with the field redefinitions~\eqref{eq:canonicalFieldRedefinitionYM} implemented as in \cref{sec:canonicalTransformationYM} and consider an action $\tilde S$ with the same fields, the same kinematic parts and identical tree-level scattering amplitudes. The discrepancies in the interaction vertices at each order are proportional to (at least) one of the terms 
        \begin{equation}\label{eq:on-shell-vanishing-YM}
            \tilde A^{\forw\,a}_\mu~,~~~
            \sqrt{\wave}\,\tilde b^a+\tilde \xi\partial^\mu \tilde A_\mu^a~,~~~
            \wave \tilde A_\mu^a~,~~~
            \wave \tilde c^a~,~~~
            \wave\tilde{\bar c}^a~,
            \eand
            \wave \tilde b^a~.
        \end{equation}
        Given the BRST invariance, we can always exclude terms proportional to $\tilde A^{\forw\,a}_\mu$, as these can be absorbed by residual gauge transformations. Terms proportional to $\sqrt{\wave}\tilde b^a+\tilde \xi\partial^\mu \tilde A_\mu^a$ can be absorbed by a field redefinition of the Nakanishi--Lautrup field due to \cref{ob:b-field-shifts}. All remaining differences are sums of terms proportional to $\wave \tilde A^a_\mu$, $\wave \tilde c^a$, $\wave\tilde{\bar c}^a$, or $\wave \tilde b^a$, and they can be absorbed by iterative field redefinitions, starting with the three-point amplitudes. There is an evident field redefinition of the relevant field, quadratic in the fields and their derivatives, such that the kinetic term of redefined Yang--Mills theory produces the difference in kinetic terms. Since such a field redefinition is a quasi-isomorphism of the corresponding $L_\infty$-algebras, it preserves the minimal model and thus the tree-level amplitudes. Moreover, such a field redefinition is clearly local.

        \subsection{Strictification of Yang--Mills theory}\label{ssec:strictified_YM}
        
        \paragraph{Generalities.}
        An important structure theorem for homotopy algebras is the strictification theorem, cf.~\cref{app:structure_theorems}. In particular, it implies that any $L_\infty$-algebra is quasi-isomorphic to a strict $L_\infty$-algebra, i.e.~an $L_\infty$-algebra with $\mu_i=0$ for $i\geq 3$, better known as a differential graded Lie algebra.
        
        From a field theory perspective, this implies that any classical field theory is equivalent to a classical field theory with interaction terms which are all cubic in the fields. Generically, a strictifying quasi-isomorphism may produce non-local terms, but there is always a  systematic choice of strictification that is entirely local. This is quite evident for the interactions of scalar fields, since we can `blow up' $n$-ary vertices to cubic graphs with edges corresponding to propagating auxiliary fields, cf.~e.g.~the discussions in~\cite{Jurco:2018sby,Macrelli:2019afx}.
        
        As a simple example of a strictification, consider the first-order formalism of Yang--Mills theory on four-dimensional Euclidean space $\IR^4$~\cite{Okubo:1979gt}, in which an additional self-dual two form $B_+\in \Omega^2_+(\IR^4)\otimes\frg$ in the adjoint representation of the gauge Lie algebra is added to the field content,
        \begin{equation}\label{eq:YM1st}
            S^{\rm YM_1}\ \coloneqq\ \int \rmd^4x\,\Big\{\tfrac12\eps^{\mu\nu\kappa\lambda}F_{a\mu\nu}B^a_{+\kappa\lambda}+\tfrac14\eps^{\mu\nu\kappa\lambda}B_{+a\mu\nu}B^a_{+\kappa\lambda}\Big\}~.
        \end{equation}
        The $L_\infty$-algebra corresponding to the full BV complex of this theory is indeed strict; see~\cite{Rocek:2017xsj,Jurco:2018sby} for a quasi-isomorphism between this $L_\infty$-algebra and that of the ordinary, second-order formulation of Yang--Mills theory.
        
        Note, however, that the full strictification of gauge theories including ghosts is a bit more involved: the equations of motion of the introduced auxiliary fields would be at least quadratic in other fields, and if these transform in the adjoint representation or as connections, the gauge transformations of auxiliary fields are at least cubic in fields and ghosts, leading to quartic or higher terms in the BV action. The strictification theorem still guarantees the existence of an equivalent formulation as a field theory with cubic interaction vertices, but we may have to extend our field space not merely by adding fields, but by switching e.g.~to its loop space. This is due to the fact that cubic gauge transformations in an $L_\infty$-algebra are encoded in a $\mu_3$, which in turn corresponds to a particular three-cocycle. The latter can be transgressed to a two-cocycle over loop space, which merely corresponds to a Lie algebra extension and thus, is turned into a higher product $\mu_2$. For fully gauge-fixed actions, however, this problem never arises.
        
        We also note that the factorisation in the double copy is most easily performed in a specific strictification\footnote{It is actually a family of strictifications.}, which is \emph{not} the first order formulation~\eqref{eq:YM1st}. Its precise form is discussed in the following.
        
        \paragraph{Colour--kinematics-dual form and cubic diagrams.} 
        Recall from \cref{ssec:BCJduality} that the tree-level Yang--Mills amplitudes can be rearranged in colour--kinematics-dual form, which is by now a well-established fact~\cite{Stieberger:2009hq,BjerrumBohr:2009rd,Jia:2010nz,BjerrumBohr:2010hn,Feng:2010my,Chen:2011jxa,Mafra:2011kj,Du:2016tbc,Mizera:2019blq}.
        \begin{observation}\label{ob:usual_double_copy}
            The tree amplitudes of Yang--Mills theory can be written in colour--kinematics-dual form.
        \end{observation}
        Explicitly, one can construct a Lagrangian whose Feynman diagrams  generate colour--kinematics-dual tree-level amplitudes of physical (transverse) gluons in Yang--Mills theory, making colour--kinematics duality manifest at the Lagrangian level. This is achieved by adding non-local interaction terms $\caO(A^n)$, for all $n>5$, to the action that vanish identically due to the colour Jacobi identity. The necessary terms were first constructed in~\cite{Bern:2010yg} up to six points. The algorithm of Tolotti--Weinzierl~\cite{Tolotti:2013caa} is a prescription of how to find the necessary terms to arbitrary order.
        
         Since the new terms are identically zero, they obviously leave the theory and amplitudes invariant, but nonetheless change the individual kinematic numerators to realise colour--kinematics duality. Moreover, the new terms can be rendered cubic and local through the introduction of auxiliary fields~\cite{Borsten:2020zgj}, as demonstrated explicitly at five points in~\cite{Bern:2010yg}. Roughly speaking, one starts from Yang--Mills theory and strictifies the already present quartic  interaction vertex by inserting an auxiliary field, redistributing the contributions to ensure colour--kinematics duality for four-point amplitudes. The colour--kinematics duality of the five-point amplitudes then requires a new interaction term $\caO(A^5)$ which vanishes due to the Jacobi identity. This vertex is then strictified by inserting further auxiliary fields, etc. The overall action is thus trivially equivalent to Yang--Mills theory. We note that the form of the strictification is encoded in the action produced by the Tolotti--Weinzierl algorithm. We shall be completely explicit below, but let us first summarise the situation.
        \begin{observation}\label{ob:ampltiudes_to_Lagrangian}
            Given tree-level physical gluon amplitudes in colour--kinematics-dual form, there is a corresponding purely cubic Lagrangian whose Feynman diagrams (summed over identical topologies) produce kinematic numerators satisfying the kinematic Jacobi identities.
        \end{observation}
        To illustrate the strictification, let us consider the four- and five-point contributions, which were already computed in~\cite{Bern:2010yg}:
        \begin{equation}\label{eq:4th5thorderAction}
            \begin{aligned}
                \scL^{(4)}\ &\sim\ \tr\big\{[A_\mu,A_\nu][A^\mu,A^\nu]\big\}\ =\ -\eta^{\mu\nu}\eta^{\kappa\rho}\eta^{\lambda\sigma}~\partial^{12}_\mu\partial_\nu^{34}~\frac{\tr\big\{[A_\kappa,A_\lambda][A_\rho,A_\sigma]\big\}}{\wave_{12}}~,\\
                \scL^{(5)}\ &\sim\ \tr\left\{[A^\nu,A^\rho]\frac{1}{\wave}\left(\left[[\partial_\mu A_\nu,A_\rho],\frac{\wave}{\wave} A^{\mu}\right]+\right.\right.
                \\
                &\kern4cm\left.\left.+\left[[A_\rho,A^\mu],\frac{\wave}{\wave}\partial_\mu A_\nu\right]+\left[[A^\mu,\partial_\mu A_\nu],\frac{\wave}{\wave}A_\rho\right]\right)\right\}.
            \end{aligned}
        \end{equation}
        We immediately note that $\scL^{(5)}$ vanishes by the colour Jacobi identity. Its presence, however, is required for the kinematic Jacobi identity to hold after factorisation. 
        
        As explained in \cref{ssec:manifest_bcj_action}, these terms reflect a `blow up' of $n$-point interaction vertices into trees with trivalent vertices and all symmetries taken into account:
        \begin{equation}
            \begin{aligned}
                &n=4:
                ~~~~
                \begin{tikzpicture}[
                    scale=1,
                    every node/.style={scale=1},
                    baseline={([yshift=-.5ex]current bounding box.center)}
                    ]
                    \matrix (m) [
                    matrix of nodes,
                    ampersand replacement=\&,
                    column sep=0.2cm,
                    row sep=0.2cm
                    ]{
                        $1$ \& {} \& {} \& {} \& $3$
                        \\
                        {} \& {} \& {} \& {} \& {}
                        \\
                        {} \& {} \& {} \& {} \& {}
                        \\
                        {} \& {} \& {} \& {} \& {}
                        \\
                        $2$ \& {}\& {} \& {} \& $4$
                        \\
                    };
                    \draw [gluon] (m-1-1) -- (m-3-2.center);
                    \draw [gluon] (m-5-1) -- (m-3-2.center);
                    \draw [aux] (m-3-2.center) -- (m-3-4.center);
                    \draw [gluon] (m-1-5) -- (m-3-4.center);
                    \draw [gluon] (m-5-5) -- (m-3-4.center);
                    \foreach \x in {(m-3-2), (m-3-4)}{
                        \fill \x circle[radius=2pt];
                    }
                \end{tikzpicture},
                ~~~~
                \begin{tikzpicture}[
                    scale=1,
                    every node/.style={scale=1},
                    baseline={([yshift=-.5ex]current bounding box.center)}
                    ]
                    \matrix (m) [
                    matrix of nodes,
                    ampersand replacement=\&,
                    column sep=0.2cm,
                    row sep=0.2cm
                    ]{
                        $1$ \& {} \& {} \& {} \& $2$
                        \\
                        {} \& {} \& {} \& {} \& {}
                        \\
                        {} \& {} \& {} \& {} \& {}
                        \\
                        {} \& {} \& {} \& {} \& {}
                        \\
                        $3$ \& {}\& {} \& {} \& $4$
                        \\
                    };
                    \draw [gluon] (m-1-1) -- (m-3-2.center);
                    \draw [gluon] (m-5-1) -- (m-3-2.center);
                    \draw [aux] (m-3-2.center) -- (m-3-4.center);
                    \draw [gluon] (m-1-5) -- (m-3-4.center);
                    \draw [gluon] (m-5-5) -- (m-3-4.center);
                    \foreach \x in {(m-3-2), (m-3-4)}{
                        \fill \x circle[radius=2pt];
                    }
                \end{tikzpicture},
                ~~~~
                \begin{tikzpicture}[
                    scale=1,
                    every node/.style={scale=1},
                    baseline={([yshift=-.5ex]current bounding box.center)}
                    ]
                    \matrix (m) [
                    matrix of nodes,
                    ampersand replacement=\&,
                    column sep=0.2cm,
                    row sep=0.2cm
                    ]{
                        $1$ \& {} \& {} \& {} \& $2$
                        \\
                        {} \& {} \& {} \& {} \& {}
                        \\
                        {} \& {} \& {} \& {} \& {}
                        \\
                        {} \& {} \& {} \& {} \& {}
                        \\
                        $4$ \& {}\& {} \& {} \& $3$
                        \\
                    };
                    \draw [gluon] (m-1-1) -- (m-3-2.center);
                    \draw [gluon] (m-5-1) -- (m-3-2.center);
                    \draw [aux] (m-3-2.center) -- (m-3-4.center);
                    \draw [gluon] (m-1-5) -- (m-3-4.center);
                    \draw [gluon] (m-5-5) -- (m-3-4.center);
                    \foreach \x in {(m-3-2), (m-3-4)}{
                        \fill \x circle[radius=2pt];
                    }
                \end{tikzpicture},
                \\
                &n=5:~~~~
                \begin{tikzpicture}[
                    scale=1,
                    every node/.style={scale=1},
                    baseline={([yshift=-.5ex]current bounding box.center)}
                    ]
                    \matrix (m) [
                    matrix of nodes,
                    ampersand replacement=\&,
                    column sep=0.2cm,
                    row sep=0.2cm
                    ]{
                        $1$ \& {} \& {} \& $3$ \& {} \& {} \& $4$
                        \\
                        {} \& {} \& {} \& {} \& {} \& {} \& {}
                        \\
                        {} \& {} \& {} \& {} \& {} \& {} \& {}
                        \\
                        {} \& {} \& {} \& {} \& {} \& {} \& {}
                        \\
                        $2$ \& {}\& {} \& {} \& {} \& {} \& $5$
                        \\
                    };
                    \draw [gluon] (m-1-1) -- (m-3-2.center);
                    \draw [gluon] (m-5-1) -- (m-3-2.center);
                    \draw [gluon] (m-3-2.center) -- (m-3-4.center);
                    \draw [gluon] (m-3-4.center) -- (m-1-4);
                    \draw [aux] (m-3-4.center) -- (m-3-6.center);
                    \draw [gluon] (m-1-7) -- (m-3-6.center);
                    \draw [gluon] (m-5-7) -- (m-3-6.center);
                    \foreach \x in {(m-3-2), (m-3-4), (m-3-6)}{
                        \fill \x circle[radius=2pt];
                    }
                \end{tikzpicture},
                ~~~~
                \begin{tikzpicture}[
                    scale=1,
                    every node/.style={scale=1},
                    baseline={([yshift=-.5ex]current bounding box.center)}
                    ]
                    \matrix (m) [
                    matrix of nodes,
                    ampersand replacement=\&,
                    column sep=0.2cm,
                    row sep=0.2cm
                    ]{
                        $1$ \& {} \& {} \& $4$ \& {} \& {} \& $3$
                        \\
                        {} \& {} \& {} \& {} \& {} \& {} \& {}
                        \\
                        {} \& {} \& {} \& {} \& {} \& {} \& {}
                        \\
                        {} \& {} \& {} \& {} \& {} \& {} \& {}
                        \\
                        $2$ \& {}\& {} \& {} \& {} \& {} \& $5$
                        \\
                    };
                    \draw [gluon] (m-1-1) -- (m-3-2.center);
                    \draw [gluon] (m-5-1) -- (m-3-2.center);
                    \draw [gluon] (m-3-2.center) -- (m-3-4.center);
                    \draw [gluon] (m-3-4.center) -- (m-1-4);
                    \draw [aux] (m-3-4.center) -- (m-3-6.center);
                    \draw [gluon] (m-1-7) -- (m-3-6.center);
                    \draw [gluon] (m-5-7) -- (m-3-6.center);
                    \foreach \x in {(m-3-2), (m-3-4), (m-3-6)}{
                        \fill \x circle[radius=2pt];
                    }
                \end{tikzpicture},
                ~~~~
                \ldots                
            \end{aligned}
        \end{equation}
        Here, an internal wavy line comes with a propagator in Feynman gauge $\frac{1}{\wave}$, while a dashed line corresponds to the identity operator $\frac{\spacer{2pt}\wave}{\wave}$.
        
        The general Lagrangian at $n$-th order is then of the form 
        \begin{equation}\label{eq:form_add_terms}
            \scL^{(n)}\ =\ f_{M_1\cdots M_k}E^{M_1}_1D_1(E^{M_2}_2D_2(E^{M_3}_3D_3\cdots))~,
        \end{equation}
        where $D_i$ stands for either $\frac{1}{\wave}$ or $\frac{\spacer{2pt}\wave}{\wave}$ and the $M_i$s are Lorentz multi-indices. Note that all the $E_i$s are polynomials of degree one or two in the fields. In the tree picture, the wave operators in the denominator correspond precisely to the edges in the trees.
        
        \paragraph{Strictification.}
        To strictify the non-local action, we now iteratively insert auxiliary fields $G^M_{n,\Gamma,i}$ and $\bar G^{n,\Gamma,i}_M$ for each operator $D_i$. If we are dealing with an operator of the form $\frac{\spacer{2pt}\wave}{\wave}$, we first use partial integration
        \begin{equation}
            \frac{E_1^{M_1}\wave_1E^{M_2}_2}{\wave_1}\ =\ -\frac{(\partial_\mu E_1^{M_1})(\partial^\mu E_2^{M_2})}{\wave_1}~,
        \end{equation}
        where $E_i^M$ is an arbitrary expression in the fields, derivatives, and auxiliary fields. We then use the fact that the Lagrangians
        \begin{subequations}
            \begin{equation}
                E_1^M\frac{1}{\wave}E^2_M
            \end{equation}
            and
            \begin{equation}
                -G^M_{n,\Gamma,i}\wave\bar G^{n,\Gamma,i}_M+G^M_{n,\Gamma,i}E^2_M+E_1^M\bar G^{n,\Gamma,i}_M
            \end{equation}
        \end{subequations}
        are physically equivalent after integrating out the auxiliary fields $G^M_{n,\Gamma,i}$ and $\bar G^{n,\Gamma,i}_M$. We iterate this process until all the inverse wave operators have been replaced in this manner. 
        
        We note that in each iteration, $E_1^M$ and $E^2_M$ are both polynomials of degree at least two in the fields. Introducing the auxiliary fields reduces the polynomial degree at least by one, and in the end, the action has indeed only cubic interaction terms and thus is a strictification of the original action. We also note that two auxiliary fields can be combined into one if they have identical equations of motion.
        
        \paragraph{Homotopy algebraic perspective.} The strictification $\frL^{\rm YM,\,st}_{\rm BRST}$ of the $L_\infty$-algebra $\frL^{\rm YM}_{\rm BRST}$ or, equivalently, of the colour--kinematics-dual action is nothing but a quasi-isomorphism (see \cref{app:hA_L_infinity})
        \begin{equation}
            \phi\,:\,\frL^{\rm YM}_{\rm BRST}\ \rightarrow\ \frL^{\rm YM,\,st}_{\rm BRST}~,
        \end{equation}
        and the map $\phi$ is given by
        \begin{equation}
            \begin{aligned}
                A^{\rm st}+\sum_{n,\Gamma,i}G_{n,\Gamma,i}\ =\ \phi_1(A)+\tfrac12\phi_2(A,A)+\cdots\ =\ \sum_{k\geq 1}\tfrac1{k!}\phi_k(A,\ldots,A)~,
            \end{aligned}
        \end{equation}
        where $A^{\rm st}$ is the gauge potential in $\frL^{\rm YM,\,st}$,
        \begin{equation}
            A^{\rm st}\ =\ \phi_1(A)~,
        \end{equation}
        and the higher maps are such that $G_{n,\Gamma,i}$ are given by their equations of motion, fully reduced to expressions in the original gauge potential $A$. 
        
        Let us work out the details for the example of the fourth- and fifth-order terms~\eqref{eq:4th5thorderAction}. The explicit form of the corresponding strictified Lagrangian is already found in~\cite{Bern:2010yg},
        \begin{subequations}
            \begin{equation}
                \begin{aligned}
                    \scL^{\rm YM,\,st}\ \coloneqq\ \tfrac12\tr\big\{A_\mu\wave A^\mu\big\}+\scL^{\rm YM,\,st}_4+\scL^{\rm YM,\,st}_5~,
                \end{aligned}
            \end{equation}
            with
            \begin{equation}
                \begin{aligned}
                    \scL^{\rm YM,\,st}_4\ &\coloneqq\ \tr\Big\{-\tfrac12 G_{4,\Gamma,1}^{\mu\nu\kappa}\wave G^{4,\Gamma,1}_{\mu\nu\kappa}-g(\partial_\mu A_\nu+\tfrac{1}{\sqrt{2}}\partial^\kappa G^{4,\Gamma,1}_{\kappa\mu\nu})[A^\mu,A^\nu]\Big\}~,
                    \\
                    \scL^{\rm YM,\,st}_5\ &\coloneqq\ \tr\Big\{G_{5,\Gamma,1}^{\mu\nu}\wave\bar G^{5,\Gamma,1}_{\mu\nu}+G_{5,\Gamma,2}^{\mu\nu\kappa}\wave \bar G^{5,\Gamma,2}_{\mu\nu\kappa}+G_{5,\Gamma,3}^{\mu\nu\kappa\lambda}\wave\bar G^{5,\Gamma,3}_{\mu\nu\kappa\lambda}\,+
                    \\
                    &\kern1.5cm+gG_{5,\Gamma,1}^{\mu\nu}[A_\mu,A_\nu]+g\partial_\mu G_{5,\Gamma,2}^{\mu\nu\kappa}[A_\nu,A_\kappa]-\tfrac g2\partial_\mu G_{5,\Gamma,3}^{\mu\nu\kappa\lambda}[\partial_{[\nu}A_{\kappa]},A_\lambda]\,+
                    \\
                    &\kern1.5cm+g\bar G_{5,\Gamma,1}^{\mu\nu}\big(\tfrac12[\partial^\kappa\bar G^{5,\Gamma,2}_{\kappa\lambda\mu},\partial^\lambda A_\nu]+[\partial^\kappa\bar G^{5,\Gamma,3}_{\kappa\lambda\nu[\mu},A^\lambda]\big)\Big\}~.
                \end{aligned}
            \end{equation}
        \end{subequations}
        Consequently, the resulting quasi-isomorphism reads as
        \begin{equation}
            \begin{aligned}
                &\phi_1(A)+\tfrac12\phi_2(A,A)+\tfrac1{3!}\phi_3(A,A,A)\ =
                \\
                &\kern2cm\ =\
                \begin{pmatrix}
                    A^s_\mu
                    \\
                    G^{4,\Gamma,1}_{\mu\nu\kappa}
                    \\
                    G_{\mu\nu}^{5,\Gamma,1}
                    \\
                    \bar G^{5,\Gamma,1}_{\mu\nu}
                    \\
                    G_{5,\Gamma,2}^{\mu\nu\kappa}
                    \\
                    \bar G^{5,\Gamma,2}_{\mu\nu\kappa}
                    \\
                    G_{5,\Gamma,3}^{\mu\nu\kappa\lambda}
                    \\
                    \bar G^{5,\Gamma,3}_{\mu\nu\kappa\lambda}
                \end{pmatrix}
                \ =\
                \begin{pmatrix}
                    A_\mu
                    \\
                    \frac{g}{2\wave}\partial_\mu[A_\nu,A_\kappa]
                    \\
                    -\frac{g^2}{2\wave}\big([[A_\lambda,A_\mu],\partial^\lambda A_\nu]-[[\partial_{[\lambda} A_{\nu]},A_\mu],A^\lambda]\big)
                    \\
                    -\frac{g}{\wave}[A_\mu,A_\nu]
                    \\
                    -\frac{g^2}{2\wave}\partial^\mu\big[\partial^\nu A_\lambda,\frac{1}{\wave}[A^\kappa,A^\lambda]\big]
                    \\
                    \frac{g}{\wave}\partial_\mu[A_\nu,A_\kappa]
                    \\
                    -\frac{g^2}{\wave}\partial^\mu\big[A^\nu,\frac{1}{\wave}[A^\lambda,A^\kappa]\big]
                    \\
                    -\frac{g}{2\wave}\partial_\mu[\partial_{[\nu}A_{\kappa]},A_\lambda]
                \end{pmatrix}~.
            \end{aligned}
        \end{equation}
        Note that the decomposition into the images of the maps $\phi_i$ corresponds to the decomposition of the image into monomials of power $i$ in the fields.
        
        \paragraph{Tree-level double copy.}
        As reviewed in \cref{ssec:double_copy}, the double copy of the kinematic numerators in the scattering amplitudes of the strictified Yang--Mills theory produces the tree-level scattering amplitudes of $\caN=0$ supergravity~\cite{Bern:2008qj,Bern:2010ue,Bern:2010yg}.
        \begin{observation}\label{ob:established_dc}
            Double copying the Yang--Mills tree-level scattering amplitudes of physical gluons in colour--kinematics-dual form yields the physical tree-level scattering amplitudes of $\caN=0$ supergravity.
        \end{observation}
        
        \paragraph{Compatibility with quantisation.}
        It is clear that quantisation does not commute with quasi-isomorphisms: classically equivalent field theories can have very different quantum field theories. A simple example making this evident is the $L_\infty$-algebra of Yang--Mills theory $\frL^{\rm YM}_{\rm BRST}$ and one of its quasi-isomorphic minimal models $\frL^{\rm YM\,\circ}_{\rm BRST}$. The vector space of $\frL^{\rm YM\,\circ}_{\rm BRST}$ is simply the free fields labelling external states in Yang--Mills scattering amplitudes, together with some irrelevant cohomological remnants in the ghosts, Nakanishi--Lautrup fields, and anti-ghosts. The tree-level scattering amplitudes of $\frL^{\rm YM}_{\rm BRST}$ are given by the higher products of $\frL^{\rm YM\,\circ}_{\rm BRST}$. They are also the tree-level scattering amplitudes of $\frL^{\rm YM\,\circ}_{\rm BRST}$ since there are no propagating degrees of freedom left. Clearly, however, there are loop-level scattering amplitudes in Yang--Mills theory which $\frL^{\rm YM}_{\rm BRST}$ can describe but which are absent in $\frL^{\rm YM\,\circ}_{\rm BRST}$. Thus, the quantum theories described by the quasi-isomorphic $L_\infty$-algebras $\frL^{\rm YM}_{\rm BRST}$ and $\frL^{\rm YM\,\circ}_{\rm BRST}$ differ.
        
        Certainly, there are quasi-isomorphisms which are compatible with quantisation. In particular, any quasi-isomorphism that corresponds to integrating out fields which appear at most quadratically in the action are of this type: we can simply complete the square in the path integral and perform the Gau{\ss}ian integral. This amounts to replacing each auxiliary field by the equation of motion.
        
        This is precisely the case in the above strictification of Yang--Mills theory, and the original formulation is quantum equivalent to its strictification. This is also clear at the level of Feynman diagrams: as the kinematic terms are all of the form $-G^M_{n,\Gamma,i}\wave \bar G^{n,\Gamma,i}_M$, each auxiliary field propagates into precisely one other auxiliary field. Moreover, each auxiliary field $G$ appears in precisely one type of vertex and then only as one leg. That is, once a propagator ends in one of the auxiliary fields, the continuation of the diagram at this end is unique until all the remaining open legs are non-auxiliaries. There are no loops consisting of purely auxiliary fields. All loops containing at least one gluon propagator are simply contracted to gluon loops. It is thus clear that the degrees of freedom running around loops in the strictified theory are the same as those running around loops in ordinary Yang--Mills theory.
        
        \subsection{Colour--kinematics duality for unphysical states}\label{ssec:ck_duality_YM_BRST_extended_Hilbertspace}
        
        The action and factorisation we have presented so far are the complete data to double copy tree-level gauge theory scattering amplitudes to gravity scattering amplitudes. For the full double copy at the loop level, however, we need to work a bit harder, as explained in our previous paper~\cite{Borsten:2020zgj}. 
        
        So far, colour--kinematics duality is only ensured for all on-shell gluon states with physical polarisation. Our goal will be to double copy arbitrary tree-level correlators, which can have unphysical polarisations of gluons as well as ghost states on external legs. We therefore need to ensure that colour--kinematics duality holds more generally. In order to establish the off-shell double copy it is sufficient to guarantee colour--kinematics duality for on-shell states in the BRST-extended Hilbert space from \cref{sec:BRSTExtension}. 
        
        \paragraph{Unphysical states.}
        Colour--kinematics duality is not affected by forward-polarised gluons, as these can be absorbed by residual gauge transformations. Furthermore, colour--kinematics duality for backward-polarised gluons can be achieved by adding new terms to the action, which are  physically irrelevant since they are introduced only through the gauge-fixing fermion. Colour--kinematics duality for ghosts is then achieved by transferring colour--kinematics duality for longitudinal gluons to the ghost sector by \cref{ob:ghost_pair_reduction} via the BRST Ward identities. We now explain the procedure in detail.
        
        We perform the corrections order by order in the degree $n$ of the vertices and for each degree order by order in the number $k$ of ghost--anti-ghost pairs. The first vertex to consider is $n=4$, and we start at $k=0$. Colour--kinematics duality for four on-shell gluons in the BRST-extended Hilbert space can only be violated by terms proportional to $\xi b^a+\partial^\mu A^a_\mu$, and we can introduce a vertex compensating these violations in the Lagrangian. 
                
        A short calculation shows that the vertex correcting the violation of colour--kinematics duality for longitudinal gluons at four points is given by 
        \begin{equation}\label{eq:4_point_long_correction_vertex}
            g^2(\partial^\rho A_{\rho}^{b})A^{c\mu}\frac{1}{\wave}\big[(\partial^\nu A_\mu^d)A_\nu^e\big]{f_{ed}}^a{f_{acb}}~.
        \end{equation}
        Unlike the vertices added to make colour--kinematics duality manifest for transverse gluons, which are identically zero, the above is non-vanishing. It cannot be simply added to the Yang--Mills Lagrangian without changing the amplitudes. However, since it is proportional to $\partial^\rho A_{\rho}$ (as had to be the case, since its contribution to the kinematic numerators  must vanish for transverse gluons) we can employ \cref{ob:b-field-shifts,ob:match_Nakanishi}. Using \cref{ob:b-field-shifts} we shift 
        \begin{equation}\label{eq:4_point_NL_shift}
            b^a\ \mapsto\ b^a+Z^a~,
            \ewith
            Z^a\ \coloneqq\ g^2A^{c\mu}\frac{1}{\wave}\big[(\partial^\nu A_\mu^d)A_\nu^e)\big]{f_{ed}}^b{f_{bc}}^a~,
         \end{equation}
        which introduces~\eqref{eq:4_point_long_correction_vertex} while leaving the amplitudes invariant. It also has the unimportant, but inconvenient with respect to the double copy, effect of adding $\xi b_a Z^a$. Using \cref{ob:match_Nakanishi} this term can be eliminated by shifting the gauge-fixing function by 
        \begin{equation}
            Y^a\ \coloneqq\ -\xi Z^a~,
        \end{equation}
        which produces 
        \begin{equation}\label{eq:CK-completion_n4}
            \begin{aligned}
                \scL^{\rm YM,\,comp}_{n=4,\,k=0}\ &=\ -\xi Q_{\rm BRST} (\bar c^aZ_a)\\
                &=\ -\xi b^bZ_b+g^2\bar c^b\,Q_{\rm BRST}\left(A^{c\mu}\frac{1}{\wave}\big[(\partial^\nu A_\mu^d)A_\nu^e\big]\right){f_{ed}}^a{f_{acb}}~,
            \end{aligned}
        \end{equation}
        The first term cancels the $\xi b_a Z^a$ term generated by~\eqref{eq:4_point_NL_shift}, while the second term generates higher-point ghost interactions that will be dealt with at the next iterative step with $n=4$, $k=1$. The total correction to the Lagrangian is then       
        \begin{equation}
            \scL^{\rm YM,\,corr}_{n=4,\,k=0}\ =\ g^2\left\{(\partial^\rho A_{\rho}^{b})A^{c\mu}\frac{1}{\wave}\big[(\partial^\nu A_\mu^d)A_\nu^e\big]+\bar c^b\,Q_{\rm BRST}\left(A^{c\mu}\frac{1}{\wave}\big[(\partial^\nu A_\mu^d)A_\nu^e\big]\right)\right\}{f_{ed}}^a{f_{acb}}~.
        \end{equation}
        where the first term enforces colour--kinematics duality for longitudinal gluons and the second ensures BRST invariance. We note that the terms in $\scL^{\rm YM,\,corr}_{n=4,\,k=0}$ come with a canonical strictification given by the colour structure. This strictification then yields colour--kinematics-dual four-point gluon scattering amplitudes. 
        
        The next case to consider is $n=4$, $k=1$. We now use \cref{ob:ghost_pair_reduction} to relate the four-gluon scattering amplitude to this scattering amplitude, and, correspondingly, the four-gluon tree-level scattering amplitude to the two gluon, one ghost-anti-ghost pair tree-level scattering amplitude. We obtain colour--kinematics duality for scattering amplitudes consisting of a ghost--anti-ghost pair as well as two physically polarised gluons. Generalising the latter to two arbitrary gluons in the BRST-extended Hilbert space, we expect colour--kinematics duality violating terms proportional to $\xi b^a+\partial^\mu A^a_\mu$. It turns out that these terms happen to vanish and there is nothing left to do. Note that if these terms had not vanished, we would have compensated for them again by inserting physically irrelevant terms to the action in a BRST-invariant fashion.
        
        \cref{ob:ghost_pair_reduction} now immediately implies that the scattering amplitudes for $n=4$, $k=2$ are colour--kinematics-dual, because those for $n=4$, $k=1$ are.
        
        So far, we constructed a strict Lagrangian for Yang--Mills theory with the same tree-level scattering amplitudes for the BRST-extended Hilbert space as ordinary Yang--Mills theory, but with a manifestly colour--kinematics-dual factorisation of the four-point scattering amplitudes.
        
        We now turn to $n=5$, $k=0$ and iterate our procedure in the evident fashion:
        \begin{enumerate}[Step 1)]\itemsep-2pt
            \item Identify the colour--kinematics duality violating terms. They are necessarily proportional to $\xi b^a+\partial^\mu A^a_\mu$.
            \item Compensate by inserting a corresponding non-local vertex. Complete the compensating term to a BRST-invariant one, which may be deduced directly via the gauge-fixing fermion.
            \item The colour structure of the vertices induces a canonical strictification, implement this strictification.
            \item Use \cref{ob:ghost_pair_reduction} to transfer colour--kinematics duality to tree-level scattering amplitudes with one more ghost--anti-ghost pair, but all other gluons physically polarised.
            \item Continue with Step 1), if there is room for backward-polarised gluons. Otherwise turn to the next higher $n$-point scattering amplitudes.
        \end{enumerate}
        
        The outcome of this construction is a strictified BRST action for Yang--Mills theory which is perturbatively quantum equivalent to ordinary Yang--Mills theory and whose scattering amplitudes come canonically factorised in colour--kinematics-dual form.
        
        We note that this action comes with a BRST operator which is cubic in the fields of the BRST-extended Hilbert space, but of higher order in its action on the auxiliary fields introduced in the strictification.
        
        \section{Double copy from factorisation of homotopy algebras}\label{sec:full_double_copy}
        
        We now turn to the factorisation of the full, interacting theories. In this case, the double copy procedure is implied by the factorisations 
        \begin{equation}
            \tilde\frL^{\rm YM,\,st}_{\rm BRST}\ =\ \frg\otimes(\frKin^{\rm st}\otimes_\tau\frScal)
            \eand
            \tilde\frL^{\caN=0,\,\rm st}_{\rm BRST}\ =\ \frKin^{\rm st}\otimes_\tau(\frKin^{\rm st}\otimes_\tau\frScal)~,
        \end{equation}
        which now hold at the level of strict homotopy algebras. In order to establish these factorisations, we use the definition of twisted tensor products of differential graded algebras we presented in~\eqref{eq:twisted_products}.
        
        \subsection{Biadjoint scalar field theory}
        
        Let us start with a brief consideration of the factorisation of biadjoint scalar field theory, cf.~\cref{sec:biadjoint}. This theory does not require any twists, and we lift the factorisation of cochain complexes~\eqref{eq:fac:diff_comp:biadjoint} to the factorisation 
        \begin{equation}\label{eq:fac:homalg:biadjoint}
            \frL^{\rm biadj}_{\rm BRST}\ =\ \frg\otimes(\bar\frg\otimes\frScal)
        \end{equation}
        into (strict) $L_\infty$-algebras. In general, a tensor product between a Lie algebra and an $L_\infty$-algebra is not well-defined; in particular, it is not a homotopy version of any of the products in the list~\eqref{eq:ex_tensor_products}. However, for nilpotent $L_\infty$-algebras, i.e.~$L_\infty$-algebras with $\mu_i\circ\mu_j=0$, the product exists and yields a $C_\infty$-algebra. The latter is then further tensored by a Lie algebra in the canonical way as explained in \cref{ssec:tensor_prod_homotopy_algebras}, leading to an $L_\infty$-algebra.
        
        \paragraph{\mathversion{bold}$L_\infty$-algebra $\frScal$.}
        Explicitly, the $L_\infty$-algebra $\frScal$ is built from the cochain complex~\eqref{eq:Scal},
        \begin{subequations}\label{eq:L_infty_scalar}
            \begin{equation}
                \frScal\ \coloneqq\ \left(~\stackrel{\tts_x}{\frF[-1]}~\xrightarrow{~\wave~}~\stackrel{\tts^+_x}{\frF[-2]}~\right),
            \end{equation}
            and the only non-vanishing higher product beyond the differential $\mu^\frScal_1$ is
            \begin{equation}
                \mu^\frScal_2\left(\int\rmd^dx_1\,\tts_{x_1}\varphi_1(x_1),\int\rmd^dx_2\,\tts_{x_2}\varphi_2(x_2)\right)\ \coloneqq\ \lambda\int\rmd^dx\,\tts^+_x\varphi_1(x)\varphi_2(x)~.
            \end{equation}
        \end{subequations}
        Evidently, $\frScal$ is nilpotent.  
        
        \paragraph{Factorisation.}
        Following the prescription for the untwisted tensor product of strict homotopy algebras from \cref{ssec:tensor_prod_homotopy_algebras}, we obtain the binary product
        \begin{equation}
            \mu_2(\tte_a\otimes\bar\tte_{\bar a}\otimes \tts_{x_1},\tte_b\otimes\bar\tte_{\bar b}\otimes \tts_{x_2})\ =\ [\tte_a,\tte_b]\otimes[\bar\tte_{\bar a},\bar\tte_{\bar b}]\otimes\lambda\delta^{(d)}(x_1-x_2)\tts_{x_1}^+~,
        \end{equation}
        which, together with the differential
        \begin{equation}
            \mu_1(\tte_a\otimes\bar\tte_{\bar a}\otimes \tts_{x_1})\ =\ \tte_a\otimes\bar\tte_{\bar a}\otimes\wave\tts^+_{x_1}~,
        \end{equation}
        and the cyclic structure
        \begin{equation}
            \inner{\varphi}{\varphi^+}\ =\ \int\rmd^dx\,\varphi^{a\bar a}(x)\varphi^+_{a\bar a}(x)~,
        \end{equation}
        forms the cyclic $L_\infty$-algebra $\frL^{\rm biadj}_{\rm BRST}$. The homotopy Maurer--Cartan action of this $L_\infty$-algebra is then the action~\eqref{eq:biadjoint:action} of biadjoint scalar field theory,
        \begin{equation}
            \begin{aligned}
                S^{\rm biadj}\ &=\ \tfrac12\inner{\varphi}{\mu_1(\varphi)}+\tfrac1{3!}\inner{\varphi}{\mu_2(\varphi,\varphi)}
                \\
                &=\ \int\rmd^dx\,\Big\{\tfrac12\varphi_{a\bar a}\wave\varphi^{a\bar a}-\tfrac{\lambda}{3!}f_{abc}f_{\bar a\bar b\bar c}\varphi^{a\bar a}\varphi^{b\bar b}\varphi^{c\bar c}\Big\}~,
            \end{aligned}
        \end{equation}
        which verifies~\eqref{eq:fac:homalg:biadjoint}.
        
        \subsection{Strictified Yang--Mills theory}
        
        \paragraph{General considerations.}
        The strictification of Yang--Mills theory formulated in \cref{ssec:strictified_YM} is now readily extended to a BV action, which can then be gauge fixed and converted into a strict $L_\infty$-algebra $\tilde\frL^{\rm YM,\,st}_{\rm BRST}$. 
        
        The full strictification of Yang--Mills theory involves an infinite number of additional auxiliary fields and corresponding interaction terms in the Lagrangian. Thus, our discussion cannot be fully explicit and has to remain somewhat conceptual, but as before, we shall give explicit lowest order terms to exemplify our discussion. Recall, however, that for computing $n$-point correlation function at the tree-level, only a finite number of auxiliary fields and interaction terms are necessary. Moreover, for computing $n$-point scattering amplitudes up to $\ell$ loops, only a finite number of correlators is necessary. Therefore, we can always truncate the Yang--Mills action to finitely many auxiliary fields to perform our computations.
        
        We note that gauge fixing of Yang--Mills theory is fully equivalent to gauge fixing of the strictified theory. Moreover, the additional interaction vertices that arise from the BV formalism are all cubic, except for the terms involving anti-fields of the auxiliary fields; the latter, however, will not contribute.
        
        The last point implies that the $L_\infty$-algebra $\tilde\frL^{\rm YM,\,st}_{\rm BRST}$ for the strictified and gauge-fixed form of Yang--Mills theory contains the cochain complex of the $L_\infty$-algebra $\tilde\frL^{\rm YM}_{\rm BRST}$ which we have computed in \cref{ssec:ex:YM}. This cochain complex is enlarged by the kinematic terms for all the auxiliary fields. We then have additional binary products encoding the cubic interactions.
        
        \paragraph{\mathversion{bold}$L_\infty$-algebra of Yang--Mills theory.}
        For concreteness, we consider the strictification up to quartic terms, as explained in \cref{ssec:strictified_YM}. By the arguments given there, however, it is clear that our discussion trivially generalises to strictifications up to an arbitrary order. The Lagrangian, including the strictification of the colour--kinematics duality producing terms~\eqref{eq:CK-completion_n4}, reads as
        \begin{equation}
            \begin{aligned}
                \scL^{\rm YM,\,st}_{{\rm BRST},\,4}\ &=\
                \tfrac12\tilde A_{a\mu}\wave\tilde A^{\mu a}-\tilde{\bar c}_a\wave\tilde c^a+\tfrac12\tilde b_a\wave\tilde b^a+\tilde\xi\,\tilde b_a\,\sqrt{\wave}\,\partial_\mu\tilde A^{\mu a}-gf_{abc}\tilde{\bar c}^a\partial^\mu(\tilde A_\mu^b\tilde c^c)\,-
                \\
                &\kern.6cm-\tfrac12\tilde G^{\mu\nu\kappa}_a\wave\tilde G^{a}_{\mu\nu\kappa}+gf_{abc}\Big(\partial_\mu\tilde A^a_\nu+\tfrac1{\sqrt{2}}\partial^\kappa\tilde G^{a}_{\kappa\mu\nu}\Big)\tilde A^{\mu b}\tilde A^{\nu c}\,-
                \\
                &\kern.6cm-\tilde K_{1a}^{\mu}\wave\tilde{\bar K}^{1a}_\mu-\tilde K_{2a}^{\mu}\wave\tilde{\bar K}^{2a}_\mu\,-
                \\
                &\kern.6cm-gf_{abc}\Big\{\tilde K^{a\mu}_1(\partial^\nu\tilde A^b_\mu)\tilde A_\nu^c+[(\partial^\kappa\tilde A_\kappa^a)\tilde A^{b\mu}+\tilde{\bar c}^a\partial^\mu\tilde c^b]\tilde{\bar K}^{1c}_\mu\Big\}\,+
                \\
                &\kern.6cm+gf_{abc}\Big\{\tilde K_2^{a\mu}\Big[(\partial^\nu\partial_\mu\tilde c^b)\tilde A_\nu^c+(\partial^\nu\tilde A_\mu^b)\partial_\nu\tilde c^c\Big]+\tilde{\bar c}^a\tilde A^{b\mu}\tilde{\bar K}^{2c}_\mu\Big\}~,
            \end{aligned}
        \end{equation}
        where $K_i^{a\mu}$ and $\bar K^{ai}_\mu$ are auxiliary $\frg$-valued one-forms, strictifying $\scL^{\rm YM,\,comp}_{{\rm BRST},\,n=4,\,k=0}$, and we used the shorthand $\tilde G^{a}_{\mu\nu\kappa}\coloneqq\tilde G^{4,\gamma,1,a}_{\mu\nu\kappa}$. Note that $K_1^{a\mu}$ and $\bar K^{a1}_\mu$ are of ghost number zero, while $K_2^{a\mu}$ and $\bar K^{a2}_\mu$ carry ghost numbers $-1$ and $+1$, respectively. The $L_\infty$-algebra $\tilde\frL^{\rm YM,\,st}_{\rm BRST}$ to quartic order has underlying cochain complex
        \begin{subequations}\label{eq:L_infty_strictified_YM}
            \begin{equation}\label{eq:YM_diff_complex_with_G}
                \begin{tikzcd}[
                    every label/.append style={scale=.95},
                    cells={nodes={scale=.95}}
                    ]
                    & \stackrel{(\tilde K_1^{a\mu},\,\tilde{\bar K}^{1a}_\mu)}{\IR^2\otimes\Omega^1(\IM^d)\otimes\frg} \arrow[r,"\wave"]   & \stackrel{(\tilde K_{1}^{+a\mu},\,\tilde{\bar K}^{1+a}_\mu)}{\IR^2\otimes\Omega^1(\IM^d)\otimes\frg} & 
                    \\
                    & \stackrel{\tilde G_{\mu\nu\kappa}^a}{\otimes^3\Omega^1(\IM^d)\otimes\frg} \arrow[r,"\wave"] & \stackrel{\tilde G^{+a}_{\mu\nu\kappa}}{\otimes^3\Omega^1(\IM^d)\otimes\frg} & 
                    \\
                    & \stackrel{\tilde A_\mu^a}{\Omega^1(\IM^d)\otimes\frg} \arrow[r,"\wave"] \arrow[start anchor=south east, end anchor= north west,rd, "\tilde\xi\sqrt{\wave}\,\partial^\mu",pos=0.02, swap] &[1cm] \stackrel{\tilde A_\mu^{+a}}{\Omega^1(\IM^d)\otimes\frg} & 
                    \\[0.8cm]
                    &\stackrel{\tilde b^a}{\scC^\infty(\IM^d)\otimes\frg} \arrow[r,"\wave",swap] \arrow[start anchor=north east, end anchor= south west,ur, "-\tilde\xi\sqrt{\wave}\,\partial_\mu",pos=0.02, crossing over] & \stackrel{\tilde b^{+a}}{\scC^\infty(\IM^d)\otimes\frg} 
                    \\
                    \stackrel{\tilde{\bar K}^{2a}_\mu}{\Omega^1(\IM^d)\otimes\frg} \arrow[r,"-\wave"]   & \stackrel{ \tilde{K}_2^{+a\mu}}{\Omega^1(\IM^d)\otimes\frg} & \stackrel{\tilde{K}^{a\mu}_{2}}{\Omega^1(\IM^d)\otimes\frg} \arrow[r,"-\wave"]   & \stackrel{\tilde{\bar K}^{2+a}_\mu}{\Omega^1(\IM^d)\otimes\frg} 
                    \\
                    \underbrace{\spacer{2ex}\stackrel{\tilde c^a}{\scC^\infty(\IM^d)\otimes\frg}}_{\eqqcolon\,\tilde\frL^{\rm YM,\,st}_{{\rm BRST},\,0}}
                    \arrow[r,"-\wave"] & \underbrace{\spacer{2ex}\stackrel{\tilde{\bar c}^{+a}}{\scC^\infty(\IM^d)\otimes\frg}}_{\eqqcolon\,\tilde\frL^{\rm YM,\,st}_{{\rm BRST},\,1}} & \underbrace{\spacer{2ex}\stackrel{\tilde{\bar c}^a}{\scC^\infty(\IM^d)\otimes\frg}\arrow[r,"-\wave"]}_{\eqqcolon\,\tilde\frL^{\rm YM,\,st}_{{\rm BRST},\,2}} & \underbrace{\spacer{2ex}\stackrel{\tilde c^{+a}}{\scC^\infty(\IM^d)\otimes\frg}}_{\eqqcolon\,\tilde\frL^{\rm YM,\,st}_{{\rm BRST},\,3}}
                \end{tikzcd}
            \end{equation}
            Besides the differentials in~\eqref{eq:YM_diff_complex_with_G}, we also have the following higher products
            \begin{equation}
                \begin{aligned}
                    \left(
                    \colvec{
                        \tilde K_1^{a\mu},
                        \tilde{\bar K}^{1a}_\mu,
                        \tilde G^a_{\mu\nu\kappa},
                        \tilde A_\mu^a,
                        \tilde b
                    },
                    \colvec{
                        \tilde{\bar K}^{2a}_{\mu},
                        \tilde c^a
                    }
                    \right)
                    \ &\mapstomap{\mu_2}\ gf_{bc}{}^a
                    \colvec{
                        (\partial^\nu \tilde A_\mu^b)\partial_\nu \tilde c^c-\tilde A_\nu^b\partial^\nu\partial_\mu\tilde c^c,
                        -\partial^\mu(\tilde A_\mu^b\tilde c^c)+\tilde{\bar K}_{1\mu}^{b}(\partial^\mu\tilde c^c)+\tilde{A}^{b\mu}\tilde{\bar K}^{2c}_{\mu}
                    }
                    \\[10pt]
                    &\kern2cm\in\ \spacee{\tilde\frL^{\rm YM,\,st}}{{\rm BRST},\,1}{\tilde K_2^+,\,\tilde{\bar c}^+}~,
                \end{aligned}
            \end{equation}
            \begin{equation}
                \begin{aligned}
                    \left(
                    \colvec{
                        \tilde{\bar K}^{2a}_{\mu},
                        \tilde c^a
                    },
                    \colvec{
                        \tilde{K}_{2}^{a\mu},
                        \tilde{\bar c}^a
                    }
                    \right)
                    \ &\mapstomap{\mu_2}\ gf_{bc}{}^a
                    \colvec{
                        (\partial^\mu \tilde c^b)\tilde{\bar c}^c,
                        -\tilde{\bar K}^{2b}_\mu\tilde{\bar c}^c +(\partial_\mu\partial_\nu\tilde c^b)\tilde K^{c\nu}_2+\partial^\nu(\partial_\nu \tilde c^b\tilde K^c_{2\mu})-\tilde c^b\partial_\mu\tilde{\bar c}^c
                    }
                    \\[10pt]
                    &\kern2cm\in \spacee{\tilde\frL^{\rm YM,\,st}}{{\rm BRST},\,1}{\tilde{\bar K}_1^+,\,\tilde A^+}~,
                \end{aligned}
            \end{equation}
            \begin{equation}
                \begin{aligned}
                    \left(
                    \colvec{
                        \tilde K_1^{a\mu},
                        \tilde{\bar K}^{1a}_\mu,
                        \tilde G^a_{\mu\nu\kappa},
                        \tilde A_\mu^a,
                        \tilde b^a
                    },
                    \colvec{
                        \tilde K_1^{a\mu},
                        \tilde{\bar K}^{1a}_\mu,
                        \tilde G^a_{\mu\nu\kappa},
                        \tilde A_\mu^a,
                        \tilde b^a
                    }
                    \right)
                    \ &\mapstomap{\mu_2}\ gf_{bc}{}^a
                    \colvec{
                        2(\partial^\nu \tilde A_\mu^b)\tilde A_\nu^c,
                        2\partial^\kappa \tilde A_\kappa^b\tilde A^{c\mu},
                        \sqrt{2}\partial_\mu(\tilde A^b_\nu\tilde A^c_\kappa),
                        R^{\tilde A^+}_{bc\mu},
                        0
                    }
                    \\[10pt]
                    &\kern2cm\in\ \spacee{\tilde\frL^{\rm YM,\,st}}{{\rm BRST},\,2}{\tilde K_1^+,\,\tilde{\bar K}^{1+},\,\tilde G^+,\,\tilde A^+,\tilde b^+}~,
                    \\[10pt]
                    R^{\tilde A^+}_{bc\mu}\ &\coloneqq\ -3!\partial^\nu(\tilde A_\nu^b\tilde A_\mu^c)-\sqrt{8}\tilde A^{\nu b}\partial^\kappa\tilde G^c_{\kappa\nu\mu}-4\tilde K_1^{b\nu}\partial_\mu \tilde A^c_\nu\,-
                    \\
                    &\kern2cm-4(\partial^\kappa \tilde A_\kappa^b)\tilde{\bar K}^{1c}_\mu~,
                \end{aligned}
            \end{equation}
            and
            \begin{equation}
                \begin{aligned}
                    \left(
                    \colvec{
                        \tilde K_1^{a\mu},
                        \tilde{\bar K}^{1a}_\mu,
                        \tilde G^a_{\mu\nu\kappa},
                        \tilde A_\mu^a,
                        \tilde b^a
                    },
                    \colvec{
                        \tilde K^{2a}_\mu,
                        \tilde{\bar c}^a
                    }
                    \right)
                    \ &\mapstomap{\mu_2}\ gf_{bc}{}^a
                    \colvec{
                        \tilde A^b_\mu\tilde{\bar c}^c,
                        -\tilde A_\mu^b\partial^\mu\tilde{\bar c}^c+\partial^\mu(\tilde{\bar K}^{1b}_\mu\tilde{\bar c}^c)+\partial^\nu\partial_\mu(\tilde A^b_\nu\tilde K_2^{c\mu})
                    }
                    \\[10pt]
                    &\kern2cm\in\ \spacee{\tilde\frL^{\rm YM,\,st}}{{\rm BRST},\,1}{\tilde{\bar K}^{2+},\,\tilde{c}^+}~,
                \end{aligned}
            \end{equation}
            and the cyclic structure is given by 
            \begin{equation}
                \begin{aligned}
                    \inner{\tilde A}{\tilde A^+}\ &\coloneqq\ \int\rmd^dx\,\tilde A^a_\mu\tilde A^{+\mu}_a~,~~~
                    &
                    \inner{\tilde b}{\tilde b^+}\ &\coloneqq\ \int\rmd^dx\,\tilde b^a\tilde b^{+}_a~,
                    \\
                    \inner{\tilde c}{\tilde c^+}\ &\coloneqq\ \int\rmd^dx\,\tilde c^a\tilde c^{+}_a~,~~~
                    &
                    \inner{\tilde{\bar c}}{\tilde{\bar c}^+}\ &\coloneqq\ -\int\rmd^dx\,\tilde{\bar c}^a\tilde{\bar c}^{+}_a~,
                    \\
                    \inner{\tilde K_{1}}{\tilde K_1^+}\ &\coloneqq\ -\int\rmd^dx\,\tilde K_{1}^{a\mu}\tilde K^{+}_{1a\mu}~,~~~
                    &
                    \inner{\tilde{\bar K}^{1}}{\tilde{\bar K}^{1+}}\ &\coloneqq\ -\int\rmd^dx\,\tilde{\bar K}^{1a}_{\mu}\tilde{\bar K}^{1+\mu}_a~,
                    \\
                    \inner{\tilde K_{2}}{\tilde K_{2}^+}\ &\coloneqq\ -\int\rmd^dx\,\tilde K^{a\mu}_2\tilde K^{+}_{2a\mu}~,~~~
                    &
                    \inner{\tilde{\bar K}^{2}}{\tilde{\bar K}^{2+}}\ &\coloneqq\ \int\rmd^dx\,\tilde{\bar K}^{2a}_{\mu}\tilde{\bar K}^{2+\mu}_a~,
                    \\
                    \inner{\tilde G}{\tilde G^+}\ &\coloneqq\ -\int\rmd^dx\,\tilde G^a_{\mu\nu\kappa}\tilde G^{+\mu\nu\kappa}_a~.
                \end{aligned}
            \end{equation}
        \end{subequations}
        
        \paragraph{Factorisation and twist datum.} 
        We factorise this $L_\infty$-algebra as 
        \begin{equation}\label{eq:full_factorisation_YM}
            \tilde\frL^{\rm YM,\,st}_{\rm BRST}\ =\ \frg\otimes(\frKin^{\rm st}\otimes_\tau\frScal)~,
        \end{equation}
        where $\frg$ is the usual colour Lie algebra, $\frKin^{\rm st}$ the graded vector space
        \begin{equation}\label{eq:kinYMst}
            \frKin^{\rm st}\ \coloneqq\ 
            \left(~
            \underbrace{
                \begin{array}{c} 
                    \stackrel{\bar \ttt^\mu_{2}}{\IM^d}
                    \\[-0.2cm]
                    \oplus
                    \\
                    \stackrel{\ttg}{\IR[1]}
                \end{array}
            }_{\eqqcolon\,\frKin^{\rm st}_{-1}}                
            ~\oplus~
            \underbrace{
                \begin{array}{c} 
                    \stackrel{\ttt_\mu^{1},~\bar \ttt^\mu_{1}}{\IM^d\oplus\IM^d}
                    \\[-0.2cm]
                    \oplus
                    \\
                    \stackrel{\ttt^{\mu\nu\kappa}_0}{\IM^d\otimes(\IM^d\wedge \IM^d)}
                    \\[-0.2cm]
                    \oplus
                    \\[-0.1cm]
                    \stackrel{\ttv^\mu}{\IM^d}
                    \\[-0.2cm]
                    \oplus
                    \\[-0.1cm]
                    \underset{\ttn}{\IR}
                \end{array}
            }_{\eqqcolon\,\frKin^{\rm st}_0}
            ~\oplus~
            \underbrace{
                \begin{array}{c} 
                    \stackrel{\ttt_\mu^{2}}{\IM^d}
                    \\[-0.2cm]
                    \oplus
                    \\
                    \stackrel{\tta}{\IR[-1]}
                \end{array}
            }_{\eqqcolon\,\frKin^{\rm st}_1}
            ~\right),
        \end{equation}
        and $\frScal$ the $L_\infty$-algebra defined in~\eqref{eq:L_infty_scalar}. This $L_\infty$-algebra is cyclic with the inner products given by~\eqref{eq:inner_product_YM_1} together with
        \begin{equation}\label{eq:def_frKin_prime}
            \begin{gathered}
                \inner{\ttt^1_\mu}{\bar\ttt_1^\nu}\ \coloneqq\ -\delta^\nu_\mu~,~~~
                \inner{\bar\ttt_1^\nu}{\ttt^1_\mu}\ \coloneqq\ -\delta^\nu_\mu~,~~~
                \inner{\ttt^2_\mu}{\bar\ttt_2^\nu}\ \coloneqq\ \delta^\nu_\mu~,~~~
                \inner{\bar\ttt_2^\nu}{\ttt^2_\mu}\ \coloneqq\ \delta^\nu_\mu~,
                \\
                \inner{\ttt^{\mu\nu\kappa}_0}{\ttt^{\lambda\rho\sigma}_0}\ \coloneqq\ -\tfrac12\eta^{\mu\lambda}(\eta^{\nu\rho}\eta^{\kappa\sigma}-\eta^{\nu\sigma}\eta^{\kappa\rho})~.
            \end{gathered}
        \end{equation}
        
        \begin{table}[ht]
            \begin{center}
                \resizebox{\textwidth}{!}{
                    \begin{tabular}{|c|c|c|c|c|c|c|c|}
                        \hline
                        \multicolumn{4}{|c|}{fields} & \multicolumn{4}{c|}{anti-fields}
                        \\
                        \hline
                        factorisation & $|-|_{\rm gh}$ & $|-|_\frL$ & dim & factorisation & $|-|_{\rm gh}$ & $|-|_\frL$ & dim
                        \\
                        \hline
                        $\tilde c=\cb_a{\ttg}\tts_x\tilde c^a(x)$ & $1$ & $0$ & $\tfrac{d}{2}-2$ & $\tilde c^+=\cb_a{\tta}\tts^+_x\tilde c^{+a}(x)$ & $-2$ & $3$ & $\tfrac{d}{2}+2$ 
                        \\
                        $\tilde A=\cb_a{\ttv^\mu}\tts_x\tilde A^a_\mu(x)$ & $0$ & $1$ & $\tfrac{d}{2}-1$& $\tilde A^+=\cb_a{\ttv^\mu}\tts^+_x\tilde A^{+a}_\mu(x)$ & $-1$ & $2$ & $\tfrac{d}{2}+1$
                        \\
                        $\tilde b=\cb_a{\ttn}\tts_x \tilde b^a(x)$ & $0$ & $1$ & $\tfrac{d}{2}-1$& $\tilde b^+=\cb_a{\ttn}\tts^+_x \tilde b^{+a}(x)$ & $-1$ & $2$ & $\tfrac{d}{2}+1$
                        \\
                        $\tilde{\bar c}=\cb_a{\tta}\tts_x\tilde{\bar c}^a(x)$ & $-1$ & $2$ & $\tfrac{d}{2}$ & $\tilde{\bar c}^+=\cb_a{\ttg}\tts^+_x\tilde{\bar c}^{+a}(x)$ & $0$ & $1$ & $\tfrac{d}{2}$
                        \\
                        \hline
                        $\tilde K_1=\cb_a\ttt_\mu^1\tts_x\tilde{K}_1^\mu(x)$ & $0$ & $1$ & $\tfrac{d}{2}-1$ & 
                        $\tilde K_1^+=\cb_a{\ttt_\mu^1}\tts^+_x\tilde{K}^{+a\mu}_1(x)$ & $-1$ & $2$ & $\tfrac{d}{2}-1$
                        \\
                        $\tilde{\bar K}^1=\cb_a{\bar \ttt^\mu_1}\tts_x\tilde{\bar K}^{1a}_\mu(x)$ & $0$ & $1$ & $\tfrac{d}{2}-1$ & 
                        $\tilde{\bar K}^{1+}=\cb_a{\bar \ttt^\mu_1}\tts^+_x\tilde{\bar K}^{1+a}_\mu(x)$ & $-1$ & $2$ & $\tfrac{d}{2}-1$
                        \\
                        $\tilde K_2=\cb_a\ttt_\mu^2\tts_x\tilde{K}_2^\mu(x)$ & $-1$ & $2$ & $\tfrac{d}{2}-1$ & 
                        $\tilde K_2^+=\cb_a{\ttt_\mu^2}\tts^+_x\tilde{K}^{+a\mu}_2(x)$ & $0$ & $1$ & $\tfrac{d}{2}-1$
                        \\
                        $\tilde{\bar K}^2=\cb_a{\bar \ttt^\mu_2}\tts_x\tilde{\bar K}^{2a}_\mu(x)$ & $1$ & $0$ & $\tfrac{d}{2}-1$ & 
                        $\tilde{\bar K}^{2+}=\cb_a{\bar \ttt^\mu_2}\tts^+_x\tilde{\bar K}^{2+a}_\mu(x)$ & $-2$ & $3$ & $\tfrac{d}{2}-1$
                        \\
                        $\tilde G=\cb_a\ttt^{\mu\nu\kappa}_0\tts_x\tilde{G}_{\mu\nu\kappa}^a(x)$ & $0$ & $1$ & $\tfrac{d}{2}-1$ & $\tilde{G}^+=\cb_a{\ttt^{\mu\nu\kappa}_0}\tts^+_x\tilde G_{\mu\nu\kappa}^{+a}(x)$ & $-1$ & $2$ & $\tfrac{d}{2}-1$
                        \\
                        \hline
                    \end{tabular}
                }
            \end{center}
            \caption{Factorisation of the fields in the $L_\infty$-algebra corresponding to the Lagrangian $\scL^{\rm YM,\,st}_{\rm BRST,\,4}$. Note that we suppressed the integrals over $x$ and the tensor products for simplicity.\label{tab:fac_vec_YM_strict}}
        \end{table}
        
        The twist datum $\tau$, see~\eqref{eq:twistDatumDGA} for the general definition, in the factorisation~\eqref{eq:full_factorisation_YM} is then given by the maps
        \begin{subequations} 
            \begin{equation}\label{eq:full_twist_1_YM}
                \tau_1(\ttg)\ \coloneqq\ \ttg\otimes\sfid~,~~~
                \begin{array}{c}
                    \tau_1(\ttt_\mu^i)\ \coloneqq\ \ttt_\mu^i\otimes\sfid~,~~~
                    \tau_1(\bar\ttt^\mu_i)\ \coloneqq\ \bar \ttt^\mu_i\otimes\sfid~,
                    \\
                    \tau_1(\ttt^{\mu\nu\kappa}_0)\ \coloneqq\ \ttt^{\mu\nu\kappa}_0\otimes\sfid~,
                    \\
                    \tau_1(\ttv^\mu)\ \coloneqq\ \ttv^\mu \otimes \sfid+\tilde\xi\ttn\otimes\wave{}^{-\frac12}\partial^\mu~,
                    \\
                    \tau_1(\ttn)\ \coloneqq\ \ttn\otimes\sfid-\tilde\xi\ttv^\mu\otimes\wave{}^{-\frac12}\partial_\mu~,
                \end{array}~~~
                \tau_1(\tta)\ \coloneqq\ \tta\otimes\sfid
            \end{equation}
            and 
            \begingroup
            \allowdisplaybreaks
            \begin{align}\label{eq:full_twist_2_YM}
                \tau_2(\ttg,\ttv^\mu)\ &\coloneqq\ -\ttg\otimes(\sfid\otimes\partial^\mu+\partial^\mu\otimes\sfid)+\ttt_2^\mu\otimes(\partial^\nu\otimes\partial_\nu-\partial^\mu\partial_\nu\otimes\sfid)~,\notag
                \\
                \tau_2(\ttv^\mu,\ttg)\ &\coloneqq\ \ttg\otimes(\sfid\otimes\partial^\mu+\partial^\mu\otimes\sfid)-\ttt_2^\mu\otimes(\partial^\nu\otimes\partial_\nu-\sfid\otimes\partial^\mu\partial_\nu)~,\notag
                \\
                \tau_2(\ttg,\bar \ttt^1_\mu)\ &\coloneqq\ \ttg\otimes\partial_\mu\otimes\sfid~,\notag
                \\
                \tau_2(\bar\ttt^1_\mu,\ttg)\ &\coloneqq\ -\ttg\otimes\sfid\otimes\partial_\mu~,\notag
                \\
                \tau_2(\bar\ttt_2^\mu,\ttv^\nu)\ &\coloneqq\ \eta^{\mu\nu}\ttg\otimes\sfid\otimes\sfid~,\notag
                \\
                \tau_2(\ttv^\mu,\bar\ttt_2^\nu)\ &\coloneqq\ -\eta^{\mu\nu}\ttg\otimes\sfid\otimes\sfid~,\notag
                \\
                \tau_2(\ttg,\tta)\ &\coloneqq\ \ttv^\mu\otimes\sfid\otimes\partial_\mu-\bar\ttt_1^\mu\otimes\partial_\mu\otimes\sfid~,\notag
                \\
                \tau_2(\tta,\ttg)\ &\coloneqq\ -\ttv^\mu\otimes\partial_\mu\otimes\sfid+\bar\ttt_1^\mu\otimes\sfid\otimes\partial_\mu~,\notag
                \\
                \tau_2(\bar\ttt_2^\mu,\tta)\ &\coloneqq\ \ttv^\mu\otimes\sfid\otimes\sfid~,\notag
                \\
                \tau_2(\tta,\bar \ttt_2^\mu)\ &\coloneqq\ -\ttv^\mu\otimes\sfid\otimes\sfid~,\notag
                \\
                \tau_2(\ttg,\ttt_2^\mu)\ &\coloneqq\ -\ttv^\nu\otimes\partial_\nu\partial^\mu\otimes\sfid-\ttv^\mu\otimes \wave\otimes\sfid-\ttv^\mu\otimes \partial_\nu\otimes\partial^\nu~,\notag
                \\
                \tau_2(\ttt_2^\mu,\ttg)\ &\coloneqq\ \ttv^\nu\otimes\sfid\otimes\partial_\nu\partial^\mu+\ttv^\mu\otimes\sfid\otimes\wave+\ttv^\mu\otimes\partial_\nu\otimes\partial^\nu~,\notag
                \\
                \tau_2(\ttv^\mu,\tta)\ &\coloneqq\ -\bar\ttt^2_\mu\otimes\sfid\otimes\sfid+\tta\otimes\sfid\otimes\partial^\mu~,\notag
                \\
                \tau_2(\tta,\ttv^\mu)\ &\coloneqq\ \bar\ttt^2_\mu\otimes\sfid\otimes\sfid-\tta\otimes\partial^\mu\otimes\sfid~,\notag
                \\
                \tau_2(\bar\ttt_1^\mu,\tta)\ &\coloneqq\ -\tta\otimes(\partial^\mu\otimes\sfid+\sfid\otimes\partial^\mu)~,
                \\
                \tau_2(\tta,\bar\ttt_1^\mu)\ &\coloneqq\ +\tta\otimes(\partial^\mu\otimes\sfid+\sfid\otimes\partial^\mu)~,\notag
                \\
                \tau_2(\ttv^\mu,\ttt_2^\nu)\ &\coloneqq\ -\tta\otimes(\partial^\mu\partial^\nu\otimes\sfid+\partial^\mu\otimes \partial^\nu+\partial^\nu\otimes\partial^\mu+\sfid\otimes\partial^\mu\partial^\nu)~,\notag
                \\
                \tau_2(\ttt_2^\mu,\ttv^\nu)\ &\coloneqq\ \tta\otimes(\partial^\mu\partial^\nu\otimes\sfid+\partial^\mu\otimes\partial^\nu+\partial^\nu\otimes\partial^\mu+\sfid\otimes\partial^\mu\partial^\nu)~,\notag
                \\
                \tau_2(\ttv^\mu,\ttv^\nu)\ &\coloneqq\ \ttt_1^\mu\otimes\partial^\nu\otimes\sfid-\ttt_1^\nu\otimes\sfid\otimes\partial^\mu\,+\notag
                \\
                &\kern1cm+\bar\ttt_1^\nu\otimes\partial^\mu\otimes\sfid-\bar\ttt_1^\mu\otimes\sfid\otimes\partial^\nu\,-\notag
                \\
                &\kern1cm-3\Big[\ttv^\nu\otimes(\partial^\mu\otimes\sfid+\sfid\otimes\partial^\mu)-\ttv^\mu\otimes(\partial^\nu\otimes\sfid+\sfid\otimes\partial^\nu)\Big]\,+\notag
                \\
                &\kern1cm+\sqrt{2}\big(\ttt^{\kappa\mu\nu}_0\otimes\partial_\kappa\otimes\sfid+\ttt^{\kappa\mu\nu}_0\otimes\sfid\otimes\partial_\kappa\big)\,,\notag
                \\
                \tau_2(\ttv^\mu,\ttt^{\nu\kappa\lambda}_0)\ &\coloneqq\ -\frac{\sqrt{2}}{2}\big(\eta^{\mu\kappa}\ttv^\lambda\otimes\sfid\otimes\partial^\nu-\eta^{\mu\lambda}\ttv^\kappa\otimes\sfid\otimes\partial^\nu\big)\,,\notag
                \\
                \tau_2(\ttt^{\nu\kappa\lambda}_0,\ttv^\mu)\ &\coloneqq\ \frac{\sqrt{2}}{2}\big(\eta^{\mu\kappa} \ttv^\lambda\otimes\partial^\nu\otimes\sfid-\eta^{\mu\lambda}\ttv^\kappa\otimes\partial^\nu\otimes\sfid\big)\,,\notag
                \\
                \tau_2(\ttt_1^\mu,\ttv^\nu)\ &\coloneqq\ -2\eta^{\mu\nu}\ttv^\kappa\otimes\sfid\otimes\partial^\kappa~,\notag
                \\
                \tau_2(\ttv^\nu,\ttt_1^\mu)\ &\coloneqq\ 2\eta^{\mu\nu}\ttv^\kappa\otimes\partial^\kappa\otimes\sfid~,\notag
                \\
                \tau_2(\ttv^\nu,\bar\ttt_1^\mu)\ &\coloneqq\ -2\ttv^\mu\otimes\partial^\nu\otimes \sfid~,\notag
                \\
                \tau_2(\bar\ttt_1^\mu,\ttv^\nu)\ &\coloneqq\ 2\ttv^\mu\otimes\sfid\otimes\partial^\nu~.\notag
            \end{align}
            \endgroup
        \end{subequations}
        
        We note that the twisted tensor product $\frKin^{\rm st}\otimes_\tau\frScal$ is a (strict) $C_\infty$-algebra, which becomes an $L_\infty$-algebra after the tensor product with the colour Lie algebra $\frg$; see \cref{ssec:tensor_prod_homotopy_algebras} for details.
        
        \subsection{BRST--Lagrangian double copy}
        
        A key to showing that our double copy prescription based on factorisations of the $L_\infty$-algebras of gauge-fixed BRST Lagrangians is that not only the action but also the BRST operator double copies. This fact guarantees that the double copy creates the appropriate gauge-fixing sectors which is crucial in considering the double copy at the loop level. In the following, we give a general discussion of what we called the \uline{BRST--Lagrangian double copy} in~\cite{Borsten:2020zgj}.
        
        \paragraph{Strictification of BRST-invariant actions.} 
        As discussed in \cref{ssec:strictified_YM}, any field theory can be strictified to a classically equivalent field theory with purely cubic interaction terms, and this equivalence extends to the quantum level. Consider a general strictified field theory
        \begin{equation}\label{eq:cubic_Lagrangian_standard_form}
            S\ =\ \frac12\Phi^I\sfg_{IJ}\Phi^J+\frac{1}{3!}\Phi^I\sff_{IJK}\Phi^J\Phi^K~,
        \end{equation}
        where $\sfg_{IJ}$ and $\sff_{IJK}$ are some structure constants. As in \cref{sec:BVMotivation}, $I,J,\ldots$ are DeWitt indices that include labels for the field species, the gauge and Lorentz representations, as well as the space--time position. 
        
        Let us now consider a theory which is invariant under a gauge symmetry. We extend the action of this theory to its BV form by including ghosts, anti-ghosts, and the Nakanishi--Lautrup field, as done in \cref{sec:Examples}. We then strictify the full BV action to an action with cubic interaction vertices. Restricting to gauge-fixing fermions which are quadratic in the fields\footnote{This is the case for all explicit gauge-fixing fermions used in this paper.} guarantees that the action remains cubic after gauge fixing. The resulting BRST operator $Q_{\rm BRST}$, given by~\eqref{eq:BRST_differential_definition}, is then automatically at most quadratic in the fields, and we can write 
        \begin{equation}\label{eq:cubic_BRST_standard_form}
            \Phi^I\ \mapstomap{Q_{\rm BRST}}\ \sfQ^I_J\Phi^J+\frac12\sfQ^I_{JK}\Phi^J\Phi^K
        \end{equation}
        for some structure constants $\sfQ^I_J$ and $\sfQ^I_{JK}$.
        
        \paragraph{Factorisation of structure constants.} 
        As indicated previously, the key to the double copy is the factorisation of the field space $\frL$ into
        \begin{equation}\label{eq:generalVectorSpaceFactorisation}
            \frL\ \coloneqq\ \frV\otimes\bar\frV\otimes\scC^\infty(\IM^d)~,
        \end{equation}
        where $\frV$ and $\bar\frV$ are two (graded) vector spaces. In our preceding discussion, we have encountered the three examples in \cref{tab:example_factorisations}. Consequently, in our formulas, we shall split the multi indices into triples, that is, $I=(\alpha,\bar\alpha,x)$, and write (see e.g.~\eqref{eq:YMsuperfieldAbstract})
        \begin{equation}\label{eq:general_field_factorisation}
            (\frL[1])^*\otimes\frL\ \ni\ \sfa\ =\ \Phi^I\otimes\cb_I\ =\ \int\rmd^dx\,\Phi^{\alpha\bar\alpha}(x)\otimes(\cb_\alpha\otimes\bar\cb_{\bar\alpha}\otimes\tts_x)~.
        \end{equation}
        We also demand that the structure constants $\sfg_{IJ}$ and $\sff_{IJK}$ that appear in the action~\eqref{eq:cubic_Lagrangian_standard_form} as well as the structure constants $\sfQ^I_J$ and $\sfQ^I_{JK}$ that appear in the BRST operator~\eqref{eq:cubic_BRST_standard_form} are local in the sense that they vanish unless all the space--time points in the multi-indices agree. 
        
        We write
        \begin{equation}
            \sfg_{IJ}\ \eqqcolon\ \sfg_{\alpha\beta}\,\bar\sfg_{\bar\alpha\bar\beta}\wave~,
        \end{equation}
        where $\sfg_{\alpha\beta}$ and $\bar\sfg_{\bar\alpha\bar\beta}$ are differential operators, mapping $\scC^\infty(\IM^d)$ to itself. In more detail, we have 
        \begin{subequations}\label{eq:cubic_Lagrangian_coefficients_g}
            \begin{equation}
                \sfg_{IJ}\Phi^J\ \equiv\ \int \rmd^d y~\sfg_{(\alpha,\bar\alpha,x);(\beta,\bar\beta,y)}\Phi^{\beta\bar\beta,y}\ =\ \int \rmd^d y\int \rmd^d z~\sfg_{\alpha\beta}(x,y)\bar\sfg_{\bar\alpha\bar\beta}(y,z)\wave{}\Phi^{\beta\bar \beta,z}~,
            \end{equation}
            where the integral kernels are of the form
            \begin{equation}
                \sfg_{\alpha\beta}(x,y)\ =\ \delta^{(d)}(x-y)\sfg_{\alpha\beta}(x)
                \eand
                \bar\sfg_{\bar\alpha\bar\beta}(y,z)\ =\ \delta^{(d)}(y-z)\bar\sfg_{\bar\alpha\bar\beta}(y)
            \end{equation}
            due to our assumption about locality, and we assume that $g_{\alpha\beta}(x)$ is invertible.
            
            Analogously, we write
            \begin{equation}\label{eq:def_fact_f}
                \sff_{IJK}\ =\ \sff_{(\alpha,\bar\alpha,x);(\beta,\bar\beta,y);(\gamma,\bar\gamma,z)}\ \eqqcolon\ \sfp\,\sff_{\alpha\beta\gamma}\,\bar\sff_{\bar\alpha\bar\beta\bar\gamma}~,
            \end{equation}
            where $\sff_{\alpha\beta\gamma}$ and $\bar\sff_{\bar\alpha\bar\beta\bar\gamma}$ are bi-differential operators $\scC^\infty(\IM^d)\otimes\scC^\infty(\IM^d)\rightarrow\scC^\infty(\IM^d)\otimes\scC^\infty(\IM^d)$ and
            \begin{equation}
                \sfp\,:\,\scC^\infty(\IM^d)\otimes\scC^\infty(\IM^d)\ \rightarrow\ \scC^\infty(\IM^d)
            \end{equation}
            is the natural diagonal product of functions. For the integral kernels of $\sff_{\alpha\beta\gamma}$ and $\bar\sff_{\bar\alpha\bar\beta\bar\gamma}$ we have again the locality condition
            \begin{equation}\label{eq:int_kernels_f}
                \begin{aligned}
                    \sff_{\alpha\beta\gamma}(x_1,x_2;y_1,y_2)\ &=\ \delta^{(d)}(x_1-y_1)\delta^{(d)}(x_2-y_2)\sff_{\alpha\beta\gamma}(y_1,y_2)~,
                    \\
                    \bar\sff_{\bar\alpha\bar\beta\bar\gamma} (x_1,x_2;y_1,y_2)\ &=\ \delta^{(d)}(x_1-y_1)\delta^{(d)}(x_2-y_2)
                    \bar\sff_{\bar\alpha\bar\beta\bar\gamma}(y_1,y_2)~.
                \end{aligned}
            \end{equation}
            We note that there is some ambiguity in the definition~\eqref{eq:def_fact_f} due to the projection onto the diagonal involved in $\sfp$, but this redundancy never arises in any formula.
        \end{subequations}
        To give a clearer picture of what the above construction is doing, we can expand the $\sff_{\alpha\beta\gamma}$ and the $\bar\sff_{\bar\alpha\bar\beta\bar\gamma}$ further in a basis of differential operators $\partial^M$ for $M$ a Lorentz multiindex, and we have
        \begin{equation}
            (\sfp\,\sff_{\alpha\beta\gamma}\,\bar\sff_{\bar\alpha\bar\beta\bar\gamma})(\Phi\otimes \Phi)\ =\ 
            \sff_{\alpha\beta M_1\gamma M_2}\bar\sff_{\bar\alpha\bar\beta N_1\bar\gamma N_2}(\partial^{M_1}\partial^{N_1}\Phi^{\beta\bar \beta})(\partial^{M_2}\partial^{N_2}\Phi^{\gamma\bar \gamma})~.
        \end{equation}
        
        For convenience, we also introduce the operators $\sff^\alpha_{\beta\gamma}$ and $\bar \sff^{\bar \alpha}_{\bar\beta\bar\gamma}$ by
        \begin{equation}
            \sfp\,\sff_{\alpha\beta\gamma}\ \eqqcolon\ \sfg_{\alpha\delta}\,\sfp\,\sff^\delta_{\beta\gamma}
            \eand
            \sfp\,\bar\sff_{\bar\alpha\bar\beta\bar\gamma}\ \eqqcolon\ \bar\sfg_{\bar\alpha\bar\delta}\,\sfp\,\bar\sff^{\bar\delta}_{\bar\beta\bar\gamma}~,
        \end{equation}
        which is possible due to the invertibility of $\sfg_{\alpha\beta}$ and $\bar\sfg_{\bar\alpha\bar\beta}$ as well as the form of the integral kernels~\eqref{eq:int_kernels_f}~. Evidently, $\sff^\alpha_{\beta\gamma}$ and $\bar \sff^{\bar \alpha}_{\bar\beta\bar\gamma}$ are again bi-differential operators, just as $\sff_{\alpha\beta\gamma}$ and $\bar\sff_{\bar\alpha\bar\beta\bar\gamma}$.
        
        \begin{table}[ht]
            \begin{center}
                \begin{tabular}{|l|c|c|}
                    \hline
                    & $\frV$ & $\bar\frV$
                    \\
                    \hline
                    Biadjoint scalar field theory & $\frg$ & $\bar\frg$ 
                    \\
                    Yang--Mills theory & $\frg$ & $\frKin$
                    \\
                    $\caN=0$ supergravity & $\frKin$ & $\frKin$ 
                    \\
                    \hline
                \end{tabular}
                \vspace{-15pt}
            \end{center}
            \caption{Factors appearing in the field space factorisation~\eqref{eq:generalVectorSpaceFactorisation} with $\frKin$ given in~\eqref{eq:kinYM} and $\frg$ and $\bar\frg$ the colour Lie algebras.\label{tab:example_factorisations}}
        \end{table}
        
        With the factorisation restriction, the action~\eqref{eq:cubic_Lagrangian_standard_form} becomes
        \begin{equation}\label{eq:decomposition_BRST_op}
            S\ =\ \int\rmd^dx\,\left\{\frac12\Phi^{\alpha\bar\alpha}\sfg_{\alpha\beta}\bar\sfg_{\bar\alpha\bar\beta}\wave\Phi^{\beta\bar\beta}+\frac1{3!}\Phi^{\alpha\bar\alpha}(\sfp\,\sff_{\alpha\beta\gamma}\,\bar\sff_{\bar\alpha\bar\beta\bar\gamma})(\Phi^{\beta\bar\beta}\otimes\Phi^{\gamma\bar\gamma})\right\}\,.
        \end{equation}
        For the BRST operator $Q_{\rm BRST}$, the factorisation of indices and the linearity of $Q_{\rm BRST}$ imply the decomposition
        \begin{equation}
            Q_{\rm BRST}\ \eqqcolon\ q_{\rm BRST}+\bar q_{\rm BRST}~,
        \end{equation}
        where $q_{\rm BRST}$ and $\bar q_{\rm BRST}$ are BRST operators acting in a non-trivial way on the factors $\frV\otimes\scC^\infty(\IM^d)$ and $\bar\frV\otimes\scC^\infty(\IM^d)$ in the factorisation~\eqref{eq:general_field_factorisation}, respectively. By this, we mean that the structure constants $\sfQ^I_J$ and $\sfQ^I_{JK}$ decompose as $\sfQ^I_J\to(\sfq^I_J,\bar\sfq^I_J)$ and $\sfQ^I_{JK}\to(\sfq^I_{JK},\bar\sfq^I_{JK})$. More explicitly,
        \begin{equation}
            \begin{aligned}
                \sfq^{(\alpha,\bar\alpha,x)}_{(\beta,\bar\beta,y)}\ &=\ \delta^{(d)}(x-y)\sfq^\alpha_\beta(x)\delta^{\bar\alpha}_{\bar\beta}~,~~~
                &\sfq^{(\alpha,\bar\alpha,x)}_{(\beta,\bar\beta,y);(\gamma,\bar \gamma,z)}\ &=\ \delta^{(d)}(x-y)\delta^{(d)}(x-z)\sfq^{\alpha}_{\beta\gamma}(x)\bar\sff^{\bar\alpha}_{\bar\beta\bar\gamma}(x)~,
                \\
                \bar\sfq^{(\alpha,\bar\alpha,x)}_{(\beta,\bar\beta,y)}\ &=\ \delta^{(d)}(x-y)\delta^\alpha_\beta\bar\sfq^{\bar\alpha}_{\bar\beta}(x)
                ~,~~~
                &\bar\sfq^{(\alpha,\bar\alpha,x)}_{(\beta,\bar\beta,y);(\gamma,\bar \gamma,z)}\ &=\ \delta^{(d)}(x-y)\delta^{(d)}(x-z)\sff^{\alpha}_{\beta\gamma}(x)\bar\sfq^{\bar\alpha}_{\bar\beta\bar\gamma}(x)~,
            \end{aligned}
        \end{equation}
        where $\sfq^\alpha_\beta$ and $\bar\sfq^{\bar \alpha}_{\bar \beta}$ are differential operators and $\sfq^\alpha_{\beta\gamma}$ and $\bar\sfq^{\bar\alpha}_{\bar\beta\bar\gamma}$ are again bi-differential operators, just as $\sff^\alpha_{\beta\gamma}$ and $\bar\sff^{\bar\alpha}_{\bar\beta\bar\gamma}$, with locality again restricting their integral kernels. Note that in this splitting, the association of terms of the form $\delta^{(d)}(x-y)\delta^\alpha_\beta\delta^{\bar\alpha}_{\bar\beta}$ and $\delta^{(d)}(x-y)\delta^{(d)}(x-z)\sff^{\alpha}_{\beta\gamma}(y,z)\bar\sff^{\bar\alpha}_{\bar\beta\bar\gamma}(y,z)$ is not unique; we assign half of each of these terms to $(\sfq^I_J,\sfq^I_{JK})$ and half to $(\bar\sfq^I_J,\bar\sfq^I_{JK})$.
        
        \paragraph{Example.}
        To make our rather abstract discussion more concrete, let us briefly consider the case of Yang--Mills theory~\eqref{eq:BVActionYM}. We refrain from discussing the details of the strictification of the BV action, but it is clear that $\frV=\frg$ and $\bar\frV=\frKin'$ with $\frKin'$ some extension of $\frKin$ allowing for auxiliary fields, similar to $\frKin^{\rm st}$ defined in~\eqref{eq:def_frKin_prime}. It is then also clear that $\sfg_{\alpha\beta}$ and $\sff^\alpha_{\beta\gamma}$ are the Killing form and the structure constants of the gauge Lie algebra $\frg$.
        
        On $\frKin'$, the integral kernel for the differential operator $\bar\sfg_{\mu\nu}$ is given by 
        \begin{equation}
            \bar\sfg_{\mu\nu}\ =\ \eta_{\mu\nu}-\frac{1}{\wave}\partial_\mu\partial_\nu~.
        \end{equation}
        We note that $\sfq^\alpha_\beta=0$ and $\bar\sfq^{\bar\alpha}_{\bar\beta}$ is only non-trivial for $\bar\alpha$ labelling ghost and Nakanishi--Lautrup fields and $\bar\beta$ labelling the gauge potential and the anti-ghost field, all colour-stripped. Working out all other structure constants is a straightforward but tedious process; since no more insights would be obtained from it, we refrain from listing them here. We only note that for Yang--Mills theory, the ambiguity in assigning terms to $q$ and $\bar q$ is absent.
        
        \paragraph{Double copy.}
        We now note that the decomposition of the Lagrangian matches precisely the decomposition of scattering amplitudes in the discussion of colour--kinematics duality, cf.~\cref{ssec:scattering_amplitudes_generalities}, which is the starting point for the double copy. We merely extended the factorisation of the interaction vertices to a factorisation of the whole BRST structure.
        
        In the usual double copy, we start from the factorisation for Yang--Mills theory and replace the colour factor by a kinematic factor. More generally, however, we can certainly replace any one of the (graded) vector spaces $\frV$ and $\bar\frV$ and the corresponding structure constants with (graded) vector spaces and structure constants from other theories. This gives us a new action, which we shall denote by $\dbl{S}^{\rm DC}_{\rm BRST}$. The corresponding BRST operator $\dbl{Q}^{\rm DC}_{\rm BRST}$ is obtained by replacing one set of kinematic structure constants in the decomposition of the BRST operator~\eqref{eq:decomposition_BRST_op} with those from the new factor.
        
        \paragraph{BRST--Lagrangian double copy.} 
        In order to obtain a consistent and quantisable theory, we demand the new BRST structure to be consistent. Specifically,
        \begin{equation}
            \dbl{Q}^{\rm DC}_{\rm BRST}\dbl{S}^{\rm DC}_{\rm BRST}\ =\ 0
            \eand
            (\dbl{Q}^{\rm DC}_{\rm BRST})^2\ =\ 0~.
        \end{equation}
        By construction, we have again a decomposition $\dbl{Q}^{\rm DC}_{\rm BRST}\eqqcolon\dbl{q}^{\rm DC}_{\rm BRST}+\dbl{\bar{q}}^{\rm DC}_{\rm BRST}$. The condition $Q_{\rm BRST}^2=0$ implies $q_{\rm BRST}^2=0$, and we decompose the latter into linear, quadratic, and cubic terms in the fields,
        \begin{equation}
            q_{\rm BRST}^2\Phi^{\cdots}\ \eqqcolon\ q^{(2,0)}_1+q^{(2,0)}_2+q^{(2,0)}_3~,
        \end{equation}
        and analogously for $\bar q_{\rm BRST}^2$, $(\dbl{q}^{\rm DC}_{\rm BRST})^2$, and $(\dbl{\bar q}^{\rm DC}_{\rm BRST})^2$, respectively. Schematically, the summands read as
        \begin{subequations}
            \begin{equation}
                \begin{aligned}
                    q^{(2,0)}_1\ &=\ \cdots\sfq^\alpha_\beta\sfq^\beta_\gamma\cdots
                    ~,
                    \\
                    q^{(2,0)}_2\ &=\ \cdots(\sfq^\alpha_\delta\sfq^\delta_{\beta\gamma}+\sfq^\delta_\beta \sfq^\alpha_{\delta\gamma}\pm\sfq^\delta_\gamma\sfq^\alpha_{\beta\delta}){\bar\sff}^{\bar \alpha}_{\bar\beta\bar\gamma}\cdots
                    ~,
                    \\
                    q^{(2,0)}_3\ &=\ \cdots(\sfq^\eps_{\beta\gamma}\sfq^\alpha_{\eps\delta}{\bar\sff}^{\bar\eps}_{\bar\beta\bar\gamma}{\bar\sff}^{\bar\alpha}_{\bar\eps\bar\delta}\pm\sfq^\eps_{\beta\gamma}\sfq^\alpha_{\delta\eps}{\bar\sff}^{\bar\eps}_{\bar\beta\bar\gamma}{\bar \sff}^{\bar\alpha}_{\bar\delta\bar\eps})\cdots
                    ~,
                \end{aligned}
            \end{equation}
            and
            \begin{equation}
                \begin{aligned}
                    \dbl{q}^{(2,0)}_1\ &=\ \cdots\sfq^\alpha_\beta\sfq^\beta_\gamma\cdots~,
                    \\
                    \dbl{q}^{(2,0)}_2\ &=\ \cdots(\sfq^\alpha_\delta\sfq^\delta_{\beta\gamma}+\sfq^\delta_\beta\sfq^\alpha_{\delta\gamma}\pm\sfq^\delta_\gamma \sfq^\alpha_{\beta\delta})\dbl{\bar\sff}^{\bar \alpha}_{\bar\beta\bar\gamma}\cdots~,
                    \\
                    \dbl{q}^{(2,0)}_3\ &=\ \cdots(\sfq^\eps_{\beta\gamma}\sfq^\alpha_{\eps\delta}\dbl{\bar\sff}^{\bar\eps}_{\bar\beta\bar\gamma}\dbl{\bar \sff}^{\bar\alpha}_{\bar\eps\bar\delta}\pm\sfq^\eps_{\beta\gamma}\sfq^\alpha_{\delta\eps}\dbl{\bar\sff}^{\bar\eps}_{\bar\beta\bar\gamma}\dbl{\bar\sff}^{\bar\alpha}_{\bar\delta\bar\eps})\cdots~,
                \end{aligned}
            \end{equation}
        \end{subequations}
        where $\tilde\sff^\alpha_{\beta\gamma}$ and $\dbl{\bar\sff}^{\bar\alpha}_{\bar\beta\bar\gamma}$ denote the kinematic constants in $\dbl{S}^{\rm DC}_{\rm BRST}$. It is now clear that $\dbl{q}^{(2,0)}_1$ and $\dbl{q}^{(2,0)}_2$ vanish if $q^2_{\rm BRST}=0$ and thus, $q^{(2,0)}_1$ and $q^{(2,0)}_2$ vanish on arbitrary fields.
        
        So far, our discussion was fairly general and nothing singled out colour--kinematics-dual theories from other theories. This changes with the condition that $q^{(2,0)}_3=0$ must imply $\dbl{q}^{(2,0)}_3=0$. Vanishing of $q^{(2,0)}_3$ relies on a transfer of the symmetry properties of the open indices of ${\bar \sff}^{\bar\eps}_{\bar\beta\bar\gamma}{\bar \sff}^{\bar\alpha}_{\bar\eps\bar\delta}$ and ${\bar \sff}^{\bar\eps}_{\bar\beta\bar\gamma}{\bar \sff}^{\bar\alpha}_{\bar\delta\bar\eps}$ via the contracting fields (in which the expression is totally symmetric) to $\sfq^\eps_{\beta\gamma}\sfq^\alpha_{\eps\delta}$ and $\sfq^\eps_{\beta\gamma}\sfq^\alpha_{\delta\eps}$. It follows that if the symmetry properties of the open indices in the terms quadratic in $\bar\sff^{\bar\alpha}_{\bar\beta\bar\gamma}$ are the same as for the terms quadratic in $\dbl{\bar\sff}^{\bar\alpha}_{\bar\beta\bar\gamma}$ then $\dbl{q}^{(2,0)}_3=0$. The colour--kinematics duality provides such a condition.
        
        The same argument shows that $(\dbl{\bar{q}}^{\rm DC}_{\rm BRST})^2=0$, and we can directly turn to the cross terms and split them again into linear, quadratic, and cubic pieces,
        \begin{subequations}
            \begin{equation}
                (q_{\rm BRST}\bar q_{\rm BRST}+\bar q_{\rm BRST} q_{\rm BRST})\Phi^{\cdots}\ \eqqcolon\ q^{(1,1)}_1+q^{(1,1)}_2+q^{(1,1)}_3~,
            \end{equation}
            and 
            \begin{equation}
                (\dbl{q}^{\rm DC}_{\rm BRST}\dbl{\bar q}^{\rm DC}_{\rm BRST}+\dbl{\bar q}^{\rm DC}_{\rm BRST}\dbl{q}^{\rm DC}_{\rm BRST})\Phi^{\cdots}\ \eqqcolon\ \tilde q^{(1,1)}_1+\tilde q^{(1,1)}_2+\tilde q^{(1,1)}_3~.
            \end{equation}
        \end{subequations}
        We note that the conditions $q^{(1,1)}_1=0$ and $\dbl{q}^{(1,1)}_1=0$ are implied directly when $q_1$ and $\bar{q}_1$ and $\tilde q_1$ and $\tilde{\bar{q}}_1$ anti-commute, respectively, which is always the case in the theories we study. Moreover, we have, again schematically, the conditions
        \begin{equation}
            \begin{aligned}
                q^{(1,1)}_2\ &=\ \cdots\sfq^\alpha_{\beta\gamma}(\bar\sfq^{\bar\alpha}_{\bar\delta}{\bar\sff}^{\bar\delta}_{\bar\beta\bar\gamma}\pm\bar\sfq^{\bar\delta}_{\bar\beta}{\bar\sff}^{\bar\alpha}_{\bar\delta\bar\gamma}\pm\bar\sfq^{\bar\delta}_{\bar\gamma}{\bar \sff}^{\bar \alpha}_{\bar\beta\bar \delta})\cdots+\cdots\bar\sfq^{\bar\alpha}_{\bar\beta\bar\gamma}(\sfq^\alpha_\delta\sff^\delta_{\beta\gamma}\pm\sfq^\delta_\beta\sff^\alpha_{\delta\gamma}\pm\sfq^{\delta}_{\gamma}{\sff}^{\alpha}_{\beta\delta})\cdots~,
                \\
                q^{(1,1)}_3\ &=\ \cdots(\sfq^\alpha_{\eps\delta}{\bar\sff}^{\bar\alpha}_{\bar\eps\bar\delta}\sff^\eps_{\beta\gamma}{\bar\sfq}^{\bar\eps}_{\bar\beta\bar\gamma}\pm\sfq^\alpha_{\beta\eps}{\bar\sff}^{\bar\alpha}_{\bar\beta\bar\eps}\sff^\eps_{\gamma\delta}{\bar\sfq}^{\bar\eps}_{\bar\gamma\bar\delta}\pm\sff^\alpha_{\eps\delta}{\bar\sfq}^{\bar\alpha}_{\bar\eps\bar\delta}\sfq^\eps_{\beta\gamma}{\bar\sff}^{\bar\eps}_{\bar\beta\bar\gamma}\pm\sff^\alpha_{\beta\eps}{\bar\sfq}^{\bar\alpha}_{\bar\beta\bar\eps}\sfq^\eps_{\gamma\delta}{\bar\sff}^{\bar\eps}_{\bar\gamma\bar\delta})\cdots~.
            \end{aligned}
        \end{equation}
        We see that $q^{(1,1)}_2=0$ splits into two separate conditions on the indices in $\frV$ and $\bar\frV$ and thus it implies $\dbl{q}^{(1,1)}_2=0$. The condition $\dbl{q}^{(1,1)}_3=0$ can, in principle, be non-trivial, but again colour--kinematics duality as well as the special form of the BRST operator in the theories in which we are interested renders $\dbl{q}^{(1,1)}_3=0$ equivalent to $q^{(1,1)}_3=0$.
        
        Finally, we have to check that $\dbl{Q}^{\rm DC}_{\rm BRST}\dbl{S}^{\rm DC}_{\rm BRST}=0$, and we consider 
        \begin{equation}
            q_{\rm BRST}S\ \eqqcolon\ s^{(1,0)}_2+s^{(1,0)}_3+s^{(1,0)}_4~,
        \end{equation}
        where $s^{(1,0)}_2$, $s^{(1,0)}_3$, and $s^{(1,0)}_4$ are quadratic, cubic, and quartic in the fields. Analogously, we have $\dbl{q}^{\rm DC}_{\rm BRST}\dbl{S}^{\rm DC}_{\rm BRST}\eqqcolon\dbl{s}^{(1,0)}_2+\dbl{s}^{(1,0)}_3+\dbl{s}^{(1,0)}_4$, and the discussion for $\bar q_{\rm BRST}$ and $\dbl{\bar q}^{\rm DC}_{\rm BRST}$ is similar. Schematically, we compute
        \begin{equation}
            \begin{aligned}
                s^{(1,0)}_2\ &=\ \int\rmd^dx\,\cdots(\sfq^\gamma_\alpha\sfg_{\gamma\beta}\bar\sfg_{\bar\alpha\bar\beta}\wave)\cdots~,
                \\
                s^{(1,0)}_3\ &=\ \int\rmd^dx\,\cdots(\sfg_{\alpha\delta}\wave\sfq^\delta_{\beta\gamma}+\sff_{\alpha\delta\gamma}\sfq^\delta_\beta+\sff_{\alpha\beta\delta}\sfq^\delta_\beta)\bar\sff_{\bar\alpha\bar\beta\bar\gamma}\cdots~,
                \\
                s^{(1,0)}_4\ &=\ \int\rmd^dx\,\cdots(\sff_{\alpha\eps\delta}\sfq^\eps_{\beta\gamma}\bar\sff_{\bar\alpha\bar\eps\bar\delta}\bar\sff^{\bar \eps}_{\bar\beta\bar\gamma}+\sff_{\alpha\beta\eps}\sfq^\eps_{\gamma\delta}\bar\sff_{\bar\alpha\bar\beta\bar\eps}\bar\sff^{\bar\eps}_{\bar\gamma\bar\delta})\cdots~,
            \end{aligned}
        \end{equation}
        where we have assumed that $q_{\rm BRST}$ commutes with the differential and bi-differential operators in the action, which is the case in all our theories. We see that $s^{(1,0)}_2=0$ and $s^{(1,0)}_3=0$ imply $\dbl{s}^{(1,0)}_2=0$ and $\dbl{s}^{(1,0)}_3=0$, respectively. The relation $\dbl{s}^{(1,0)}_4=0$ can, in principle, lead to additional conditions. In a theory with colour--kinematics duality, however, the contraction of the kinematic structure constants $\bar\sff^{\bar\alpha}_{\bar\beta\bar\gamma}$ appears as in the Jacobi identity, and $s^{(1,0)}_4$ as well as $\dbl{s}^{(1,0)}_4$ vanish automatically. 
        
        \paragraph{Partial BRST--Lagrangian double copy.}
        There are few theories where we expect the BRST--Lagrangian double copy to work perfectly. The reason is that in most formulations, colour--kinematics duality will not hold. In Yang--Mills theory, for example, it is not known if colour--kinematics duality can be made manifest for off-shell fields.\footnote{Recall that we only extended colour--kinematics to the BRST-extended Hilbert space in~\cref{ssec:ck_duality_YM_BRST_extended_Hilbertspace}, but with all fields still on-shell.}
        
        Now if colour--kinematics duality fails to hold up to certain terms, say the ideal of functions of the fields vanishing on-shell as in the case of Yang--Mills theory, then the equation $\dbl{Q}^{\rm DC}_{\rm BRST}\dbl{S}^{\rm DC}_{\rm BRST}=0$ will also fail to hold up to the same ideal. Consequently, $\dbl{Q}^{\rm DC}_{\rm BRST}\dbl{S}^{\rm DC}_{\rm BRST}$ is a product of factors whose vanishing amounts to the equations of motion possibly multiplied by other fields and their derivatives.
        
        \subsection{BRST--Lagrangian double copy of Yang--Mills theory}\label{ssec:construction_double_copied_action}
        
        Let us now make the abstract discussion above concrete by working out the example of the BRST--Lagrangian double copy
        \begin{equation}\label{eq:TP_Full_N=0}
            \tilde\frL^{\rm DC}_{\rm BRST}\ \coloneqq\ \frKin^{\rm st}\otimes_\tau(\frKin^{\rm st}\otimes_\tau\frScal)~,
        \end{equation}
        where $\frKin^{\rm st}$ is given in~\eqref{eq:kinYMst} and $\frScal$ in~\eqref{eq:L_infty_scalar}, respectively.
        
        \paragraph{Field content.}
        From the discussion in \cref{ssec:Factorisation_CC_N=0}, we already know that the double-copied field content of the BRST-extended Hilbert space of Yang--Mills theory agrees with the field content of the BRST-extended Hilbert space. We shall continue to use the field labels introduced in \cref{tab:fac_vec_N0}.
        
        In the interactive case of the full homotopy algebras, however, we have infinitely many additional auxiliary fields, arising from the infinitely many additional auxiliary fields of strictified and colour--kinematics duality preserving Yang--Mills theory. In the previous section, we made five of the auxiliary fields in Yang--Mills theory explicit,
        \begin{equation}
            \tilde K^{a\mu}_1~,~~~
            \tilde{\bar K}^{a\mu}_1~,~~~
            G^a_{\mu\nu\kappa}~,~~~
            \tilde K^{2a}_\mu~,~~~
            \tilde{\bar K}^{2a}_\mu~,
        \end{equation}
        which correspond to the additional basis elements
        \begin{equation}
            \sft^1_\mu~,~~~
            \bar \sft^\mu_1~,~~~
            \sft^{\mu\nu\kappa}_0~,~~~
            \sft^2_\mu~,~~~
            \bar\sft^\mu_2
        \end{equation}
        in $\frKin^{\rm st}$. In the tensor product~\eqref{eq:TP_Full_N=0} this gives rise to 40~auxiliary fields involving one auxiliary kinetic factor and another 25~auxiliary fields involving two auxiliary kinetic factors. Instead of giving these auxiliary fields individual labels, we shall collectively denote them by $\dcf{\ttk_1}{\ttk_2}$, where $\ttk_1$ and $\ttk_2$ denote the first and second kinematic factors. For example,
        \begin{equation}
            \begin{aligned}
                \dcf{\ttg}{\ttg}\ &\coloneqq\ \ttg\otimes \ttg\otimes\left(\int\rmd^dx\,\tts_x\varphi^{\ttg\ttg}(x)\right)\ =\ \tilde \lambda~,
                \\
                \dcf{\ttv}{\ttv}\ &\coloneqq\ \cb_a\otimes\ttv^\mu\otimes\ttv^\nu\otimes \left(\int\rmd^dx\,\tts_x\varphi^{\ttv\ttv}{}_{\mu\nu}(x)\right)\ =\ \tilde h+\tilde B~,
                \\
                \dcf{\ttt^1}{\ttt_0}\ &\coloneqq\ \ttt^1_\mu\otimes\ttt_0^{\nu\kappa\lambda}\otimes\left(\int\rmd^dx\,\tts_x\varphi^{\ttt^1\ttt_0}{}^\mu_{\nu\kappa\lambda}(x)\right)~.
            \end{aligned}
        \end{equation}
        
        \paragraph{Higher products.}
        Next, we use the twist~\eqref{eq:full_twist_1_YM} and~\eqref{eq:full_twist_2_YM} to compute the higher products $\mu_1$ and $\mu_2$ between the elements of $\tilde\frL^{\rm DC}_{\rm BRST}$. The formulas from \cref{ssec:twisted_tp_homotopy_algebras} with all the appropriate signs included read as
        \begin{equation}
            \begin{aligned}
                &\mu_1(\ttx_1\otimes \tty_1\otimes \varphi_1)\ \coloneqq\ (-1)^{|\tau^{(1)}_1(\ttx_1)|+|\tau^{(1)}_1(\tty_1)|}~
                \tau^{(1)}_1(\ttx_1)\otimes \tau^{(1)}_1(\tty_1)\otimes \big(\tau^{(2)}_1(\ttx_1)(\tau^{(2)}_2(\tty_1)(\varphi_1))\big)\,,
                \\
                &\mu_2(\ttx_1\otimes \tty_1\otimes \varphi_1\,,\,\ttx_2\otimes \tty_2\otimes \varphi_2)\ \coloneqq
                \\
                &~~\ \coloneqq\ (-1)^{(|\tty_1|+|\varphi_1|)|\ttx_2|+|\varphi_1|\,|\tty_2|}\times
                \\
                &\hspace{2cm}\times\tau_2^{(1)}(\ttx_1,\ttx_2)\otimes \tau_2^{(1)}(\tty_1,\tty_2)\otimes \big(\tau_2^{(2)}(\ttx_1,\ttx_2)\varphi_1(x)\big)\big(\tau_2^{(2)}(\tty_1,\tty_2)\varphi_2(x)\big)~.
            \end{aligned}
        \end{equation}
        Note that there are no additional signs because our $\tau_i^{(2)}$ are always even. While the computation is readily performed, listing the higher products for all 81~fields is not particularly helpful. 
        
        \paragraph{Action.} 
        The factorisation~\eqref{eq:TP_Full_N=0} induces the following cyclic structure:
        \begin{equation}
            \begin{aligned}
                &\inner{\ttx_1\otimes\tty_1\otimes\varphi_1}{\ttx_2\otimes\tty_2\otimes\varphi_2}\ \coloneqq
                \\
                &\hspace{1cm}\coloneqq \ (-1)^{|\ttx_2|_\frKin(|\tty_1|_\frKin+|\varphi_1|_\frScal)+|\tty_2|_\frKin|\varphi_1|_\frScal}\inner{\ttx_1}{\ttx_2}\,\inner{\tty_1}{\tty_2}\,\inner{\varphi_1}{\varphi_2}~.
            \end{aligned}
        \end{equation}
        Together with the formulas for the super homotopy Maurer--Cartan action~\eqref{eq:shMC_translation_formulas}, we can compute the (gauge-fixed) BRST action corresponding to the $L_\infty$-algebra $\tilde\frL^{\rm DC}_{\rm BRST}$. Again, listing all the terms would not provide much insight, but we stress that we obtain all the expected terms, in particular the lowest terms of the Fierz--Pauli version of the $\caN=0$ supergravity action as well as the evident terms involving ghosts.
        
        \paragraph{Double copy of the BRST operator.} 
        Let us now also consider the double copy of the BRST operator to a BRST operator $\dbl{Q}^{\rm DC}_{\rm BRST}$. For our purposes, the double copy of the linearised part without considering the auxiliary fields will be sufficient. We start from Yang--Mills theory with the factors $\frV\coloneqq\frg$ and $\bar\frV\coloneqq\frKin$ in~\eqref{eq:generalVectorSpaceFactorisation} and the usual BRST relations in terms of coordinate functions on $\tilde\frL^{\rm YM}_{\rm BRST}$,
        \begin{equation}
            \begin{aligned}
                \tilde A_\mu^a\ &\mapstomap{Q_{\rm BRST}^{\rm YM,\,lin}}\ \delta^a_b \partial_\mu\tilde c^b~,~~~& \tilde b^a\ &\mapstomap{Q_{\rm BRST}^{\rm YM,\,lin}} \ \delta^a_b\frac{1-\sqrt{1-\xi}}{\sqrt{\xi}}\sqrt{\wave}\,\tilde c^b~,
                \\
                \tilde c^a\ &\mapstomap{Q_{\rm BRST}^{\rm YM,\,lin}}\ 0~,~~~
                & \tilde{\bar c}^a\ &\mapstomap{Q_{\rm BRST}^{\rm YM,\,lin}}\ \delta^a_b \left(\sqrt{\frac{\wave}{\xi}}\tilde b^b-\frac{1-\sqrt{1-\xi}}{\xi}\partial^\mu\tilde A_\mu^b\right).
            \end{aligned}
        \end{equation}
        We thus have $\sfq^\alpha_\beta=\delta^\alpha_\beta$, and the non-vanishing components of $\bar\sfq^{\bar\alpha}_{\bar\beta}$ are given by 
        \begin{equation}
            \bar\sfq^{\bar\alpha}_{\bar\beta}\ =\ 
            \begin{cases}
                \partial_\mu & \efor\bar\alpha\ =\ \ttg^*~,~~\bar\beta\ =\ \ttv_\mu^*~,
                \\
                \frac{1-\sqrt{1-\xi}}{\sqrt{\xi}}\sqrt{\wave} & \efor\bar\alpha\ =\ \ttg^*~,~~\bar\beta\ =\ \ttn^*~,
                \\
                \sqrt{\frac{\spacer{2pt}\wave}{\xi}} & \efor\bar\alpha\ =\ \ttn^*~,~~\bar\beta\ =\ \tta^*~,
                \\
                -\frac{1-\sqrt{1-\xi}}{\xi}\partial^\mu & \efor\bar\alpha\ =\ \ttv^*_\mu~,~~\bar\beta\ =\ \tta^*~.
            \end{cases}
        \end{equation}
        After the double copy, we have $\frV\coloneqq\frKin\eqqcolon\bar\frV$ and, correspondingly, $\sfq^{\alpha}_{\beta}=\bar \sfq^{\alpha}_{\beta}$. The linearisation of the double-copied BRST operator is then non-trivial on a field containing a factor of $\ttv^\mu$ or $\tta$, and we have in the anti-symmetrised sector
        \begin{subequations}\label{eq:double_copied_BRST}
            \begin{equation}
                \begin{aligned}
                    \tilde\lambda \ &\mapstomap{\dbl{Q}^{\rm DC,\,lin}_{\rm BRST}}\ 0~,
                    \\
                    \tilde\Lambda_\mu\ &\mapstomap{\dbl{Q}^{\rm DC,\,lin}_{\rm BRST}}\ \partial_\mu \tilde\lambda~,
                    \\
                    \tilde\gamma\ &\mapstomap{\dbl{Q}^{\rm DC,\,lin}_{\rm BRST}}\ \frac{1-\sqrt{1-\xi}}{\sqrt{\xi}}\sqrt{\wave}\,\tilde\lambda ~,
                    \\
                    \tilde B_{\mu\nu}\ &\mapstomap{\dbl{Q}^{\rm DC,\,lin}_{\rm BRST}}\ \partial_\mu\tilde\Lambda_\nu-\partial_\nu\tilde\Lambda_\mu~,
                    \\
                    \tilde\alpha_{\mu}\ &\mapstomap{\dbl{Q}^{\rm DC,\,lin}_{\rm BRST}}\ \frac{1-\sqrt{1-\xi}}{\sqrt{\xi}}\sqrt{\wave}\,\tilde\Lambda_\mu-\partial_\mu\tilde\gamma~,
                    \\
                    \tilde\eps\ &\mapstomap{\dbl{Q}^{\rm DC,\,lin}_{\rm BRST}}\ \sqrt{\frac{\wave}{\xi}}\tilde\gamma-\frac{1-\sqrt{1-\xi}}{\xi}\partial^\mu\tilde \Lambda_\mu~,
                    \\
                    \tilde{\bar\Lambda}_\mu\ &\mapstomap{\dbl{Q}^{\rm DC,\,lin}_{\rm BRST}}\ \partial_\mu \tilde \eps+\sqrt{\frac{\wave}{\xi}}\tilde \alpha_\mu-\frac{1-\sqrt{1-\xi}}{\xi}\partial^\nu\tilde B_{\mu\nu}~,
                    \\
                    \tilde{\bar\gamma}\ &\mapstomap{\dbl{Q}^{\rm DC,\,lin}_{\rm BRST}}\ \frac{1-\sqrt{1-\xi}}{\sqrt{\xi}}\sqrt{\wave}\,\tilde\eps+\frac{1-\sqrt{1-\xi}}{\xi}\partial^\mu\tilde\alpha_\mu~,
                    \\
                    \tilde{\bar\lambda}\ &\mapstomap{\dbl{Q}^{\rm DC,\,lin}_{\rm BRST}}\
                    \sqrt{\frac{\wave}{\xi}}\tilde{\bar\gamma}-\frac{1-\sqrt{1-\xi}}{\xi}\partial^\mu\tilde{\bar\Lambda}_\mu
                    ~,
                \end{aligned}
            \end{equation}
            and in the symmetrised sector
            \begin{equation}
                \begin{aligned}
                    \tilde X^\mu\ &\mapstomap{\dbl{Q}^{\rm DC,\,lin}_{\rm BRST}}\ 0~,
                    \\
                    \tilde\beta\ &\mapstomap{\dbl{Q}^{\rm DC,\,lin}_{\rm BRST}}\ 0~,
                    \\
                    \tilde h_{\mu\nu}\ &\mapstomap{\dbl{Q}^{\rm DC,\,lin}_{\rm BRST}}\ \partial_\mu \tilde X_\nu+\partial_\nu \tilde X_\mu~,
                    \\
                    \tilde\varpi^\mu\ &\mapstomap{\dbl{Q}^{\rm DC,\,lin}_{\rm BRST}}\ -\frac{1-\sqrt{1-\xi}}{\sqrt{\xi}}\sqrt{\wave}\,\tilde X^\mu-\partial^\mu\tilde\beta~,
                    \\
                    \tilde\pi\ &\mapstomap{\dbl{Q}^{\rm DC,\,lin}_{\rm BRST}}\ 2\frac{1-\sqrt{1-\xi}}{\sqrt{\xi}}\sqrt{\wave}\,\tilde\beta~,
                    \\
                    \tilde\delta\ &\mapstomap{\dbl{Q}^{\rm DC,\,lin}_{\rm BRST}}\ \sqrt{\frac{\wave}{\xi}}\tilde\beta-\frac{1-\sqrt{1-\xi}}{\xi}\partial_\mu\tilde X^\mu~,
                    \\
                    \tilde{\bar X}^\mu\ &\mapstomap{\dbl{Q}^{\rm DC,\,lin}_{\rm BRST}}\ -\partial^\mu\tilde\delta-\sqrt{\frac{\wave}{\xi}}\tilde\varpi^\mu-\frac{1-\sqrt{1-\xi}}{\xi}\partial_\nu\tilde h^{\nu\mu}~,
                    \\
                    \tilde{\bar\beta}\ &\mapstomap{\dbl{Q}^{\rm DC,\,lin}_{\rm BRST}} \ -\frac{1-\sqrt{1-\xi}}{\sqrt{\xi}}\sqrt{\wave}\,\tilde\delta+\frac{1-\sqrt{1-\xi}}{\xi}\partial_\mu\tilde\varpi^\mu+\sqrt{\frac{\wave}{\xi}}\tilde\pi~.
                \end{aligned}
            \end{equation}
            This BRST operator is related to the usual linearised BRST operator for $\caN=0$ supergravity,~\eqref{eq:BVOperatorKR} and~\eqref{eq:BVOperatorEH},
            \begin{equation}
                \begin{aligned}
                    \lambda\ &\mapstomap{Q^{\caN=0,\,\rm lin}_{\rm BRST}}\ 0~,~~~
                    &\varphi\ &\mapstomap{Q^{\caN=0,\,\rm lin}_{\rm BRST}}\ 0~,
                    \\
                    \Lambda_\mu\ &\mapstomap{Q^{\caN=0,\,\rm lin}_{\rm BRST}} \ \partial_\mu\lambda~,~~~
                    & X^\mu\ &\mapstomap{Q^{\caN=0,\,\rm lin}_{\rm BRST}}\ 0~,
                    \\
                    \gamma\ &\mapstomap{Q^{\caN=0,\,\rm lin}_{\rm BRST}}\ 0~,~~~
                    & \beta\ &\mapstomap{Q^{\caN=0,\,\rm lin}_{\rm BRST}}\ 0~,
                    \\
                    B_{\mu\nu}\ &\mapstomap{Q^{\caN=0,\,\rm lin}_{\rm BRST}}\ \partial_\mu\Lambda_\nu-\partial_\nu\Lambda_\mu
                    ~,~~~ 
                    & h_{\mu\nu}\ &\mapstomap{Q^{\caN=0,\,\rm lin}_{\rm BRST}}\ \partial_\mu X_\nu+\partial_\nu X_\mu~,
                    \\
                    \alpha_\mu\ &\mapstomap{Q^{\caN=0,\,\rm lin}_{\rm BRST}}\ 0
                    ~,~~~ 
                    & \varpi^\mu\ &\mapstomap{Q^{\caN=0,\,\rm lin}_{\rm BRST}}\ 0~,
                    \\
                    \eps\ &\mapstomap{Q^{\caN=0,\,\rm lin}_{\rm BRST}}\ \gamma~,~~~
                    &\delta\ &\mapstomap{Q^{\caN=0,\,\rm lin}_{\rm BRST}}\ \beta~,
                    \\
                    \bar\Lambda_\mu\ &\mapstomap{Q^{\caN=0,\,\rm lin}_{\rm BRST}}\ \alpha_\mu~,~~~ 
                    & \bar X^\mu\ &\mapstomap{Q^{\caN=0,\,\rm lin}_{\rm BRST}}\ \varpi^\mu~,
                    \\
                    \bar\gamma\ &\mapstomap{Q^{\caN=0,\,\rm lin}_{\rm BRST}}\ 0~,~~~
                    &\bar\beta\ &\mapstomap{Q^{\caN=0,\,\rm lin}_{\rm BRST}}\ \pi~,
                    \\
                    \bar\lambda\ &\mapstomap{Q^{\caN=0,\,\rm lin}_{\rm BRST}}\ \bar\gamma~,~~~
                    & \pi\ &\mapstomap{Q^{\caN=0,\,\rm lin}_{\rm BRST}}\ 0
                \end{aligned}
            \end{equation}
        \end{subequations}
        by the field redefinitions~\eqref{eq:canonicalFieldRedefinitionKR} and~\eqref{eq:canonicalFieldRedefinitionEH}, respectively.
        
        \subsection{Equivalence of the double copied action and \texorpdfstring{$\caN=0$}{N=0} supergravity}\label{ssec:quantum_equivalence}
        
        Let us complete the argument by showing that the double copied action $\dbl{S}^{\rm DC}_{\rm BRST}$ we constructed in \cref{ssec:construction_double_copied_action} is fully perturbatively quantum equivalent to the suitably gauge-fixed version of the BV action of $\caN=0$ supergravity, $S^{\caN=0}_{\rm BRST}$, defined in \cref{ssec:N=0SUGRA}. For this, we have to show that up to a field redefinition, both theories have the same tree-level correlation functions. A crucial point in our discussion will be the BRST--Lagrangian double copy formalism developed in the previous section.
        
        In the following, we shall speak of `auxiliary fields connected to a field $\phi$' by which we mean all auxiliary fields which appear together with $\phi$ in Feynman diagrams containing only propagators of auxiliary fields. Put differently, an auxiliary field $\psi$ connected to a field $\phi$ can have an interaction vertex with $\phi$ or interact with an auxiliary field that propagates to an auxiliary field that non-trivially interacts with $\phi$, etc.:
        \begin{equation}
            \begin{tikzpicture}[
                scale=1,
                every node/.style={scale=1},
                baseline={([yshift=-.5ex]current bounding box.center)}
                ]
                \matrix (m) [
                matrix of nodes,
                ampersand replacement=\&,
                column sep=0.1cm,
                row sep=0.4cm
                ]{
                    {} \& $\ldots$ \& {} 
                    \\
                    {} \& {} \& {} 
                    \\
                    $\psi$ \& \& $\phi$
                    \\
                };
                \draw (m-1-1) -- (m-2-2.center);
                \draw (m-1-3) -- (m-2-2.center);
                \draw [aux] (m-3-1) -- (m-2-2.center);
                \draw  (m-3-3) -- (m-2-2.center);
                \foreach \x in {(m-2-2)}{
                    \fill \x circle[radius=2pt];
                }
            \end{tikzpicture}
            ,~~~
            \begin{tikzpicture}[
                scale=1,
                every node/.style={scale=1},
                baseline={([yshift=-.5ex]current bounding box.center)}
                ]
                \matrix (m) [
                matrix of nodes,
                ampersand replacement=\&,
                column sep=0.1cm,
                row sep=0.4cm
                ]{
                    {} \& $\ldots$ \& {} \& {} \& $\ldots$ \& {} \& {} 
                    \\
                    {} \& {} \& {} \& {} \& {} \& {} \& {} 
                    \\
                    $\psi$ \& {} \& {} \& {} \& {}  \& $\phi$
                    \\
                };
                \draw (m-1-1) -- (m-2-2.center);
                \draw (m-1-3) -- (m-2-2.center);
                \draw (m-1-6) -- (m-2-5.center);
                \draw (m-1-4) -- (m-2-5.center);
                \draw [aux] (m-3-1) -- (m-2-2.center);
                \draw [aux] (m-2-2.center) -- (m-2-5.center);
                \draw (m-3-6) -- (m-2-5.center);
                \foreach \x in {(m-2-2), (m-2-5)}{
                    \fill \x circle[radius=2pt];
                }
            \end{tikzpicture}
            ,~~~
            \begin{tikzpicture}[
                scale=1,
                every node/.style={scale=1},
                baseline={([yshift=-.5ex]current bounding box.center)}
                ]
                \matrix (m) [
                matrix of nodes,
                ampersand replacement=\&,
                column sep=0.1cm,
                row sep=0.4cm
                ]{
                    {} \& $\ldots$ \& {} \& {} \& $\ldots$ \& {} \& {} \& $\ldots$ \& {} \& {} 
                    \\
                    {} \& {} \& {} \& {} \& {} \& {} \& {} \& {} \& {} 
                    \\
                    $\psi$ \& {} \& {} \& {} \& {} \& {} \& {} \& {}  \& $\phi$
                    \\
                };
                \draw (m-1-1) -- (m-2-2.center);
                \draw (m-1-3) -- (m-2-2.center);
                \draw (m-1-6) -- (m-2-5.center);
                \draw (m-1-4) -- (m-2-5.center);
                \draw (m-1-9) -- (m-2-8.center);
                \draw (m-1-7) -- (m-2-8.center);
                \draw [aux] (m-3-1) -- (m-2-2.center);
                \draw [aux] (m-2-2.center) -- (m-2-5.center);
                \draw [aux] (m-2-5.center) -- (m-2-8.center);
                \draw  (m-3-9) -- (m-2-8.center);
                \foreach \x in {(m-2-2), (m-2-5), (m-2-8)}{
                    \fill \x circle[radius=2pt];
                }
            \end{tikzpicture}
            ,~\ldots~,
        \end{equation}
        where a dashed line denotes an auxiliary field. We also use the terms physical and unphysical fields/interaction terms/scattering amplitudes. The unphysical fields are all ghosts, anti-ghosts, and Nakanishi--Lautrup fields as well as auxiliary fields connected to these. Physical fields are the remaining fields, consisting of the metric, the Kalb--Ramond field and the dilaton as well as a number of auxiliary fields. Physical interaction vertices are those consisting exclusively of physical fields. Physical scattering amplitudes are those with physical fields as external labels.
        
        \paragraph{Physical tree-level scattering amplitudes.} 
        We first note that the auxiliary fields in the double copied action $\dbl{S}^{\rm DC}_{\rm BRST}$ can be integrated out, after which the field content and the kinematic terms in both actions fully agree, up to the field redefinitions we discussed in \cref{sec:factorisation_free_field_theories}. Implementing these field redefinitions on $S^{\caN=0}_{\rm BRST}$, we obtain the action $S^{\caN=0}_{{\rm BRST},\,0}$.
        
        Moreover, the physical tree-level scattering amplitudes computed from the interaction vertices of the action $\dbl{S}^{\rm DC}_{\rm BRST}$ are by design precisely those arising in the usual double copy prescription for the construction of $\caN=0$ supergravity tree amplitudes from a factorisation of Yang--Mills amplitudes. The tree-level double copy has been demonstrated to hold, cf.~\cref{ob:established_dc}, and therefore the physical tree-level scattering amplitudes of $\dbl{S}^{\rm DC}_{\rm BRST}$ and $S^{\caN=0}_{{\rm BRST},\,0}$ agree. 
        
        If we put all unphysical fields to zero, the resulting theories $\dbl{S}^{\rm DC}_{\rm BRST,\,phys}$ and $S^{\caN=0}_{\rm BRST,\,phys}$ are classically equivalent by \cref{ob:classical_equivalence}. In the homotopy algebraic picture, this corresponds to a restriction $\frL^{\caN=0}_{\rm BRST,\,phys}$ and $\tilde\frL^{\rm DC}_{\rm phys}$ to two quasi-isomorphic $L_\infty$-subalgebras.
        
        In order to improve this restricted classical equivalence to a full perturbative quantum equivalence, we need to adjust and modify the actions or, equivalently, the corresponding $L_\infty$-algebras. We shall do this now in a sequence of steps, expanding the discussion in~\cite{Borsten:2020zgj}.
        
        \paragraph{Auxiliary fields of ghost number zero.} The reformulation of the tree-level scattering amplitudes of $\caN=0$ supergravity used in the double copy defines a local strictification of the physical part of the action~$S^{\caN=0}_{\rm BRST}$ to the action $S^{\caN=0}_{\rm BRST,\,1}$ by promoting all cubic interaction vertices to cubic interaction terms. This is fully analogous to the strictification implied by the manifestly colour--kinematics-dual form of the Yang--Mills action explained in~\cref{ssec:strictified_YM}. 
        
        By construction, the actions $S^{\caN=0}_{\rm BRST,\,1}$ and $\dbl{S}^{\rm DC}_{\rm BRST,\,phys}$ have the same field content, the same kinematical terms for the physical and auxiliary fields and identical tree-level scattering amplitudes for the physical fields. 
        
        Let us now consider the tree-level scattering amplitudes which have auxiliary fields of ghost number zero on their external legs. Such amplitudes are fully determined by the (iterated) collinear limits of physical tree-level scattering amplitudes. Because, again, the physical tree-level scattering amplitudes of $S^{\caN=0}_{\rm BRST,\,1}$ and $\dbl{S}^{\rm DC}_{\rm BRST,\,phys}$ agree, the tree-level scattering amplitudes with physical and auxiliary fields of ghost number zero on external legs agree.
        
        By \cref{ob:classical_equivalence}, we then have a local field redefinition of $S^{\caN=0}_{\rm BRST,\,1}$ to $S^{\caN=0}_{\rm BRST,\,2}$ such that both actions agree after all fields except for physical ones and auxiliary fields of ghost number zero are put to zero. If we integrated out all auxiliary fields in both actions, the resulting actions would agree in their purely physical parts.
        
        \paragraph{Nakanishi--Lautrup fields.}
        In the next step, we deal with the difference between $\dbl{S}^{\rm DC}_{\rm BRST}$ and $S^{\caN=0}_{{\rm BRST},\,2}$ proportional to any of the Nakanishi--Lautrup fields ($\tilde{\bar\beta},\tilde{\varpi}_\mu,\tilde\pi,\tilde\gamma,\tilde\alpha_\mu,\tilde{\bar\gamma}$); we shall come to the ghost field $\beta$ later. After integrating out all auxiliary fields, this difference can be compensated by \cref{ob:match_Nakanishi}. That is, we can modify the gauge-fixing fermion and perform a field redefinition of the Nakanishi--Lautrup fields such that this difference is removed. We note that neither of these two processes modifies the physical parts of the action and both preserve quantum equivalence. We can thus replace all terms in $S^{\caN=0}_{{\rm BRST},\,2}$ containing Nakanishi--Lautrup fields by the terms in $\dbl{S}^{\rm DC}_{\rm BRST}$ containing Nakanishi--Lautrup fields as well as auxiliary fields connected to Nakanishi--Lautrup fields. We call the resulting action $S^{\caN=0}_{{\rm BRST},\,3}$.
        
        Recall that there is a ghost number $-2$ field $\bar \lambda$ which is paired with the Nakanishi--Lautrup-type field $\gamma$ in the gauge-fixing fermion~\eqref{eq:gaugeFixingFermionKR}, allowing us to absorb any term proportional to $\gamma$ in a different gauge choice. This is not the case for the corresponding Nakanishi--Lautrup-type field in the gravity sector, $\beta$. Any discrepancy proportional to $\beta$ between $S^{\caN=0}_{{\rm BRST},\,3}$ and $\dbl{S}^{\rm DC}_{\rm BRST}$ (again, after integrating out all the auxiliary fields) should instead be absorbed by shifting the gauge-fixing fermion $\Psi$ from~\eqref{eq:gaugeFixingFermionEH} by a term $\delta P$, where $\beta P$ is the discrepancy. This will generate the desired corrections. This will also lead to new ghost terms, which we will treat separately in the next step.
        
        \paragraph{Ghost sector.}
        Let us now examine the ghost interactions. We know that the action $S^{\caN=0}_{{\rm BRST},\,3}$ comes with a BRST operators $Q^{\caN=0}_{\rm BRST,\,3}$ which satisfies 
        \begin{equation}
            (Q^{\caN=0}_{\rm BRST,\,3})^2\ =\ 0
            \eand
            Q^{\caN=0}_{\rm BRST,\,3}S^{\caN=0}_{{\rm BRST},\,3}\ =\ 0~.
        \end{equation}
        From our discussion in the previous section, we know that the double-copied BRST operator $\dbl{Q}^{\rm DC}_{\rm BRST}$ satisfies
        \begin{equation}
            (\dbl{Q}^{\rm DC}_{\rm BRST})^2\ \in\ \scI
            \eand
            \dbl{Q}^{\rm DC}_{\rm BRST}\dbl{S}^{\rm DC}_{\rm BRST}\ \in\ \scI~,
        \end{equation}
        where $\scI$ is the ideal of polynomials in the fields and their derivatives which vanishes for on-shell fields. We also know from the discussion around~\eqref{eq:double_copied_BRST} that the linearisations of the BRST operators satisfy
        \begin{equation}
            \dbl{Q}^{\rm DC,\,lin}_{\rm BRST}\ =\ Q^{\caN=0,\,\rm lin}_{\rm BRST,\,3}~.
        \end{equation}
        After integrating out all auxiliary fields, these BRST operators link the physical tree-level scattering amplitudes to tree-level scattering amplitudes containing ghosts by the on-shell Ward identities, cf.~\cref{ob:ghost_pair_reduction}. 
        
        At the level of the BRST operators $\dbl{Q}^{\rm DC,\,lin}_{\rm BRST}$ and $Q^{\caN=0,\,\rm lin}_{\rm BRST,\,3}$ the situation is more involved, but we still end up with similar on-shell Ward identities. The BRST doublets in the BRST-extended Hilbert space of Yang--Mills theory double copy to BRST doublets of auxiliary and non-auxiliary fields.
        
        Therefore, the tree-level scattering amplitudes for the BRST-extended Hilbert spaces of $S^{\caN=0}_{{\rm BRST},\,3}$ and $\dbl{S}^{\rm DC}_{\rm BRST}$ are fully determined via on-shell Ward identities by the tree-level scattering amplitudes of the physical and auxiliary fields of ghost number zero. We conclude that all these tree-level scattering amplitudes agree between both theories.

        \paragraph{Full quantum equivalence.} 
        For full quantum equivalence, it remains to show that there is a local field redefinition that links the action $S^{\caN=0}_{{\rm BRST},\,3}$ to $\dbl{S}^{\rm DC}_{\rm BRST}$. Both actions fully agree in their kinematic terms and their interaction vertices that contain exclusively fields of ghost number zero. Moreover, they have identical tree-level scattering amplitudes on their BRST-extended (i.e.~full) Hilbert spaces. By \cref{ob:off-shell-Ward}, the tree-level ghost correlators are related to the correct tree-level correlators of the physical sector. Moreover, we can invoke  \cref{ob:classical_equivalence} one final time in order to provide us with a field redefinition that shifts the discrepancies between both actions to interaction vertices of arbitrarily high degree. While this field redefinition may generically be non-local, it becomes local if we leave in all auxiliary fields. This renders the actions fully quantum equivalent from the perspective of perturbative quantum field theory.
        
        \appendix
        \addappheadtotoc
        \appendixpage 
        
        \section{Definitions and conventions for homotopy algebras}\label{app:homotopyAlgebras}
        
        The homotopy algebras that appear naturally in the context of field theories, namely $A_\infty$-, $C_\infty$-, and $L_\infty$-algebras are homotopy versions of associative, commutative and Lie algebras. In particular, associativity and the Jacobi identity only hold up to coherent homotopies.\footnote{But graded commutativity (in the case of $C_\infty$-algebras) and graded anti-symmetry (in the case of $L_\infty$-algebras) are not relaxed.} In the following, we list relevant definitions and our conventions. For more details on $L_\infty$-algebras and some of the calculations detailed in this appendix, see e.g.~\cite{Jurco:2018sby,Jurco:2019bvp}; our conventions match the ones in these references. Other helpful references with original results listed in this section are~\cite{Fukaya:2001uc,Kajiura:2001ng,Kajiura:2003ax}. A unifying description of all the homotopy algebras and their cyclic structures listed below is given by operads, but we refrain from introducing this additional layer of abstraction.
        
        \subsection{\texorpdfstring{$A_\infty$}{A-infty}-algebras}
        
        \paragraph{\mathversion{bold}$A_\infty$-algebras.}
        An \uline{$A_\infty$-algebra} or \uline{strong homotopy associative algebra} is a graded vector space $\frA=\bigoplus_{i\in\IZ}\frA_i$ together with \uline{higher products} which are $i$-linear maps $\sfm_i:\frA\times\cdots\times\frA\rightarrow\frA$ of degree~$2-i$ that satisfy the homotopy associativity relation
        \begin{equation}\label{eq:hJI_associative}
            \sum_{i_1+i_2+i_3=i}(-1)^{i_1i_2+i_3}\sfm_{i_1+i_3+1}(\sfid^{\otimes i_1}\otimes\sfm_{i_2}\otimes\sfid^{\otimes i_3})\ =\ 0
        \end{equation}
        for all $i\in\IN^+$. The lowest identities read as
        \begin{equation}
            \begin{gathered}
                \sfm_1(\sfm_1(\ell_1))\ =\ 0~,
                \\
                \sfm_1(\sfm_2(\ell_1,\ell_2))\ =\ \sfm_2(\sfm_1(\ell_1),\ell_2)+(-1)^{|\ell_1|_\frA}\sfm_2(\ell_1,\sfm_1(\ell_2))~,
                \\
                \sfm_1(\sfm_3(\ell_1,\ell_2,\ell_3))+\sfm_3(\sfm_1(\ell_1),\ell_2,\ell_3)+(-1)^{|\ell_1|_\frA}\sfm_3(\ell_1,\sfm_1(\ell_2),\ell_3)\,+\hspace{2cm}
                \\
                \hspace{2cm}+\,(-1)^{|\ell_1|_\frA+|\ell_2|_\frA}\sfm_3(\ell_1,\ell_2,\sfm_1(\ell_3))\ =\ \sfm_2(\sfm_2(\ell_1,\ell_2),\ell_3)-\,\sfm_2(\ell_1,\sfm_2(\ell_2,\ell_3))~,
                \\
                \vdots
            \end{gathered}
        \end{equation}
        for all $\ell_1,\ldots,\ell_i\in\frA$. We thus see that the unary product $\sfm_1$ is a differential and a derivation for the binary product $\sfm_2$. Furthermore, the ternary product $\sfm_3$ captures the failure of the binary product $\sfm_2$ to be associative.
        
        \paragraph{\mathversion{bold}Cyclic $A_\infty$-algebras.}
        A \uline{cyclic $A_\infty$-algebra} $(\frA,\inner{-}{-}_\frA)$ is an $A_\infty$-algebra $\frA$ equipped with a non-degenerate graded-symmetric bilinear form $\inner{-}{-}_\frA:\frA\times\frA\rightarrow\IR$ such that
        \begin{equation}\label{eq:cyclicity_A}
            \inner{\ell_1}{\sfm_i(\ell_2,\ldots,\ell_{i+1})}_\frA\ =\ (-1)^{i+i(|\ell_1|_\frA+|\ell_{i+1}|_\frA)+|\ell_{i+1}|_\frA\sum_{j=1}^{i}|\ell_j|_\frA}\inner{\ell_{i+1}}{\sfm_i(\ell_1,\ldots,\ell_{i})}_\frA
        \end{equation}
        for all $\ell_i\in\frA$. When it is clear from the context, we shall suppress the subscript $\frA$ on the inner products.
        
        \paragraph{Homotopy Maurer--Cartan theory.}
        Each $A_\infty$-algebra comes with a homotopy Maurer--Cartan theory, where the \uline{gauge potential} is an element $a\in \frA_1$ whose \uline{curvature} $f\in\frA_2$ is defined as
        \begin{equation}\label{eq:Curvature_A}
            f\ \coloneqq\ \sfm_1(a)+\sfm_2(a,a)+\cdots\ =\ \sum_{i\geq 1}\sfm_i(a,\ldots,a)
        \end{equation}
        and satisfies the Bianchi identity
        \begin{equation}\label{eq:BianchiIdentity_A}
            \sum_{i\geq 0}\sum_{j=0}^{i}(-1)^{i+j}\sfm_{i+1}(\underbrace{a,\ldots,a}_{j},f,\underbrace{a,\ldots,a}_{i-j})\ =\ 0~.
        \end{equation}
        
        If the \uline{homotopy Maurer--Cartan equation}
        \begin{equation}
            f\ =\ 0
        \end{equation}
        holds, we say that $a$ is a \uline{homotopy Maurer--Cartan element}. Provided $\frA$ is cyclic with pairing of degree~$-3$, homotopy Maurer--Cartan elements are the stationary points of the \uline{homotopy Maurer--Cartan action}
        \begin{equation}\label{eq:hMCAction_A}
            S^{\rm hMC}[a]\ \coloneqq\ \sum_{i\geq1}\frac{1}{i+1}\inner{a}{\sfm_i(a,\ldots,a)}_\frA~.
        \end{equation}
        Infinitesimal gauge transformations are mediated by elements $c_0\in\frA_0$ and are given by
        \begin{equation}\label{eq:GaugeTrafo_A}
            \delta_{c_0}a\ \coloneqq\ \sum_{i\geq0}\sum_{j=0}^i(-1)^{i+j} \sfm_{i+1}(\underbrace{a,\ldots,a}_{j}, c_0,\underbrace{a,\ldots,a}_{i-j})~.
        \end{equation}
        One may check that the action~\eqref{eq:hMCAction_A} is invariant under the transformations~\eqref{eq:GaugeTrafo_A}, and the curvature~\eqref{eq:Curvature_A} transforms as
        \begin{equation}
            \delta_{c_0}f\ =\ \sum_{i\geq0}\sum_{j=0}^i\sum_{k=0}^{i-j}(-1)^k\sfm_{i+2}(\underbrace{a,\ldots,a}_{j},f,\underbrace{a,\ldots,a}_{i-j},c_0,\underbrace{a,\ldots,a}_{i-j-k})~.
        \end{equation}
        To verify these statements, one makes use of~\eqref{eq:hJI_associative}.
        
        \subsection{\texorpdfstring{$C_\infty$}{C-infty}-algebras}\label{app:CinftyAlgebras}

        \paragraph{Permutations, shuffles, and unshuffles.}
        Let $S_n$ be the permutation group of degree $n\in\IN^+$. We shall write for a permutation $\sigma\in S_n$
        \begin{equation}
            \sigma\ \coloneqq\ 
                \begin{pmatrix}
                    1 & 2 & \cdots & n
                    \\
                    \sigma(1) & \sigma(2) & \cdots & \sigma(n)
                \end{pmatrix}.
        \end{equation}

        A \uline{$(p,q)$-shuffle} for $p,q\in\IN^+$ is a permutation $\sigma\in S_{p+q}$ which satisfies the condition that if $1\leq\sigma(i)<\sigma(j)\leq p$ or $p+1\leq\sigma(i)<\sigma(j)\leq p+q$ then $i<j$. We denote the set of all $(p,q)$-shuffles in $S_{p+q}$ by ${\rm Sh}(p;p+q)$. Consider, for instance, $S_3$. We have the permutations
        \begin{equation}
            S_3\ =\ \left\{
                \begin{pmatrix}
                    1 & 2 & 3
                    \\
                    1 & 2 & 3
                \end{pmatrix},
                \begin{pmatrix}
                    1 & 2 & 3
                    \\
                    1 & 3 & 2
                \end{pmatrix},
                \begin{pmatrix}
                    1 & 2 & 3
                    \\
                    2 & 1 & 3
                \end{pmatrix},
                \begin{pmatrix}
                    1 & 2 & 3
                    \\
                    2 & 3 & 1
                \end{pmatrix},
                \begin{pmatrix}
                    1 & 2 & 3
                    \\
                    3 & 1 & 2
                \end{pmatrix},
                \begin{pmatrix}
                    1 & 2 & 3
                    \\
                    3 & 2 & 1
                \end{pmatrix}
            \right\}.
        \end{equation}
        Then, the sets of $(1,2)$- and $(2,1)$-shuffles are given by
        \begin{equation}\label{eq:shuffles}
            \begin{aligned}
                {\rm Sh}(1;3)\ &=\ \left\{
                    \begin{pmatrix}
                        1 & 2 & 3
                        \\
                        1 & 2 & 3
                    \end{pmatrix},
                    \begin{pmatrix}
                        1 & 2 & 3
                        \\
                        2 & 1 & 3
                    \end{pmatrix},
                    \begin{pmatrix}
                        1 & 2 & 3
                        \\
                        2 & 3 & 1
                    \end{pmatrix}
                \right\},
                \\
                {\rm Sh}(2;3)\ &=\ \left\{
                    \begin{pmatrix}
                        1 & 2 & 3
                        \\
                        1 & 2 & 3
                    \end{pmatrix},
                    \begin{pmatrix}
                        1 & 2 & 3
                        \\
                        1 & 3 & 2
                    \end{pmatrix},
                    \begin{pmatrix}
                        1 & 2 & 3
                        \\
                        3 & 1 & 2
                    \end{pmatrix}
                \right\}.
            \end{aligned}
        \end{equation}
        Likewise, a \uline{$(p,q)$-unshuffle} for $p,q\in\IN^+$ is a permutation $\sigma\in S_{p+q}$ which satisfies the condition that $\sigma(1)<\cdots<\sigma(p)$ and $\sigma(p+1)<\cdots<\sigma(p+q)$. We denote the set of all $(p,q)$-unshuffles in $S_{p+q}$ by $\overline{{\rm Sh}}(p;p+q)$. For instance, the sets of $(1,2)$- and $(2,1)$-unshuffles in $S_3$ are given by
        \begin{equation}\label{eq:unshuffles}
            \begin{aligned}
                \overline{{\rm Sh}}(1;3)\ &=\ \left\{
                    \begin{pmatrix}
                        1 & 2 & 3
                        \\
                        1 & 2 & 3
                    \end{pmatrix},
                    \begin{pmatrix}
                        1 & 2 & 3
                        \\
                        2 & 1 & 3
                    \end{pmatrix},
                    \begin{pmatrix}
                        1 & 2 & 3
                        \\
                        3 & 1 & 2
                    \end{pmatrix}
                \right\},
                \\
                \overline{{\rm Sh}}(2;3)\ &=\ \left\{
                    \begin{pmatrix}
                        1 & 2 & 3
                        \\
                        1 & 2 & 3
                    \end{pmatrix},
                    \begin{pmatrix}
                        1 & 2 & 3
                        \\
                        1 & 3 & 2
                    \end{pmatrix},
                    \begin{pmatrix}
                        1 & 2 & 3
                        \\
                        2 & 3 & 1
                    \end{pmatrix}
                \right\}.
            \end{aligned}
        \end{equation}

        It follows from the above definitions, and it is evident from the explicit examples~\eqref{eq:shuffles} and~\eqref{eq:unshuffles}, that a permutation is a $(p,q)$-shuffle if and only if its inverse is a $(p,q)$-unshuffle, and vice versa.

        \paragraph{\mathversion{bold}$C_\infty$-algebras.}
        A \uline{$C_\infty$-algebra} or \uline{strong homotopy commutative algebra} is an $A_\infty$-algebra $\frC=\bigoplus_{i\in\IZ}\frC_i$ where the higher products $\sfm_i$, in addition to~\eqref{eq:hJI_associative}, also satisfy the \uline{homotopy commutativity relations}
        \begin{equation}\label{eq:homotopy_commutativity}
            \sum_{\sigma\in{\rm Sh}(i_1;i)}\chi(\sigma;\ell_1,\ldots,\ell_i)\,\sfm_i(\ell_{\sigma(1)},\ldots,\ell_{\sigma(i_1)},\ell_{\sigma(i_1+1)},\ldots,\ell_{\sigma(i)})\ =\ 0
        \end{equation}
        for all $0<i_1<i$ and for all $\ell_1,\ldots,\ell_i\in\frC$. Here, $\chi(\sigma;\ell_1,\ldots,\ell_i)$ is the \uline{Koszul sign} for total graded anti-symmetrisation defined by 
        \begin{equation}\label{eq:Koszul-chi}
            \ell_1\wedge\ldots\wedge\ell_i\ =\ \chi(\sigma;\ell_1,\ldots,\ell_i)\,\ell_{\sigma(1)}\wedge\ldots\wedge\ell_{\sigma(i)}~.
        \end{equation}
        
        The lowest four homotopy commutativity relations are
        \begin{equation}
            \begin{gathered}
                \sfm_2(\ell_1,\ell_2)-(-1)^{|\ell_1|_\frC\,|\ell_2|_\frC}\sfm_2(\ell_2,\ell_1)\ =\ 0~,
                \\
                \sfm_3(\ell_1,\ell_2,\ell_3)-(-1)^{|\ell_2|_\frC\,|\ell_3|_\frC}\sfm_3(\ell_1,\ell_3,\ell_2)+(-1)^{(|\ell_1|_\frC+|\ell_2|_\frC)|\ell_3|_\frC}\sfm_3(\ell_3,\ell_1,\ell_2)\ =\ 0~,
                \\
                \sfm_4(\ell_1,\ell_2,\ell_3,\ell_4)-(-1)^{|\ell_1|_\frC\,|\ell_2|_\frC}\sfm_4(\ell_2,\ell_1,\ell_3,\ell_4)\,+\hspace{6.5cm}
                \\
                \hspace{0.5cm}+\,(-1)^{|\ell_1|_\frC(|\ell_2|_\frC+|\ell_3|_\frC)}\sfm_4(\ell_2,\ell_3,\ell_1,\ell_4)-(-1)^{|\ell_1|_\frC(|\ell_2|_\frC+|\ell_3|_\frC+|\ell_4|_\frC)}\sfm_4(\ell_2,\ell_3,\ell_4,\ell_1)\ =\ 0~,
                \\
                \sfm_4(\ell_1,\ell_2,\ell_3,\ell_4)-(-1)^{|\ell_2|_\frC\,|\ell_3|_\frC}\sfm_4(\ell_1,\ell_3,\ell_2,\ell_4)\,+\hspace{4.5cm}
                \\
                +\,(-1)^{|\ell_2|_\frC(|\ell_3|_\frC+|\ell_4|_\frC)}\sfm_4(\ell_1,\ell_3,\ell_4,\ell_2)+(-1)^{(|\ell_1|_\frC+|\ell_2|_\frC)|\ell_3|_\frC}\sfm_4(\ell_3,\ell_1,\ell_2,\ell_4)\,-\hspace{0cm}
                \\-(-1)^{(|\ell_1|_\frC+|\ell_2|_\frC)|\ell_3|_\frC+|\ell_2|_\frC\,|\ell_4|_\frC}\sfm_4(\ell_3,\ell_1,\ell_4,\ell_2)\,+\hspace{4.5cm}
                \\
                \hspace{1cm}+\,(-1)^{(|\ell_1|_\frC+|\ell_2|_\frC)|\ell_3|_\frC+(|\ell_1|_\frC+|\ell_2|_\frC)|\ell_4|_\frC}\sfm_4(\ell_3,\ell_4,\ell_1,\ell_2)\ =\ 0~,
            \end{gathered}
        \end{equation}
        and we see that the product $\sfm_2$ is indeed graded commutative. Note that, a priori, there are two relations for $\sfm_3$ given by the $(2,1)$- and $(1,2)$-shuffles. However, the $(1,2)$-shuffles for $(\ell_1,\ell_2,\ell_3)$ are the same as the $(2,1)$-shuffles for $(\ell_3,\ell_1,\ell_2)$. Since $\ell_1$, $\ell_2$, and $\ell_3$ are arbitrary elements of $\frC$, the two relations thus reduce to one relation. Generally, the number of independent relations for $\sfm_i$ is $\lfloor\frac{i}{2}\rfloor$.
        
        \paragraph{\mathversion{bold}Cyclic $C_\infty$-algebras.}
        A cyclic $C_\infty$-algebra is a cyclic $A_\infty$-algebra satisfying the homotopy commutativity relations~\eqref{eq:homotopy_commutativity}.
        
        \subsection{\texorpdfstring{$L_\infty$}{L-infty}-algebras}\label{app:hA_L_infinity}
        
        \paragraph{\mathversion{bold}$L_\infty$-algebras.}
        An \uline{$L_\infty$-algebra} or \uline{strong homotopy Lie algebra} is a graded vector space $\frL=\bigoplus_{i\in \IZ}\frL_i$ together with \uline{higher products} which are graded anti-symmetric $i$-linear maps $\mu_i:\frL\times\cdots\times\frL\rightarrow\frL$ of degree~$2-i$ that satisfy the \uline{homotopy Jacobi identities}
        \begin{equation}\label{eq:Jacobi_L}
            \sum_{i_1+i_2=i} \sum_{\sigma\in{\rm \overline{Sh}}(i_1;i)}(-1)^{i_2}\chi(\sigma;\ell_1,\ldots,\ell_i)\mu_{i_2+1}(\mu_{i_1}(\ell_{\sigma(1)},\ldots,\ell_{\sigma(i_1)}),\ell_{\sigma(i_1+1)},\ldots,\ell_{\sigma(i)})\ =\ 0~.
        \end{equation}
        for all $\ell_1,\ldots,\ell_i\in\frL$ and $i\in\IN^+$; see \cref{app:CinftyAlgebras} for the definitions of the unshuffles $\overline{\rm Sh}(i_1;i)$ and of the Koszul sign $\chi(\sigma;\ell_1,\ldots,\ell_i)$. The lowest homotopy Jacobi identities, slightly rewritten, read as
        \begin{equation}\label{eq:Jacobi_L_details}
            \begin{gathered}
                \mu_1(\mu_1(\ell_1))\ =\ 0~,
                \\
                \mu_1(\mu_2(\ell_1,\ell_2))\ =\ \mu_2(\mu_1(\ell_1),\ell_2)+(-1)^{|\ell_1|_\frL}\mu_2(\ell_1,\mu_1(\ell_2))~,
                \\
                \mu_2(\mu_2(\ell_1,\ell_2),\ell_3)+(-1)^{|\ell_1|_\frL\,|\ell_2|_\frL}\mu_2(\ell_2,\mu_2(\ell_1,\ell_3))-\mu_2(\ell_1,\mu_2(\ell_2,\ell_3))\ =\ 
                \\
                \hspace{2cm}\ =\ \mu_1(\mu_3(\ell_1,\ell_2,\ell_3))+\mu_3(\mu_1(\ell_1),\ell_2,\ell_3)+(-1)^{|\ell_1|_\frL}\mu_3(\ell_1,\mu_1(\ell_2),\ell_3)\,+
                \\
                \hspace{5cm}+\,(-1)^{|\ell_1|_\frL+|\ell_2|_\frL}\mu_3(\ell_1,\ell_2,\mu_1(\ell_3))~,
                \\
                \vdots
            \end{gathered}
        \end{equation}
        and we can interpret them as follows. The unary product $\mu_1$ is a differential and a derivation with respect to the binary product $\mu_2$. In addition, the ternary product $\mu_3$ captures the failure of the binary product $\mu_2$ to satisfy the standard Jacobi identity.
        
        We note that any $A_\infty$-algebra yields an $L_\infty$-algebra with higher products obtained from total anti-symmetrisation, 
        \begin{equation}\label{eq:A_infty_to_L_infty}
            \mu_i(\ell_1,\ldots,\ell_i)\ =\ \sum_{\sigma \in S_i}\chi(\sigma;\ell_1,\ldots,\ell_i)\,\sfm_i(\ell_{\sigma(1)},\ldots,\ell_{\sigma(i)})~.
        \end{equation}
        In particular, the Lie algebra arising from the commutator on any matrix algebra is an $L_\infty$-algebra. Likewise, the anti-symmetrisation of a $C_\infty$-algebra is an $L_\infty$-algebra with $\mu_i=0$ for $i\geq 2$ due to the homotopy commutativity relations~\eqref{eq:homotopy_commutativity}.
        
        We call an $L_\infty$-algebra \uline{nilpotent}, if all nested higher products vanish, i.e.
        \begin{equation}
            \mu_i(\mu_j(-,\ldots,-),\ldots,-)\ =\ 0~~~\mbox{for all}~i,j\geq 1~.
        \end{equation}
        
        \paragraph{\mathversion{bold}Cyclic $L_\infty$-algebras.}
        A \uline{cyclic $L_\infty$-algebra} $(\frL,\inner{-}{-}_\frL)$ is an $L_\infty$-algebra $\frL$ equipped with a non-degenerate graded-symmetric bilinear form $\inner{-}{-}_\frL:\frL\times\frL\rightarrow\IR$ such that 
        \begin{equation}\label{eq:cyclicity_L}
            \inner{\ell_1}{\mu_i(\ell_2,\ldots,\ell_{i+1})}_\frL\ =\ (-1)^{i+i(|\ell_1|_\frL+|\ell_{i+1}|_\frL)+|\ell_{i+1}|_\frL\sum_{j=1}^{i}|\ell_j|_\frL}\inner{\ell_{i+1}}{\mu_i( \ell_1,\ldots, \ell_{i})}_\frL
        \end{equation}
        for all $\ell_i\in\frL$. As before, when it is clear from the context, we shall suppress the subscript $\frL$ on the inner products.
        
        \paragraph{Homotopy Maurer--Cartan theory.}
        Similar to $A_\infty$-algebras, any $L_\infty$-algebra $(\frL,\mu_i)$ comes with its homotopy Maurer--Cartan theory. In particular, a \uline{gauge potential} is an element $a\in\frL_1$, and its \uline{curvature} is 
        \begin{equation}\label{eq:Curvature_L}
            f\ \coloneqq\ \mu_1(a)+\tfrac12 \mu_2(a,a)+\cdots\ =\ \sum_{i\geq 1}\frac{1}{i!}\mu_i(a,\ldots,a)\ \in\ \frL_2~.
        \end{equation}
        The Bianchi identity reads here as
        \begin{equation}\label{eq:BianchiIdentity_L}
            \sum_{i\geq0}\frac{1}{i!}\mu_{i+1}(a,\ldots,a,f)\ =\ 0~.
        \end{equation}
        \uline{Homotopy Maurer--Cartan elements}, i.e.~gauge potentials with vanishing curvature $f=0$, are the stationary points of the homotopy Maurer--Cartan action
        \begin{equation}
            S^{\rm hMC}[a]\ \coloneqq\ \sum_{i\geq1}\frac{1}{(i+1)!}\inner{a}{\mu_i(a,\ldots,a)}_\frL
        \end{equation}
        provided $\frL$ comes with a cyclic pairing $\inner{-}{-}_\frL$ of degree~$-3$. Similarly to~\eqref{eq:GaugeTrafo_A}, infinitesimal gauge transformations are of the form
        \begin{equation}\label{eq:GaugeTrafo_L}
            \delta_{c_0}a\ \coloneqq\ \sum_{i\geq0}\frac{1}{i!}\mu_{i+1}(a,\ldots,a,c_0)
        \end{equation}
        and are parametrised by elements $c_0\in\frL_0$. The action is invariant under such transformations, and the curvature behaves as
        \begin{equation}
            \delta_{c_0}f\ =\ \sum_{i\geq0}\frac{1}{i!}\mu_{i+2}(a,\ldots,a,f,c_0)~.
        \end{equation}
        To verify these statements, one makes use of~\eqref{eq:Jacobi_L}. 
        
        \paragraph{Covariant derivative.}
        Given an $L_\infty$-algebra $(\frL,\mu_i)$, consider $\varphi\in\frL_k$ for some $k\in\IZ$ and require that under infinitesimal gauge transformations, $\varphi$ transforms adjointly, that is,
        \begin{equation}\label{eq:GaugeTrafoPhi_L}
            \delta_{c_0}\varphi\ \coloneqq\ \sum_{i\geq0}\frac{1}{i!}\mu_{i+2}(a,\ldots,a,\varphi,c_0)
        \end{equation}
        for $c_0\in\frL_0$. We then define the \uline{covariant derivative} $\nabla:\frL_k\rightarrow\frL_{k+1}$ by
        \begin{equation}
            \nabla\varphi\ \coloneqq\ \mu_1(\varphi)+\mu_2(a,\varphi)+\cdots\ =\ \sum_{i\geq0}\frac{1}{i!}\mu_{i+1}(a,\ldots,a,\varphi)
        \end{equation}
        for $a\in\frL_1$. Using~\eqref{eq:Jacobi_L}, one can show that under infinitesimal gauge transformations~\eqref{eq:GaugeTrafo_L} and~\eqref{eq:GaugeTrafoPhi_L}, $\nabla\varphi$ transforms as
        \begin{equation}
            \delta_{c_0}(\nabla\varphi)\ =\ \sum_{i\geq0}\frac{1}{i!}\mu_{i+2}(a,\ldots,a,\nabla\varphi,c_0)+\sum_{i\geq0}\frac{1}{i!}\mu_{i+3}(a,\ldots,a,f,\varphi,c_0)~,
        \end{equation}
        where $f$ is the curvature~\eqref{eq:Curvature_L} of $a$. Thus, for homotopy Maurer--Cartan elements $a$, the covariant derivative transforms adjointly as well.\footnote{It will always transform adjointly when $\mu_i=0$ for all $i>2$, that is, for differential graded Lie algebras also known as strict $L_\infty$-algebras, cf.~\cref{app:structure_theorems}.} Using~\eqref{eq:Jacobi_L} again, we obtain in addition
        \begin{equation}\label{eq:nableSquared}
            \nabla^2\varphi\ =\ \sum_{i\geq0}\frac{1}{i!}\mu_{i+2}(a,\ldots,a,f,\varphi)~.
        \end{equation}        
        
        \paragraph{Curved \mathversion{bold}morphisms of $L_\infty$-algebras.}
        Morphisms between Lie algebras are maps preserving the Lie bracket. In the context of $L_\infty$-algebras, this notion generalises and one obtains what is known as a \uline{curved morphism (of $L_\infty$-algebras)}. Specifically, a curved morphism $\phi:(\frL,\mu_i)\rightarrow(\tilde\frL,\tilde\mu_i)$ between two $L_\infty$-algebras $(\frL,\mu_i)$ and $(\tilde\frL,\tilde\mu_i)$ is a collection of $i$-linear totally graded anti-symmetric maps $\phi_i:\frL\times\cdots\times\frL\rightarrow\tilde\frL$ of degree~$1-i$ such that
        \begin{subequations}\label{eq:morphism_L}
            \begin{equation}
                \begin{aligned}
                    &\sum_{i_1+i_2=i}\sum_{\sigma\in{\rm \overline{Sh}}(i_1;i)}(-1)^{i_2}\chi(\sigma;\ell_1,\ldots,\ell_i)\phi_{i_2+1}(\mu_{i_1}(\ell_{\sigma(1)},\ldots,\ell_{\sigma(i_1)}),\ell_{\sigma(i_1+1)},\ldots,\ell_{\sigma(i)}) \ =\\\
                    \ & = \ \sum_{j\geq1}\frac{1}{j!}\sum_{k_1+\cdots+k_j=i}\sum_{\sigma\in{\rm \overline{Sh}}(k_1,\ldots,k_{j-1};i)}\chi(\sigma;\ell_1,\ldots,\ell_i)\zeta(\sigma;\ell_1,\ldots,\ell_i)\,\times\\
                    &\kern1cm\times\tilde\mu_j\Big(\phi_{k_1}\big(\ell_{\sigma(1)},\ldots,\ell_{\sigma(k_1)}\big),\ldots,\phi_{k_j}\big(\ell_{\sigma(k_1+\cdots+k_{j-1}+1)},\ldots,\ell_{\sigma(i)}\big)\Big)
                \end{aligned}
            \end{equation}
            for $i\in\IN^+\cup\{0\}$ with $\chi(\sigma;\ell_1,\ldots,\ell_i)$ the Koszul sign~\eqref{eq:Koszul-chi} and $\zeta(\sigma;\ell_1,\ldots,\ell_i)$ given by
            \begin{equation}\label{eq:zeta-sign}
                \zeta(\sigma;\ell_1,\ldots,\ell_i)\ \coloneqq\ (-1)^{\sum_{1\leq m<n\leq j}k_mk_n+\sum_{m=1}^{j-1}k_m(j-m)+\sum_{m=2}^j(1-k_m)\sum_{k=1}^{k_1+\cdots+k_{m-1}}|\ell_{\sigma(k)}|_\frL}~.
            \end{equation}
        \end{subequations}
        Note that $\phi_0:\IR\to\tilde\frL_1$ is the constant map. Explicitly, the lowest expressions of~\eqref{eq:morphism_L} read as
        \begin{equation}\label{eq:explicitMorphism_L}
            \begin{gathered}
                0\ =\ \sum_{i\geq1}\frac{1}{i!}\tilde\mu_i(\phi_0,\ldots,\phi_0)~,
                \\
                \phi_1(\mu_1(\ell_1))\ =\ \tilde\mu_1(\phi_1(\ell_1))+\sum_{i\geq1}\frac{1}{i!}\tilde\mu_{i+1}(\phi_0,\ldots,\phi_0,\phi_1(\ell_1))~,
                \\
                \phi_1(\mu_2(\ell_1,\ell_2))-\phi_2(\mu_1(\ell_1),\ell_2)+(-1)^{|\ell_1|_\frL|\ell_2|_\frL}\phi_2(\mu_1(\ell_2),\ell_1)\ =\ 
                \\
                \ =\ \tilde\mu_1(\phi_2(\ell_1,\ell_2))+\tilde\mu_2(\phi_1(\ell_1),\phi_1(\ell_2))\,+
                \\
                +\,\sum_{i\geq1}\frac{1}{i!}\tilde\mu_{i+1}(\phi_0,\ldots,\phi_0,\phi_2(\ell_1,\ell_2))+\sum_{i\geq1}\frac{1}{i!}\tilde\mu_{i+2}(\phi_0,\ldots,\phi_0,\phi_1(\ell_1),\phi_1(\ell_2))~,
                \\
                \vdots
            \end{gathered}
        \end{equation}
        It is easily seen that this definition reduces to the standard definition of a Lie algebra morphism in the context of Lie algebras. Note that a curved morphism is simply called an \uline{(uncurved) morphism (of $L_\infty$-algebras)} whenever $\phi_0=0$, and this notion of morphisms is usually used in the literature when discussing $L_\infty$-algebras. As we will see below, we shall need the more general notion of curved morphisms to reinterpret gauge transformations as morphisms of $L_\infty$-algebras.
        
        Evidently, the first equation of~\eqref{eq:explicitMorphism_L} implies that $\phi_0$ is necessarily a homotopy Maurer--Cartan element of $\tilde\frL$. For such $\phi_0$, we now set
        \begin{equation}\label{eq:twistedMu_L}
            \tilde\mu_i^{\phi_0}(\tilde\ell_1,\ldots,\tilde\ell_i)\ \coloneqq\ \sum_{j\geq0}\frac{1}{j!}\tilde\mu_{i+j}(\phi_0,\ldots,\phi_0,\tilde\ell_1,\ldots,\tilde\ell_i)
        \end{equation}
        for all $\tilde\ell_1,\ldots,\tilde\ell_i\in\tilde\frL$ and $i\in\IN^+$. By virtue of~\eqref{eq:nableSquared}, we immediately have that $\tilde\mu_1^{\phi_0}=\tilde \mu_1$ is a differential. In fact, one can show that $(\tilde\frL,\tilde\mu_i^{\phi_0})$ forms an $L_\infty$-algebra, that is, the $\tilde\mu_i^{\phi_0}$ satisfy the homotopy Jacobi identities~\eqref{eq:Jacobi_L} thus defining another $L_\infty$-structure on $\tilde\frL$. From~\eqref{eq:morphism_L} we may then conclude that any curved morphism between two $L_\infty$-algebras $(\frL,\mu_i)$ and $(\tilde\frL,\tilde\mu_i)$ can be viewed as an uncurved morphism between $(\frL,\mu_i)$ and $(\tilde\frL,\tilde\mu_i^{\phi_0})$.
        
        \paragraph{Maurer--Cartan elements and curved morphisms.}
        Consider $a\in\frL_1$ and let $f\in\frL_2$ be its curvature~\eqref{eq:Curvature_L}. We define the image of a gauge potential under a curved morphism $\phi:(\frL,\mu_i)\rightarrow(\tilde\frL,\tilde\mu_i)$ as
        \begin{equation}\label{eq:morphismA_L}
            \tilde a\ \coloneqq\ \phi_0+\phi_1(a)+\tfrac12\phi_2(a,a)+\cdots\ =\ \sum_{i\geq0}\frac{1}{i!}\phi_i(a,\ldots,a)\ \in\ \tilde\frL_1~.
        \end{equation}
        The curvature of $\tilde a$ is then 
        \begin{equation}
            \tilde f\ =\ \sum_{i\geq 1}\frac{1}{i!}\tilde\mu_i(\tilde a,\ldots,\tilde a)\ =\ \sum_{i\geq0}\frac{1}{i!}\phi_{i+1}(a,\ldots,a,f)\ \in\ \tilde\frL_2~,
        \end{equation}
        which one can verify using~\eqref{eq:Jacobi_L} and~\eqref{eq:morphism_L}. Hence, homotopy Maurer--Cartan elements in $\frL$ are mapped to homotopy Maurer--Cartan elements in $\tilde\frL$. 
        
        Let us extend the above observation to gauge orbits. Consider gauge transformations~\eqref{eq:GaugeTrafo_L} $a\mapsto a+\delta_{c_0}a$ and $\tilde a\mapsto\tilde a+\delta_{\tilde c_0}\tilde a$ with the image of the gauge parameter $c_0\in \frL_0$ given by
        \begin{equation}\label{eq:morphismC_L}
            \tilde c_0\ \coloneqq\ \phi_1(c_0)+\phi_2(a,c_0)+\cdots\ =\ \sum_{i\geq0}\frac{1}{i!}\phi_{i+1}(a,\ldots,a,c_0)\ \in\ \tilde\frL_0~.
        \end{equation}
        A short calculation involving~\eqref{eq:Jacobi_L} reveals that
        \begin{equation}\label{eq:infinitesimalGaugeTrafoMorphism}
            \delta_{\tilde c_0}\tilde a\ =\ -\sum_{i\geq 0}\frac{1}{i!}\phi_{i+2}(a,\ldots,a,f,c_0)+\sum_{i\geq 0}\frac{1}{i!}\phi_{i+1}(\delta_{c_0}a,a,\ldots,a)~.
        \end{equation}
        This immediately yields
        \begin{equation}
            \begin{aligned}
                \sum_{i\geq0}\frac{1}{i!}\phi_{i}(a+\delta_{c_0}a,\ldots,a+\delta_{c_0}a)\ &= \ \sum_{i\geq0}\frac{1}{i!}\phi_{i}(a,\ldots,a)+\sum_{i\geq 0}\frac{1}{i!}\phi_{i+1}(\delta_{c_0}a,a,\ldots,a)
                \\
                &=\ \tilde a+\delta_{\tilde c_0}\tilde a+\sum_{i\geq 0}\frac{1}{i!}\phi_{i+2}(a,\ldots,a,f,c_0)~.
            \end{aligned}
        \end{equation}
        Consequently, gauge equivalence classes of homotopy Maurer--Cartan elements in $\frL$ are mapped to gauge equivalence classes of homotopy Maurer--Cartan elements in $\tilde\frL$ under curved morphisms.
        
        \paragraph{Morphisms of cyclic \mathversion{bold}$L_\infty$-algebras.}
        Consider an uncurved morphism between two $L_\infty$-algebras $(\frL,\mu_i)$ and $(\tilde\frL,\tilde\mu_i)$, that is, a curved morphism with $\phi_0=0$. If, in addition, we have inner products $\inner{-}{-}_\frL$ on $\frL$ and $\inner{-}{-}_{\tilde\frL}$ on $\tilde\frL$, then a morphism of cyclic $L_\infty$-algebras has to satisfy
        \begin{subequations}\label{eq:cyclicMorphism_L}
            \begin{equation}
                \inner{\ell_1}{\ell_2}_\frL\ =\ \inner{\phi_1(\ell_1)}{\phi_1(\ell_2)}_{\tilde\frL}
            \end{equation}
            for all $\ell_{1,2}\in\frL$ and for all $i\geq3$ and $\ell_1,\ldots,\ell_i\in\frL$
            \begin{equation}
                \sum_{\substack{i_1+i_2=i\\i_1,i_2\geq1}}\inner{\phi_{i_1}(\ell_1,\ldots,\ell_{i_1})}{\phi_{i_2}(\ell_{i_1+1},\ldots,\ell_{i})}_{\tilde\frL}\ =\ 0~.
            \end{equation}
        \end{subequations}
        We note that the morphisms of cyclic $L_\infty$-algebras defined here require $\phi_1$ to be injective. More general notions of such morphisms can be defined using Lagrangian correspondences, cf.~\cite{Weinstein:1977aa}.
        
        Suppose now that the inner product $\inner{-}{-}_\frL$ on $\frL$ and $\inner{-}{-}_{\tilde\frL}$ on $\tilde\frL$ of degree~$-3$ so that the homotopy Maurer--Cartan equations, $f=0$ and $\tilde f=0$, are variational. Then, under a morphism $\phi:(\frL,\mu_i)\rightarrow(\tilde\frL,\tilde\mu_i)$, we obtain
        \begin{equation}
            \begin{aligned}
                \sum_{i\geq1}\frac{1}{(i+1)!}\inner{a}{\mu_i(a,\ldots,a)}_\frL\ &= \ S^{\rm hMC}[a]
                \\
                &=\ \tilde S^{\rm hMC}[\tilde a]\ =\ \sum_{i\geq1}\frac{1}{(i+1)!}\inner{\tilde a}{\tilde\mu_i(\tilde a,\ldots,\tilde a)}_{\tilde\frL}
            \end{aligned}
        \end{equation}
        by virtue of~\eqref{eq:cyclicMorphism_L} and~\eqref{eq:morphismA_L}.
        
        \paragraph{Curved quasi-isomorphisms of \mathversion{bold}$L_\infty$-algebras.}
        Recall that the homotopy Jacobi identities~\eqref{eq:Jacobi_L} (see also~\eqref{eq:Jacobi_L_details}) imply that $\mu_1^2=0$. Hence, we may consider the cohomology 
        \begin{equation}
            H^\bullet_{\mu_1}(\frL)\ =\ \bigoplus_{k\in\IZ}H^k_{\mu_1}(\frL)
            \ewith
            H^k_{\mu_1}(\frL)\ \coloneqq\ \ker(\mu_1|_{\frL_k})/\im(\mu_1|_{\frL_{k-1}})~.
        \end{equation}
        A curved morphism of $L_\infty$-algebras $\phi:(\frL,\mu_i)\rightarrow(\tilde\frL,\tilde\mu_i)$ is called a \uline{curved quasi-isomorphism (of $L_\infty$-algebras)} whenever $\phi_1$ induces an isomorphism $H^\bullet_{\mu_1}(\frL)\cong H^\bullet_{\tilde\mu_1}(\tilde\frL)$. There is a bijection between the moduli spaces of gauge equivalence classes of homotopy Maurer--Cartan elements of $\frL$ and $\tilde\frL$. A curved quasi-isomorphism is called an \uline{(uncurved) quasi-isomorphism} whenever $\phi_0=0$.
        
        \paragraph{\mathversion{bold}Gauge transformations as curved morphisms.}
        Let us revisit the infinitesimal gauge transformations~\eqref{eq:GaugeTrafo_L} and first explain how they arise from partially flat homotopies. In particular, set $I\coloneqq[0,1]\subseteq\IR$ and consider the tensor product (see also \cref{ssec:tensor_prod_homotopy_algebras})
        \begin{equation}
            \frL_\Omega\ \coloneqq\ \Omega^\bullet(I)\otimes\frL\ =\ \bigoplus_{k\in\IZ}(\frL_\Omega)_k
            \ewith
            (\frL_\Omega)_k\ =\ \scC^\infty(I)\otimes\frL_k\oplus\Omega^1(I)\otimes\frL_{k-1}
        \end{equation}
        between the de~Rham complex $(\Omega^\bullet(I),\rmd)$ on the interval $I$ and an $L_\infty$-algebra $(\frL,\mu_i)$. Furthermore, the higher products $\mu_i^{\frL_\Omega}$ on $\frL_\Omega$ are given by
        \begin{equation}
            \begin{aligned}
                \mu_1^{\frL_\Omega}(\omega_1\otimes\ell_1)\ &\coloneqq\ \rmd\omega_1\otimes\ell_1+(-1)^{|\omega_1|_{\Omega^\bullet(I)}}\omega_1\otimes\mu_1(\ell_1)~,
                \\
                \mu_i^{\frL_\Omega}(\omega_1\otimes\ell_1,\ldots,\omega_i\otimes\ell_i)\ &\coloneqq\ (-1)^{i\sum_{j=1}^i|\omega_j|_{\Omega^\bullet(I)}+\sum_{j=0}^{i-2}|\omega_j|_{\Omega^\bullet(I)}\sum_{k=1}^{i-j-1}|\ell_k|_\frL}\,\times
                \\
                &\kern1cm\times(\omega_1\wedge\ldots\wedge\omega_i)\otimes\mu_i(\ell_1,\ldots,\ell_i)
            \end{aligned}
        \end{equation}
        for all $\omega_1,\ldots,\omega_i\in\Omega^\bullet(I)$ and for all $\ell_1,\ldots,\ell_i\in\frL$. Hence, a general element $\sfa\in(\frL_\Omega)_1$ is of the form $\sfa(t)=a(t)+\rmd t\otimes c_0(t)$ with $a(t)\in\scC^\infty(I)\otimes\frL_1$ and $c_0(t)\in\scC^\infty(I)\otimes\frL_0$. Its curvature $\sff\in(\frL_\Omega)_2$ is then
        \begin{equation}
            \sff(t)\ =\ f(t)+\rmd t\otimes\left\{\parder[a(t)]{t}-\sum_{i\geq0}\frac{1}{i!}\mu_{i+1}(a(t),\ldots,a(t),c_0(t))\right\}, 
        \end{equation}
        where $f(t)\in\scC^\infty(I)\otimes\frL_2$ is the curvature of $a(t)$. The requirement of partial flatness of $\sff(t)$ amounts to saying that $\sff(t)$ has no components along $\rmd t$. Thus,
        \begin{equation}\label{eq:gaugeFlow}
            \parder[a(t)]{t}\ =\ \sum_{i\geq0}\frac{1}{i!}\mu_{i+1}(a(t),\ldots,a(t),c_0(t))
        \end{equation}
        and we recover the gauge transformations~\eqref{eq:GaugeTrafo_L} from
        \begin{equation}
            \delta_{c_0}a\ =\ \left.\parder[a(t)]{t}\right|_{t=0}
        \end{equation}
        with $a=a(0)$ and $c_0=c_0(0)$. Furthermore, upon solving the ordinary differential equation~\eqref{eq:gaugeFlow}, we will obtain finite gauge transformations. Let us now explain how one can understand this as a curved morphism that preserves the products $\mu_i$.
        
        Concretely, we consider~\eqref{eq:morphismA_L} and~\eqref{eq:morphismC_L} and make the ansatz
        \begin{equation}\label{eq:AnsatzGaugeMorphism}
            a(t)\ \coloneqq\ \sum_{i\geq0}\frac{1}{i!}\phi_i(t)(a,\ldots,a)
            \eand
            c_0(t)\ \coloneqq\ \sum_{i\geq0}\frac{1}{i!}\phi_{i+1}(t)(a,\ldots,a,c_0)~.
        \end{equation}
        Here, we again set $a=a(0)$ and $c_0=c_0(0)$ which, in turn, translates to the conditions $\phi_i(0)=0$ for all $i\neq1$ and $\phi_1(0)=1$. Upon substituting the ansatz~\eqref{eq:AnsatzGaugeMorphism} into~\eqref{eq:gaugeFlow} and remembering~\eqref{eq:infinitesimalGaugeTrafoMorphism}, we obtain
        \begin{equation}\label{eq:gaugeFlowMorphism}
            \begin{aligned}
                \parder[a(t)]{t}\ &= \ \sum_{i\geq1}\frac{1}{i!}\parder[\phi_i(t)]{t}(a,\ldots,a)
                \\
                &=\ -\sum_{i\geq 0}\frac{1}{i!}\phi_{i+2}(t)(a,\ldots,a,f,c_0)+\sum_{i\geq 0}\frac{1}{i!}\phi_{i+1}(t)(\delta_{c_0}a,a,\ldots,a)~,
            \end{aligned}
        \end{equation}
        where $f$ is the curvature of $a$. Thus, solving the ordinary differential equation~\eqref{eq:gaugeFlow} for gauge transformations is equivalent to solving the ordinary differential equation~\eqref{eq:gaugeFlowMorphism} for a curved morphism $\phi_i$ on the $L_\infty$-algebra that preserves the $L_\infty$-algebra structure. Put differently, finite gauge transformations are given by curved morphisms that arise as solutions to~\eqref{eq:gaugeFlowMorphism}. 
        
        Let us exemplify these discussions by considering a standard Lie algebra valued one-form gauge potential on Minkowski space $\IM^d$. Here, $a=A\in\Omega^1(\IM^d)\otimes\frg$ and $c_0=c\in\scC^\infty(\IM^d)\otimes\frg$ for a Lie algebra $\frg$. Moreover, in this case it is enough to consider $\phi_0(t)$ and $\phi_1(t)$ and set $\phi_i(t)=0$ for all $i>1$. Consequently, the ordinary differential equation~\eqref{eq:gaugeFlowMorphism} reduces to
        \begin{equation}
            \parder[A(t)]{t}\ =\ \parder[\phi_0(t)]{t}+\parder[\phi_1(t)]{t}(A)\ =\ \phi_1(t)(\rmd c+[A,c])
        \end{equation}
        and is solved by $A(t)=\phi_0(t)+\phi_1(t)(A)$ and $c(t)=\phi_1(t)(c)$ with\footnote{We can also consider the more general case $\phi_0(t)=g^{-1}(t)\,\rmd g(t)$, $\phi_1(t)(A)=g^{-1}(t)\,A\,g(t)$, and $\phi_1(t)(c)=g^{-1}(t)\,\partial_tg(t)$ for $g\in\scC^\infty(I,\sfG)$ with $g(0)=1$, that is, $g$ solves the ordinary differential equation $\partial_tg(t)=g(t)\,c(t)$; note that $\partial_tg(t)|_{t=0}=c$.}
        \begin{equation}
            \begin{gathered}
                \phi_0(t)\ =\ t\rmd c+\tfrac{t^2}{2!}[\rmd c,c]+\tfrac{t^3}{3!}[[\rmd c,c],c]+\cdots\ =\ \rme^{-tc}\,\rmd\rme^{tc}~,
                \\
                \phi_1(t)(A)\ =\ A+t[A,c]+\tfrac{t^2}{2!}[[A,c],c]+\tfrac{t^3}{3!}[[[A,c],c],c]+\cdots\ =\ \rme^{-tc}\,A\,\rme^{tc}~,
                \\
                \phi_1(t)(c)\ =\ c
            \end{gathered}
        \end{equation}
        as a short calculation reveals; recall from~\eqref{eq:explicitMorphism_L} that $\phi_0(t)$ must be a homotopy Maurer--Cartan element.
        
        \subsection{Structure theorems}\label{app:structure_theorems}
        
        In the following, the term `homotopy algebra' refers to either an $A_\infty$-, $C_\infty$-, or $L_\infty$-algebra. Note that the unary higher product is a differential for any homotopy algebra. We call a homotopy algebra \uline{minimal} provided the unary product vanishes. A homotopy algebra is called \uline{strict} if only the unary and binary products are non-vanishing. Moreover, a homotopy algebra is called \uline{linearly contractible} if only the unary product is nonvanishing, and it has trivial cohomology. Above we have introduced different notions of $L_\infty$-algebras. Likewise, there are similar notions of morphisms for $A_\infty$- and $C_\infty$-algebras. In addition, there is a stricter notion of quasi-isomorphisms known as \uline{isomorphisms}. In the context of $L_\infty$-algebras, those are morphisms for which the lowest map $\phi_1$ is invertible.
        
        \paragraph{Structure theorems.}
        We now have the following structure theorems:
        \begin{enumerate}[(i)]\itemsep-2pt
            \item The \uline{decomposition theorem}: any homotopy algebra is isomorphic to the direct sum of a minimal and a linearly contractible one; see e.g.~\cite{Kajiura:2003ax} for the case of $A_\infty$-algebras.
            \item The \uline{minimal model theorem}: any homotopy algebra is quasi-isomorphic to a minimal one. This follows directly from the decomposition theorem, see also~\cite{kadeishvili1982algebraic,Kajiura:2003ax} for the case of $L_\infty$-algebras.
            \item The \uline{strictification theorem}: any homotopy algebra is quasi-isomorphic to a strict one~\cite{igor1995,Berger:0512576}.
        \end{enumerate}
        We note that strict $A_\infty$-, $C_\infty$-, and $L_\infty$-algebras are simply differential graded associative, differential graded commutative, and differential graded Lie algebras, respectively. We also note that mathematicians would probably use the term `rectify' over `strictify'; we found the latter term more descriptive.
        
        \begin{remark}
            We also would like to make a few remarks on the relations between the homotopy algebras:
            \begin{enumerate}[(i)]\itemsep-2pt
                \item As we saw above in~\eqref{eq:A_infty_to_L_infty}, any $A_\infty$-algebra carries an $L_\infty$-structure by (graded) anti-symmetrisation the higher products.
                \item All higher products of a $C_\infty$-algebra (which is also in particular an $A_\infty$-algebra) except for the differential vanish after anti-symmetrisation.
            \end{enumerate}
        \end{remark}
        
        \section{Inverses of wave operators}\label{app:analytical_details}
        
        In this paper, we have glossed over some of the finer analytical details as not to hide the simplicity of our constructions (too much) behind arcane notation. In particular, we mostly ignored the difference between $\scC^\infty(\IM^d)$ and the actual function space
        \begin{equation}
            \frF\ \coloneqq\ \frF_{\rm int}\oplus\frF_{\rm free}\ =\ \scS(\IM^d)\oplus \ker_\scS(\wave)~,
        \end{equation}
        cf.~\eqref{eq:properFunctionSpace}. This is unproblematic, but the mathematically minded reader may wonder about the definition of inverses of the operator $\wave$ which appear throughout our discussion. Below, we give an answer to this point.
        
        A first point to note is that only gluons can label scattering amplitudes (we are not talking about correlation functions) and therefore they are the only relevant object in the minimal model consisting of (several copies of) the kernel of the wave operator. However, we do want to have gauge symmetries also at the level of free fields, and we therefore allow also the ghosts to have free components. This is not a problem for the scattering amplitudes, as gluons and ghosts live in homogeneously differently graded spaces. The anti-ghosts and Nakanishi--Lautrup field are only relevant as interacting fields, and we arrive at an $L_\infty$-algebra with underlying cochain complex
        \begin{equation}
            \begin{tikzcd}
                & \stackrel{\tilde A_\mu^a}{\frF_{\Omega^1}\otimes\frg} \arrow[r,"-\wave"] \arrow[start anchor=south east, end anchor= north west,rd, "-\tilde\xi\sqrt{\wave}\,\partial^\mu",pos=0.02, swap] &[1cm] \stackrel{-\tilde A_\mu^{+a}}{\frF_{\Omega^1}\otimes\frg} & 
                \\[0.8cm]
                &\stackrel{\tilde b^a}{\scS(\IM^d)\otimes\frg} \arrow[r,"-\wave",swap] \arrow[start anchor=north east, end anchor= south west,ur, "\tilde\xi\sqrt{\wave}\,\partial_\mu",pos=0.02, crossing over] & \stackrel{\tilde b^{+a}}{\scS(\IM^d)\otimes\frg} 
                \\
                \spacer{2ex}\stackrel{\tilde c^a}{\frF\otimes\frg}
                \arrow[r,"\wave"] & \spacer{2ex}\stackrel{\tilde{\bar c}^+}{\scS(\IM^d)\otimes\frg} & \stackrel{\tilde{\bar c}^a}{\scS(\IM^d)\otimes\frg}\arrow[r,"\wave"] & \stackrel{\tilde c^{+a}}{\frF\otimes\frg}
            \end{tikzcd}
        \end{equation}
        where $\frF_{\Omega^1}$ are one-forms on $\IM^d$ with coefficients in $\frF$ and $\wave$ (as well as $\sqrt{\wave}$) vanishes on $\ker_\scS(\wave)\otimes\frg$. In this complex, expressions such as $\frac{1}{\sqrt{\wave}}\tilde b^a$ are clearly well-defined.
        
        We may feel slightly uncomfortable with this restriction of fields as the tensor product of scalar fields with elements in $\frKin$ that we used to construct the Yang--Mills fields does not make any distinction between gluons and ghost on the one side and anti-ghosts and Nakanishi--Lautrup fields on the other side. To resolve this issue, we can consider a quasi-isomorphic $L_\infty$-algebra with underlying complex 
        \begin{equation}
            \begin{tikzcd}
                & \stackrel{\tilde A_\mu^a}{\frF_{\Omega^1}\otimes\frg} \arrow[r,"-\wave"] \arrow[start anchor=south east, end anchor= north west,rd, "-\tilde\xi\sqrt{\wave}\,\partial^\mu",pos=0.02, swap] &[1cm] \stackrel{-\tilde A_\mu^{+a}}{\frF_{\Omega^1}\otimes\frg} & 
                \\[0.8cm]
                &\stackrel{\tilde b^a}{\scS(\IM^d)\otimes\frg} \arrow[r,"-\wave",swap] \arrow[start anchor=north east, end anchor= south west,ur, "\tilde\xi\sqrt{\wave}\,\partial_\mu",pos=0.02, crossing over] & \stackrel{\tilde b^{+a}}{\scS(\IM^d)\otimes\frg} 
                \\
                &\stackrel{\tilde b^a}{\ker_\scS(\wave)\otimes\frg} \arrow[r,"\sfid"] & \stackrel{\tilde b^{+a}}{\ker_\scS(\wave)\otimes\frg} 
                \\
                \spacer{2ex}\stackrel{\tilde c^a}{\frF\otimes\frg}
                \arrow[r,"\wave"] & \spacer{2ex}\stackrel{\tilde{\bar c}^+}{\scS(\IM^d)\otimes\frg} & \stackrel{\tilde{\bar c}^a}{\scS(\IM^d)\otimes\frg}\arrow[r,"\wave"] & \stackrel{\tilde c^{+a}}{\frF\otimes\frg}
                \\
                & \spacer{2ex}\stackrel{\tilde{\bar c}^+}{\ker_\scS(\wave)\otimes\frg} \arrow[r,"\sfid"] & \stackrel{\tilde{\bar c}^a}{\ker_\scS(\wave)\otimes\frg}
            \end{tikzcd}
        \end{equation}
        We note that the identity component is readily implemented with a modification of the Yang--Mills twist $\tau_1$ in~\eqref{eq:twistYM}. However, the required modification is technically a bit cumbersome, and had we inserted it, it would have distracted from the main point of the twist. Finally, note that the twists $\tau_2$ governing interactions can be left unmodified and merely need to be restricted to the Schwartz parts $\scS(\IM^d)\otimes\frg$ for anti-ghosts and Nakanishi--Lautrup fields. This is also evident, as scattering amplitudes will never lead to interactions involving non-propagating anti-ghosts or Nakanishi--Lautrup fields.
                        
    \end{body}

\end{document}